\documentclass[twocolumn]{raa}        
\usepackage{graphicx,times}
\usepackage{natbib}
\usepackage{amssymb,amsmath}
\usepackage{indentfirst} %调整表格为垂直
\usepackage{lscape} %调整表格为垂直
\usepackage{siunitx} %手动添加，为了使arcsecond的单位能够显示出来。
\bibpunct{(}{)}{;}{a}{}{,}

\usepackage[OT1]{fontenc}  %%%自己添加，修复不能加粗的问题
\usepackage[toc]{appendix} %%为了加入appendix

\usepackage[a4paper=true,pagebackref=true]{hyperref}  

\hypersetup{pdftitle = The title of my PDF, pdfauthor = My name, pdfsubject= The subject, pdfkeywords = keyword1 keyword2 keyword3} 
\hypersetup{colorlinks = true, linkcolor = green, anchorcolor = red, citecolor = blue, filecolor = red, pagecolor = red, urlcolor = red}

\begin{document}

   \title{Molecular Clouds Associated with {H \small{II}} regions and Candidates within l = 106.65$^\circ$ to 109.50$^\circ$ and b = ${-}$1.85$^\circ$ to 0.95$^\circ$
\footnotetext{\small $*$ Supported by the National Natural Science Foundation of China.}
}

 \volnopage{ {\bf 20XX} Vol.\ {\bf X} No. {\bf XX}, 000--000}
   \setcounter{page}{1}

   \author{Chun-Xue Li\inst{1,2}, Hong-Chi Wang\inst{1,2}, Yue-Hui Ma\inst{1}, Miao-Miao Zhang\inst{1}, Chong Li \inst{3,4}, and Yu-Qing Zheng\inst{1}
   }
%% Here is an example of three authors come from different institutes.
%% For single author or all the authors from an institute, use "\inst{}" only

   \institute{ Purple Mountain Observatory and Key Laboratory of Radio Astronomy, 
   Chinese Academy of Sciences, 10 Yuanhua Road, Qixia District, Nanjing 210033, 
   P. R. China; {\it hcwang@pmo.ac.cn}\\
%% Please give the E-mail address of the author, to whom future correspondence and
%% offprint requests will be sent.
        \and
        School of Astronomy and Space Science, University of Science and Technology of China, 96 Jinzhai Road, Hefei 230026, P. R. China\\
        \and
         School of Astronomy and Space Science, Nanjing University, 163 Xianlin Avenue, Nanjing 210023, P. R. China \\
         \and
         Key Laboratory of Modern Astronomy and Astrophysics (Nanjing University), Ministry of Education, Nanjing 210023, P. R. China\\
\vs \no
   {\small Received 20XX Month Day; accepted 20XX Month Day}
}

\abstract{  We present a large-scale simultaneous survey of the CO isotopologues ($\rm {}^{12}{CO}$, $\rm{}^{13}{CO}$, and $\rm{C}{}^{18}{O}$) J = 1 ${-}$ 0 line emission toward the Galactic plane region of l = 106.65$^\circ$ to 109.50$^\circ$ and b = ${-}$1.85$^\circ$ to 0.95$^\circ$ using the Purple Mountain Observatory 13.7 m millimeter-wavelength telescope. Except for the molecular gas in the solar neighborhood, the emission from the molecular gas in this region is concentrated in the velocity range of [${-}$60, ${-}$35] $\rm km~s^{-1}$. The gas in the region can be divided into four clouds, with mass in the range of $\sim$10$^{3}$ to 10$^{4}$\,${M_{\sun}}$. We have identified 25 filaments based on the $\rm {}^{13}{CO}$ data. The median excitation temperature, length, line mass, line width, and virial parameter of the filaments are 10.89 K, 8.49 pc, 146.11 ${M}_{\odot}~ \rm pc^{-1}$, 1.01 $\rm km~s^{-1}$, and 3.14, respectively. Among these filaments, eight have virial parameters of less than 2, suggesting that they are gravitationally bound and can lead to star formation. Nineteen {H \small {II}} regions or candidates have previously been found in the region and we investigate the relationships between these {H \small {II}} regions/candidates and surrounding molecular clouds in detail. Using morphology similarity and radial velocity consistency between {H \small {II}} regions/candidates and molecular clouds as evidence for association, and raised temperature and velocity broadening as signatures of interaction, we propose that 12 {H \small {II}} regions/candidates are associated with their surrounding molecular clouds. In the case of the {H \small {II}} region of S142, the energy of the {H \small {II}} region is sufficient to maintain the turbulence in the surrounding molecular gas.
}

   \authorrunning{Chun-Xue Li et al. }            %author_head in even pages
   \titlerunning{Molecular Clouds Associated with HII regions and Candidates }  % title_head in odd pages
   \maketitle

%________________________________________________ sections below
% 
\section{Introduction}           %% first-level sections will be auto-capitalized
Dense molecular clouds are the cradle of star formation \citep{1981MNRAS.194..809L,1997ApJ...476..166W,2015ARA&A..53..583H}. Giant molecular clouds (GMCs) have masses in the range from 10$^{4}$ to 10$^{7}$ M$_{\odot}$ and sizes in the range from $\sim$10 to $\sim$150 pc \citep{2013PASA...30...44B}. The mass function of molecular clouds in the Milky Way follows a power-law distribution. For the inner Galaxy, the power-law index of the mass distribution is ${-}$1.6, while it is ${-}$2.2 for the outer Galaxy \citep{2016ApJ...822...52R}. \cite{2001ApJ...551..852H} suggested that GMCs are under the self-gravitational equilibrium, while molecular clouds with masses less than 10$^{3}$ M$_{\odot}$ are far from self-gravitational equilibrium.  \par
Observations have shown that filamentary structures are widespread in molecular clouds and that star formation prefers to take place at the junctions of cold filaments in molecular clouds \citep{1979ApJS...41...87S,1987ApJ...312L..45B,2008PASJ...60..407T,2010A&A...518L.102A,2010A&A...518L.100M,2010A&A...518L.104M,2010A&A...518L..92W,2011A&A...529L...6A,2014prpl.conf...27A,2016A&A...592A..54A,2019A&A...623A.142S}. Observationally, filaments have a wide range of masses and widths, with masses from 1 to 10$^{5}$ M$_{\odot}$, and lengths from 0.1 to 100 pc. Filaments with the highest mass and length are considered to be the ``bones" of the molecular gas in the spiral arms of the Milky Way \citep{2013A&A...554A..55H,2018A&A...610A..77H,2014ApJ...797...53G,2015MNRAS.450.4043W,2016ApJS..226....9W,2015ApJ...815...23Z,2016A&A...591A...5L,2018ApJ...864..152Z,2019A&A...622A..52Z,2020MNRAS.492.5420S}. 
Sub-pc scale filaments within molecular clouds are considered as an essential structure for the transportation of molecular gas onto molecular clumps/cores, which will lead to new born stars \citep{2010A&A...518L.102A,2011A&A...529L...6A,2013A&A...554A..55H}.      \par
The structure and physical properties of molecular clouds can be significantly influenced by feedback from massive stars (${>}$8 M$_{\odot}$), such as outflows, radiation pressure, stellar winds, and supernova explosions \citep{2007ARA&A..45..481Z,2017ApJ...850..112R,2018MNRAS.474..647C}. Radiation from {H \small {II}} regions and expansion of supernova remnants can create expanding shells in their surroundings and the expanding shells can sweep up diffuse molecular gas into dense shells which may subsequently undergo fragmentation and form a new generation of stars \citep{1994MNRAS.268..291W,2017ApJS..231....9K}. Besides, the momentum feedback due to outflows, stellar winds, and expanding shells can create high-velocity components in molecular clouds, and therefore enhance the internal turbulence of molecular clouds \citep{2004RvMP...76..125M,2014PhR...539...49K}. The above feedback processes can cause several features that can be observed in molecular clouds, for example, raised temperature and broadened linewidths \citep{1998ApJ...505..286S,2006ApJ...642L.149S,2010ApJ...712.1147J}. \par

 In this study, we conduct a detailed investigation on the possible associations between {H \small {II}} regions and molecular clouds within the Galactic plane region of l = [106.65${-}$109.50]$^\circ$ and b = [${-}$1.85 ${-}$ 0.95]$^\circ$. In this region, there are two SNRs, G108.2 ${-}$ 0.6 and CTB109. SNR G108.2 ${-}$ 0.6 is a faint and large shell-type structure as seen at 1420 MHz. The angular size of this SNR is 62$\arcmin$, and its distance is assumed to be 3.2 kpc \citep{2007A&A...465..907T}. CTB 109 also has a shell-type morphology in radio and X-ray emission \citep{1980Natur.287..805G,1981ApJ...246L.127H}. Moreover, its interaction with a dense cloud has been confirmed \citep{2006ApJ...642L.149S}. Within the  longitude range l = 100$^\circ$ ${-}$ 120$^\circ$, \cite{1978A&A....66....1C} have found two groups of {H \small {II}} regions, S147 / S148 / S149 (l = 108.35$^\circ$, b = ${-}$1.0$^\circ$) and S152 / S153 (l = 108.8$^\circ$, b = ${-}$0.1$^\circ$). \cite{1984A&A...133...99C} and \cite{1985A&A...146..325K} suggested that these {H \small {II}} regions are associated with molecular gas. Two wide-field surveys of molecular gas, the FCRAO OGS survey \citep{1998ApJS..115..241H} and the CfA-Chile survey \citep{2001ApJ...547..792D}, have been conducted toward this region. However, the spatial resolution or sensitivity of the two surveys is relatively low. \par

In this work, we present a large-scale (2.88$^\circ$ $\times$ 2.84$^\circ$) survey of molecular gas toward the S152 {H \small {II}} region, which is part of the MWISP project \citep{2019ApJS..240....9S}. The structure of the paper is arranged as follows. The observation is described in Section \ref{sect:Obs} and the results are presented in Section \ref{sect:Resluts}. We discuss the gravitational stability of the filaments in Section \ref{sect:discussion} and give our summary in Section \ref{Summary}.

% Authors can give a citation as `\citealt{Michel+etal+1992}'.
% You may also use \cite, \citep and \citet for citation, and use Table~1
% or Figure~1 and so forth. Using \ref and \label for cross-references of
% Tables/Figures is a good way in adjusting/adding/removing text, tables or
% figures.

\section{Observations}
\label{sect:Obs}
%\subsection{Observations}
The observations of this work are part of the Milky Way Imaging Scroll Painting (MWISP)\footnote{http://www.radioast.nsdc.cn/mwisp.php} project, which is a large-filed survey of the $\rm {}^{12}{CO}$, $\rm {}^{13}{CO}$, and $\rm {C}{}^{18}{O}$ J = 1 $-$ 0 emission toward the Galactic plane. The observations of the three isotopologues toward the Galactic region of 106.65$^\circ$ $\leqslant$ l $\leqslant$ 109.50 $^\circ$ and ${-}$1.85$^\circ$ $\leqslant$ b $\leqslant$ 0.95$^\circ$ were carried out between 2015 and 2017 using the PMO-13.7m millmeter-wavelength telescope at Delingha, China. The telescope is equipped with a nine-beam Superconducting Spectroscopic Array Receiver (SSAR) \citep{2012ITTST...2..593S}, working as the front end, and a Fast Fourier Transform Spectrometer (FFTS) with 16384 frequency channels and a full-bandwidth of 1 GHz, working as the back end, which provides a velocity resolution of 0.17 $\rm km~s^{-1}$ at 110 GHz. The data were collected via position-switch on-the-fly (OTF) mapping mode. The observations were conducted along the Galactic longitude and Galactic latitude directions with a scanning rate of 50$\arcsec$ per second. The half-power beamwidth (HPBW) of the telescope is $\SI{52}{\arcsecond}$ at 110 GHz and $\SI{50}{\arcsecond}$ at 115 GHz, and the pointing accuracy is about  $\SI{5}{\arcsecond}$. Calibration of the antenna temperature to the main beam temperature ($\rm {T}_{mb}$) was done by $\rm {T}_{mb}$ = $\rm {T}_{A}^{*}$ / $\rm {\eta}_{mb}$, where the main beam efficiency $\rm {\eta}_{mb}$ is 0.55 at 110 GHz and 0.5 at 115 GHz, according to the status report of the PMO-13.7 m telescope. The entire survey area was divided into cells of sizes of $\SI{30}{\arcminute}$ ${\times}$ $\SI{30}{\arcminute}$. Each cell is regridded into pixels of sizes of $\SI{30}{\arcsecond}$ ${\times}$ $\SI{30}{\arcsecond}$. The data were reduced using the GILDAS\footnote{https://www.iram.fr/IRAMFR/GILDAS/} software package. The final median RMS noise level is $\sim$0.5 K per channel and $\sim$0.3 K per channel for the $\rm {}^{12}{CO}$ and $\rm{}^{13}{CO}$ data, respectively.

\label{Observations}

\section{Results}

\label{sect:Resluts}

\subsection{Overview of Molecular Clouds in the Region }

\subsubsection{Spatial Distribution}

\begin{figure} 
  \centering
  \includegraphics[width=14.0cm, angle=0]{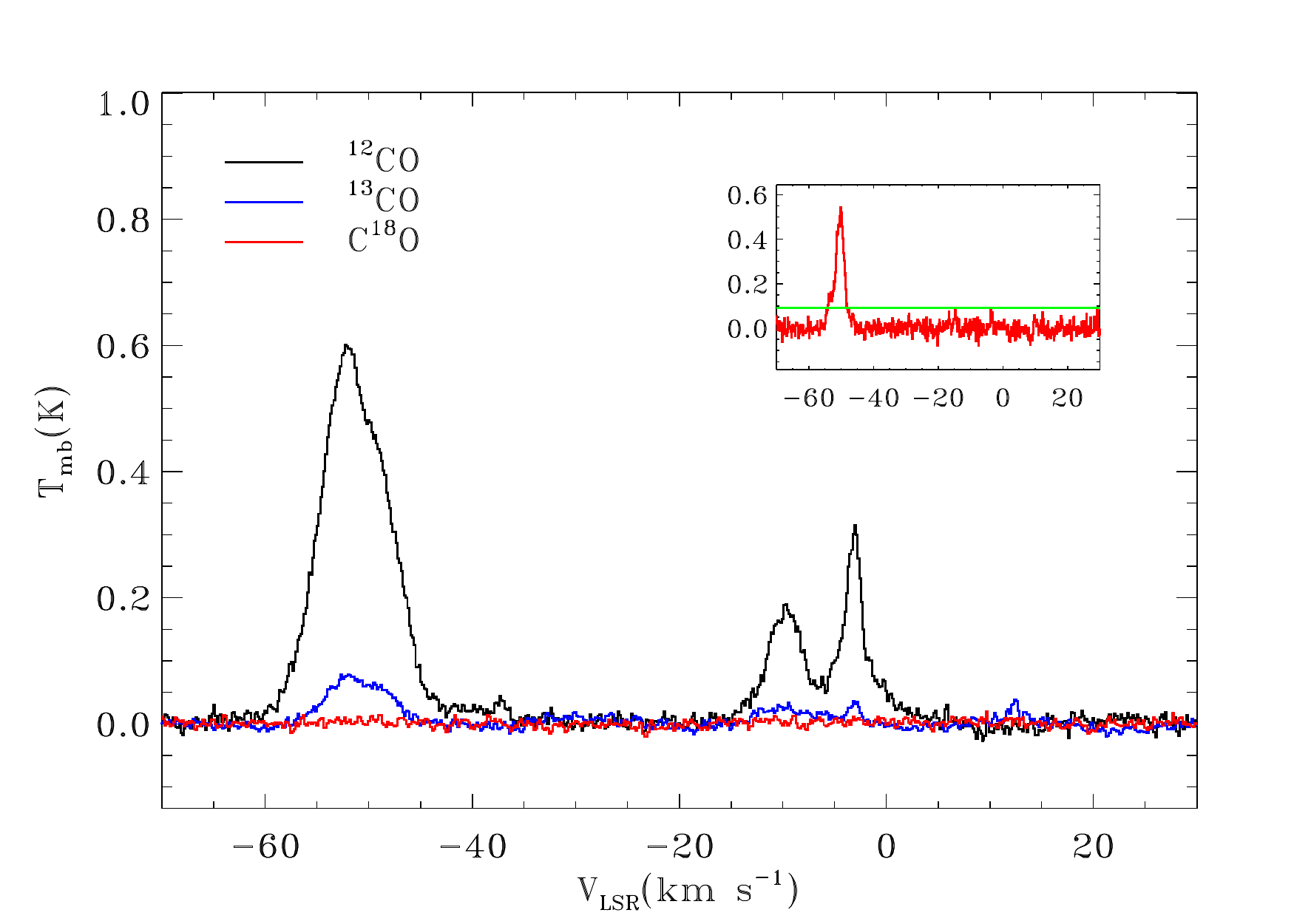}
 % \begin{minipage}[]{85mm}
  \caption{Average spectra of $\rm {}^{12}{CO}$, $\rm {}^{13}{CO}$, and $\rm {C}{}^{18}{O}$ in the surveyed area. The black spectrum is $\rm {}^{12}{CO}$ emission, while the blue is $\rm {}^{13}{CO}$ emission, and the red is $\rm {C}{}^{18}{O}$  emission. The inset panel is the average of the $\rm {C}{}^{18}{O}$  spectra in the region that have at least four contiguous channels with $\rm {C}{}^{18}{O}$ emission above 1.5$\sigma$. The green line shows the 3$\sigma$ noise level of the average spectrum.  }
%\end{minipage}
    \label{Fig1}
  \end{figure}

Figure~\ref{Fig1} displays the average spectra of $\rm {}^{12}{CO}$, $\rm {}^{13}{CO}$, and $\rm {C}{}^{18}{O}$ emission in the galactic region of l = [106.65 , 109.50]$^\circ$ and b = [${-}$1.85 , 0.95]$^\circ$. There are two velocity components in this region, one between ${-}$60 and ${-}$35 $\rm km~s^{-1}$, and the other between ${-}$15 and 7 $\rm km~s^{-1}$. The first velocity component shows higher intensity than the second one. The molecular gas with the second velocity component is located in the solar neighborhood. In this work, we focus on the molecular gas with the first velocity component, and the study on the molecular gas with the second velocity component is deferred to a future paper.  \par
    
\begin{figure}
    \centering
    \includegraphics[width=14.0cm, trim=3.5cm 3.5cm 3.5cm 3.5cm ]{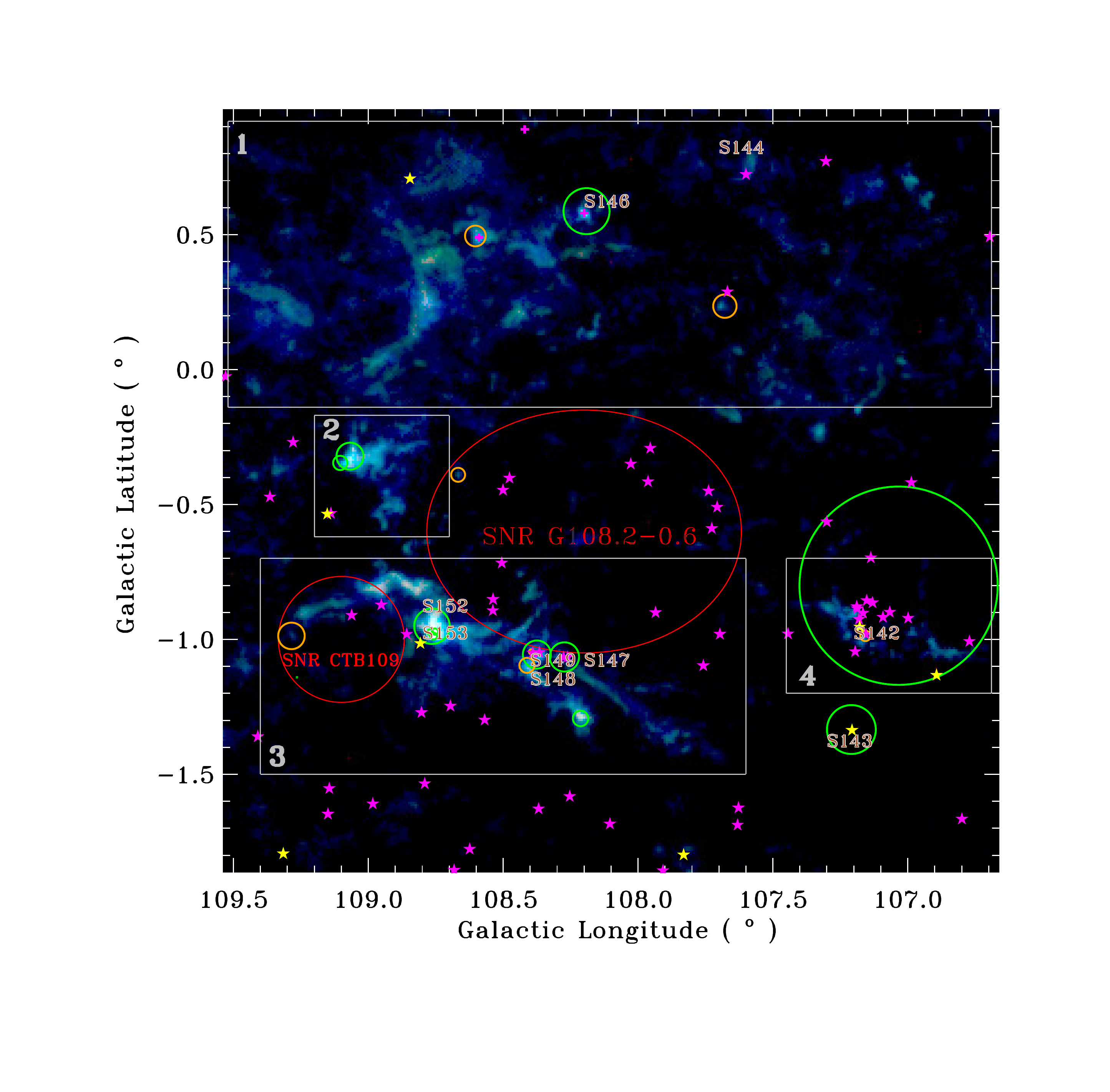}
    \caption{Three-color image of the $\rm {}^{12}{CO}$, $\rm {}^{13}{CO}$, and $\rm {}{C}^{18}{O}$ integrated intensity of the surveyed area. The blue, green, and red colors show the $\rm {}^{12}{CO}$, $\rm {}^{13}{CO}$, and $\rm {}{C}^{18}{O}$ emissions, respectively. For $\rm {}^{12}{CO}$ and $\rm {}^{13}{CO}$ emission, the integrated velocity range is from ${-}$60 $\rm km~s^{-1}$ to ${-}$35 $\rm km~s^{-1}$, while for $\rm {}{C}^{18}{O}$ emission, the integrated velocity range is from ${-}$55 $\rm km~s^{-1}$ to ${-}$45 $\rm km~s^{-1}$. Only spectra having at least five contiguous channels with a brightness temperature above 1.5$\rm {\sigma}$ are used for the intensity integration of $\rm {}^{12}{CO}$ and $\rm {}^{13}{CO}$ emission, while only spectra having at least four contiguous channels with a brightness temperature above 1.5$\rm {\sigma}$ are used for the intensity integration of $\rm {}{C}^{18}{O}$. The red circle and the red ellipse show the locations of the known SNRs in the region CTB 109 and G108.2 ${-}$  0.6 from  \cite{2019JApA...40...36G}. The orange circles and the green circles show the locations of {H \small {II}} regions/candidates from the WISE catalog \citep{2014ApJS..212....1A}, respectively, and the magenta crosses indicate the maser locations from \cite{2019ApJ...885..131R}. The yellow pentagrams indicate the locations of  O-type stars and the magenta for the B-type stars. The white squares with numbers 1${-}$4 show the division of molecular gas in the region into four structures, GMCs 1${-}$4.     }
    \label{Fig2}
  \end{figure}

    The spatial distribution of the integrated intensity of $\rm {}^{12}{CO}$, $\rm {}^{13}{CO}$, and $\rm {C}{}^{18}{O}$ emission is shown in Figure~\ref{Fig2}. The black, blue, and red colors show the intergrated intensities of the $\rm {}^{12}{CO}$, $\rm {}^{13}{CO}$, and $\rm {}{C}^{18}{O}$ emission, respectively. For $\rm {}^{12}{CO}$ and $\rm {}^{13}{CO}$ emission, the integrated velocity range is from ${-}$60 $\rm km~s^{-1}$ to ${-}$35 $\rm km~s^{-1}$, while for $\rm {}{C}^{18}{O}$ emission, the integrated velocity range is from ${-}$55 $\rm km~s^{-1}$ to ${-}$45 $\rm km~s^{-1}$. Only spectra having at least five contiguous channels with brightness temperature above 1.5$\rm {\sigma}$ are used for the intensity integration of $\rm {}^{12}{CO}$ and $\rm {}^{13}{CO}$, while only spectra having at least four contiguous channels with brightness temperature above 1.5$\rm {\sigma}$ are used for the intensity integration of $\rm {}{C}^{18}{O}$. The red circle and the red ellipse show the locations of the two known SNRs in the region, CTB 109 and G108.2 ${-}$ 0.6 \citep{2019JApA...40...36G}. The magenta crosses represent the maser locations from \cite{2019ApJ...885..131R}. The orange circles and the green circles show the {H \small {II}} regions/candidates (nineteen in total) from the WISE catalog \citep{2014ApJS..212....1A}, respectively. In Figure~\ref{Fig2}, the $\rm {C}{}^{18}{O}$ emission is quite weak compared with $\rm {}^{12}{CO}$, $\rm {}^{13}{CO}$ emission, so we only use $\rm {}^{12}{CO}$ and $\rm {}^{13}{CO}$ data in this work. The molecular gas shows four individual structures in Figure~\ref{Fig2}. The first structure is the northern one which occupies one-third of the entire area and consists of a relatively bright part with a bone shape on its east side and a diffuse part to the south of the {H \small {II}} region S144. This northern structure is spatially coincident with three {H \small {II}} regions (S146 and two others) and the three known masers in the region. The second structure is the eastern one with a cometary shape which has an intensity peak approximately at l = 109$^ \circ$ and b = ${-}$0.4$^ \circ$ and is spatially associated with two {H \small {II}} regions/candidates. The third structure is the southeastern one close to SNRs CTB 109 and G108.2 ${-}$ 0.8 and is associated with {H \small {II}} regions/candidates S147, S149, S152, and S153. The above three molecular gas structures form a broken shell surrounding the SNR G108.2 ${-}$ 0.6. The boundary of the SNR G108.2${-}$0.6, i.e., the circle in Fig.2, is delimited by the 1420 MHz continuum emission. The southeast edge of the SNR nearly reaches the northwest edge of GMC 3. The SNR G108.2${-}$0.6 is among the faintest supernova remnants discovered in the Milky Way \citep{2007A&A...465..907T}. As one of the faintest SNRs at a late evolutionary stage, it can hardly provide much kinetic energy to the molecular clouds. The fourth distinct structure is a filamentary cloud spatially coincident with the largest {H \small {II}} region, S142, in the surveyed area.    \par
    
\begin{figure}
    \centering
    \includegraphics[width=14.0cm,trim=1.0cm 1.0cm 1.0cm 1.0cm]{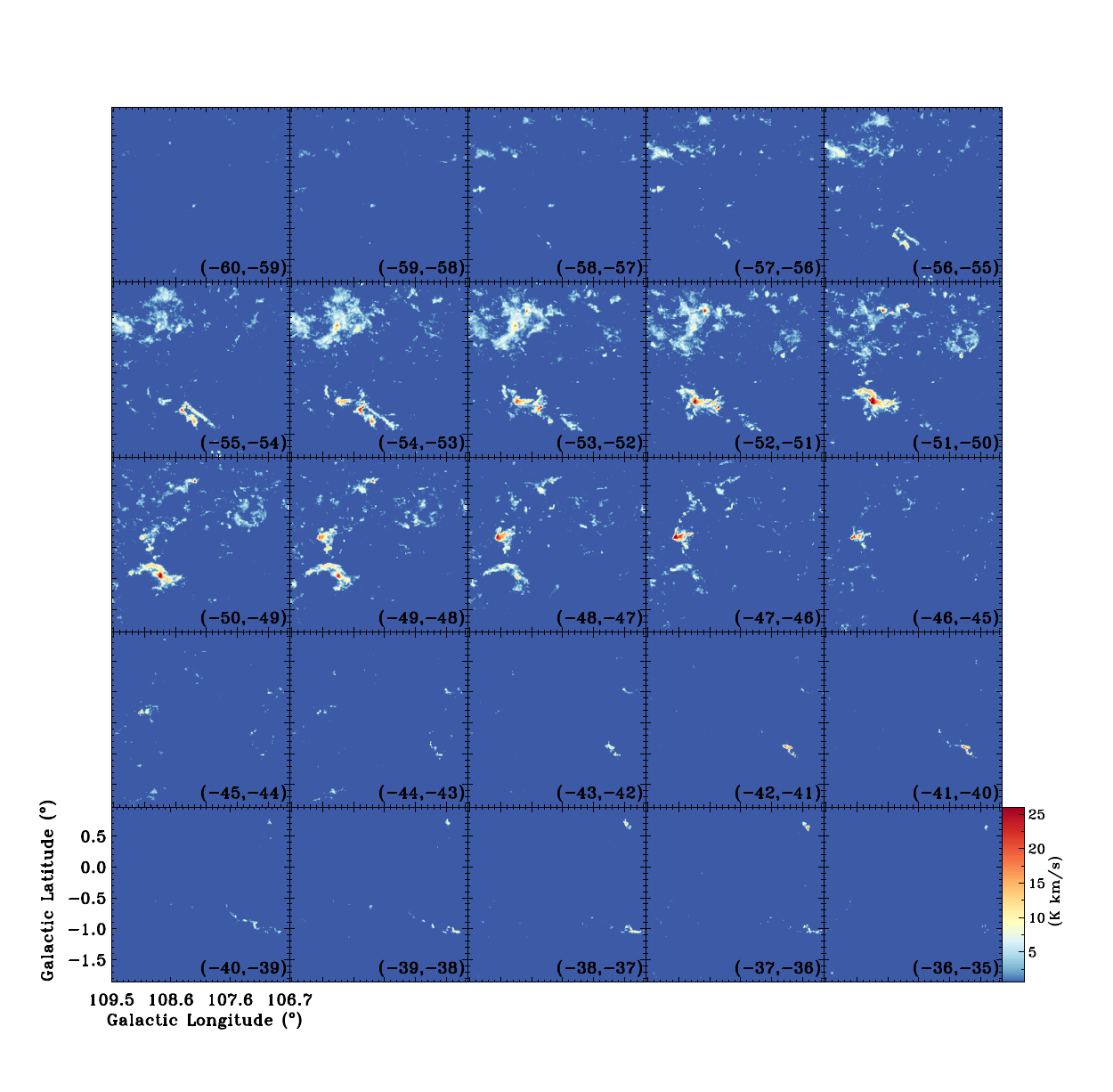}
    \caption{Velocity channel maps of $\rm {}^{12}{CO}$ from ${-}$60 to ${-}$35 $\rm km~s^{-1}$ of the region. The velocity range of each channel is shown at the bottom-right corner of each panel.}
    \label{Fig3}
  \end{figure}

    Figure~\ref{Fig3} shows the velocity channel maps of the $\rm {}^{12}{CO}$ emission within the velocity range from ${-}$60 to ${-}$35 $\rm km~s^{-1}$. The first structure seen in Figure~\ref{Fig2} first shows up at the channels with the bluest velocities and then the third structure associated with S152 {H \small {II}} region begins to show up. The two structures can be identified clearly in the velocity channel of [${-}$53, ${-}$52] $\rm km~s^{-1}$. Moreover, the molecular cloud associated with S152 has a systematic velocity gradient along its intensity ridge. The emission of this structure is mainly contained in a filamentary structure in its southwest part when v $<$ ${-}$52 $\rm km~s^{-1}$, and an arc structure (see the velocity channel [${-}$49, ${-}$48] $\rm km~s^{-1}$) in its northeastern part when v $>$ ${-}$51 $\rm km~s^{-1}$. The second structure seen in Figure~\ref{Fig2} shows up in the velocity range of [${-}$51, ${-}$44] $\rm km~s^{-1}$, and the fourth structure associated with S142 shows up in the channels of velocity above ${-}$44 km s$^{-1}$. According to the spatial distribution and the velocities shown in Figures~\ref{Fig2} and \ref{Fig3}, we divide the molecular gas in this region into four clouds: the first structure seen in [${-}$60, ${-}$46] $\rm km~s^{-1}$ which is referred to as GMC 1, the cometary cloud in [${-}$51, ${-}$44] $\rm km~s^{-1}$ as GMC 2, the cloud near the S152 {H \small {II}} region in  [${-}$58, ${-}$46] $\rm km~s^{-1}$ as GMC 3, and the filamentary cloud associated with S142 in [${-}$44, ${-}$35] $\rm km~s^{-1}$ as GMC 4. The velocity channel map of the $\rm {}^{13}{CO}$ emission in the region is given in Figure~\ref{FigA.1} in the Appendix.   \par
    
\begin{figure}[b!]
    \centering
    \begin{subfigure}[b]{0.75\linewidth}
    \includegraphics[trim=6cm 1cm 7cm 0cm, width=\linewidth]{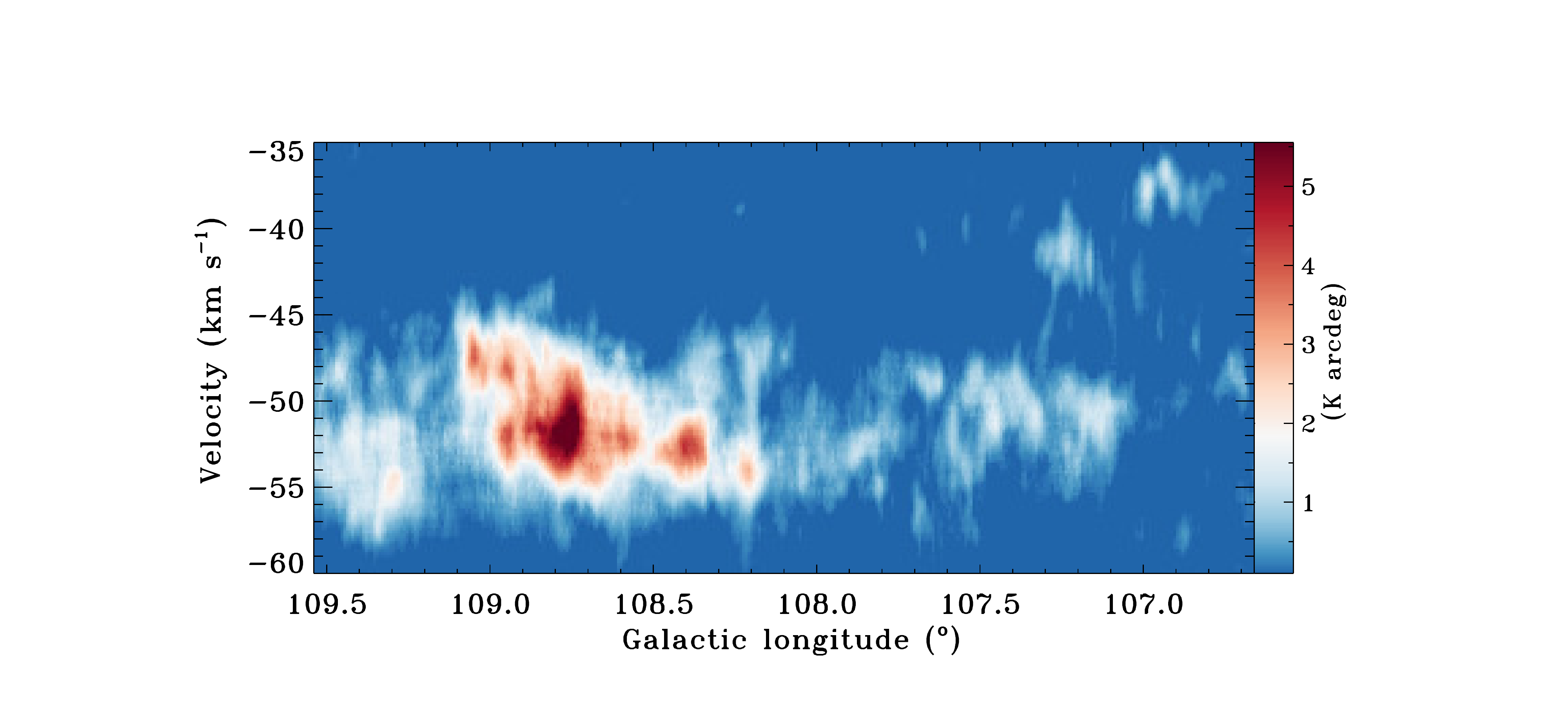}
    \caption{}
   \end{subfigure}
    \begin{subfigure}[b]{0.45\linewidth}
    \includegraphics[trim=1cm 8cm 5cm 15cm,width=\linewidth]{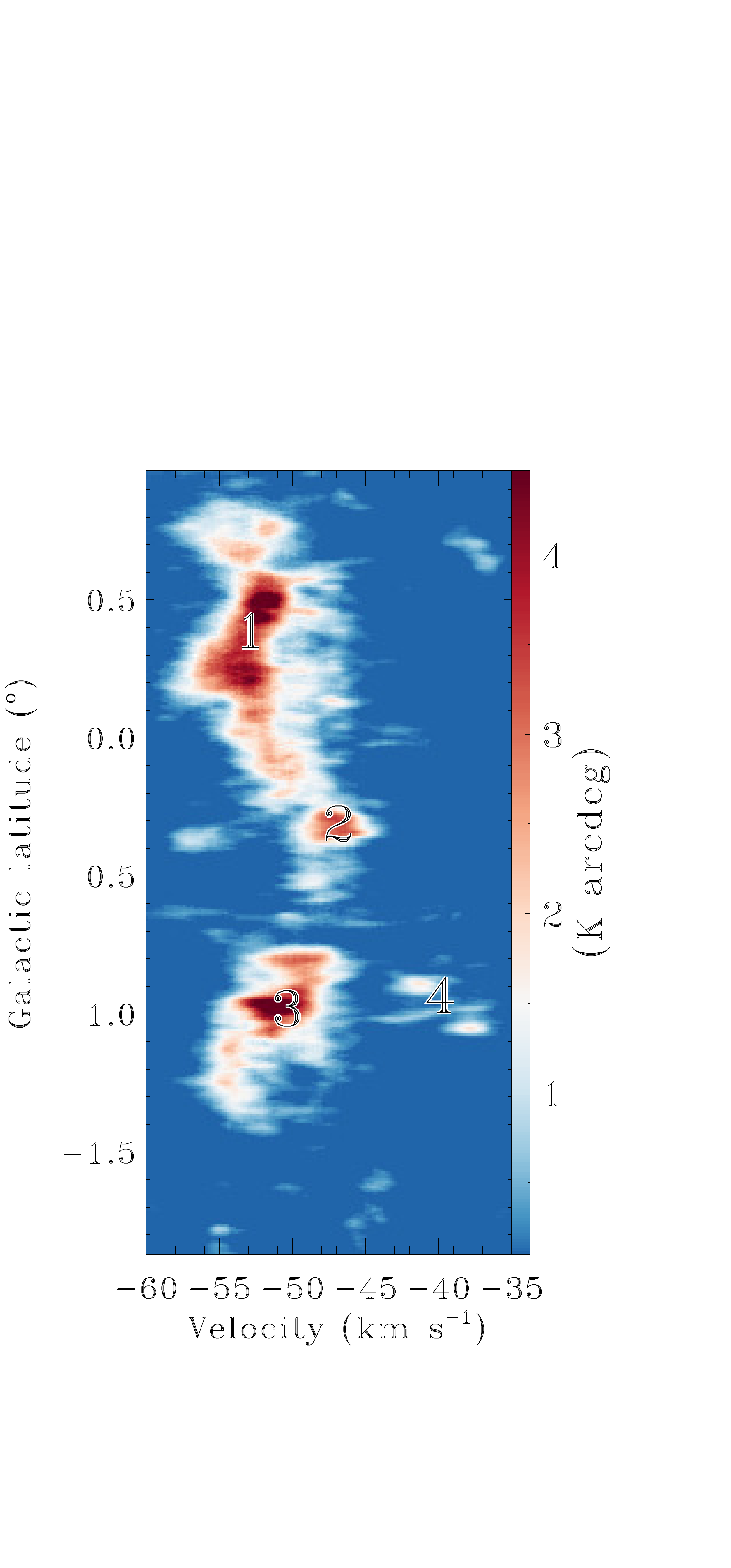}
     \caption{}
    \end{subfigure}
    \caption{(a), $\rm {}^{12}{CO}$ l-v diagram. (b), $\rm {}^{12}{CO}$ b-v diagram.The numbers 1, 2, 3 and 4 in (b) represent the four GMCs in the region. }
   \label{Fig4}
  \end{figure}

    The position-velocity maps of the $\rm {}^{12}{CO}$ emission are shown in Figure~\ref{Fig4}. Panels (a) and (b) in Figure~\ref{Fig4} are the l-v diagram and b-v diagram, respectively. In panel (a), we find that the GMCs 1 ${-}$ 3 have similar velocities which are within the range of [${-}$60, ${-}$44] $\rm km~s^{-1}$. The GMC 4 is located approximately at l = 107.20$^ \circ$, v = ${-}$40  km s$^{-1}$, which is quite separate from GMCs 1 ${-}$ 3. In panel (b), GMCs 1 ${-}$ 4 can be distinguished clearly and are indicated them by numbers 1 ${-}$ 4. The p-v diagrams of the $\rm {}^{13}{CO}$ emission of the region are presented in Figure~\ref{FigA.2} in the Appendix. \par

\begin{figure}[htbp]
    \centering
    \begin{subfigure}[b]{0.4\linewidth}
    \includegraphics[width=7.0cm, trim= 6cm 3cm -1cm 2.5cm]{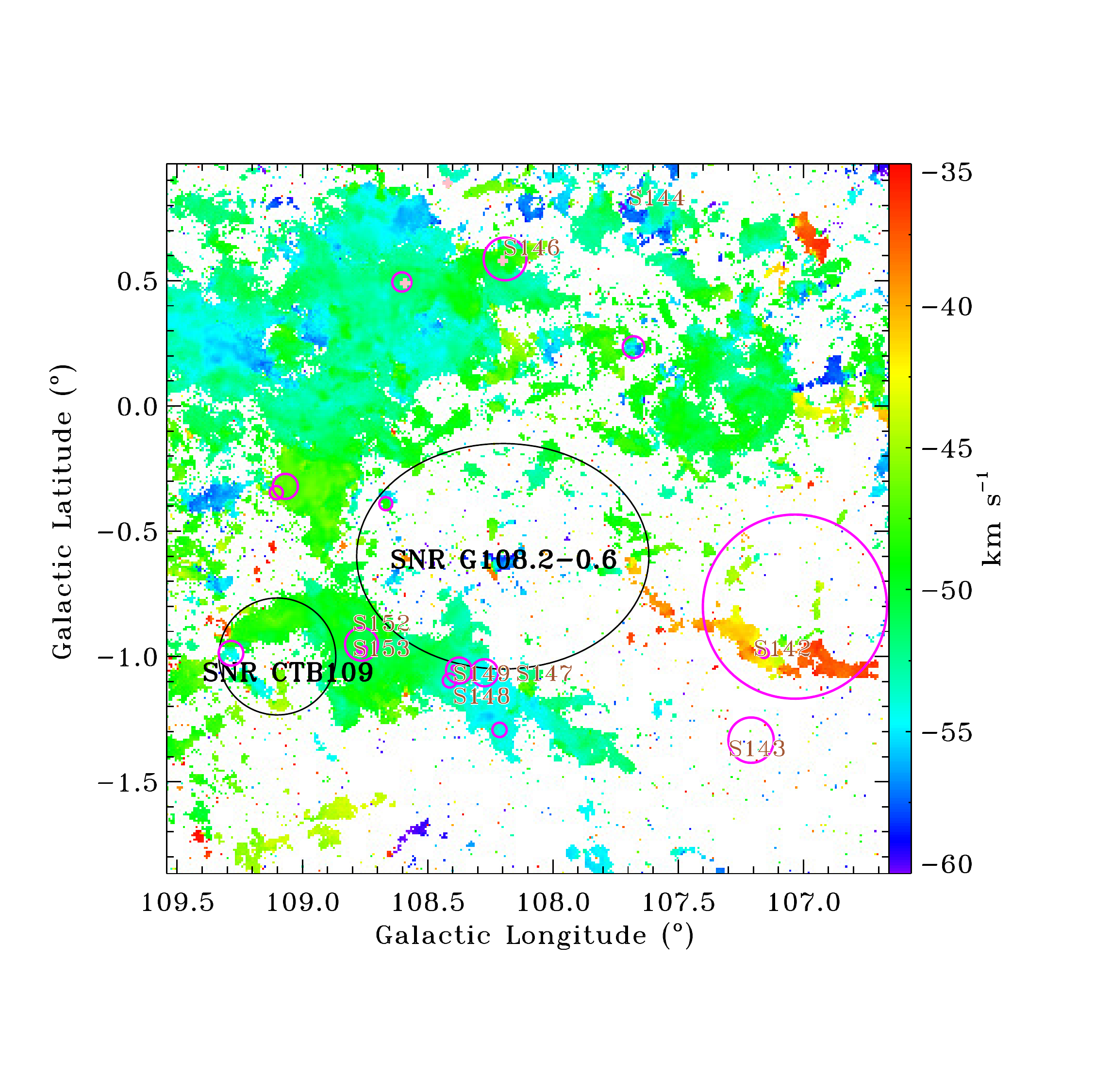}
    \caption{}
   \end{subfigure}
   \begin{subfigure}[b]{0.4\linewidth}
    \includegraphics[width=7.0cm, trim= 1cm 3cm 4cm 2.5cm]{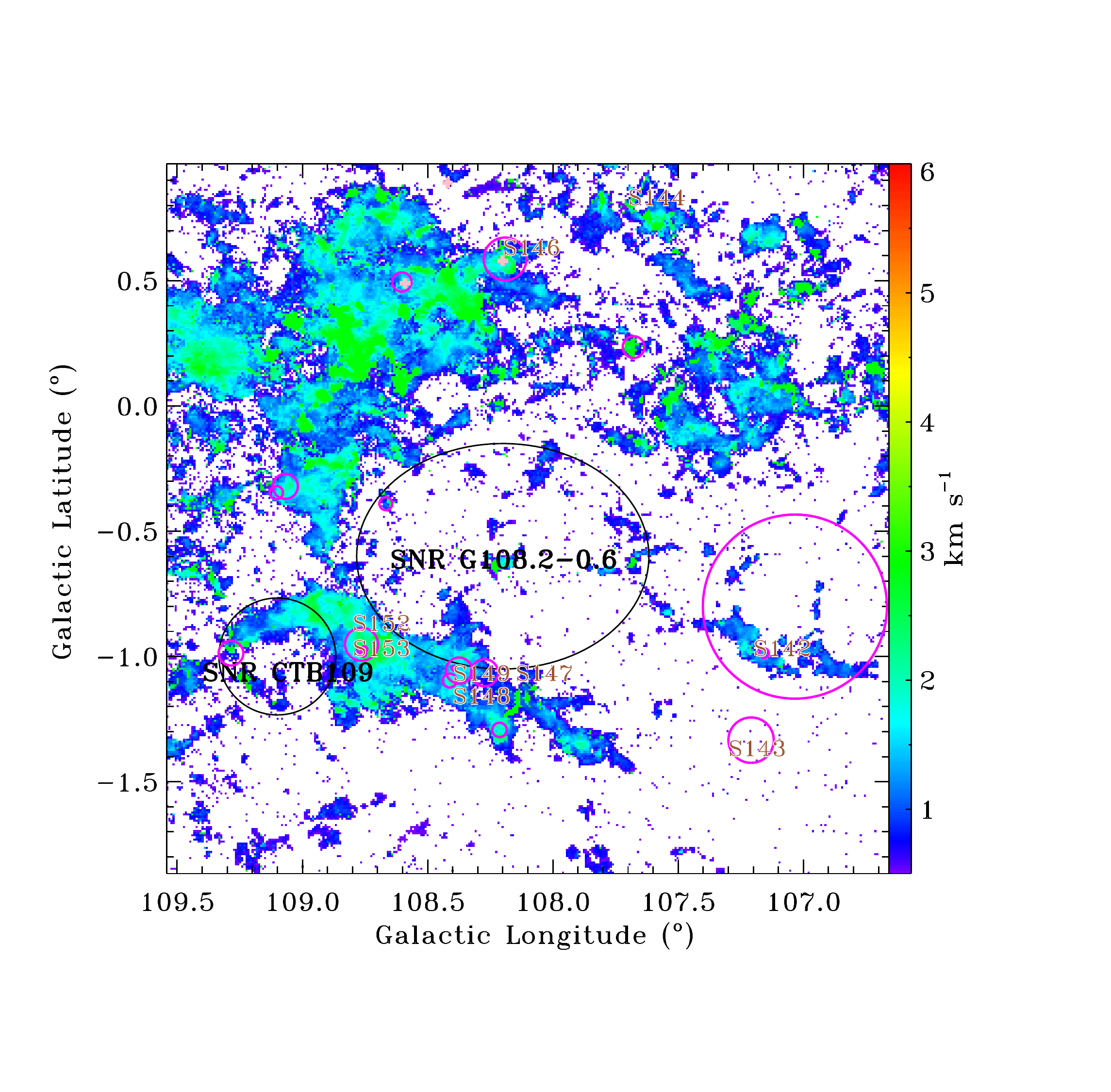}
    \caption{}
    \end{subfigure}
    \caption{(a) Intensity-weighted centroid velocity map and (b) velocity dispersion map of $\rm {}^{12}{CO}$ emission in the velocity range between ${-}$60 and ${-}$35 $\rm km~s^{-1}$. }
    \label{Fig5}
\end{figure}

    Figure~\ref{Fig5} presents the intensity-weighted centroid velocity and velocity dispersion maps of the $\rm {}^{12}{CO}$ emission. In Figure~\ref{Fig5} (a), GMC 4 shows a much-redshifted centroid velocity compared with GMCs 1  ${-}$ 3. Except for this feature, there is no apparent large-scale velocity gradient in the centroid velocity map. A systematic velocity gradient is generally considered to be an indication of expanding shells around {H \small {II}} regions or SNRs. In Figure~\ref{Fig5} (b), the highest velocity dispersions in GMC 1 and GMC 3 reach $\sim$4 km s$^{-1}$,  which is about three times the typical velocity dispersion of the molecular clouds in the Outer Galaxy as obtained by \cite{2010ApJ...723..492R}. The high velocity dispersions in GMCs 1 and 3 are probably caused by multi-velocity components along the line of sight. The highest velocity dispersion of GMC 3 appears at the location of the S152 {H \small {II}} region, which signifies possible interactions between the {H \small {II}} region and the GMC.
      
\subsubsection{Physical Properties}

    To estimate the physical properties of the molecular clouds such as mass, size, and density, it is necessary to obtain the distances to the clouds. In this region, there are three known $\rm {H}_{2}{O}$ masers \citep{2019ApJ...885..131R}. The trigonometric parallax of these masers are 0.227, 0.400, and 0.405 mas, corresponding to distances of 4.405, 2.500, and 2.470 kpc, respectively \citep{2012PASJ...64..142I,2014ApJ...790...99C}. Considering the significant discrepancy among the parallaxes of the above $\rm {H}_{2}{O}$ masers, they may not represent well the distances to the molecular clouds in the surveyed region. The kinematic method is widely used to estimate the distances of molecular clouds. However, the surveyed region in this work is toward the portion of the Perseus arm, where large peculiar motions that can result in significant uncertainties are known \citep{2019ApJ...885..131R}. Some previous works \citep{2015ApJ...810...25G,2021ApJS..254....3M} used the 3D dust extinction map to estimate the distances to molecular clouds. However, given the relatively large distances to the clouds at the Perseus arm in this region ($>$2\,kpc), the available 3D extinction maps, such as that presented by \citep{2019ApJ...887...93G} toward this region, also have large uncertainties.\par
         
   Massive stars with spectral types earlier than B2 can be the excitation sources of {H \small {II}} regions/candidates. \cite{2015AJ....150..147F} used the 21 cm continuum data from the Canadian Galactic Plane Survey (CGPS) to identify bright {H \small {II}} regions. They found 103 {H \small {II}} regions and estimated the mean distances of the OB stars associated with the {H \small {II}} regions.  \cite{2015AJ....150..147F} expanded the list of associated OB stars through a search for additional OB stars that lie in or near the outermost 21 cm continuum or H$\rm{\alpha}$ emission boundary of the {H \small {II}} regions. We use a method similar to \cite{2015AJ....150..147F} to search for OB stars that are associated with the {H \small {II}} regions in the observed region in this work. The distances of the OB stars are calculated using the parallax measurements from \cite{2018AJ....156...58B}. \par 
   
    We firstly search for OB stars with spectral types of B2 or earlier in this region from the literature \citep{2003AJ....125.2531R,2013msao.confE.198M,2019MNRAS.487.1400C,2019ApJS..241...32L,2021ApJS..253...22X,2021ApJS..253...54L}. \cite{2003AJ....125.2531R} published the Alma Luminous Star (ALS) catalog, which contains over 18, 000 known Galactic luminous  stars with available spectral type and/or photometric data. We crossmatched the ALS catalog with the Gaia EDR3 catalog \citep{2021A&A...649A...9G} using the method suggested by \cite{2020MNRAS.495..663W} and obtained the distances of about 11, 000 sources from \cite{2021AJ....161..147B}. \cite{2019MNRAS.487.1400C} compiled an OB star catalog by adopting three catalogs, i.e., the GOSC catalog \citep{2013msao.confE.198M}, the OB star catalog selected from \cite{2014yCat....1.2023S} and that selected from the Large Sky Area Multi-Object Fiber Spectroscopic Telescope \citep[LAMOST,][]{2012RAA....12.1197C} Spectroscopic Survey of the Galactic Anti-centre \citep[LSS-GAC;][]{2015MNRAS.448..855Y}. \cite{2019MNRAS.487.1400C} crossmatched their compiled OB star catalog with the Gaia DR2 catalog \citep{2018A&A...616A..14G} and obtained Gaia distance measurements for about 8, 800 O-B2 stars. \cite{2019ApJS..241...32L} presented over 16, 000 OB stars identified from the LAMOST DR5 data \citep{2015RAA....15.1095L} and \cite{2021ApJS..253...22X} combined them with the Gaia DR2 parallaxes to estimate their distances using a data-driven machine-learning method. \cite{2021ApJS..253...54L} also published 209 O-type stars identified with LAMOST data. We combined the above OB star catalogs together and excluded the duplicates. Finally, we obtained 8849 O-B2 stars with spectral type classification.

    Figure~\ref{Fig2} shows the 22 OB stars that are located within or near the {H \small {II}} regions. The S148 and S149 {H \small {II}} regions in GMC 3 each contain two OB stars. The S142 {H \small {II}} region in GMC 4 has seventeen associated OB stars, while the S143 {H \small {II}} region to the south of GMC 4 has only one. We propose that the 22 OB stars are candidates of  the excitation sources of the four {H \small {II}} regions, and Table~\ref{table A.1} shows the detailed information of the 22 OB stars. Three of the above four {H \small {II}} regions are associated with molecular clouds (for details, see Section~\ref{subsect:Giant Molecular Cloud Complex Associated with the HII Regions/Candidates}). Based on the parallax distances of the OB stars associated with the S148 and S149 {H \small {II}} regions, which are associated with GMC 3 (see Section~\ref{subsect:Giant Molecular Cloud Complex Associated with the HII Regions/Candidates} for detailed discussion), we obtain a mean distance of 2.92 kpc for GMC 3. The distance of GMC 4 is adopted to be 2.63 kpc, which is the mean distance of the OB stars within the boundary of S142. Since GMCs 1 ${-}$ 3 are velocity coherent  in the p-v diagram with velocity differences less than 4 $\rm km~s^{-1}$, we adopt the  distance of 2.92 kpc of GMC 3 for GMCs 1 ${-}$ 2.   

    The masses of the clouds are calculated using the following formula, 
\begin{equation}
M = \mu m_{H} D^{2} \int N_{H{}_{2}} d\Omega ,
\end{equation} 
where $\mu$ = 2.83 is the mean molecular weight, m$\rm _{H}$ is the mass of the hydrogen atom, D is the distance, and $\Omega$ is the solid angle. Here, we adopt two methods to calculate the H$_2$ column density of the clouds. Firstly, we assume that the $^{12}$CO and $^{13}$CO molecules are under local thermodynamic equilibrium (LTE) conditions and the $\rm {}^{12}{CO}$ emission is optically thick. Then the excitation temperature can be obtained from the peak intensity of the $\rm {}^{12}{CO}$ line \citep{2018ApJS..238...10L,2020RAA....20...60M},

\begin{equation}
T_{\rm ex} = \frac{5.532 {\rm K}} { [\ln(1 + \frac{5.532 {\rm K} }{{\rm T_{peak}(\rm {}^{12}{CO})} + 0.819})]} ,
\end{equation}
where $\rm T_{peak}(\rm {}^{12}{CO})$ is the peak brightness temperature of the $\rm {}^{12}{CO}$ line. The optical depth and the column density of the $\rm {}^{13}{CO}$ molecule are derived according to the following formula \citep{2010ApJ...721..686P,2018ApJS..238...10L,2020RAA....20...60M},

\begin{equation}
\tau_{\rm \nu}^{\rm {}^{13}{CO}} =  - \ln[1 - \frac{\rm T_{mb}(\rm {}^{13}{CO})}{5.29[\rm J_{\nu}(\rm T_{ex}) - 0.164]}] ,
\end{equation}

\begin{table}
\bc
\begin{minipage}[]{100mm}
\caption[]{Physical Parameters of the Clouds in the Region\label{tab1}}\end{minipage}
\setlength{\tabcolsep}{4pt}
\small
 \begin{tabular}{ccrccccrrc}
  \hline\noalign{\smallskip}
 (1)& (2)& (3)&(4)& (5)& (6)& (7)& (8)& (9)\\
GMC&$\rm {l}_{c}$& $\rm {b}_{c}$&$\rm {v}_{c}$&d& $\rm {M}_{LTE}$ & $\rm {M}_{{X}_{CO}}$& $\rm {T}_{ex}$& Number of Filaments& \\
&(degree) &(degree)&($\rm km~s^{-1}$)&(kpc)&(${M}_{\odot}$)&(${M}_{\odot}$)&(${K}$)&\\
  \hline\noalign{\smallskip}
GMC1&108.69&0.16&${-}$50.98&2.92&${2.71}\times{10}^4$&${1.64}\times{10}^5$&6.7&11\\
GMC2&109.00&${-}$0.35&${-}$47.17&2.92&${1.43}\times{10}^4$&${2.29}\times{10}^4$&11.3&3\\
GMC3&108.68&${-}$0.97&${-}$51.07&2.92&${4.78}\times{10}^4$&${8.35}\times{10}^4$&9.7&9\\
GMC4&107.12&${-}$0.96&${-}$40.20&2.63&${1.49}\times{10}^3$&${5.77}\times{10}^3$&8.8&2\\

  \noalign{\smallskip}\hline
\end{tabular}
\ec
%% place \tablecomments and \tablerefs below \end{center| and \end{center}:
%% you may leave the table-width parameter to editors or set to your actual size
\tablecomments{0.95\textwidth}{Columns 2 ${-}$ 4 are the centroid position. The distance, mass, and mean excitation temperature are given in columns 5 ${-}$ 8. The last column gives the numbers of filaments in each cloud.}
\end{table}

\begin{equation}
  N_{(\rm {}^{13}{CO})} = 2.42 \times 10^{14} \frac{\rm T_{ex} + 0.88}{1 - e^{\frac{-5.29}{\rm T_{ex}}}} \int \tau_{\rm \nu}^{\rm {}^{13}{CO}}{\rm d\upsilon} ,
\end{equation}
where $\rm J_{\nu}(\rm T_{ex})$ = $[\rm exp(5.29/\rm T_{ex}) - 1]^{-1}$, and $\rm T_{ex}$ is the excitation temperature. We use the abundance ratios [$\rm {}^{12}{C}$/$\rm {}^{13}{C}$] = 6.21d$\rm _{GC}$ + 18.71 \citep{2005ApJ...634.1126M} (d$\rm _{GC}$ is the distance in kpc from the cloud to the Galactic center) and H$_{2}$/$^{12}$CO = 1.1 $\times$ 10$^4$ \citep{1982ApJ...262..590F} to convert the $^{13}$CO column density to the H$_2$ column density. For the clouds studied in this work, d$\rm _{GC}$ is about 9.7 kpc and therefore [$^{12}$C/$^{13}$C] = 80. We also use the conversion factor X$_{\rm CO}$ = 2 $\times$ 10$^{20}$ cm$^{-2}$ (K km s$^{-1}$)$^{-1}$  \citep{2013ARA&A..51..207B} to convert the integrated intensity of $^{12}$CO to the column density of molecular hydrogen. The excitation temperatures of the clouds and the masses derived from the above two methods are presented in Table~\ref{tab1}. 
\label{physical properties}

\begin{figure}[htbp]
    \centering
    \begin{subfigure}[b]{1\linewidth}
    \includegraphics[width=14cm, trim= -1cm 0cm 0cm 0cm]{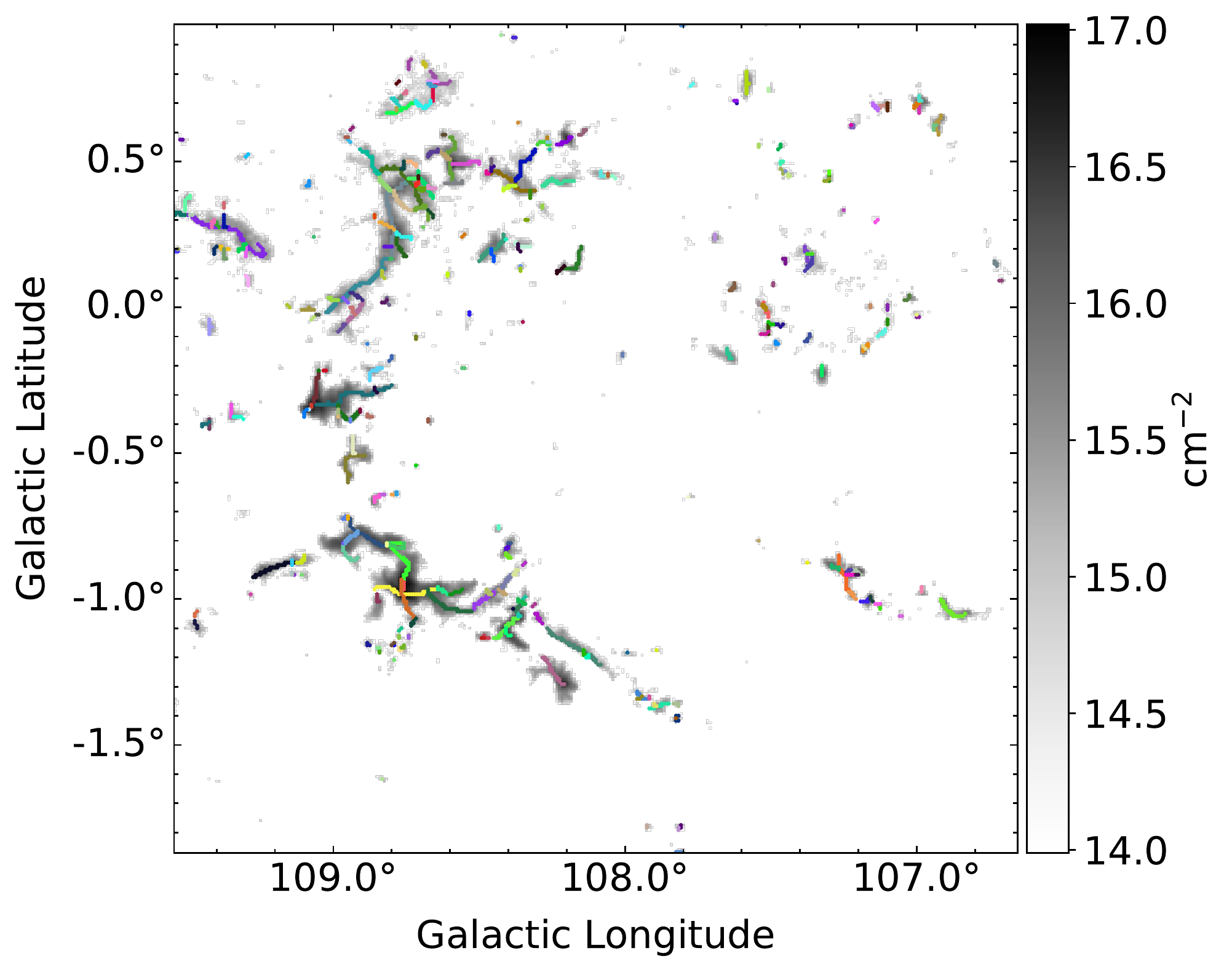}
    \caption{}
    \end{subfigure}
    \begin{subfigure}[b]{1\linewidth}
    \includegraphics[width=14cm, trim= -1cm 0cm 0cm 0cm]{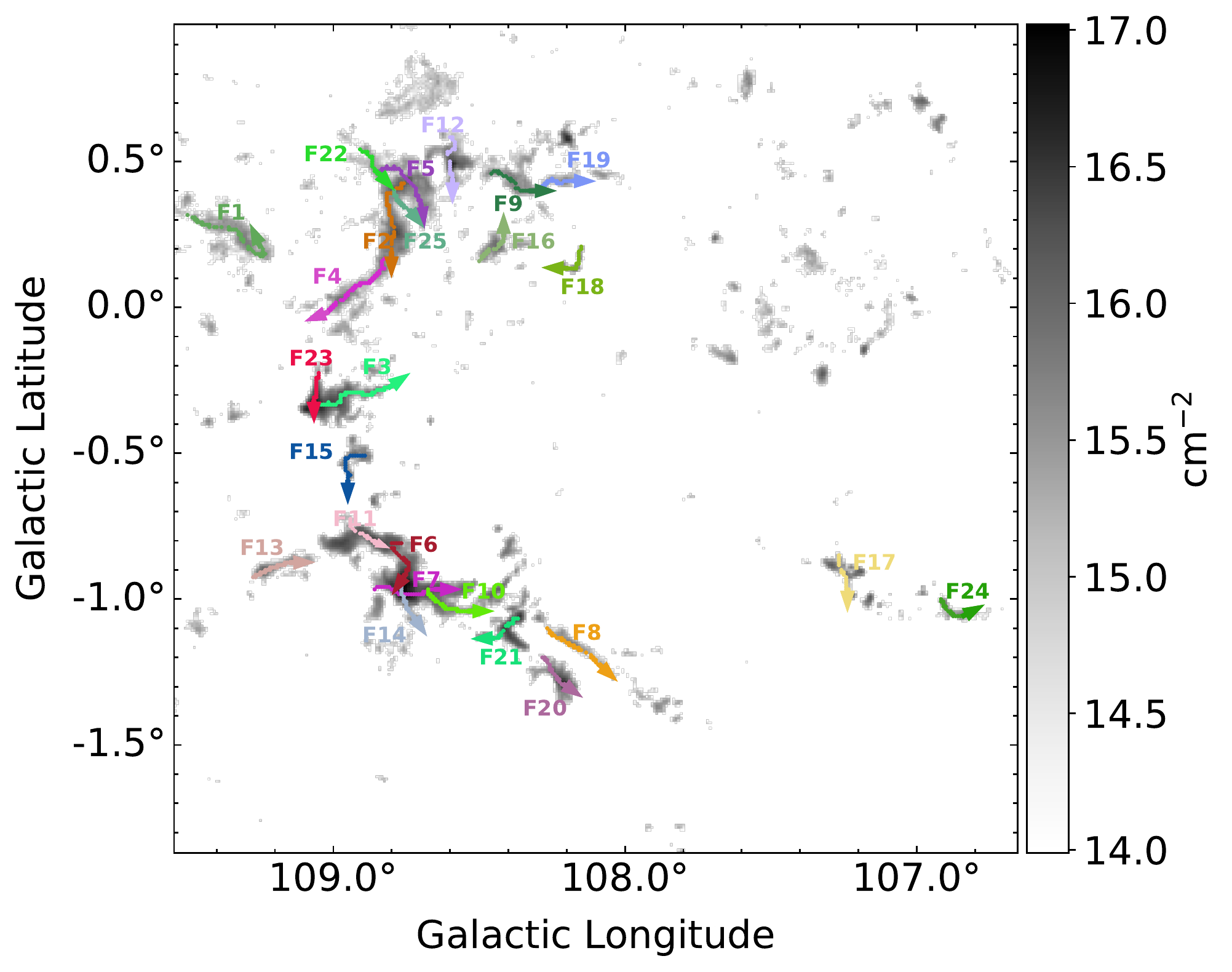}
    \caption{}
    \end{subfigure}
    \caption{ 299 skeletons identified by DisPerSE which are overlaid onto the column density map of $\rm {}^{13}{CO}$ (a) and the 25 skeletons with lengths larger than 15 pixels (b). In panel (b), the color lines represent the positions of the skeletons and the arrows indicate the directions of the position-velocity diagram shown in Figure~\ref{Fig7}. }
    \label{Fig6}
\end{figure}

\label{subsect:Overview of Molecular Cloud in the Region}
\subsection{Filamentary Contents of the Molecular Clouds}

In this section, we identify filaments in the $^{13}$CO column density map using the publicly-available algorithm Discrete Persistent Structures Extractor (DisPerSE) \citep{2011MNRAS.414..350S}. This algorithm can identify persistent topological features such as walls, sheets, and especially filamentary structures within a three-dimensional space, such as a position-position-position (PPP) data product from simulations or position-position-velocity data product from observations
 \citep{2011A&A...529L...6A,2012A&A...540L..11S,2013A&A...550A..38P,2014MNRAS.444.2507P,2014MNRAS.445.2900S,2016MNRAS.455.3640S,2018A&A...610A..62C,2019A&A...623A.142S,2021RAA....21..188Z}.
\par

\begin{figure}
\centering
    \includegraphics[width=.5\textwidth]{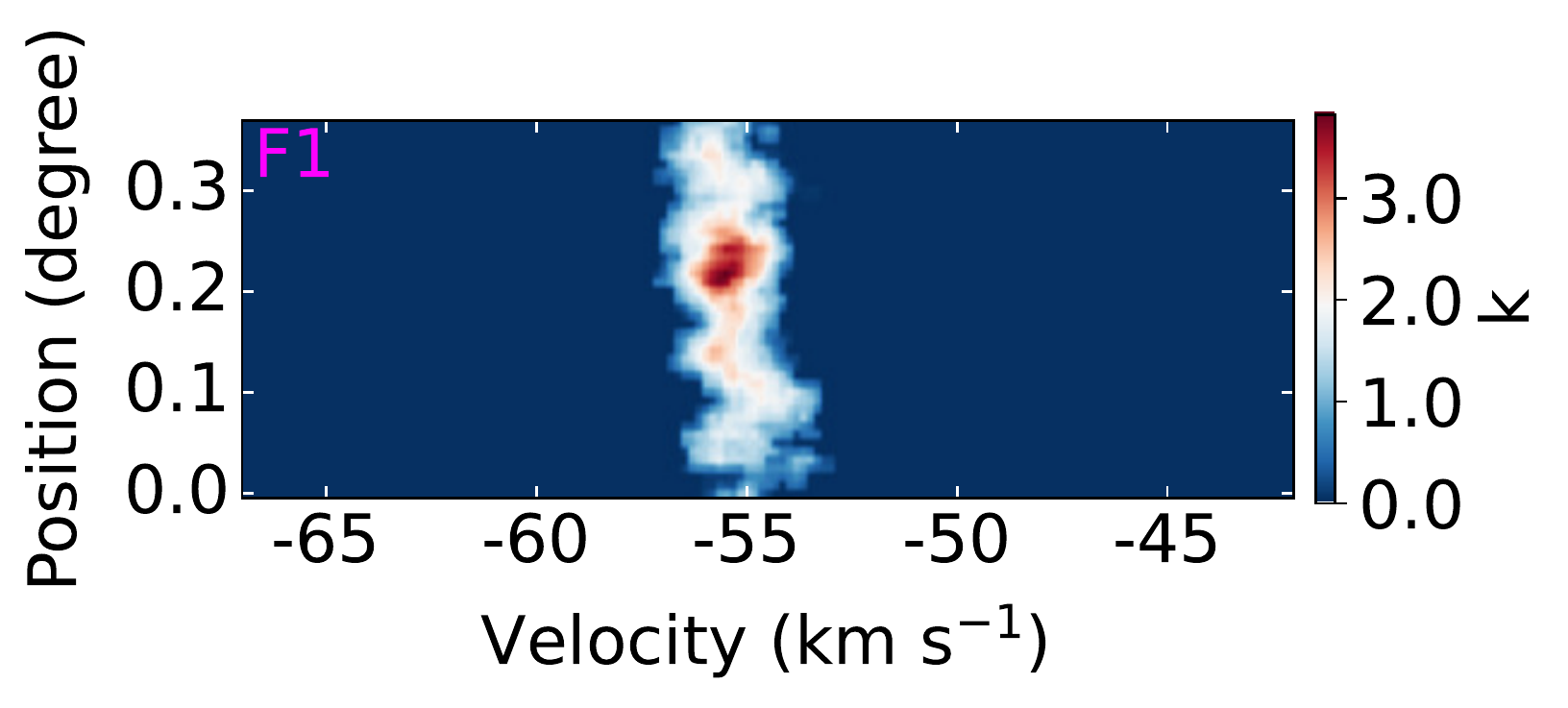}\hfill
    \includegraphics[width=.5\textwidth]{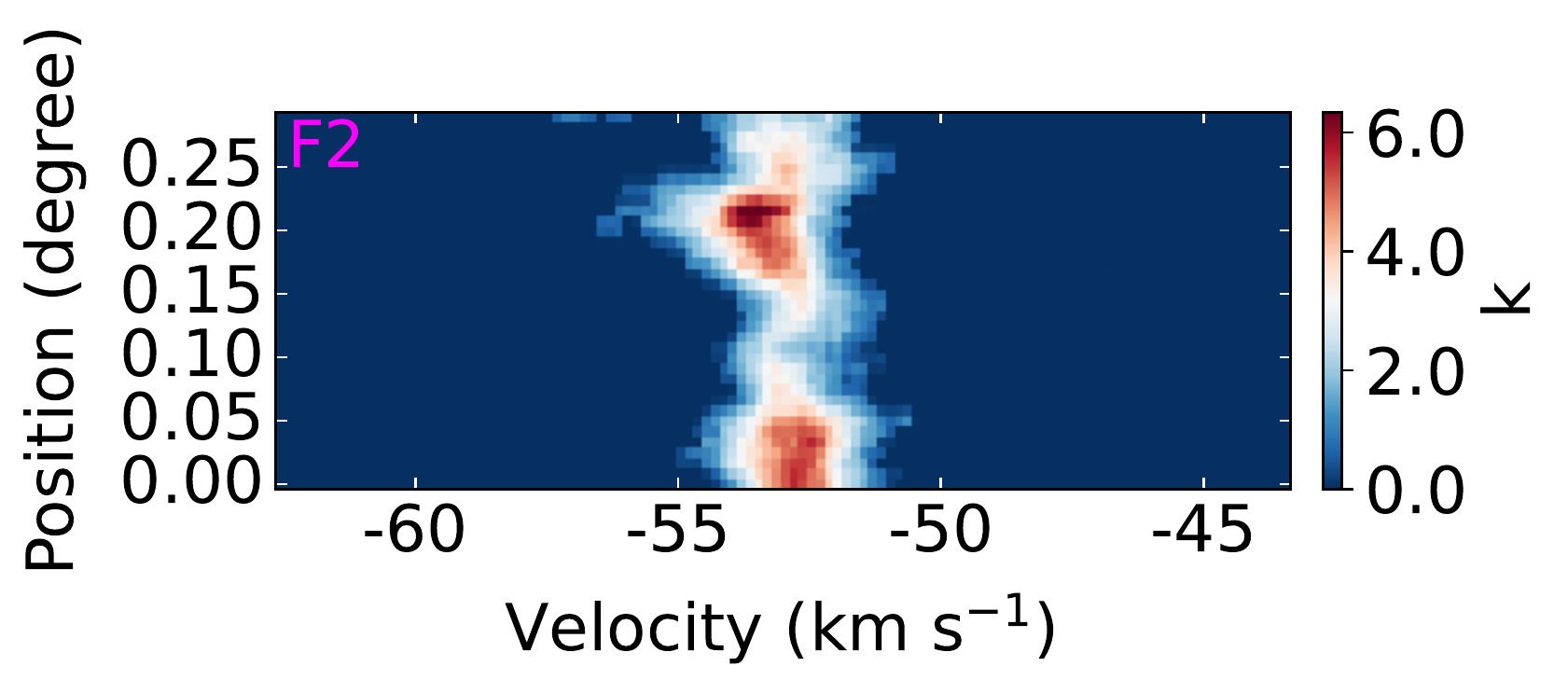}\hfill
    
    \includegraphics[width=.5\textwidth]{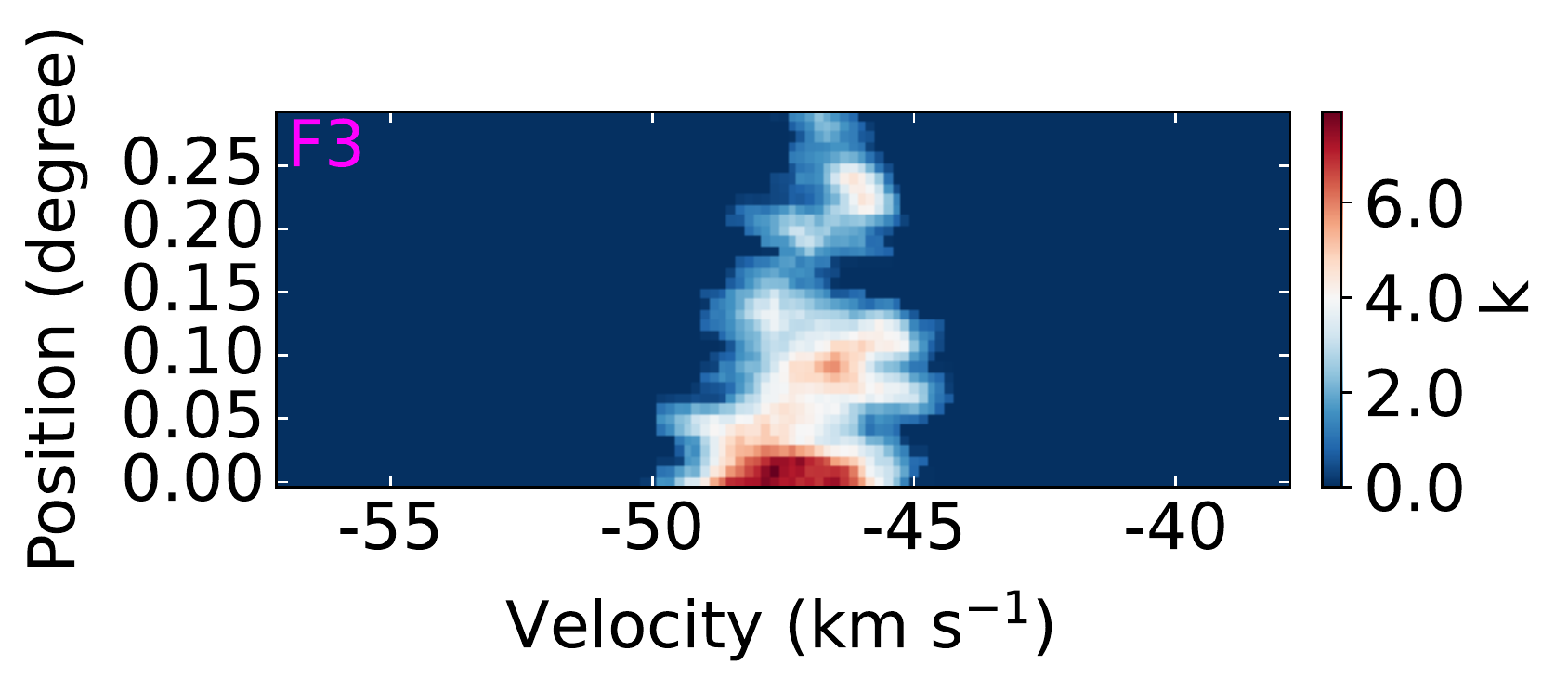}\hfill
    \includegraphics[width=.5\textwidth]{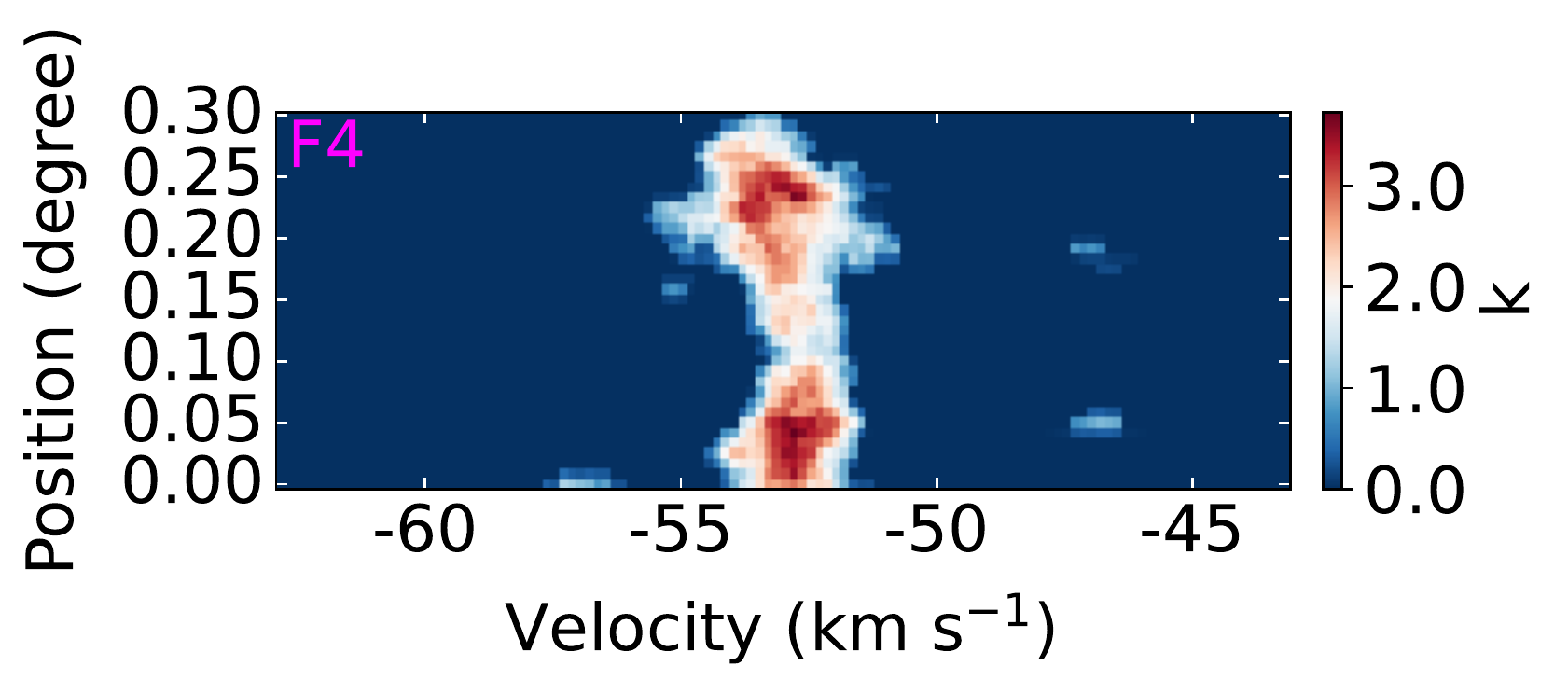}\hfill
    
    \includegraphics[width=.5\textwidth]{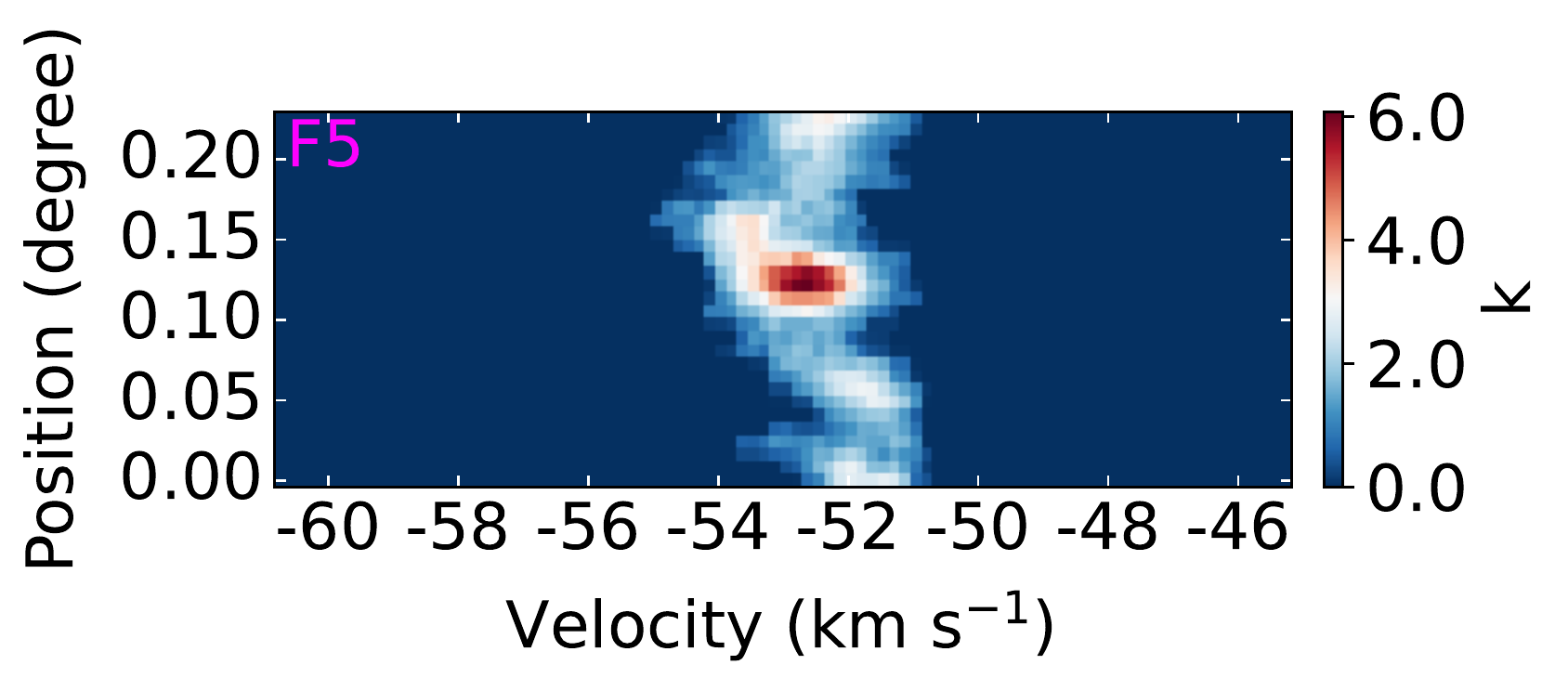}\hfill
    \includegraphics[width=.5\textwidth]{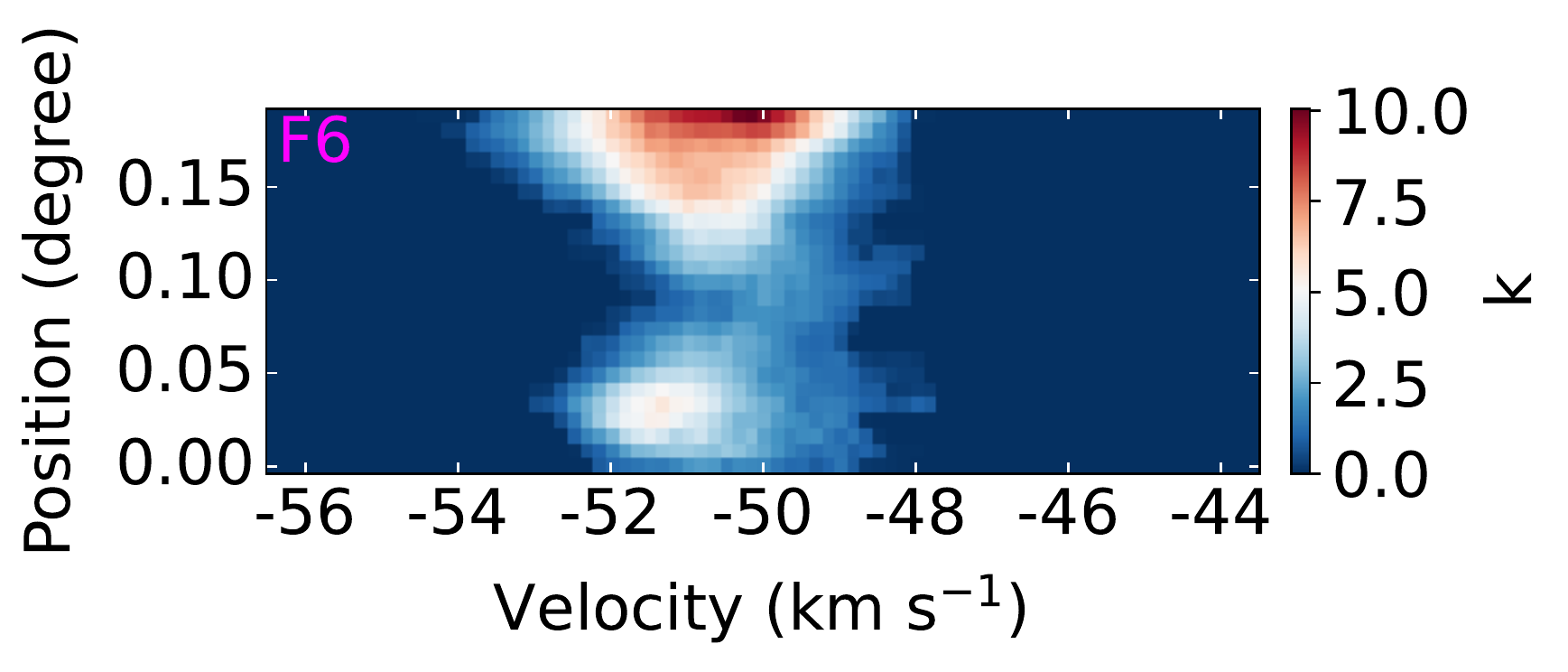}\hfill
    
    \includegraphics[width=.5\textwidth]{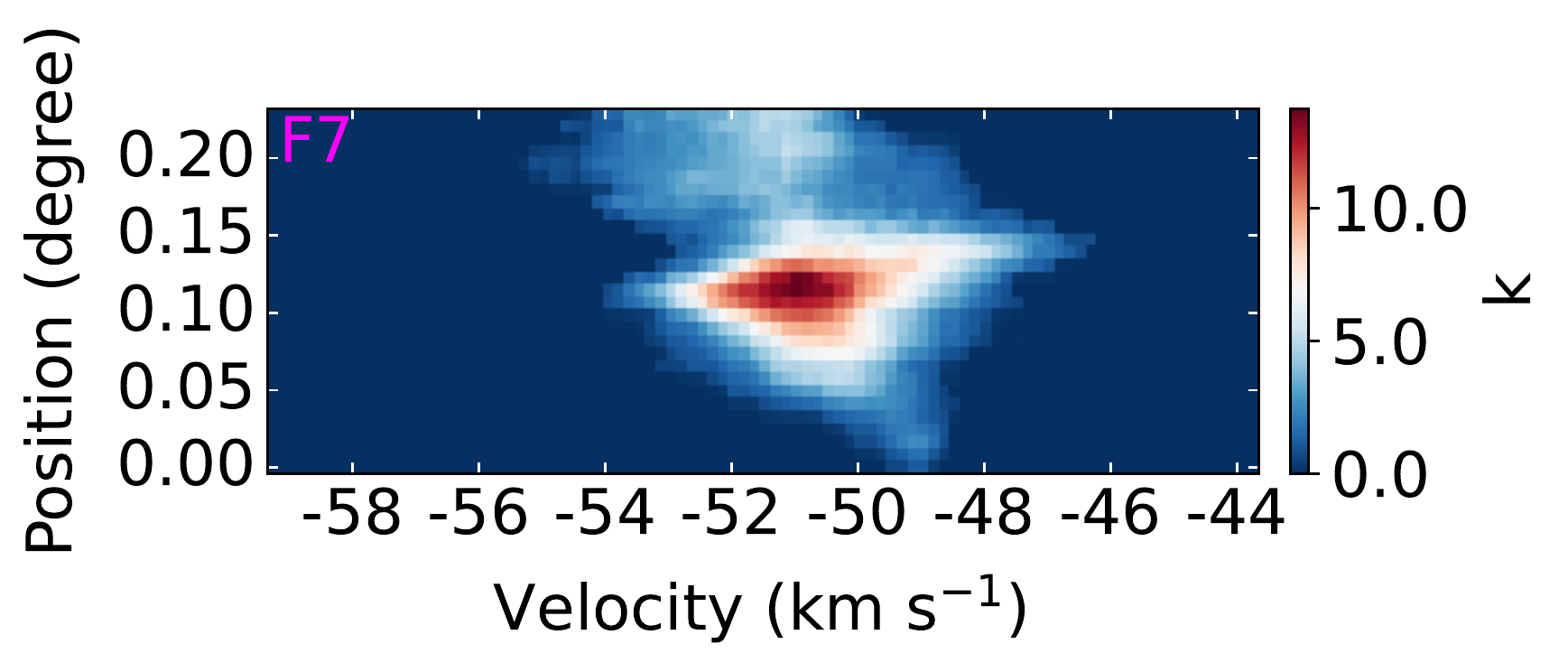}\hfill
    \includegraphics[width=.5\textwidth]{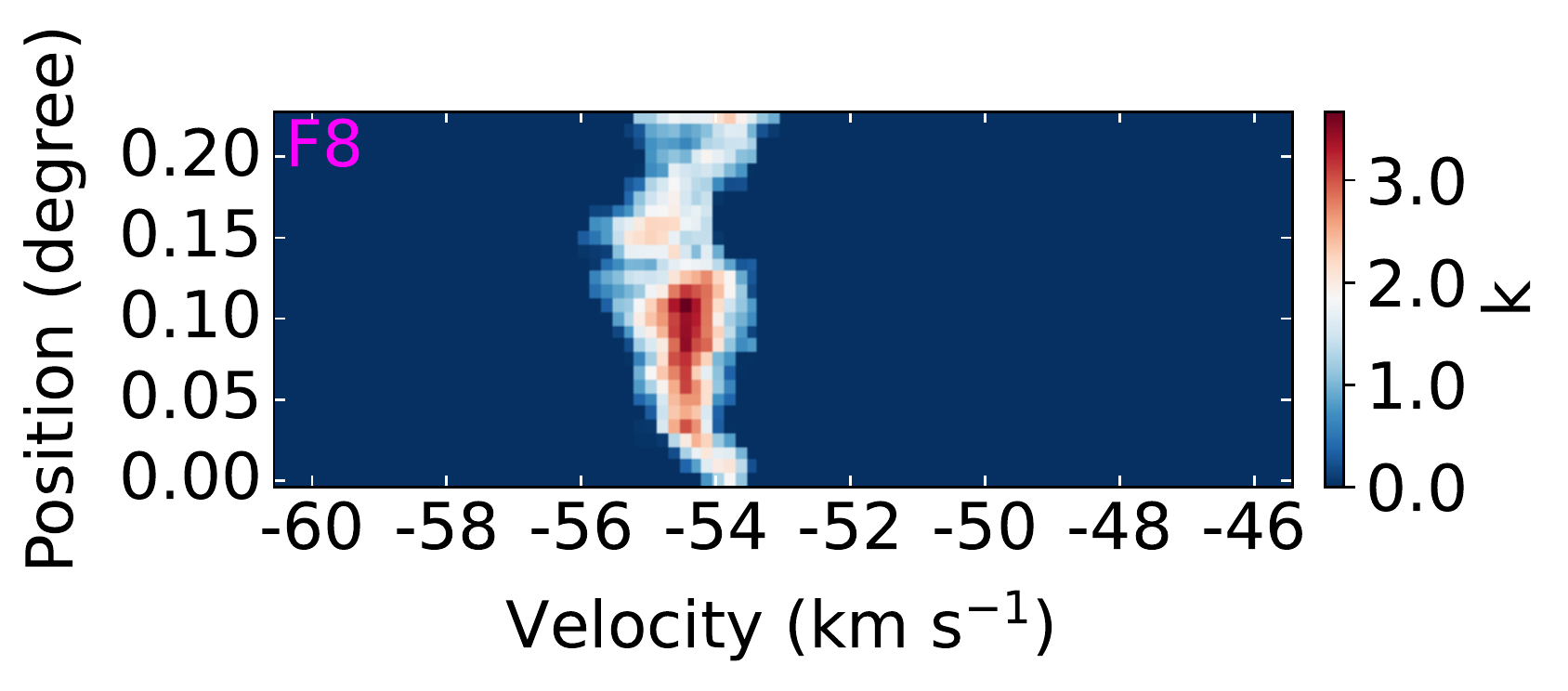}\hfill
    
    \includegraphics[width=.5\textwidth]{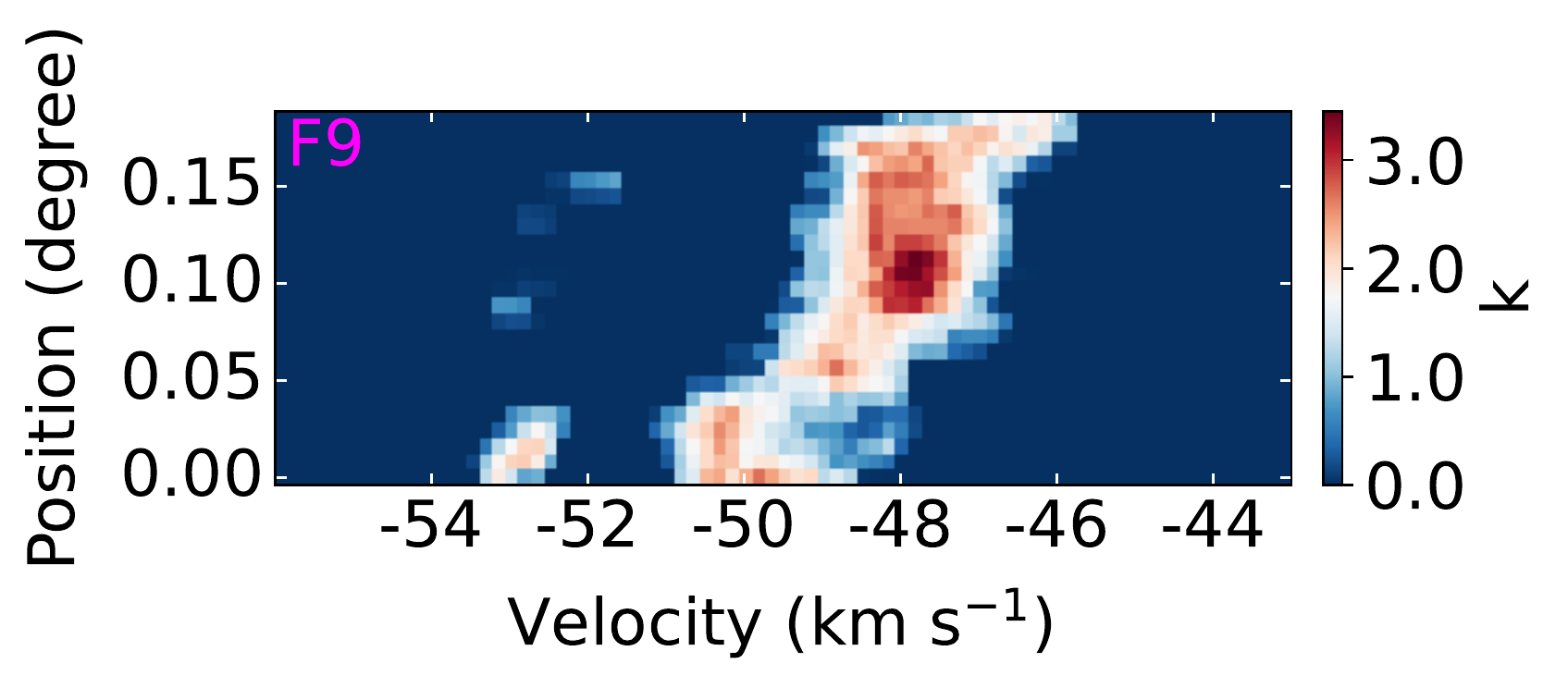}\hfill
    \includegraphics[width=.5\textwidth]{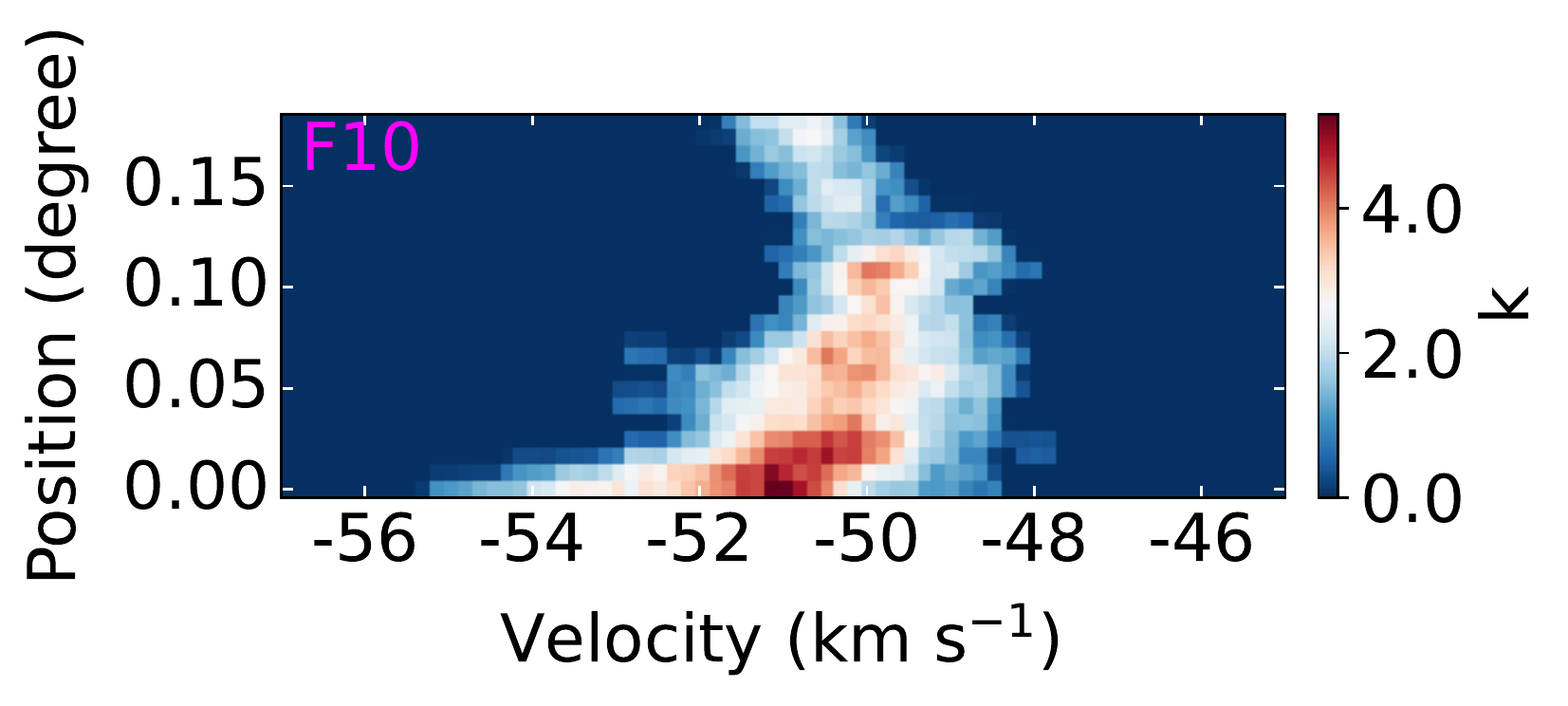}\hfill
    
    \includegraphics[width=.5\textwidth]{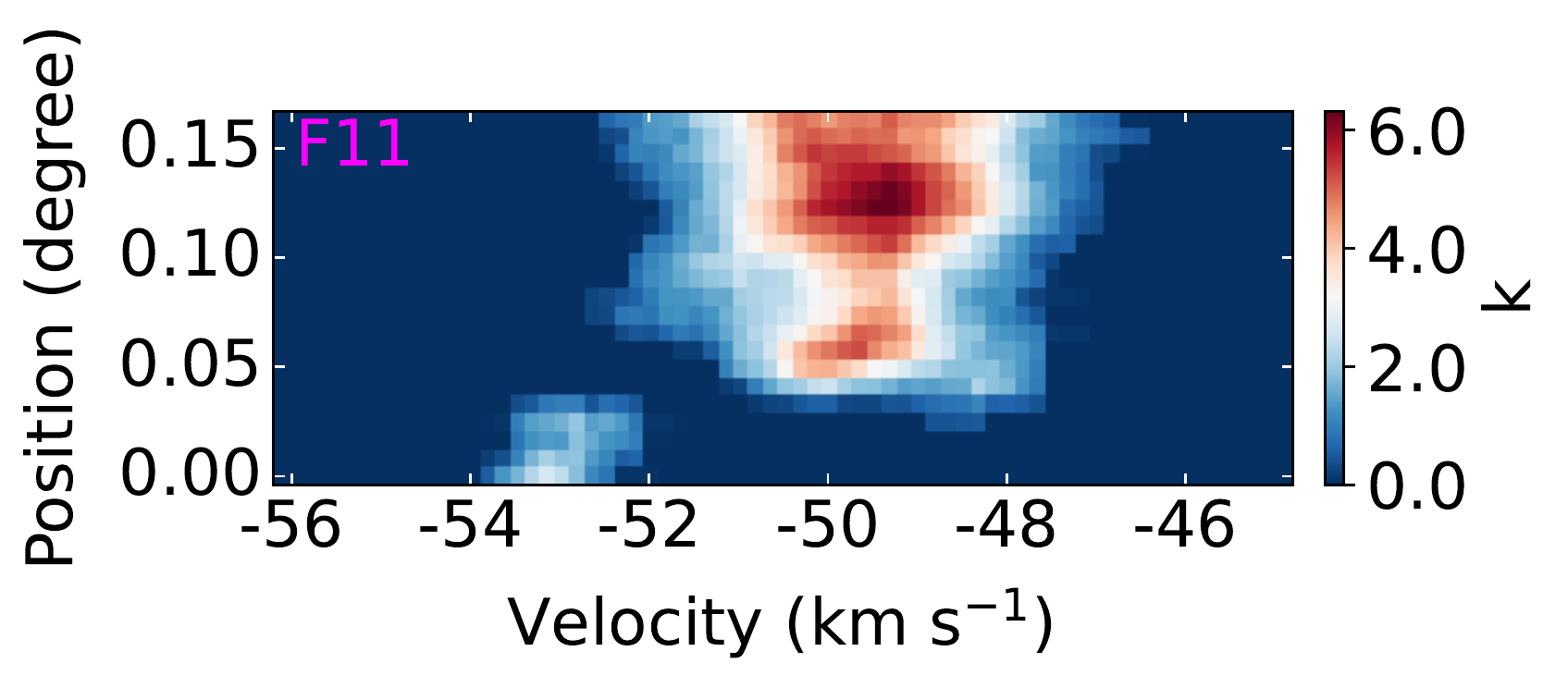}\hfill
    \includegraphics[width=.5\textwidth]{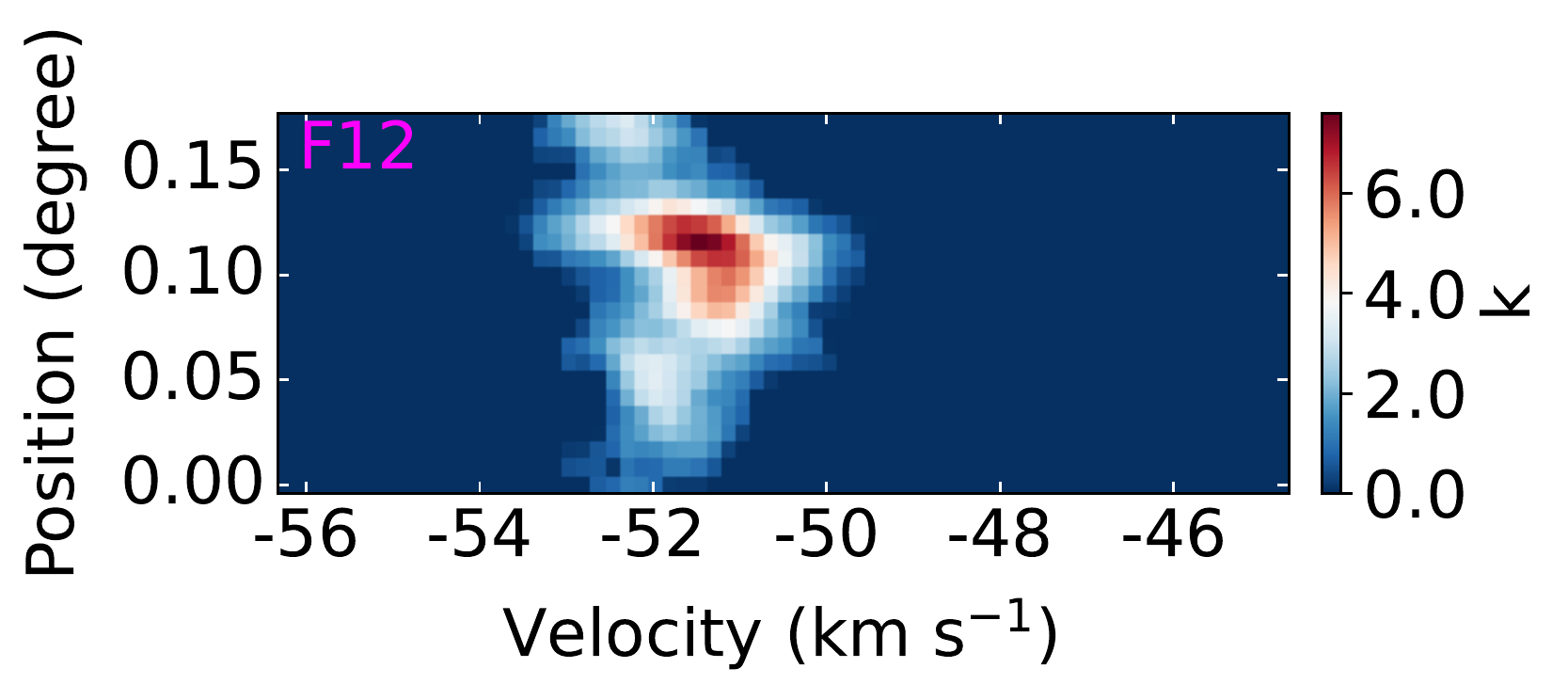}\hfill
    
    \includegraphics[width=.5\textwidth]{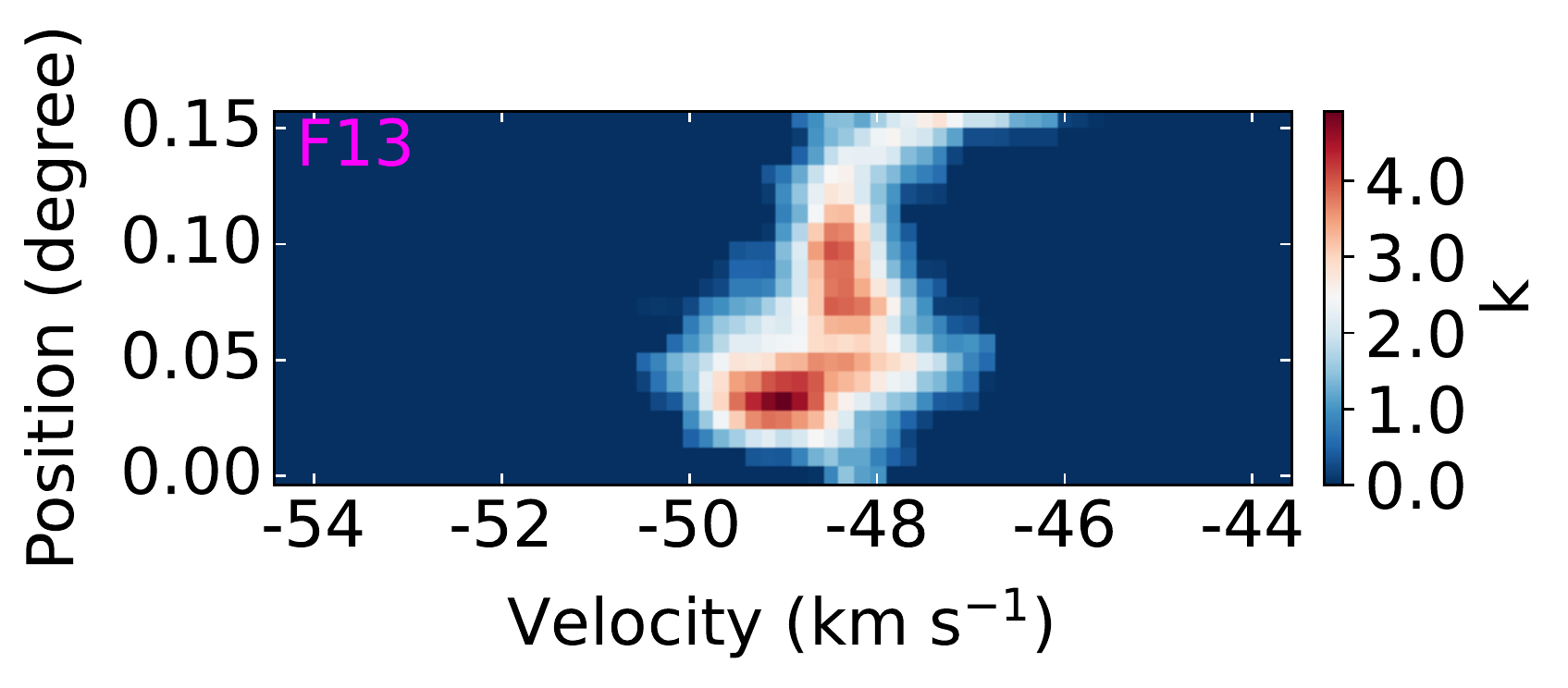}\hfill
    \includegraphics[width=.5\textwidth]{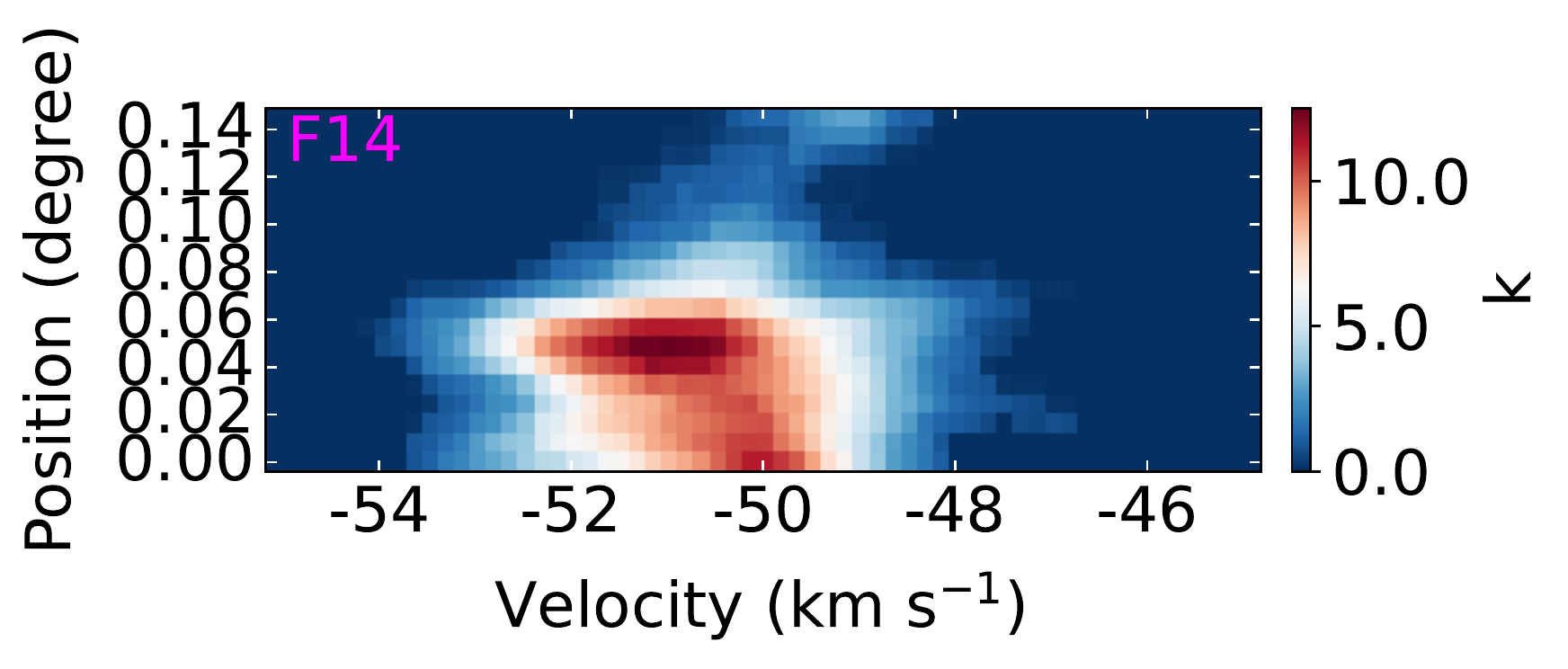}\hfill
    \\[\smallskipamount]
    \label{Fig7}
    \end{figure}
    
    \begin{figure}
    \centering
    \includegraphics[width=.5\textwidth]{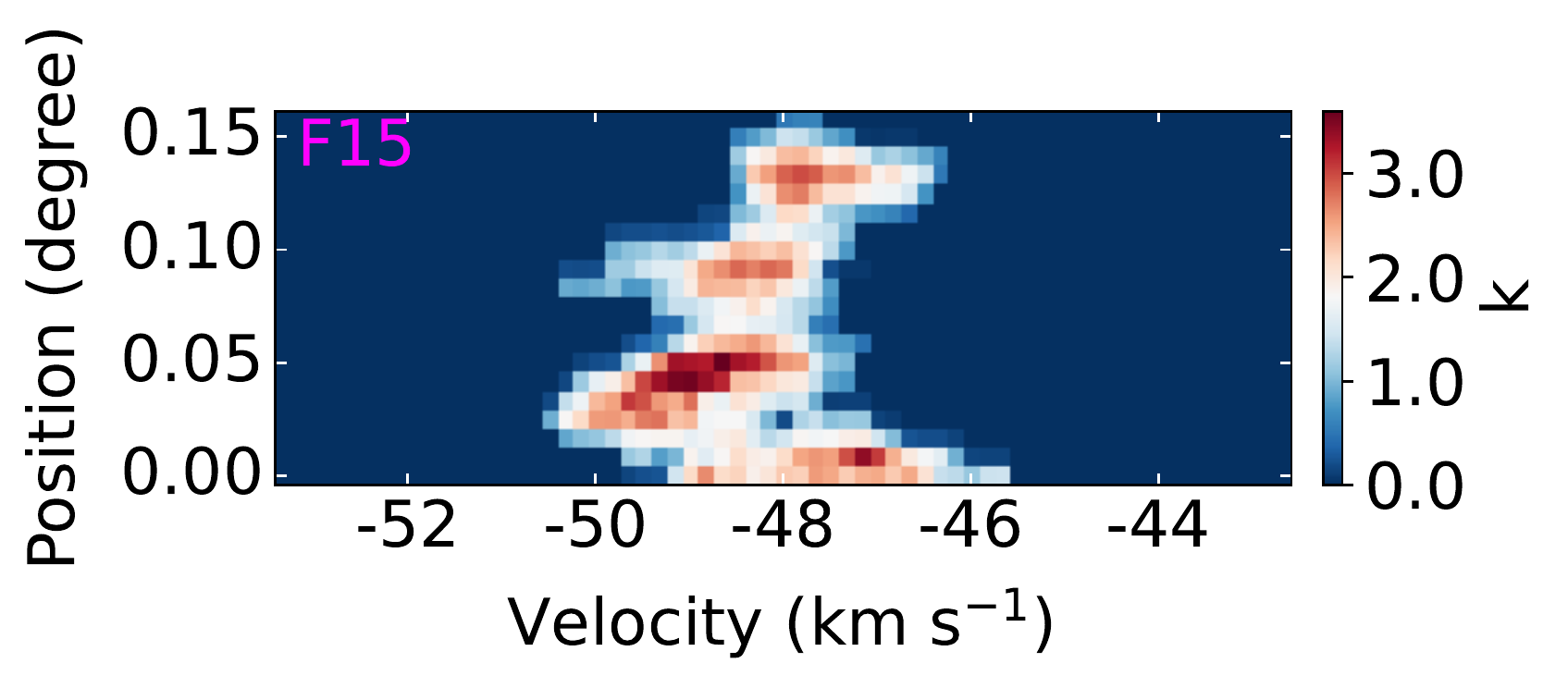}\hfill
    \includegraphics[width=.5\textwidth]{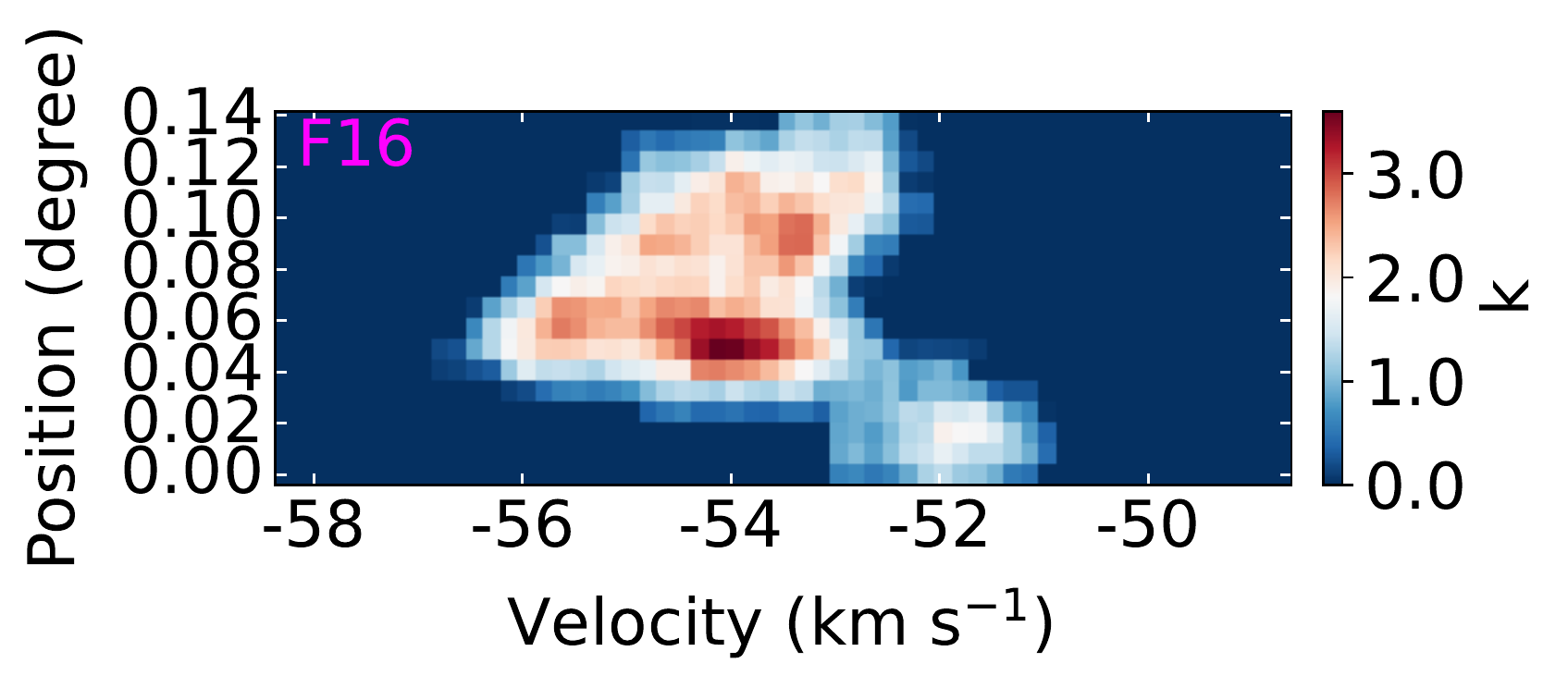}\hfill    
    
    \includegraphics[width=.5\textwidth]{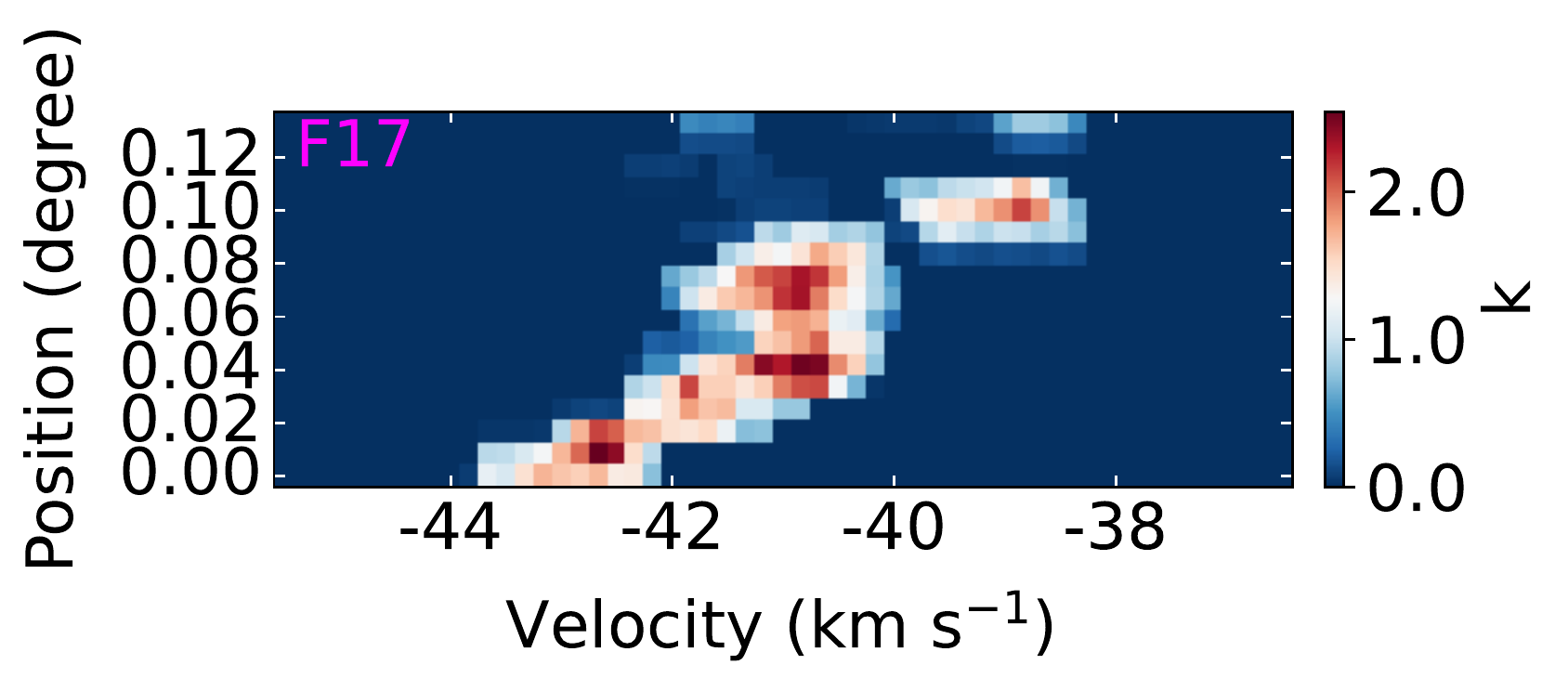}\hfill
    \includegraphics[width=.5\textwidth]{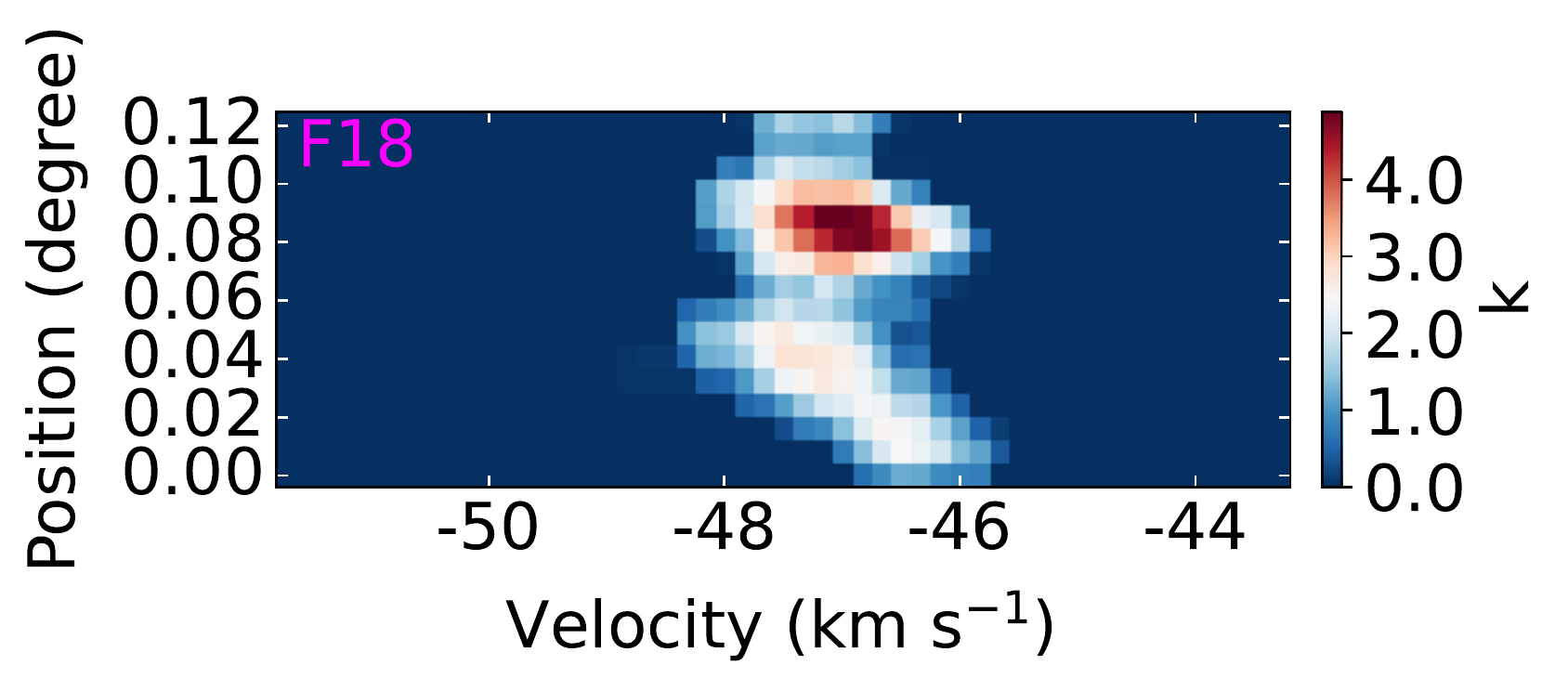}\hfill
    
    \includegraphics[width=.5\textwidth]{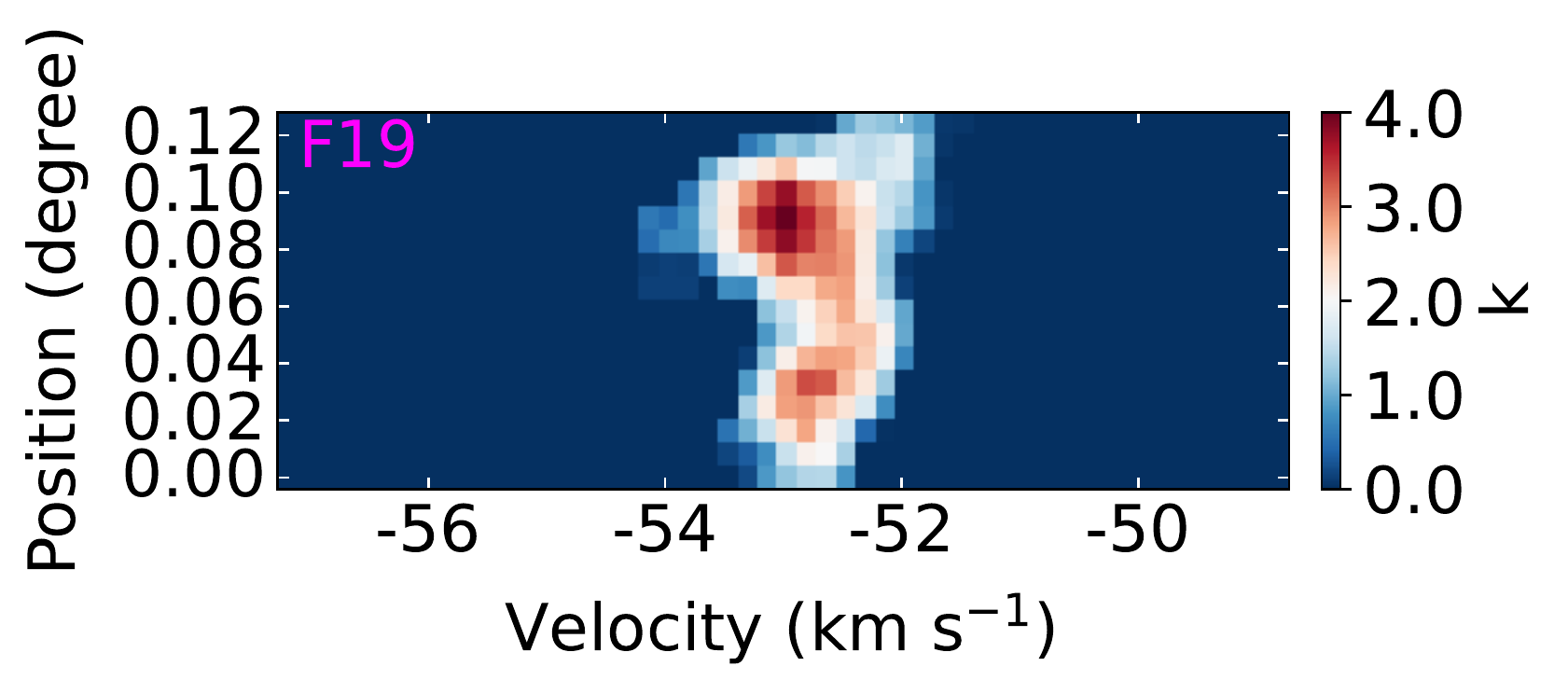}\hfill
    \includegraphics[width=.5\textwidth]{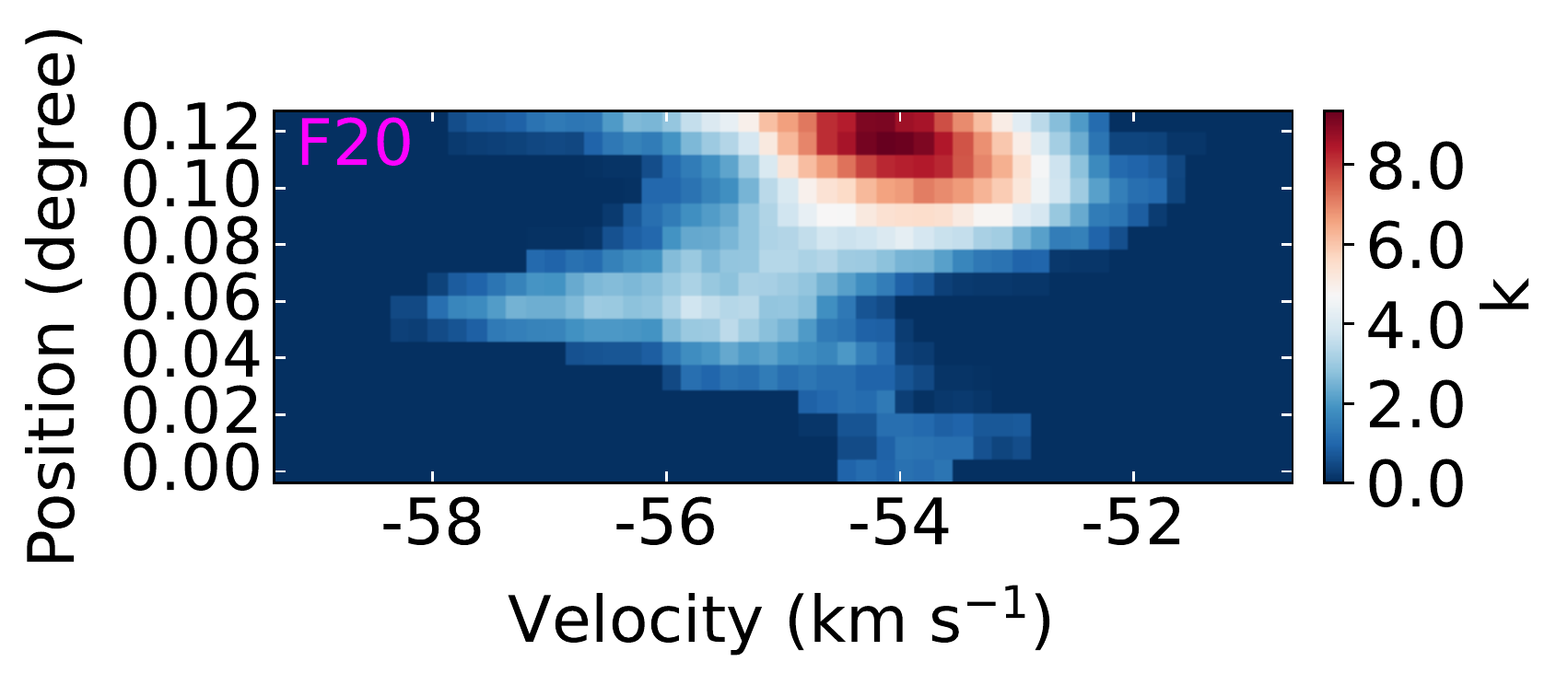}\hfill
    
   \includegraphics[width=.5\textwidth]{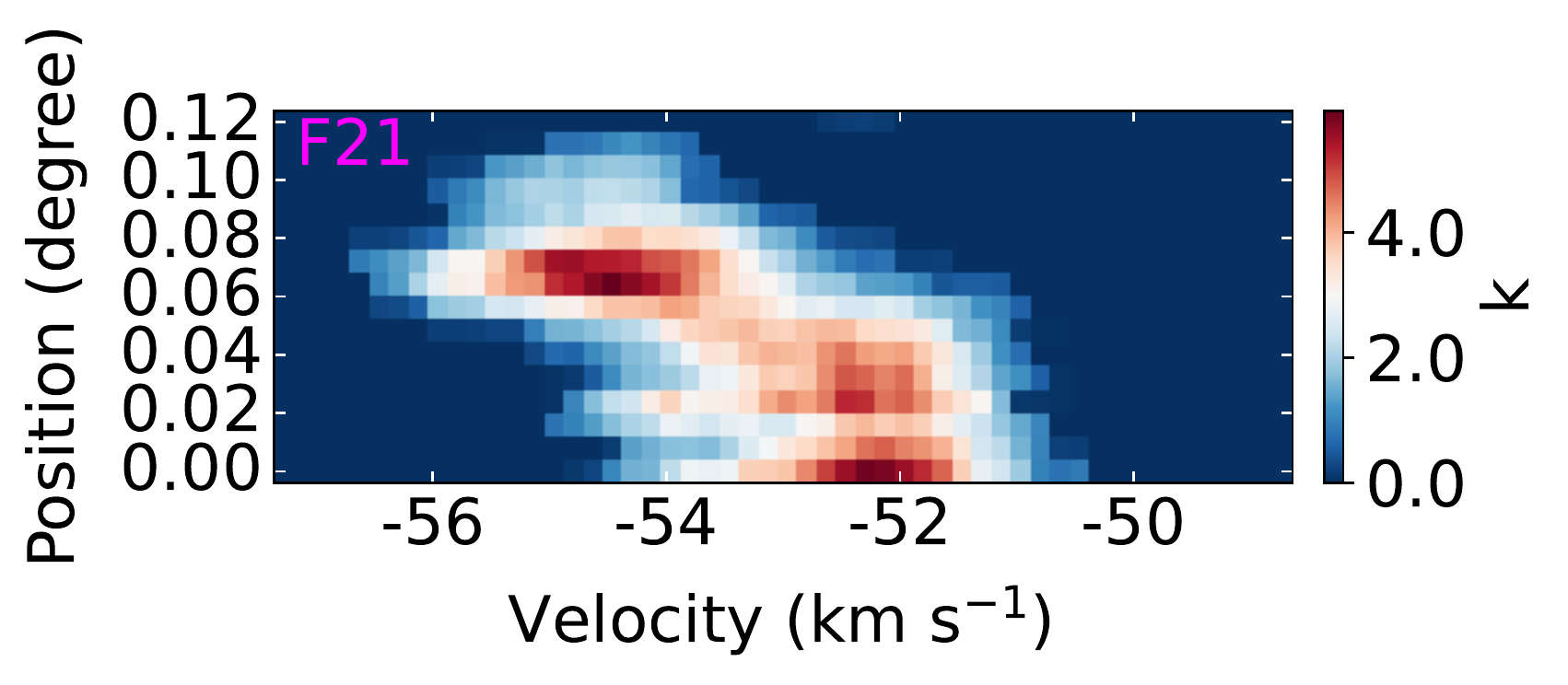}\hfill
   \includegraphics[width=.5\textwidth]{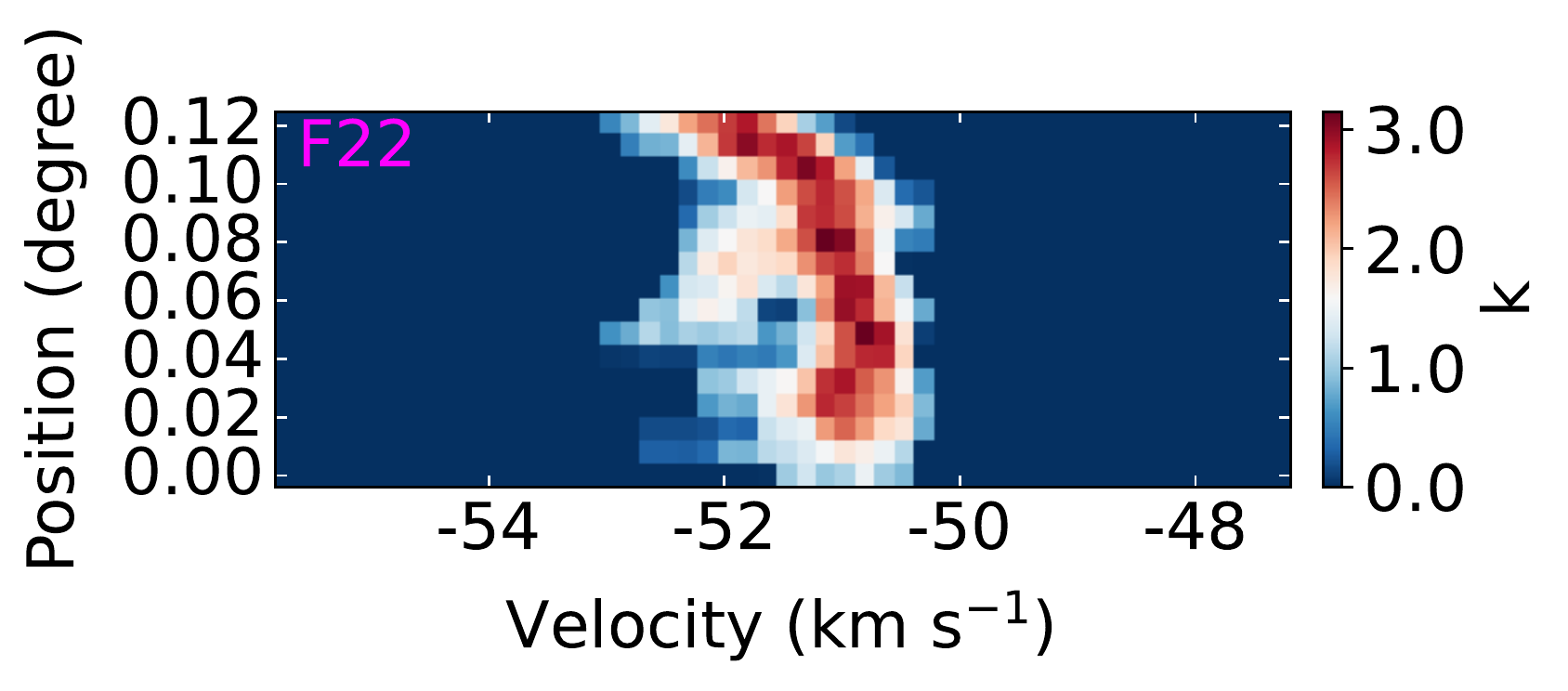}\hfill
   
   \includegraphics[width=.5\textwidth]{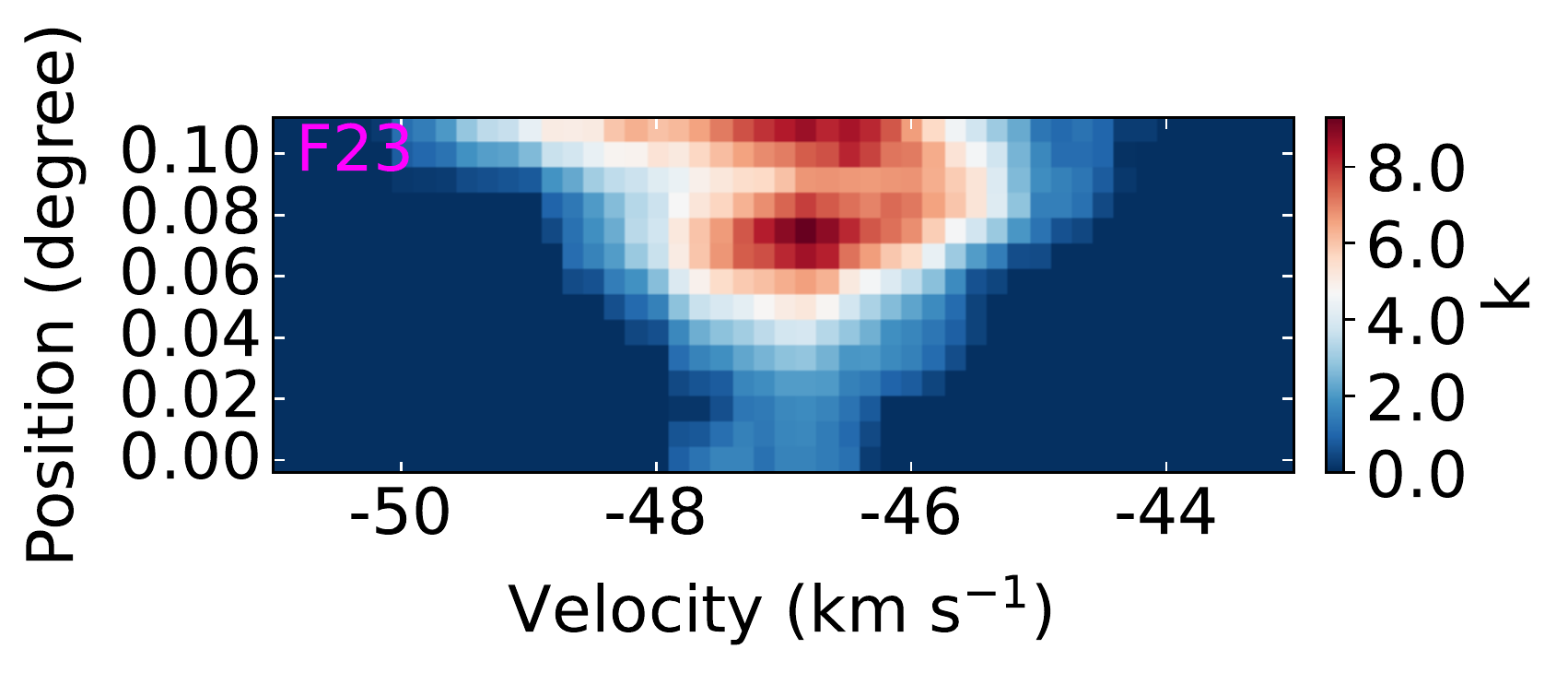}\hfill
   \includegraphics[width=.5\textwidth]{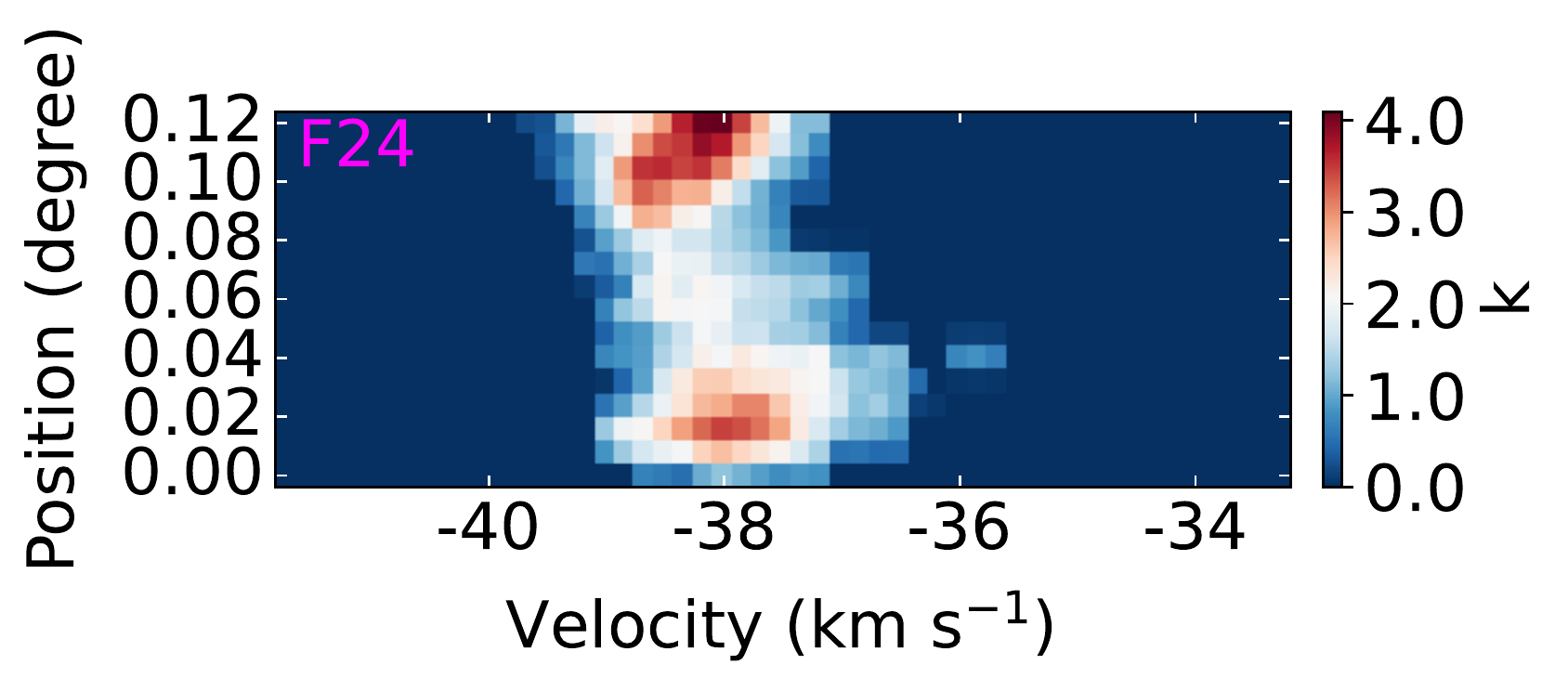}\hfill
   
   \includegraphics[width=.5\textwidth]{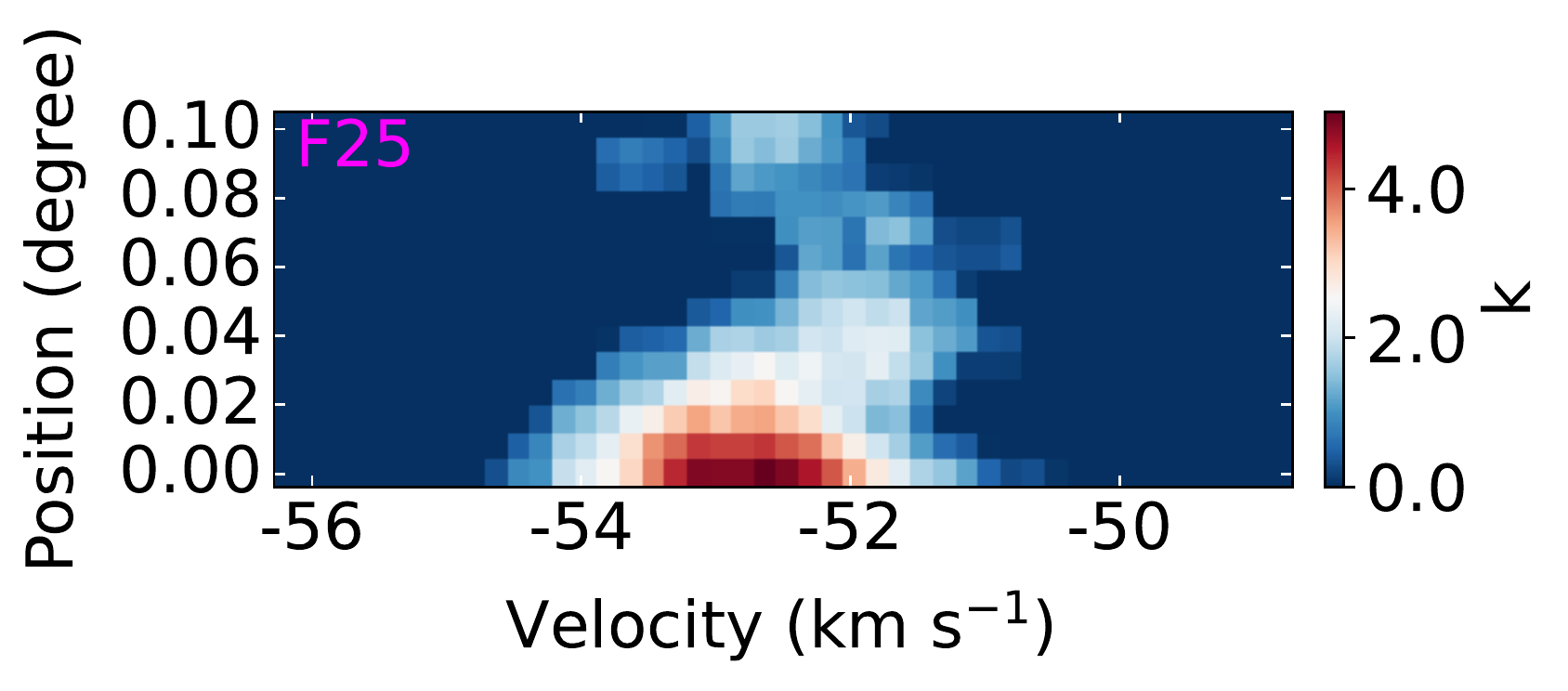}\hfill
   \\[\smallskipamount] 
   
\caption{ $\rm {}^{13}{CO}$ position-velocity diagrams of the 25 filaments shown in Figure~\ref{Fig6} (b), extracted along the solid arrowed lines with width of 1 pixel.}
\label{Fig7}
\end{figure}

Two threshold parameters, i.e., ``persistence'' and ``field${\_}$value'', are used in the DisPerSE algorithm to identify filaments. The persistence is generally defined as the column density difference of the critical points in a pair. The ``field${\_}$value'' is a threshold. The structures identified by the DisPerSE algorithm that have column density below this threshold are trimmed. The ``persistence'' and ``filed value'' in this work are set to be 5.83$\times$ 10$^{14}$ cm$^{-2}$ and 8.83$\times$ 10$^{14}$ cm$^{-2}$, which is seven and ten times the noise level of the $\rm {}^{13}{CO}$ column density, respectively. With the above two parameters, we identified 299 filaments, whose skeletons are shown in Figure~\ref{Fig6} (a). We can see from Figure~\ref{Fig6} (a) that most of the skeletons have  small aspect ratios and they look like molecular cores or clumps. Real filaments should be elongated structures with aspect ratios larger than 3. We made a rough estimation of the averaged width of the filaments in this region, which is $\sim$5 pixels, corresponding to $\sim$2.12 pc at the distance of 2.92 kpc. After the skeletons with lengths shorter than 15 pixels are excluded, 25 skeletons  remain and they are shown in Figure~\ref{Fig6} (b). We only study these 25 filaments in the following analysis.

\begin{table}
\bc
\begin{minipage}[]{100mm}
\caption[]{Parameters of the Filaments\label{tab2}}\end{minipage}
\setlength{\tabcolsep}{17pt}
\small
 \begin{tabular}{crrrrcc}
  \hline\noalign{\smallskip}
  (1)& (2)& (3)&(4)& (5)& (6)&(7)\\
Filament &  $T\rm _{ex}$ & Length & Mass & M/l &${\Delta}{\upsilon }$& $\rm{\alpha}_{vir}$  \\
 & $(\rm K)$  & ($\rm pc)$ & ($M_\odot$) & ($M_\odot~\rm pc^{-1}$)&($\rm km~s^{-1}$)&   \\
  \hline\noalign{\smallskip}
F1& 10.60 &  18.26 &  2441.70&  133.70& 1.03 &  3.81 \\
F2& 11.45 &  14.86 &  6536.53&  439.74&  0.96 &1.01  \\
F3& 13.91 &  16.99 &  9090.42&  535.11& 1.33 &1.58  \\
F4& 9.87  &  18.26 &  2204.32&  120.71&  0.84  &2.81 \\
F5& 9.57  &  12.32 &  1750.07&  142.09&  0.95  & 3.06 \\
F6& 14.30 & 10.62  &  5055.32&  476.14&  1.23  &1.53  \\
F7& 18.74 & 11.04  & 13990.77& 1267.04& 1.43   &0.78\\
F8& 11.33 & 12.74  & 717.88  & 56.34&  0.59     &3.16 \\
F9& 10.95 &  8.92  & 1351.12& 151.50&   1.17    &4.33\\
F10& 14.41 & 11.47 & 2502.31& 218.22&  1.36   &4.01\\
F11& 12.77 & 8.49  & 3370.26& 396.79&  1.36    &2.20\\
F12& 13.32 & 8.92  & 2405.66& 269.74&  0.88    &1.41\\
F13& 11.64 & 9.77 & 1028.09& 105.25&  0.82   &3.14\\
F14& 23.41 & 8.07 & 11092.26& 1374.64&  1.42  &0.71\\
F15& 13.08 & 8.50 & 1088.09& 128.10&  1.34    &6.69\\
F16& 10.51 & 8.07 & 1179.06& 146.12&  1.30    &5.45\\
F17& 11.40 & 6.12 & 464.72& 75.93&  0.94      &5.67\\
F18& 8.58  & 6.80 & 433.41& 63.78&     0.66    &3.42\\
F19& 9.63  & 7.64 & 431.17& 56.40&    0.57     &2.95\\
F20& 14.25 & 7.22 & 2238.29& 351.57& 1.12    &1.71\\
F21& 20.32 & 6.80 & 2259.42& 332.51&  1.50    &3.22\\
F22& 10.12 & 6.80 & 510.71& 75.16&     0.83   &4.43\\
F23& 16.67 & 6.37 & 2346.24& 368.30&   1.01   &1.36\\
F24& 9.25  & 6.50 & 450.06& 69.21&     0.73   &3.79\\
F25& 10.89 & 6.37 & 642.43& 100.85&  0.95     &4.30\\
  \noalign{\smallskip}\hline
\end{tabular}
\ec
\tablecomments{0.95\textwidth}{Properties of the identified filaments. The excitation temperature, length, mass, line mass, line width, and virial parameter are given in Columns 2 ${-}$ 7, respectively. }
\end{table}

The names of the filaments are composed of ``F'' and a number from ``1'' to ``25''. The position-velocity diagrams of the filaments are shown in Figure~\ref{Fig7}. The position-velocity diagrams are extracted along the spines of the filaments which are marked with the arrowed lines in Figure~\ref{Fig6} (b). As shown in Figure~\ref{Fig7}, all filaments are velocity coherent structures. The majority of the filaments show velocity broadening.   
The velocity broadening and gradient in the position-velocity diagrams possibly arise from the feedback of {H \small {II}} regions. F12 in GMC 1 is associated with the {H \small {II}} region G108.603 ${-}$ 00.494 (for details see section \ref{subsect:Giant Molecular Cloud Complex Associated with the HII Regions/Candidates}). We find a velocity broadening of about 4 $\rm km~s^{-1}$ in the p-v diagram of F12 in Figure~\ref{Fig7}. Similarly, F3 and F23 in GMC 2 are located in the vicinity of the {H \small {II}} regions G109.104 ${-}$ 00.347 and G109.068 ${-}$ 00.322, and they also show velocity broadenings of ${~}$ $\sim$5 $\rm km~s^{-1}$. There may be an interaction between these {H \small {II}} regions and filaments. For F6, F7, and F14 in GMC 3, which are associated with the {H \small {II}} region cluster G108.764 ${-}$ 00.952 (for details, see section \ref{subsect:Giant Molecular Cloud Complex Associated with the HII Regions/Candidates}), only F7 has an obvious velocity broadening ($\sim$ 8 $\rm km~s^{-1}$ ). F13 in GMC 3 is surrounded by CTB 109 and its position-velocity shows a velocity broadening ($\sim$ 3 $\rm km~s^{-1}$). This possibly implies that the F13 is associated with CTB 109. F21 in GMC 3 is possibly interacting with the S149 {H \small {II}} region, as shown by a velocity gradient of 3 $\rm km~s^{-1}$. The S142 {H \small {II}} region is spatially coincident with two filaments, F17 and F24 in GMC 4. We can see a velocity gradient of $\sim$ 6 $\rm km~s^{-1}$ in F17. Therefore, we suggest that F17 is physically associated with the S142 {H \small {II}} region.   \par

For each filament, we calculate the H$_{2}$ column densities within its boundary which is carefully selected so that most of the $\rm {}^{13}{CO}$ emission of the filament is included. The calculation of the mass of the filaments is with the LTE method as we did in Section~\ref{subsect:Overview of Molecular Cloud in the Region}. We also calculate the lengths of the filaments and their line mass (mass per unit length). The virial mass of the filament is calculated according to \citep{2000MNRAS.311...85F}
\begin{equation}
\rm {M}_{line, vir} = \frac{2 \rm {\sigma}_{tot}^2}{G} ,
\end{equation}
where $\rm {\sigma}_{tot}$ is the total velocity dispersion, which consists of the non-thermal velocity dispersion ($\rm {\sigma}_{NT}$ = $\sqrt{\rm {\sigma}_{obs}^{2} - \rm {\sigma}_{T({obs})}^{2}}$) and thermal velocity dispersion $\rm {\sigma}_{T}$ = $\sqrt{\rm {k}_{B}T / \mu{m}_{H}}$:

\begin{equation}
{\rm {\sigma}_{tot}} = \sqrt{\rm {\sigma}_{NT}^{2} + \rm {\sigma}_{T}^{2}} = \sqrt{\rm {\sigma}_{obs}^{2} - \rm {\sigma}_{T(obs)}^{2} + \frac{\rm {k}_{B}T}{\mu{m}_{H}} } ,
\end{equation}

where $\rm \sigma_{obs}$ = $\rm \Delta{\upsilon}/\sqrt{8\ln2}$, $\rm {\sigma}_{T(obs)}$ = $\sqrt{\rm {k}_{B}T / \mu_{13}{m}_{H}}$, $\rm \Delta{\upsilon}$ is the $\rm {}^{13}{CO}$ line width, $\mu$ is the mean weight of molecular hydrogen and is taken to be 2.37 \citep{2008A&A...487..993K}, $\rm {k}_{B}$ is the Boltzmann constant, ${\rm \mu_{13}}$ is the atomic weight of the $\rm {}^{13}{CO}$ molecule and its value is 29, and T is approximately equal to the excitation temperature T$\rm _{ex}$. The virial parameter of filaments is given by:

\begin{equation}
{\rm {\alpha}_{vir}} = \frac{\rm {M}_{line, vir}}{M/l} .
\end{equation}

All the parameters derived above, including the excitation temperature, length, mass, line mass, line width,mand virial parameter are listed in Table~\ref{tab2}. The median excitation temperature, length, line mass, line width, and virial parameter of the filaments  are  10.89 K, 8.49 pc, 146.12 $\rm {M}_{\odot}~ \rm pc^{-1}$, 1.01 $\rm km~s^{-1}$, and 3.14, respectively.
\label{subsect:Filamentary Contents of the Molecular Clouds}

\subsection{Molecular Clouds Associated with {H \small {II}} Regions/Candidates}

   \begin{figure}[h!]
    \centering
     \begin{subfigure}[b]{0.4\linewidth}
     \centering
    \includegraphics[trim=13cm 0cm 13cm 0cm, width=.4\textwidth]{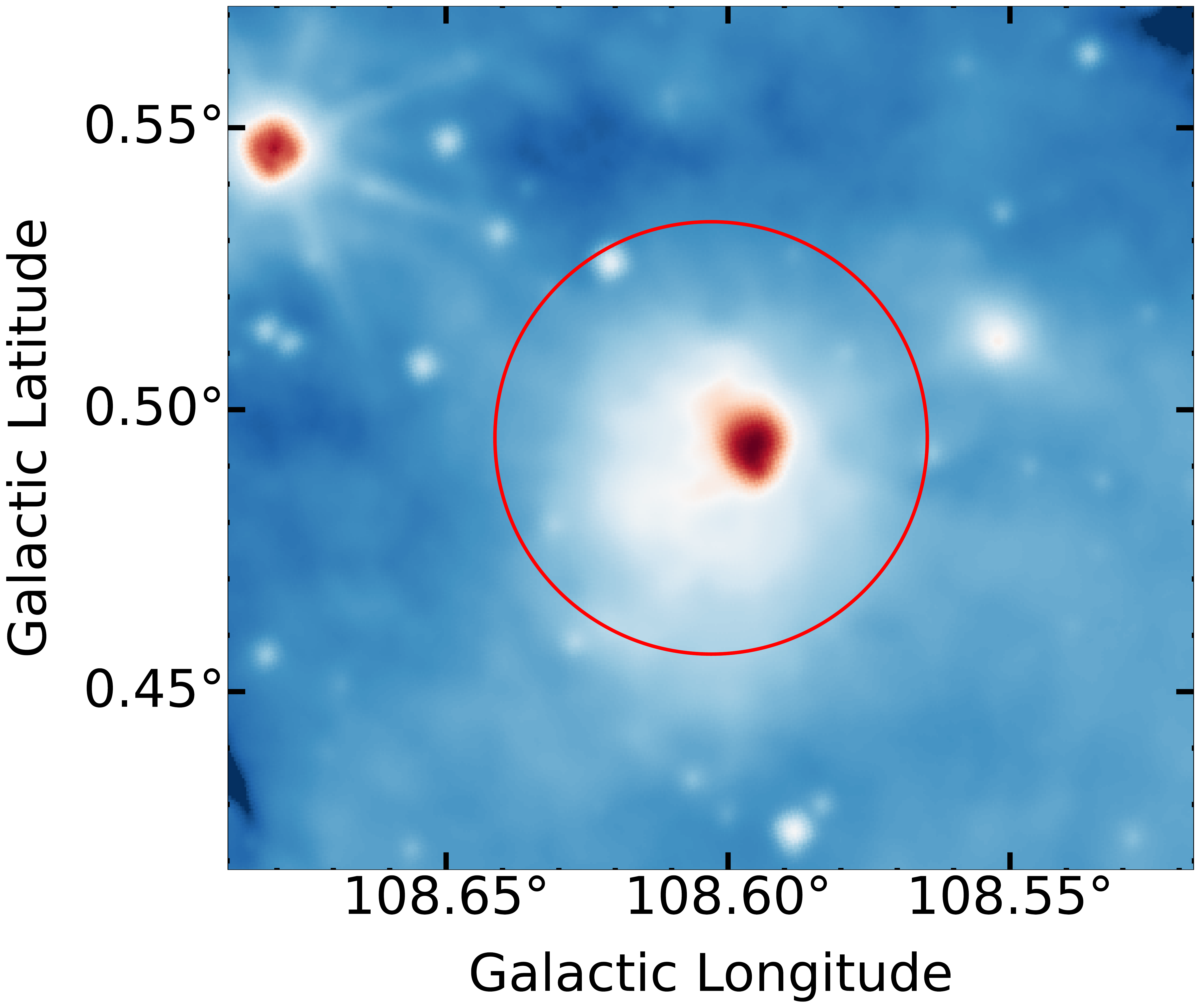}
    \\(a)
   \end{subfigure}
   \begin{subfigure}[b]{0.4\linewidth}
   \centering
    \includegraphics[trim=13cm 0cm 13cm 0cm,width=.4\textwidth]{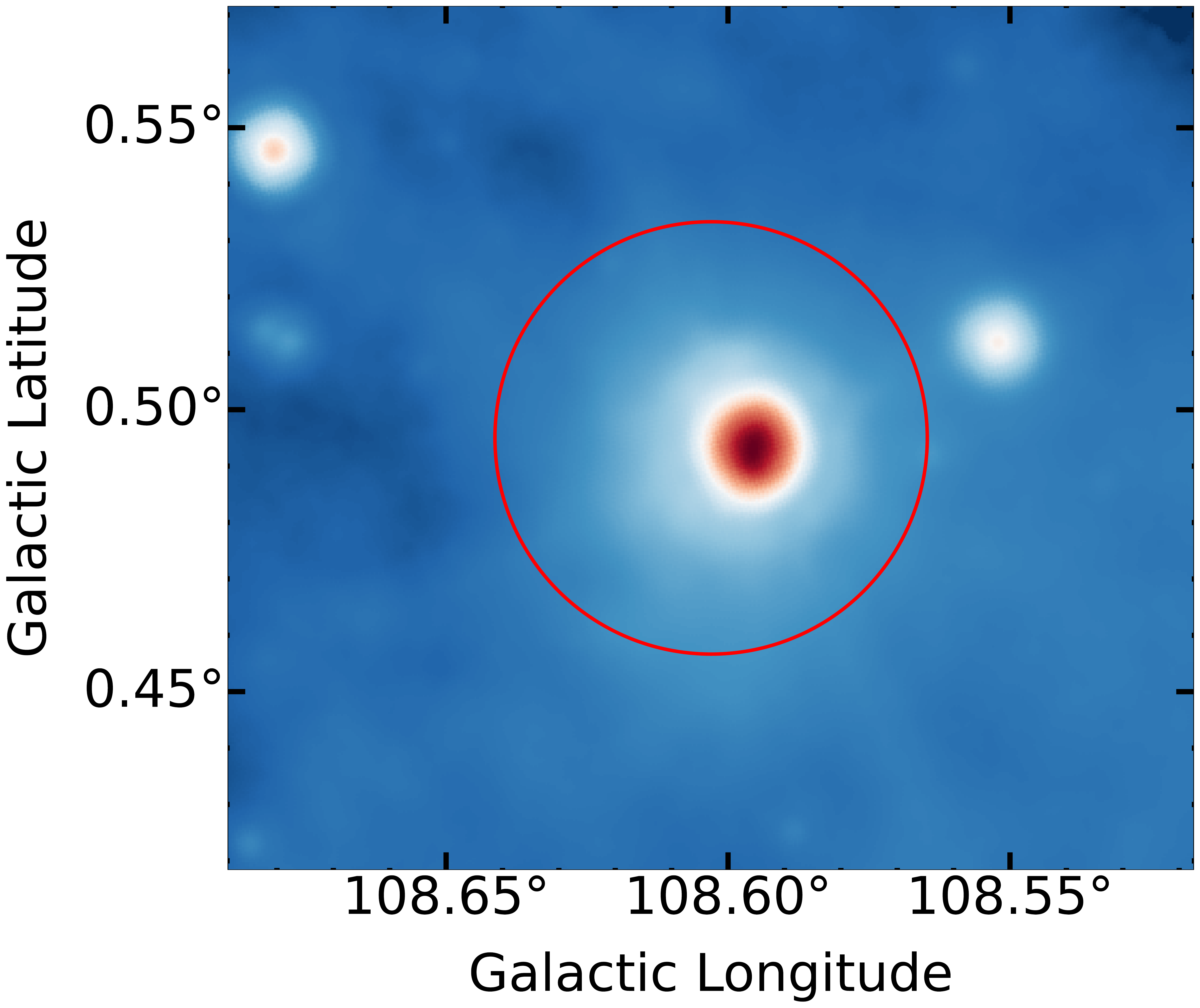}
     \\(b)
    \end{subfigure}
   \begin{subfigure}[b]{0.4\linewidth}
   \centering
    \includegraphics[trim=13cm 0cm 13cm 0cm,width=.4\linewidth]{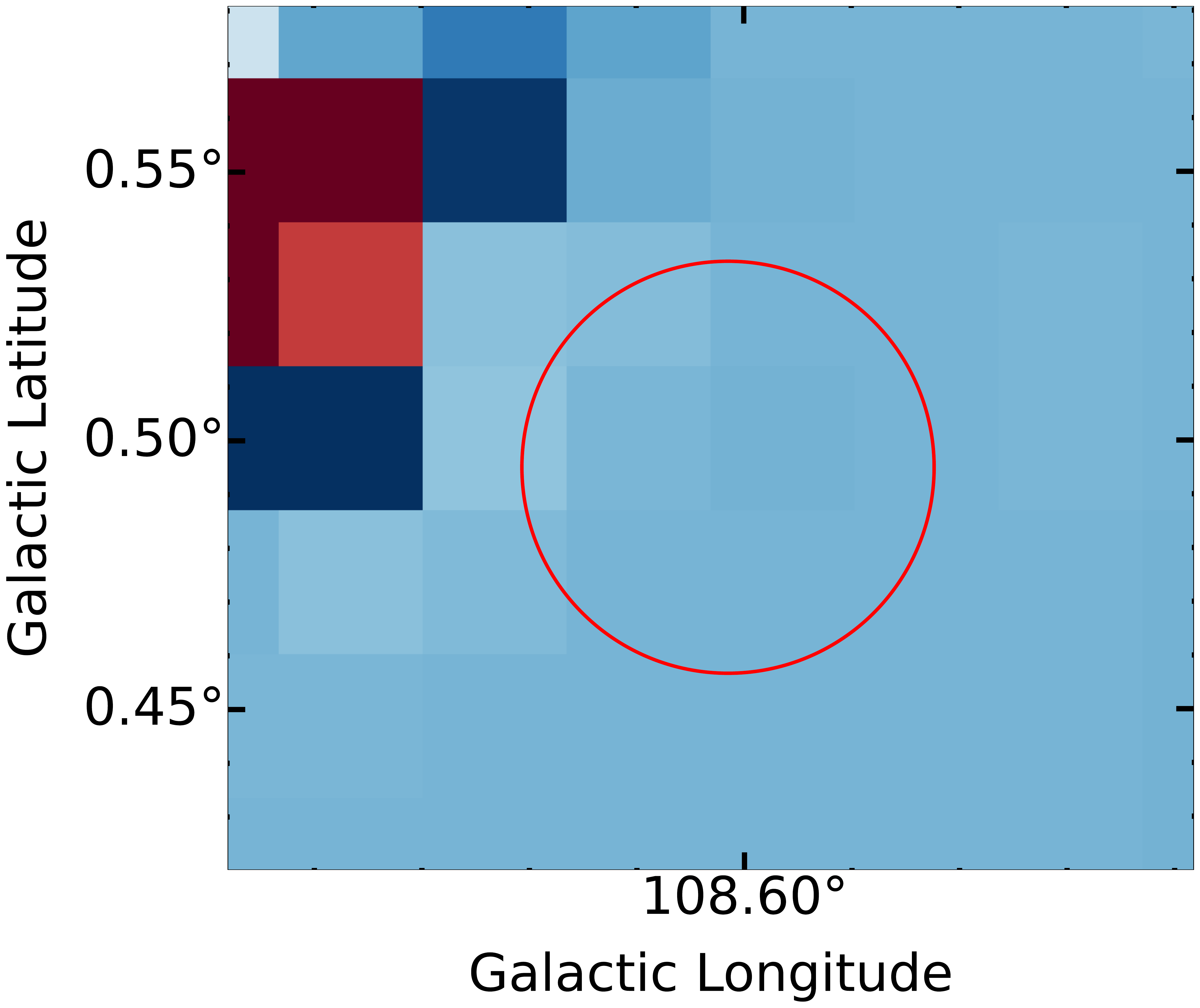}
     \\(c)
    \end{subfigure}
    \begin{subfigure}[b]{0.4\linewidth}
   \centering
    \includegraphics[trim=14cm 0cm 21cm 0cm,width=.4\textwidth]{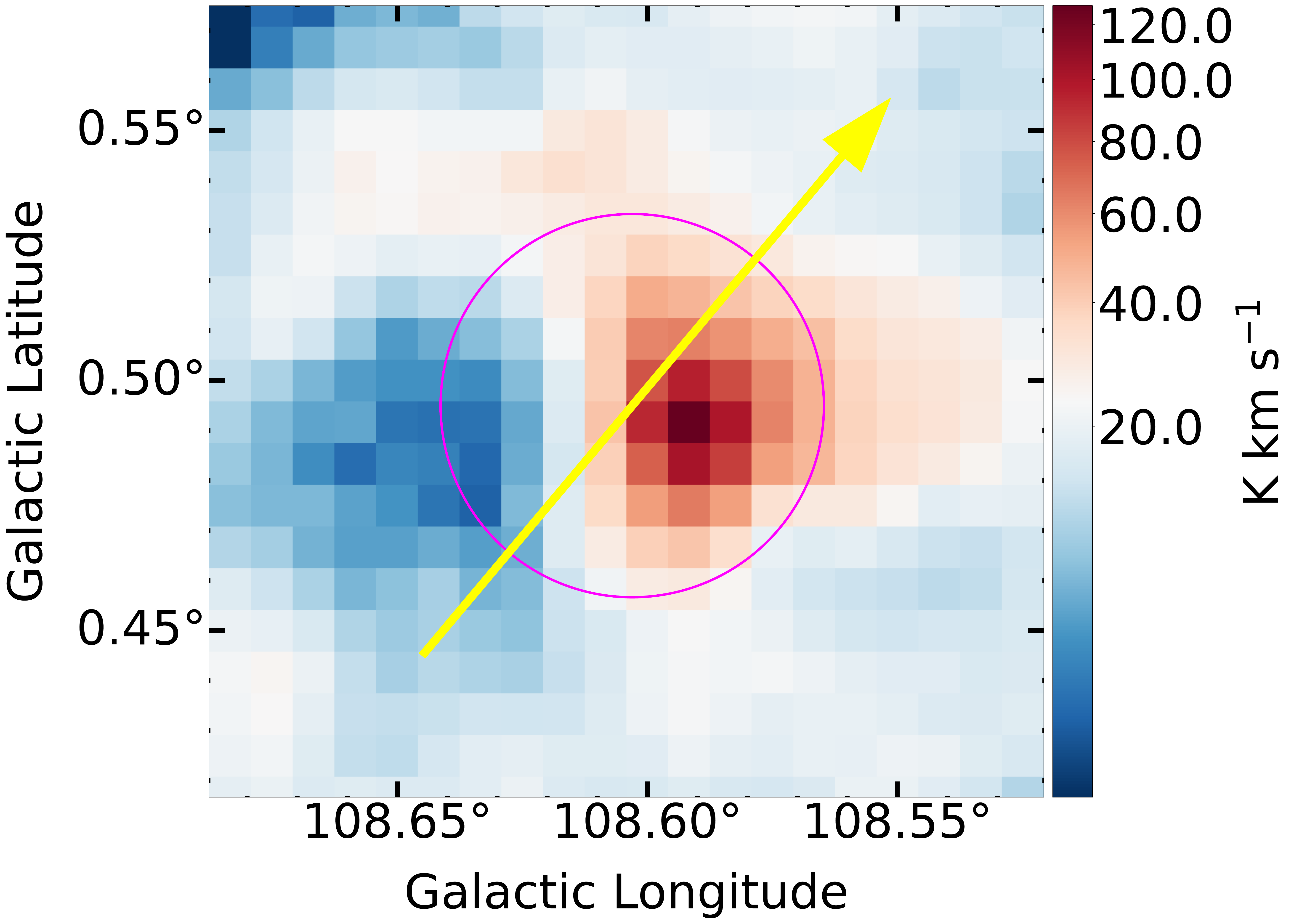}\hfill
    \\(d)
    \end{subfigure}
    \begin{subfigure}[b]{0.4\linewidth}
   \centering
    \includegraphics[trim=13.5cm 0cm 21.5cm 0cm,width=.4\textwidth]{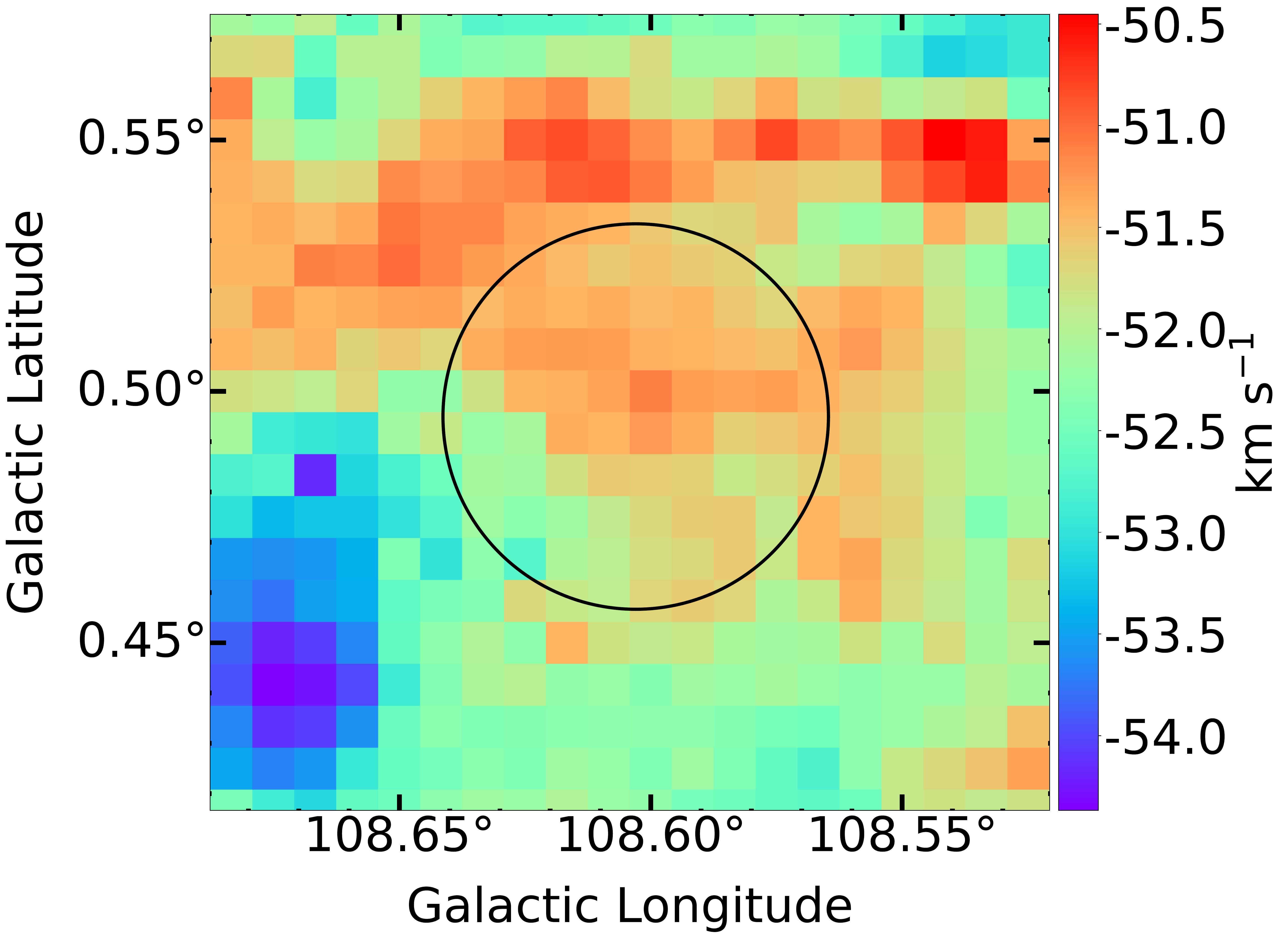}\hfill
    \\(e)
    \end{subfigure}
    \begin{subfigure}[b]{0.4\linewidth}
   \centering
    \includegraphics[trim=9cm 0cm 24cm 0.7cm,width=.4\textwidth]{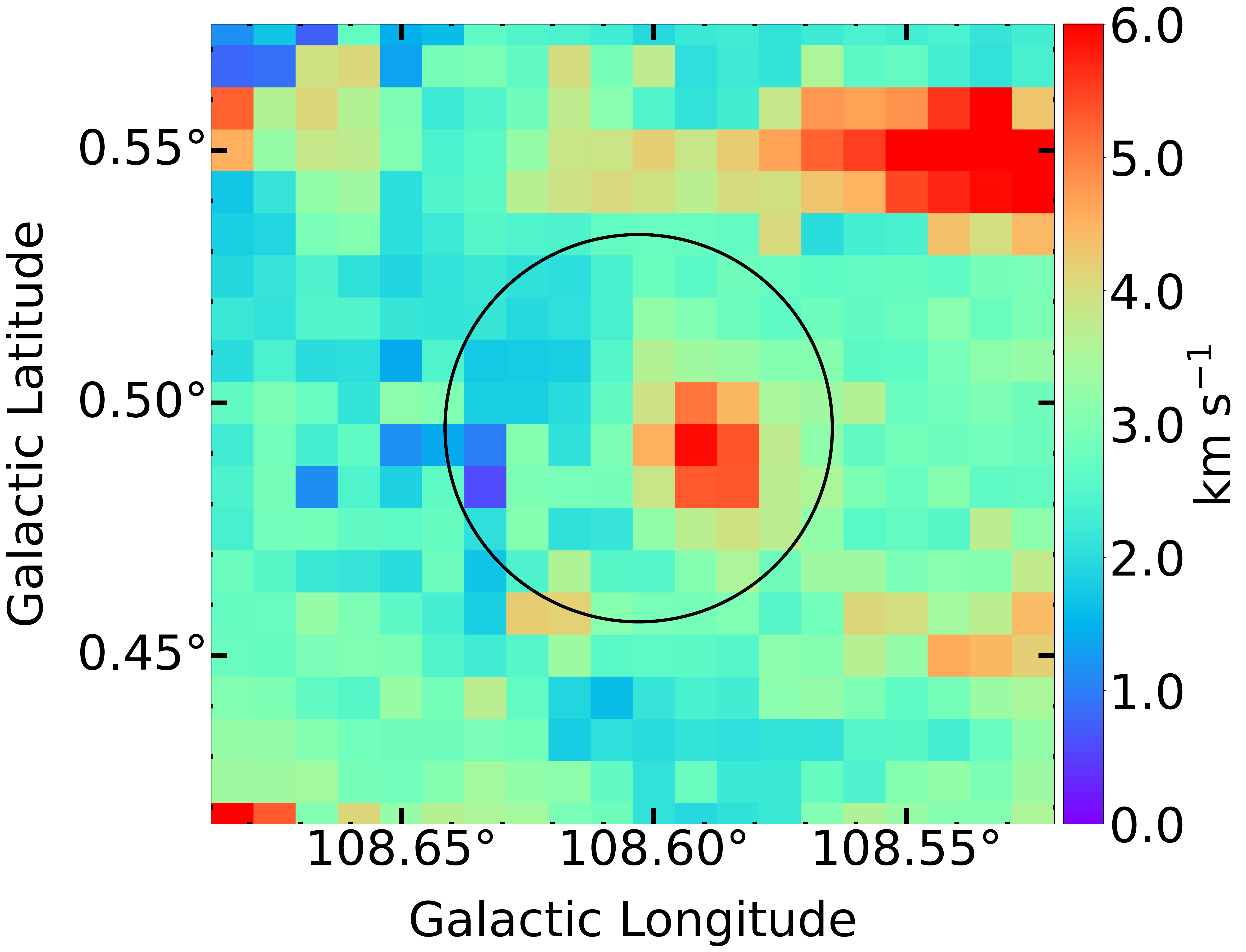}\hfill
    \\(f)
    \end{subfigure}
    \begin{subfigure}[b]{0.4\linewidth}
   \centering
    \includegraphics[trim=13.5cm 0cm 21cm 0cm, width=.4\textwidth]{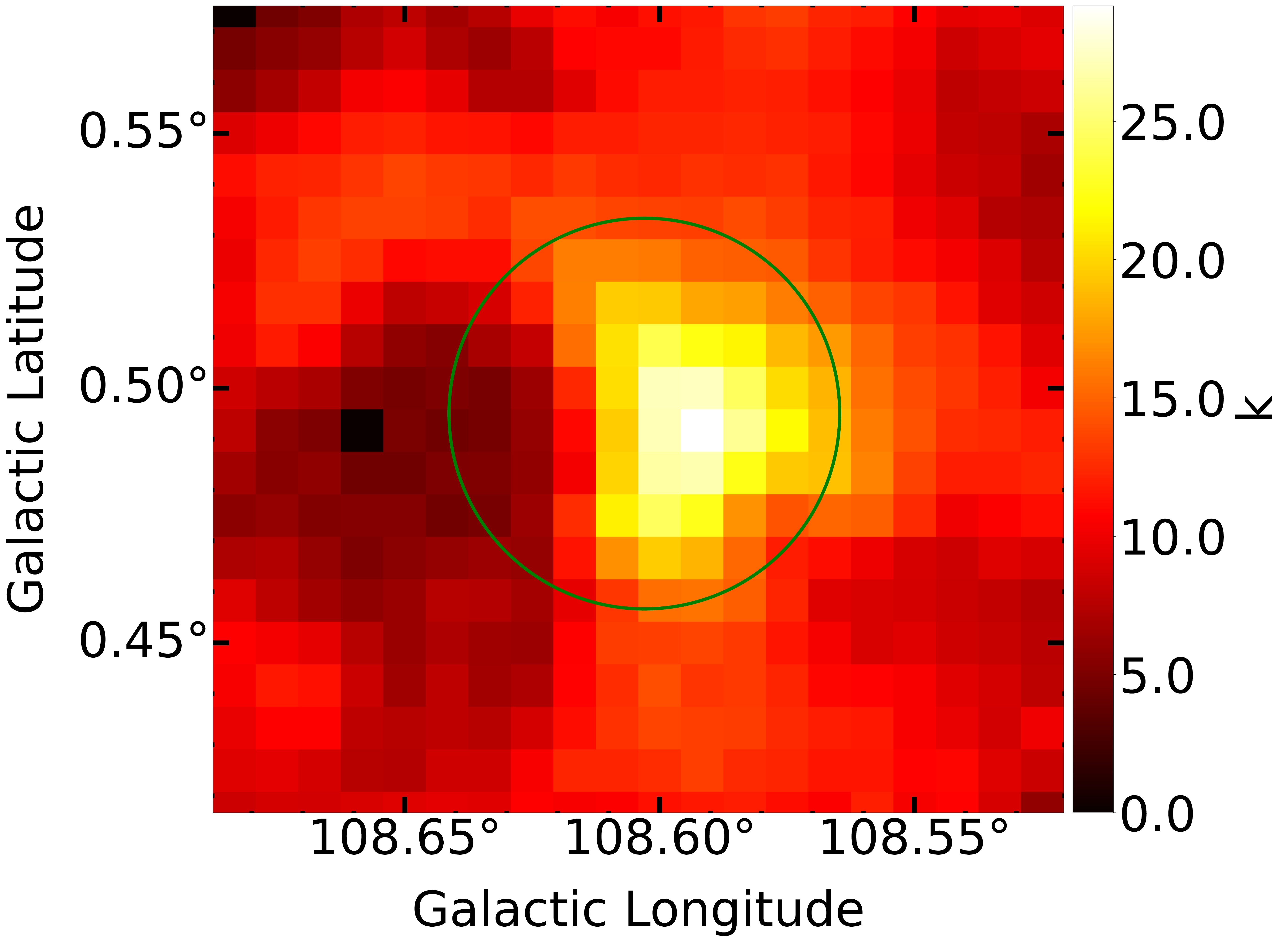}\hfill
    \\(g)
    \end{subfigure}
    \begin{subfigure}[b]{0.4\linewidth}
   \centering
    \includegraphics[trim=1cm -2cm 15cm 0cm, width=.4\textwidth]{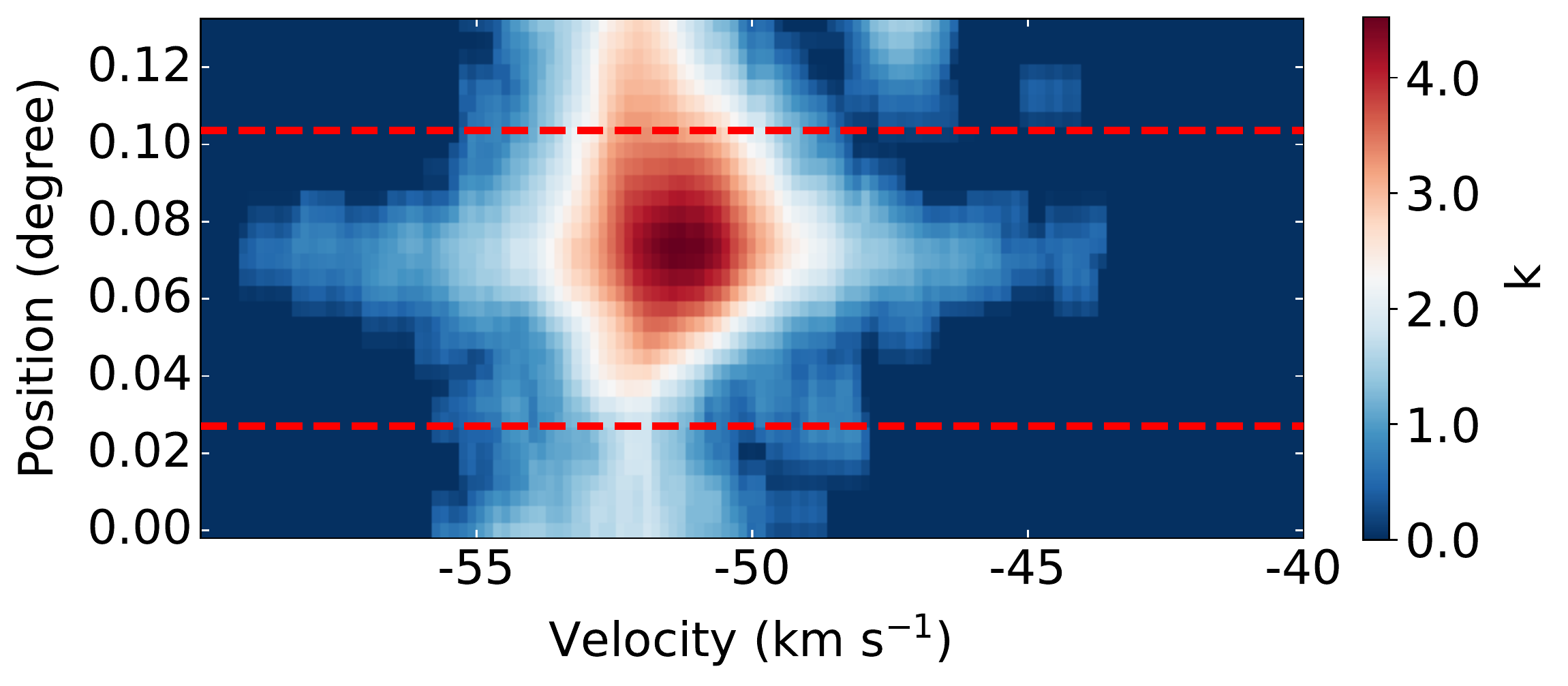}\hfill
    \\(h)
    \end{subfigure}
    \\[\smallskipamount]
    \caption{
    The WISE 3 (12$\mu$m), WISE 4 (22$\mu$m), and H$\alpha$ emission images of the {H \small {II}} region G108.603 + 00.494 from (a) to (c). The H$\alpha$ image is obtained from the archive data of the VTSS survey (http://www1.phys.vt.edu/~halpha/). The zeroth, the first, the second moment maps, and the excitation temperature maps of the $^{12}$CO emission from (d) to (e). Panel (h) shows the position-velocity map of $\rm {}^{12}{CO}$ extracted along the arrow marked in panel (d). The horizontal dashed lines indicate the position range of the {H \small {II}} region.  All panels from (a) to (g) cover the same sky area.}  
    \label{Fig8}
    
\end{figure}

In this section, we investigate the possible physical association between the molecular clouds and the {H \small {II}} regions/candidates in the surveyed region.  \cite{2014ApJS..212....1A} summarized the spatial relations between these {H \small {II}} regions/candidates and molecular line emission in their Table 5. \cite{2015AJ....150..147F} conducted detailed analysis on the distances and velocities of {H \small {II}} regions in the II and III quadrants of the Milky Way and also examined the spatial correlation between {H \small {II}} regions and $^{12}$CO emission. In this work, in addition to spatial correlation, we also use signatures of interaction, including raised temperature, broadened line width, and large velocity gradient of the molecular line emission to examine the relations between {H \small {II}} regions/candidates and molecular clouds. The relationship between {H \small {II}} regions/candidates and molecular clouds in the region is examined in the following four aspects, \par
 1. Spatial correlation between the morphology of the {H \small {II}} regions/candidates at the WISE 3 / 4 bands and that of $^{12}$CO line emission. The WISE 3 and 4 bands cover the polycyclic aromatic hydrocarbon (PAH) emission, which can trace the periphery of {H \small {II}} regions, and the emission from warm dust heated by star-formation activities. If the morphology of the {H \small {II}} regions/candidates in the WISE 3 / 4 band images are similar to that in the $^{12}$CO integrated intensity map, we consider the {H \small {II}} regions/candidates to be spatially associated with molecular clouds. \par
 2. Consistency between the velocities of the radio recombination line of the {H \small {II}} regions/candidates and that of the $^{12}$CO emission line. Similar to \cite{2014ApJS..212....1A}, we adopt the criterion that the velocity difference between RRL and the $^{12}$CO emission line should less than 10 km s$^{-1}$. \par
 3. Raised excitation temperature of the molecular gas in the vicinity of the {H \small {II}} regions/candidates. As predicted by thermal equilibrium models, physical processes like photoionization can heat up molecular clouds to kinetic temperatures of about 20 K. Therefore, the critical temperature, T$_{\rm k}$ = 20 K, has been used as a criterion in the observational study of the energetics of the molecular clouds around S235 \citep{1981ApJ...246..394E}. Therefore, we choose T$_{\rm k}$ $\geq$ 20 K as the heating criterion. \par
 4. Line broadenings of the $^{12}$CO emission of the molecular clouds near {H \small {II}} regions/candidates. Line broadenings above 6 km s$^{-1}$ are considered as a signature of physical association between {H \small {II}} regions/candidates and molecular clouds. \cite{2014ApJ...796..144K} used line broadenings larger than 6 km s$^{-1}$ as a criterion to search for the interactions between molecular clouds and supernova remnants. SNRs are a more powerful source than {H \small {II}} regions to influence the kinematics of molecular gas. Therefore, 6 km s$^{-1}$ as a criterion for velocity broadening in this work should be reasonable.

  \par
We ``scored'' the  relation between {H \small {II}} regions/candidates and molecular clouds according to the above four criteria. The first and second criteria each correspond to 1 point, while the third and fourth criteria each correspond to 0.5 points. A ``score'' equal to or higher than 2 means that the {H \small {II}} regions/candidates are physically associated with molecular clouds. The results are summarized in Table~\ref{tab3}. Among the 19 {H \small {II}} regions/candidates, twelve are physically associated with molecular clouds. \par

\begin{figure}[hb!]
    \centering
     \begin{subfigure}[b]{.4\linewidth}
     \centering
    \includegraphics[trim=14cm 0cm 14cm 0cm, width=.4\textwidth]{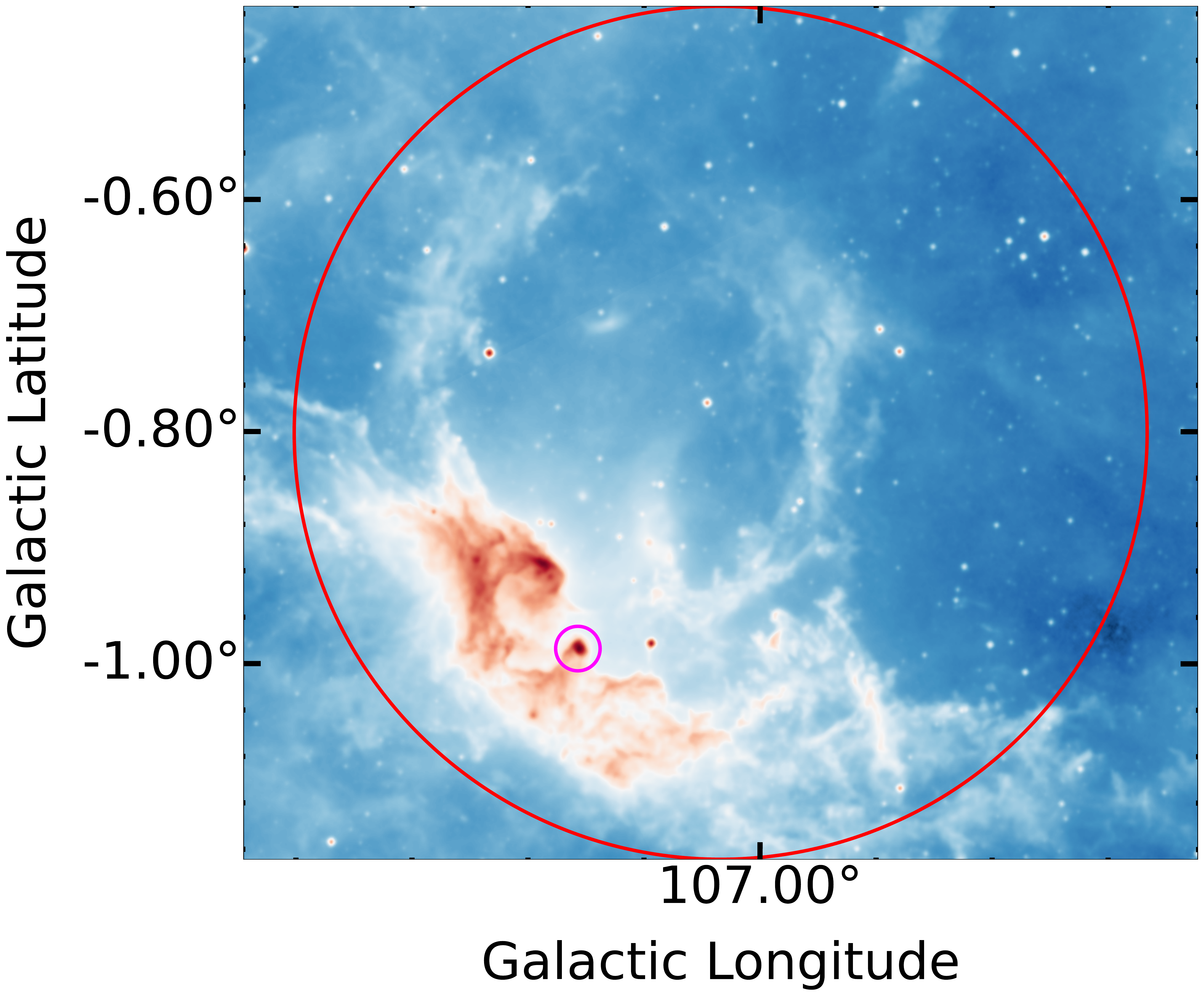}
    \\(a)
   \end{subfigure}
   \begin{subfigure}[b]{.4\linewidth}
   \centering
    \includegraphics[trim=14cm 0cm 14cm 0cm,width=.4\textwidth]{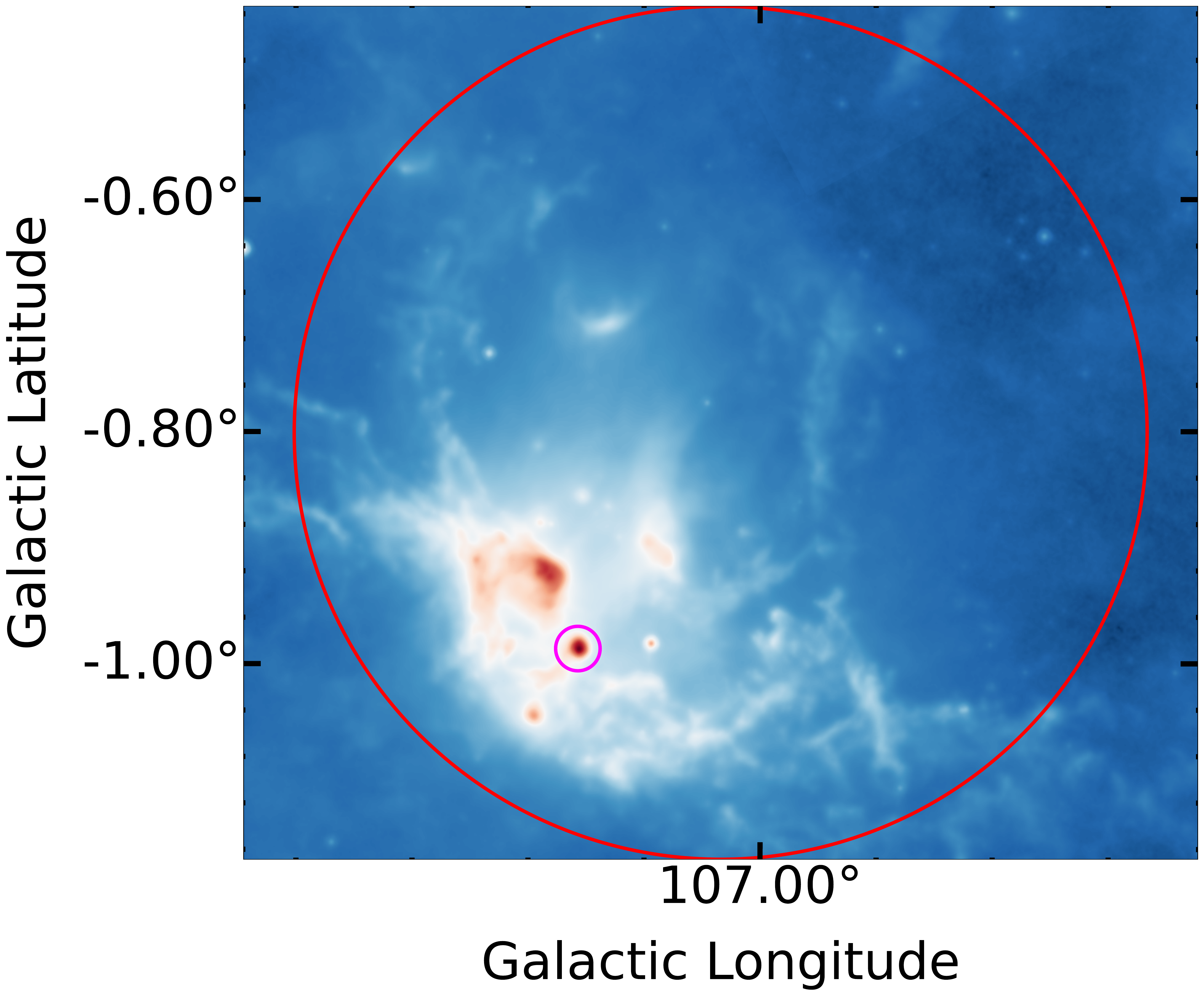}
     \\(b)
    \end{subfigure}
   \begin{subfigure}[b]{0.4\linewidth}
   \centering
    \includegraphics[trim=14cm 0cm 14cm 0cm,width=.4\textwidth]{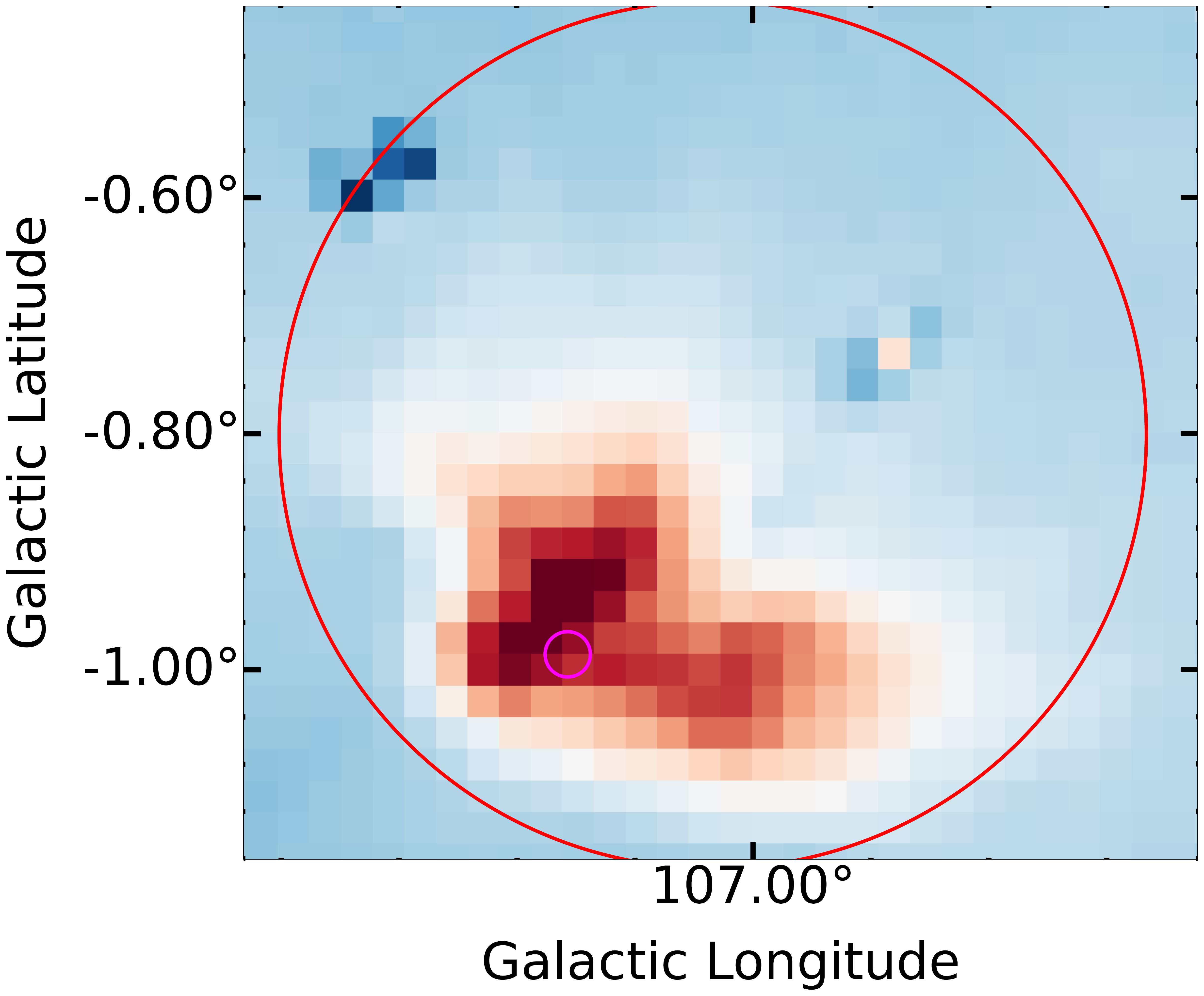}
     \\(c)
    \end{subfigure}
    \begin{subfigure}[b]{0.4\linewidth}
   \centering
    \includegraphics[trim=14cm 0cm 21.5cm 0cm,width=.4\textwidth]{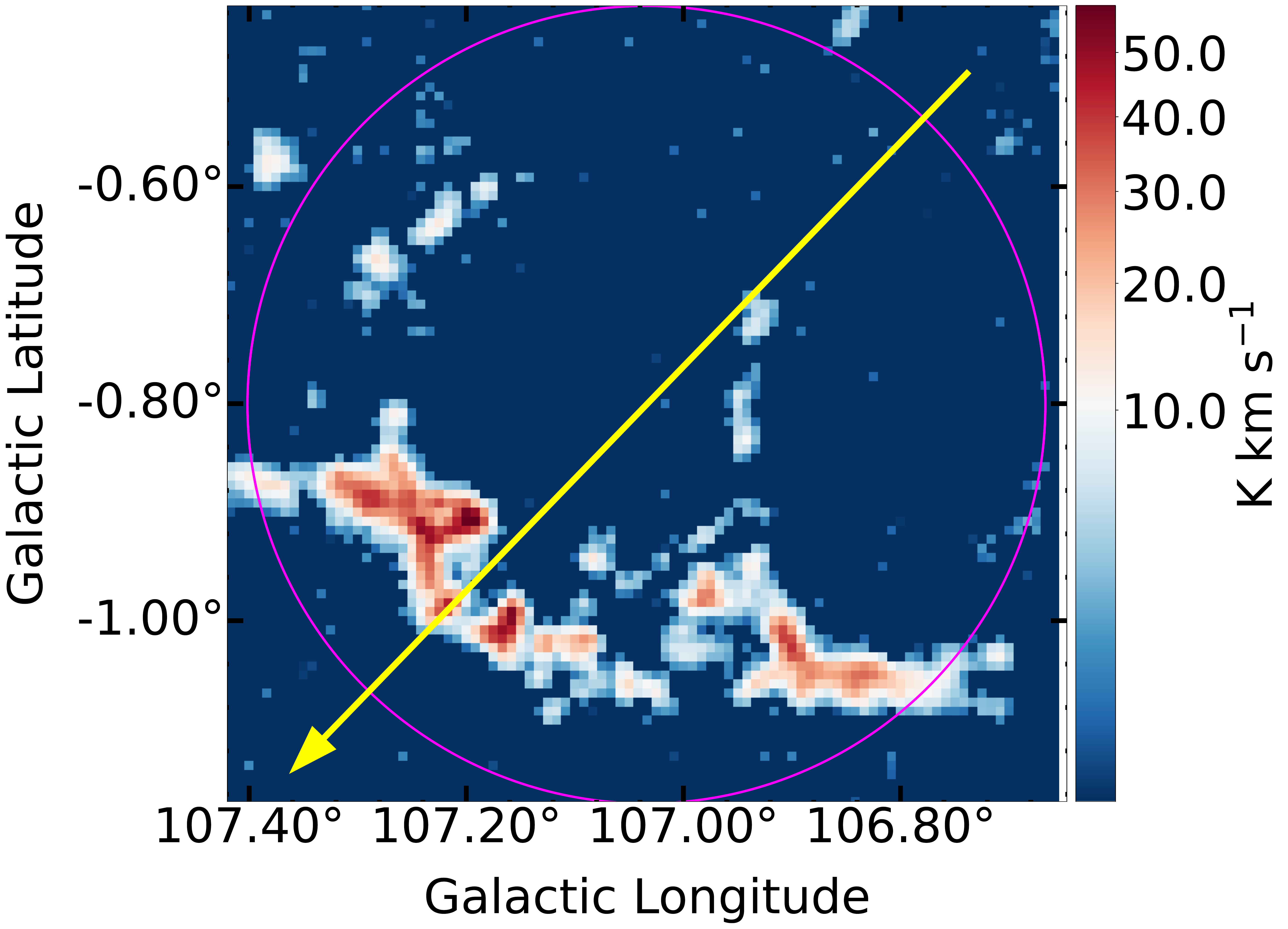}\hfill
    \\(d)
    \end{subfigure}
    \begin{subfigure}[b]{0.4\linewidth}
   \centering
    \includegraphics[trim=13.5cm 0cm 23cm 0cm,width=.4\textwidth]{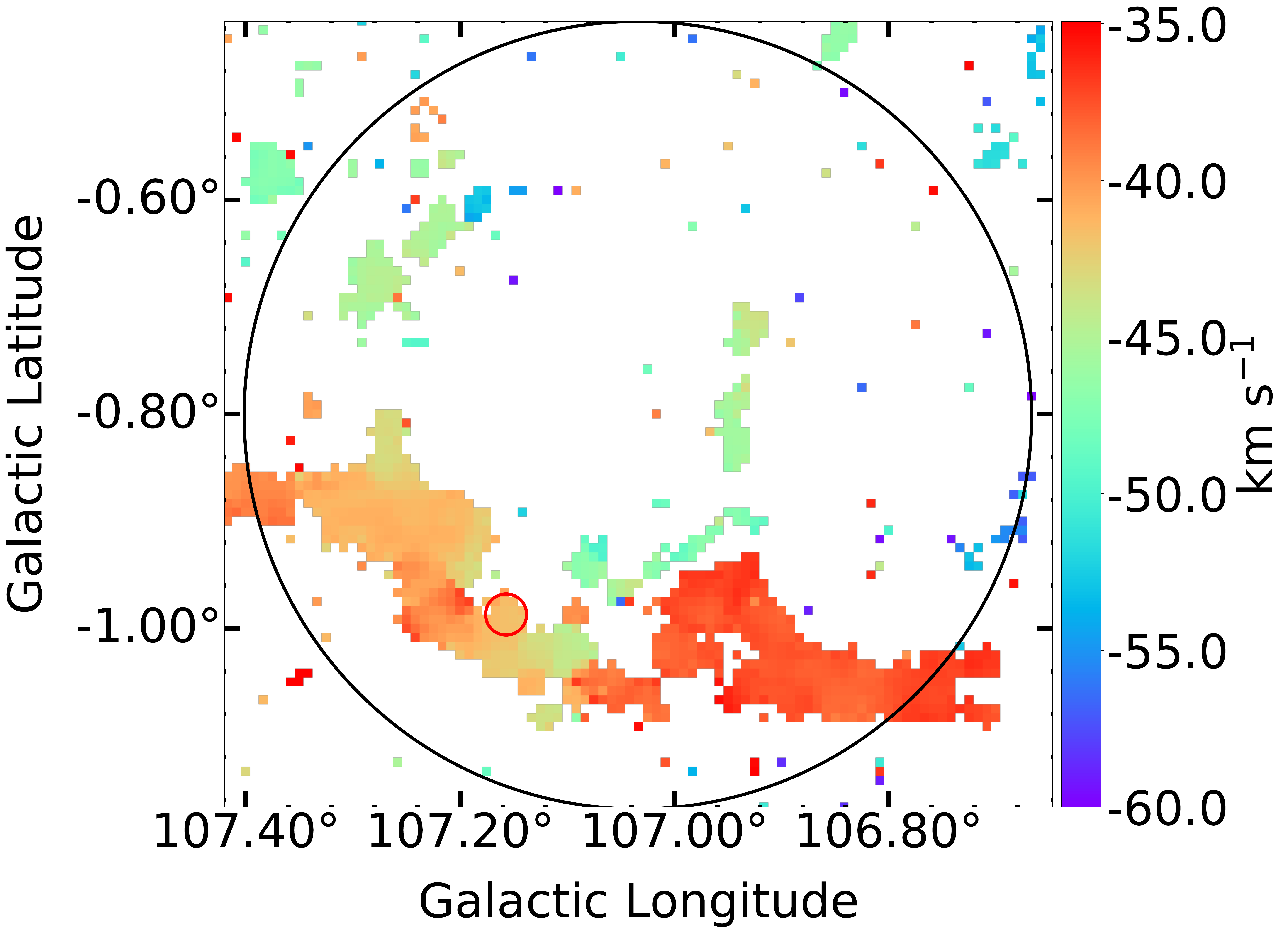}\hfill
    \\(e)
    \end{subfigure}
    \begin{subfigure}[b]{0.4\linewidth}
   \centering
    \includegraphics[trim=9cm 0cm 25cm 0.7cm,width=.4\textwidth]{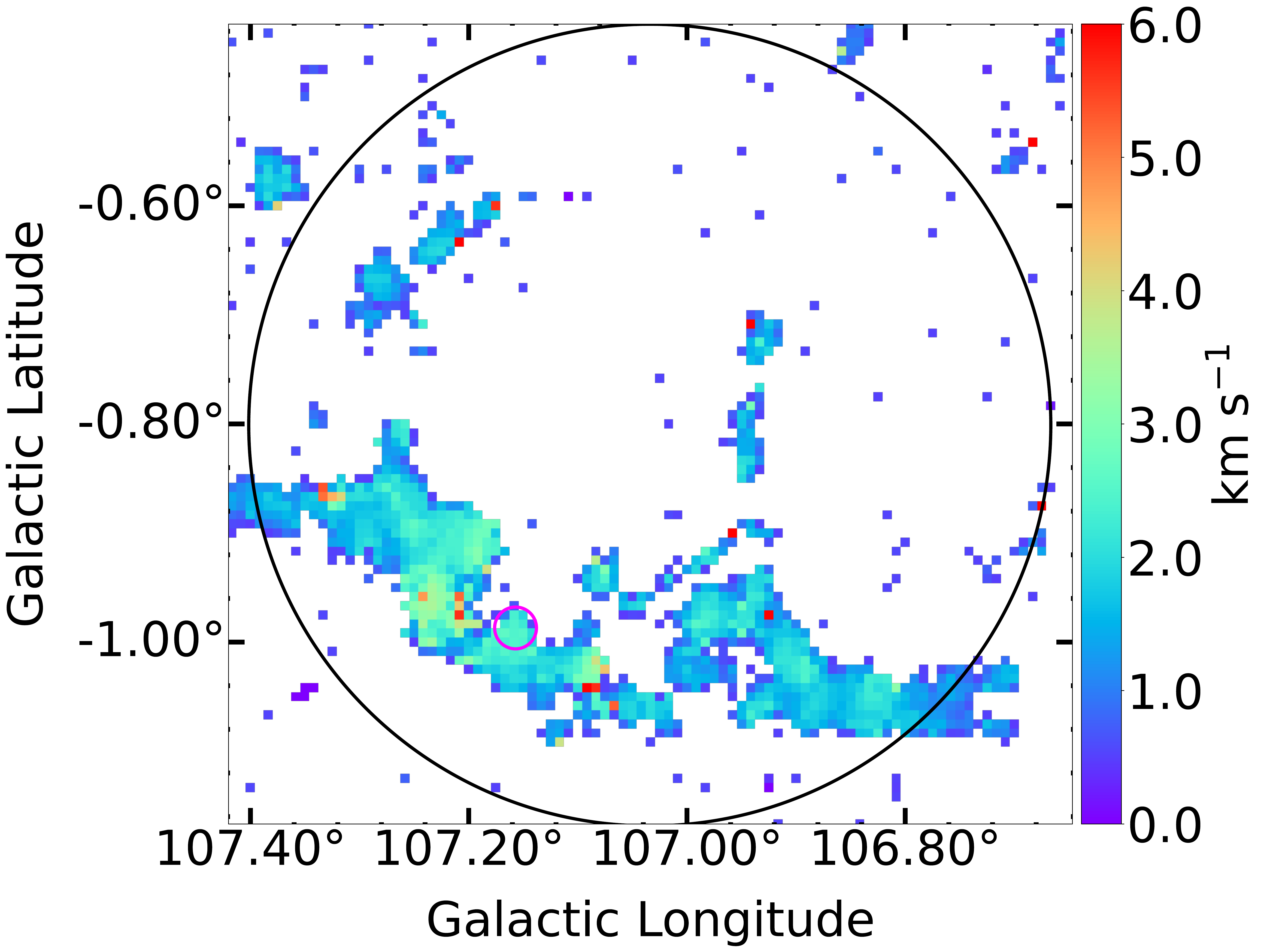}\hfill
    \\(f)
    \end{subfigure}
    \begin{subfigure}[b]{0.4\linewidth}
   \centering
    \includegraphics[trim=14cm 0cm 22cm 0cm, width=.4\textwidth]{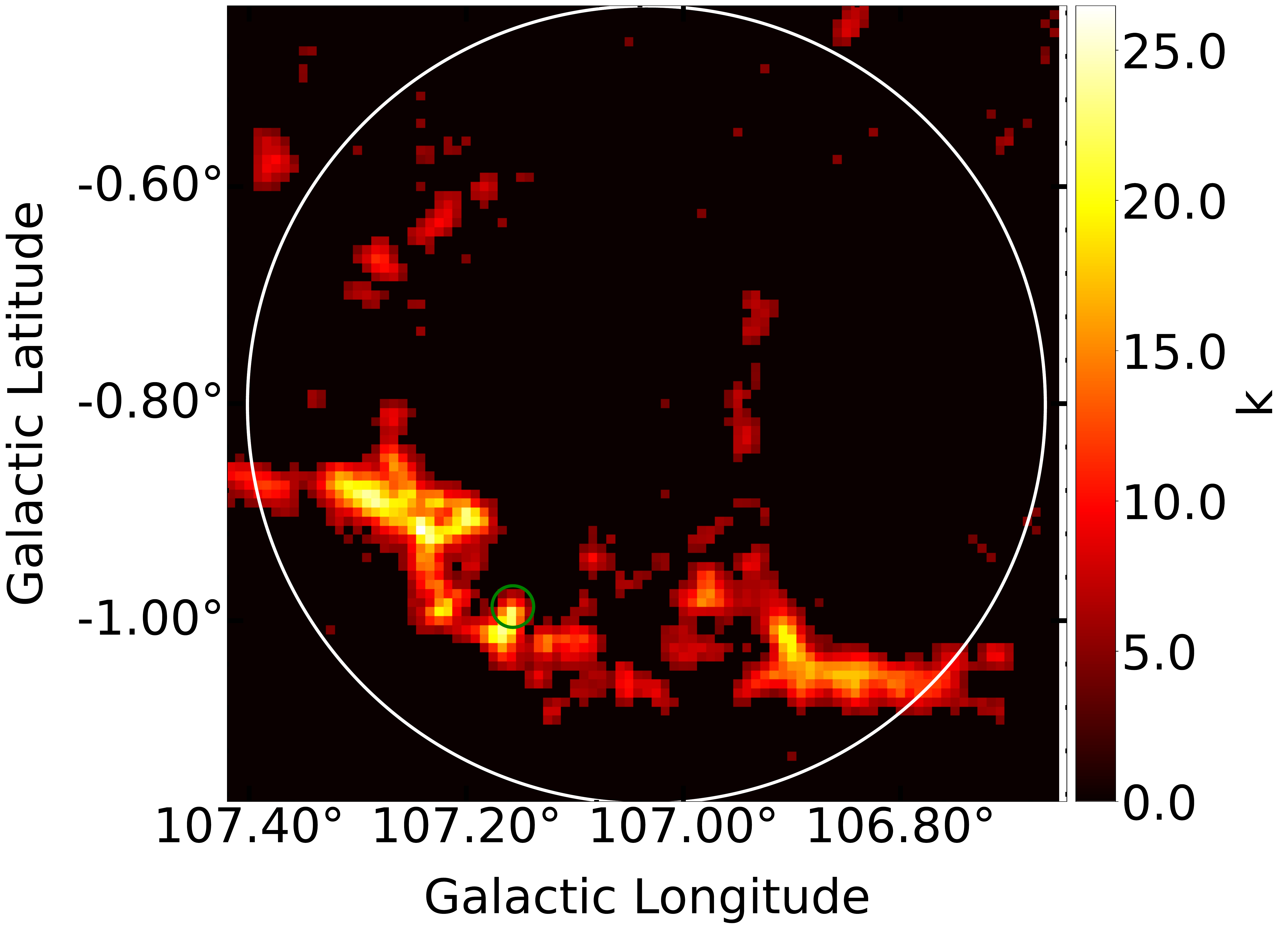}\hfill
    \\(g)
    \end{subfigure}
    \begin{subfigure}[b]{0.4\linewidth}
   \centering
    \includegraphics[trim=1cm -2cm 15cm 0cm, width=.4\textwidth]{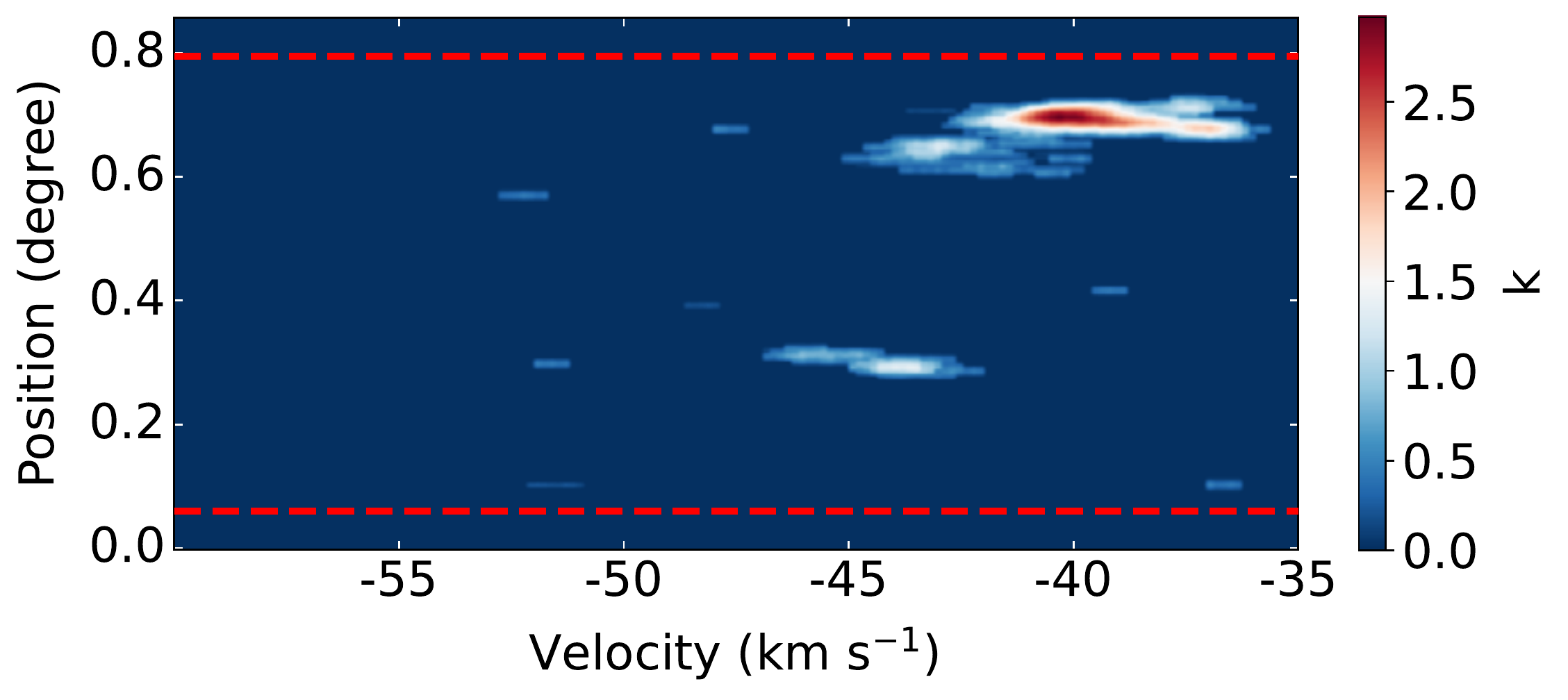}\hfill
    \\(h)
    \end{subfigure}
    \\[\smallskipamount]
    \caption{Same as Figure~\ref{Fig8} but for S142 and the G107.156 ${-}$ 00.988 {H \small {II}} regions. The large and the small circles show the radius of S142 and G107.156 ${-}$ 00.988, respectively. }
    \label{Fig9}
    
\end{figure}

The {H \small {II}} regions G108.603 + 00.494 and S142 are taken as two examples to show our processes of the examination of the possible physical association between {H \small {II}} regions/candidates and molecular clouds. The images of other {H \small {II}} regions are given in Figures~\ref{FigA.4} to \ref{FigA.15} in the Appendix. The WISE 3/4 and H$\rm _{\alpha}$ images of the {H \small {II}} region G108.603 + 00.494 are presented in Figures~\ref{Fig8}, panels (a)-(c), respectively, while its moment maps, excitation temperature distribution, and position-velocity diagram are given in panels (d)-(h), respectively. From panels (a), (b), and (d) in Figure~\ref{Fig8}, we can see that  the molecular gas in this region shows similar morphology to that of the mid-infrared emission of the {H \small {II}} region. The centroid velocity of $^{12}$CO emission within G108.603 + 00.494 is ${-}$ 51 km s$^{-1}$, which is consistent with the velocity of NH$_{3}$ gas (${-}$ 51.4 km s$^{-1}$) associated with the {H \small {II}} region \citep{2011MNRAS.418.1689U}. The excitation temperature of the molecular gas at the intensity peaks in the WISE 3/4 images is 25K, which is significantly higher than that of the molecular gas surrounding the {H \small {II}} region ($\sim$10 K). The p-v diagram in panel (h) is extracted with a width of five pixels along the yellow arrowed line in panel (d). The two red dashed lines in panel (h) indicate the positions of the boundary of the {H \small {II}} region. We can see from panel (h) that there is obvious velocity broadening at the position near the intensity peak of $^{12}$CO, which extends from ${-}$59 to ${-}$43 km s$^{-1}$. This velocity broadening of about $\sim$16 km s$^{-1}$ is significantly larger than the velocity span of about 5 km s$^{-1}$ seen in other parts of the molecular gas in this region. Therefore, we propose that the molecular cloud is influenced by the {H \small {II}} region G108.603 + 00.494. The WISE, H$\rm _{\alpha}$, and $^{12}$CO maps of the S142 region are presented in Figure~\ref{Fig9}. The morphology of the molecular cloud in the integrated intensity map (panel(d)) shows filamentary structures that resemble those seen in the WISE 3/4 band images. The velocity of the ionized gas from the RRL emission in the {H \small {II}} region is ${-}$36 km s$^{-1}$, which lies within the range of the velocities of the $^{12}$CO gas ( from ${-}$45 to ${-}$35 km s$^{-1}$ ). The excitation temperatures of the molecular gas at the region of bright WISE 3/4 emission are higher than 20 K, which is significantly higher than that of other parts of the molecular gas ($\sim$10 K), as shown in Figure~\ref{Fig9} (g). On the basis of spatial and velocity coincidence and raised temperature at the peaks of WISE 3/4 emission, we propose that the {H \small {II}} region S142 is physically associated with the molecular cloud. The {H \small {II}} region G107.156 ${-}$ 00.988, shown in Figure~\ref{Fig9} with the small circle, has a small radius of $\SI{69}{\arcsecond}$, and is located to the southwest of the WISE 3/4 intensity peak of S142. We also propose that this small {H \small {II}} region is associated with the surrounding molecular cloud. 

In summary, the {H \small {II}} regions S146, S149, S152, G109.068 $-$ 00.322, and G109.104 $-$ 00.347 are found to be associated with the surrounding molecular clouds, as shown in Figures~\ref{FigA.6}, \ref{FigA.9}, \ref{FigA.12}, \ref{FigA.13}, and \ref{FigA.14}, respectively. Among these HII regions, S149 meets the criteria 1, 2, and 4, while others meet all the criteria. Seven of the nineteen HII regions, as presented in Table~\ref{tab3} and Figures~\ref{FigA.4}, \ref{FigA.5}, \ref{FigA.7}, \ref{FigA.8}, \ref{FigA.10}, \ref{FigA.11}, and \ref{FigA.15}, are not physically associated with molecular clouds according to our criteria.

%\end{figure}

\label{subsect:Giant Molecular Cloud Complex Associated with the HII Regions/Candidates}

\begin{landscape}
\begin{table}
\bc
\begin{minipage}[]{150mm}
\caption[]{Parameters of {H \small {II}} regions/Candidates in the Surveyed Area\label{tab3}}\end{minipage}
\setlength{\tabcolsep}{2.0pt}
\small
 \begin{tabular}{rccrrrrcccccccc}
  \hline\noalign{\smallskip}
(1)& (2)& (3)&(4)& (5)& (6)& (7)& (8)& (9)&(10)&(11)& (12)&(13)&(14)&(15)\\
Name& Category& Glon&Glat& Radius& V$\rm _{lsr}$(RRL)& V$\rm _{lsr}$(molecular)& Spatial& Velocity &Temperature&Velocity Broadening&A(2014)& F(2015)&Association&Exciting Star$^{a}$\\
&& $\rm deg$ & $\rm deg$ & $\rm arcsec$ &  $\rm km~s^{-1}$& $\rm km~s^{-1}$&&&&&&&&\\
  \hline\noalign{\smallskip}
S142&K&107.034&${-}$0.801&1323&${-}$36&${-}$&Y&Y&Y&Y&N&N&Y&$\rm \alpha$\\
G107.156 ${-}$ 00.988&Q&107.157&${-}$0.990&69&${-}$36&${-}$&Y&Y&N&N&N&N&Y&${-}$\\
S143&K&107.209&${-}$1.330&326&${-}$36&${-}$&N&Y&N&N&N&Y&N&$\rm \gamma$\\
G107.678 ${+}$ 00.235&Q&107.678&0.235&156&${-}$&${-}$&Y&N&N&N&N&N&N&${-}$\\
S146&K&108.191&0.587&307&${-}$55.9&${-}$&Y&Y&Y&Y&N&N&Y&${-}$\\
G108.213 ${-}$ 01.293&C&108.213&${-}$1.290&103&${-}$&${-}$54($\rm {NH}_{3}$)&Y&N&N&N&Y&N&N&${-}$\\
S148&K&108.273&${-}$1.070&195&${-}$53.1&${-}$&N&Y&N&N&N&Y&N&$\rm \delta$\\
S149&K&108.375&${-}$1.056&187&${-}$59.1&${-}$52.4($\rm {}^{12}{CO}$)&Y&Y&Y&N&Y&Y&Y& $\rm \beta$ \\
G108.394 ${-}$ 01.046&Q&108.395&${-}$1.046&34&${-}$59.1&${-}$&N&Y&N&N&N&Y&Y&${-}$\\
G108.412 ${-}$ 01.097&Q&108.412&${-}$1.100&100&${-}$&${-}$&Y&N&N&Y&N&N&N&${-}$\\
G108.603 ${+}$ 00.494&Q&108.603&0.495&138&${-}$&${-}$51.4($\rm {NH}_{3}$)&Y&Y&Y&Y&Y&N&Y&${-}$\\
G108.666 ${-}$ 00.391&Q&108.667&${-}$0.390&93&${-}$&${-}$&N&N&N&Y&N&N&N&${-}$\\
G108.752 ${-}$ 00.972&K&108.753&${-}$0.971&42&${-}$49.1&${-}$51.5($\rm {NH}_{3}$)&N&Y&N&N&Y&N&Y&${-}$ \\
G108.758 ${-}$ 00.989&K&108.758&${-}$0.988&46&${-}$49.1&${-}$51.5($\rm {NH}_{3}$)&N&Y&N&N&Y&N&Y&${-}$\\
S152&K&108.764&${-}$0.951&237&${-}$49.1&${-}$50.3(CS)&Y&Y&Y&Y&Y&Y&Y&${-}$\\
G108.770 ${-}$ 00.974&K&108.771&${-}$0.973&38&${-}$49.1&${-}$&Y&N&N&N&N&Y&Y&${-}$ \\
G109.068 ${-}$ 00.322&K&109.068&${-}$0.321&180&${-}$&${-}$&Y&Y&Y&Y&N&N&Y&${-}$ \\
G109.104 ${-}$ 00.347&K&109.104&${-}$0.346&95&${-}$&${-}$45.8($\rm {NH}_{3}$)&Y&Y&Y&Y&Y&N&Y&${-}$ \\
G109.285 ${-}$ 00.987&Q&109.285&${-}$0.990&177&${-}$&${-}$&N&N&N&N&N&N&N&${-}$\\
  \noalign{\smallskip}\hline
\end{tabular}
\ec
%% place \tablecomments and \tablerefs below \end{center| and \end{center}:
%% you may leave the table-width parameter to editors or set to your actual size
\tablecomments{1.4\textwidth}{Column 2 gives the category of the {H \tiny{II}} regions, with K, Q, and C indicating the known, radio-quiet, and candidate categories. The radius of the {H \tiny {II}} regions is presented in column 5. Column 6 gives the LSR velocity from RRL or H ${\alpha}$, and column 7 gives the LSR velocity from molecular line observations. Columns 8 ${-}$ 11 give the possible signatures of interaction between {H \small {II}} regions/candidates and molecular clouds. Column 12 ${-}$ 14 gives the relations between {H \tiny {II}} regions/candidates and molecular clouds from \cite{2014ApJS..212....1A}, \cite{2015AJ....150..147F}, and this work, respectively. $^{a}$ $\rm \alpha$ = (1, 8, 10, 11, 12, 12, 14, 15, 16, 17, 20, 21, 22); $\rm \beta$ = (6, 7); $\rm \gamma$ = 3; $\rm \delta$ = 4. The numbers are the serial numbers of the stars in Appendix Table~\ref{table A.1}.      }
\end{table}
\end{landscape}

\section{Discussion}

\label{sect:discussion}

\subsection{Gravitational Stability of Molecular Filaments}

\begin{figure}
    \centering
    \includegraphics[width=\textwidth]{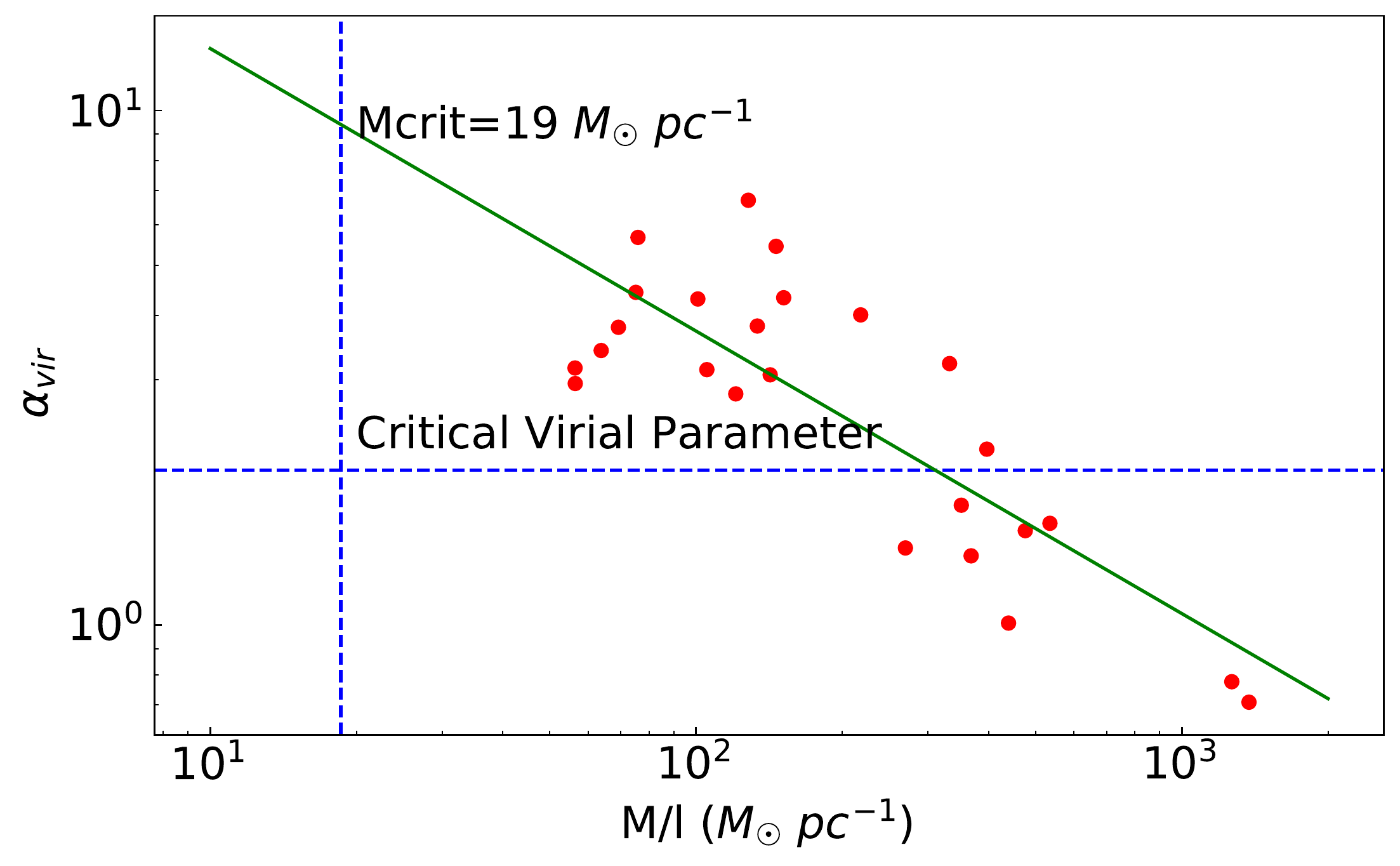}\hfill
    \caption{Relation between the virial parameter and the LTE line mass for the identified filaments. The vertical and horizontal blue dashed lines show the critical line mass $\rm {M}_{line, crit}$ ${\approx}$ 19 ${M}_{\odot}~ \rm pc^{-1}$ and the critical virial parameter of a value of two, respectively. The green line is the power-law fitting result of the correlation with an index of ${-}$0.54 $\pm$ 0.04. }
    \label{Fig10}
\end{figure}

Molecular filaments are unstable against axisymmetric perturbations. A filament supported solely by thermal pressure will collapse when its line mass exceeds the critical value, $\rm {M}_{line, crit}$ = 2$\rm {c}_{s}^2$ / ${G}$ \citep{1964ApJ...140.1056O}, where $\rm {c}_{s}$ is the thermal sound speed, and G is the gravitational constant. Taking the temperature to be 10 K, which is the typical excitation temperature of the filaments identified in this work, the value of the thermal $\rm {M}_{line,crit}$ is around 19 $\rm {M}_{\odot}$ $\rm {pc}^{-1}$. This value is much smaller than the line masses of the filaments as list in Table~\ref{tab2}. The minimum line mass of the filaments in Table~\ref{tab2} is 56.3 $\rm {M}_{\odot}$ $\rm {pc}^{-1}$. The above fact does not mean that all the filaments are gravitationally unstable. From Table~\ref{tab2}, we can see that the $\rm {\sigma}_{tot}$ of our sample of filaments are much larger than isothermal sound speed (${\sim}$0.2 $\rm km~s^{-1}$), which suggests that turbulence motion is dominant in the filaments. The relationship of the virial parameter $\rm {\alpha}_{vir}$ with the line mass M/l is given in Figure~\ref{Fig10}, which can be fitted as $\rm \alpha{_{vir}}$ $\propto$ (M/l)$^{-0.54 \pm 0.04}$. As shown in Figure~\ref{Fig10}, the filaments can be divided into two groups by the critical virial parameter, $\rm {\alpha}_{vir}$ = 2. F2, F3, F6, F7, F12, F14, F20, and F23 have viral parameters smaller than 2, indicating that they are gravitationally bound and tend to collapse and form stars. The other filaments have virial parameters greater than 2, indicating that external pressure is playing a role in binding them or that they are in an expanding state \citep{2017ApJ...838...49X}.

\subsection{Feedback from H {\small{II}} Regions}

As shown in Table~\ref{tab2}, the virial parameters of $\sim70\%$ of the filaments are less than 2, suggesting that non-thermal motions dominate the dynamics of these filaments. Feedback from massive star formation, such as expanding shells of H {\small{II}} regions and stellar winds, are possible regimes that can inject momentum and thermal energy into ambient molecular clouds, which helps maintain turbulence in these clouds. In this section, we discuss the amount of energy that can be provided by the H {\small{II}} regions associated with the molecular clouds and whether it can help maintain the turbulence in the clouds.

H {\small{II}} regions contain three forms of energy, i.e., kinetic energy, ionization energy, and thermal energy, which can be estimated through the following formula \citep{1967ApJ...149...23L,2003ApJ...594..888F},
\begin{equation}
{\rm E_{k}} = \dfrac{4}{3}{\pi}{\rm n_{i}}{\rm m_{H}}{\rm c_{s}^{2}}{\rm R_{s}^{{3/2}}}[(\dfrac{7}{4}{\rm c_{s}}{\rm R_{s}^{{3/4}}}{\rm t_{HII}} + {\rm R_{s}^{{7/4}}})^{{6/7}} - {\rm R_{s}^{{3/2}}}] ,
\end{equation}

\begin{equation}
{\rm E_{i}} = \dfrac{4}{3}{\pi}{\rm n_{i}}(\dfrac{7}{4}{\rm c_{s}}{\rm R_{s}^{{5/2}}}{\rm t_{HII}} + {\rm R_{s}^{{7/2}}})^{{6/7}}{\chi_{0}} ,
\end{equation}

\begin{equation}
{\rm E_{t}} = \dfrac{4}{3}{\pi}{\rm n_{i}}(\dfrac{7}{4}{\rm c_{s}}{\rm R_{s}^{{5/2}}}{\rm t_{HII}} + {\rm R_{s}^{{7/2}}})^{{6/7}}{\rm k}{\rm T_{HII}} ,
\end{equation}
where k is the Boltzmann constant, t$_{\rm HII}$ is the dynamical age of the H {\small{II}} region, ${\rm n_{i}}$ = 10$^{3}$ cm$^{-3}$ is the initial number density of the molecular gas \citep{2017ApJ...845...34D}, m$_{\rm H}$ is the mass of atomic hydrogen, c$_{\rm s}$ = 11 km s$^{-1}$, is the isothermal sound speed of ionized gas \citep{2009A&A...497..649B}, R$_{\rm s}$ is the radius of the Stromgren sphere which can be calculated through ${\rm R_{s}}$ = (3${\rm N_{ly}}$/4$\pi$${\rm n_{i}^{2}}$${\rm \alpha_{B}}$)$^{1/3}$, where N$_{\rm ly}$ is the ionizing luminosity and $\alpha_B = 2.6\times10^{13}$ cm$^{3}$ s$^{-1}$ is the hydrogen recombination coefficient to all levels above the ground level \citep{1980pim..book.....D}. The ionization potential $\chi_0$ in the formula is 13.6 eV, and the effective electron temperature, T$_{\rm HII}$, is adopted to be 10$^{4}$ K. According to \cite{1980pim..book.....D} and \cite{2012AJ....144..173D}, the dynamical age t$_{\rm HII}$ can be estimated through
\begin{equation}
{\rm t_{HII}} =  7.2{\times}10^{4}(\dfrac{\rm R_{HII}}{\rm pc})^{4/3}(\dfrac{\rm N_{\rm ly}}{10^{49}{\rm s^{-1}}})^{-1/4}(\dfrac{\rm n_{\rm i}}{10^{3} {\rm cm}^{-3}})^{-1/2} {\rm yr} ,
\end{equation}
where ${\rm R_{HII}}$ is the radius of the H {\small{II}}. In Section~\ref{subsect:Giant Molecular Cloud Complex Associated with the HII Regions/Candidates}, we found twelve H {\small{II}} regions are associated with their surrounding molecular gas. However, we only found possible excitation stars for two H {\small{II}} regions (S142, S149). According to Table II of \cite{1973AJ.....78..929P} and the spectral type of the excitation stars of S142 and S149, the ionization luminosities of the two H {\small{II}} regions are 2.5$\times$ 10$^{48}$ s$^{-1}$ and 2.3$\times$10$^{47}$ s$^{-1}$, respectively. The distances of S142 and S149 are 2.63 kpc and 2.92 kpc, respectively. Therefore, the R$_{\rm HII}$ of S142 and S149 are 16.8 and 2.6 pc, respectively, and the corresponding t$_{\rm HII}$ are 4.3$\times$10$^{6}$ yr and 6.6$\times$10$^{5}$ yr, respectively. Finally, we get E$_{\rm k}$ = 1.7$\times$10$^{48}$ erg, E$_{\rm i}$ = 1.9$\times$10$^{49}$ erg, and E$_{\rm t}$ = 1.2$\times $10$ ^ {48}$ erg for S142,  E$_{\rm k}$ = 6.5$\times$10$^{46}$ erg, E$_{\rm i}$ = 7.2$\times$10$^{47}$ erg, and E$_{\rm t}$ = 4.5$\times $10$^{46}$ erg for S149.

The turbulent energy of GMC 4 and F21, which are found to be interacting with H {\small{II}} regions S142 and S149, can be estimated by
\begin{equation}
{\rm E_{turb}} = \dfrac{1}{2}M{\sigma^{2}}
\end{equation}
where ${\sigma}$ $\approx$ $\rm\sqrt{3} \sigma_{\upsilon}$ is the three-dimensional turbulent velocity dispersion, which is related to $\Delta{\upsilon}$ in Table~\ref{tab2} through $\sigma = \sqrt{3}\Delta{\upsilon}/(2\sqrt{2\ln2})$.

The kinetic energy, ionization energy, and thermal energy of H {\small{II}} region S142, which is associated with GMC 4, are all larger than the turbulent energy of GMC 4 (7.8$\times$10$^{46}$ erg). Only 4${-}$6\% of the kinetic or the thermal energy, or four-thousandth of the ionization energy provided by S142 is enough to maintain the turbulence in GMC 4. The kinetic and thermal energy of H {\small{II}} region S149 is only about 30${-}$40\% of the turbulent energy of F21 ($1.6\times10^{47}$  erg), while the ionization energy of S149 is 4.5 times the turbulent energy of F21. Therefore, only the ionization energy of S149 can maintain the turbulence in F21, which is similar to the  cases studied by \cite{2018A&A...609A..43X,2019A&A...627A..27X}. We also calculated the turbulent energy for the entire GMC 3 which contains F21. The ionization energy from S149, however, is not enough to affect significantly the dynamics and structures of the GMC 3. The dynamic age of S142, 4.3$\times$10$^{6}$ yr, is about an order of magnitude higher than that of S149, 6.6$\times$10$^{5}$ yr. Most of the energy of S142 may have been converted into the kinetic energy and thermal energy of its ambient interstellar medium. However, H {\small{II}} regions with dynamical age like S149 may be too young to rigorously change the dynamics of the molecular clouds on large scales. Therefore, we presume other energy sources are needed to sustain the turbulence in GMC 1${-}$3.

\section{Summary}
We have performed a large-filed simultaneous survey of the $\rm {}^{12}{CO}$, $\rm {}^{13}{CO}$, and $\rm {C}{}^{18}{O}$ (J = 1 ${-}$ 0) line emission toward the region with 106.65$^\circ$ $\leq$ l $\leq$109.50 $^\circ$ and ${-}$1.85$^\circ$ $\leq$ b $\leq$ 0.95$^\circ$. The spatial distribution, kinematics, and filamentary structure of the molecular clouds in the surveyed region are analyzed in detail. The relationships between the known {H \small {II}} regions/candidates in this region and the molecular clouds revealed by our survey are investigated. The main results are summarized as follows.\par

1. Based on its spatial distribution and velocity structure, the molecular gas in this region is divided into four individual clouds. The distances of clouds 1 ${-}$ 3 are assigned to be 2.92 kpc and that for cloud 4 is 2.62 kpc, based on the trigonometric parallaxes of the OB stars in this region. The physical parameters derived for the four clouds are listed in Table~\ref{tab1}. \par

2. We have identified 25 filaments, F1 to F25, in the $\rm {}^{13}{CO}$ column density map. All filaments show coherent velocity along their main axes. Calculated from $^{13}$CO data, the median excitation temperature, length, line mass, line width, and virial parameter of the filaments are 10.89 K, 8.49 pc, 146.11 $\rm {M}_{\odot}~ \rm pc^{-1}$, 1.01 $\rm km~s^{-1}$, and 3.14, respectively. Eight filaments are gravitationally bound in view of their virial parameters being less than two. \par

3. Combining the data of dust and ionized gas emission of {H \small {II}} regions/candidates with our data of $\rm {}^{12}{CO}$ line emission of the molecular gas, we investigated the relationships between {H \small {II}} regions/candidates and the surrounding molecular clouds. We propose that twelve of the nineteen {H \small {II}} regions in this region are associated with their surroundings molecular clouds.

4. The kinetic, ionization, and thermal energy of the {H \small {II}} region S142 can help maintain turbulence in GMC 4. Small {H \small {II}} regions with relatively short dynamic ages like S149 do not significantly affect the dynamics and structures of molecular clouds.

\label{Summary}

\normalem
\begin{acknowledgements}
This research made use of the data from the Milky Way Imaging Scroll Painting (MWISP) project, which is a multi-line survey in $\rm {}^{12}{CO}$/$\rm{}^{13}{CO}$/$\rm{C}{}^{18}{O}$ along the northern galactic plane with PMO-13.7m telescope. We are grateful to all the members of the MWISP working group, particularly the staff members at PMO-13.7m telescope, for their long-term support. MWISP was sponsored by National Key R${\rm \&}$D Program of China with grant 2017YFA0402701 and  CAS Key Research Program of Frontier Sciences with grant QYZDJ-SSW-SLH047. We acknowledge the support of NSFC grants 11973091, 12073079, and 12103025. C.L. acknowledges the supports by China Postdoctoral Science Foundation No. 2021M691532, and Jiangsu Postdoctoral Research Funding Program No. 2021K179B. This publication makes use of data products from the Wide-field Infrared Survey Explorer, which is a joint project of the University of California, Los Angeles, and the Jet Propulsion Laboratory/California Institute of Technology, funded by the National Aeronautics and Space Administration. This publication makes use of data products from the Virginia Tech Spectral-Line Survey (VTSS), which is supported by the National Science Foundation. This work also makes use of data from the European Space Agency (ESA) space mission Gaia. Gaia data are being processed by the Gaia Data Processing and Analysis Consortium (DPAC). Funding for the DPAC is provided by national institutions, in particular, the institutions participating in the Gaia MultiLateral Agreement (MLA). 

\end{acknowledgements}
  
\appendix
\section{figures}
\label{Appendix}

\begin{figure}
\centering
    \includegraphics[width=\textwidth]{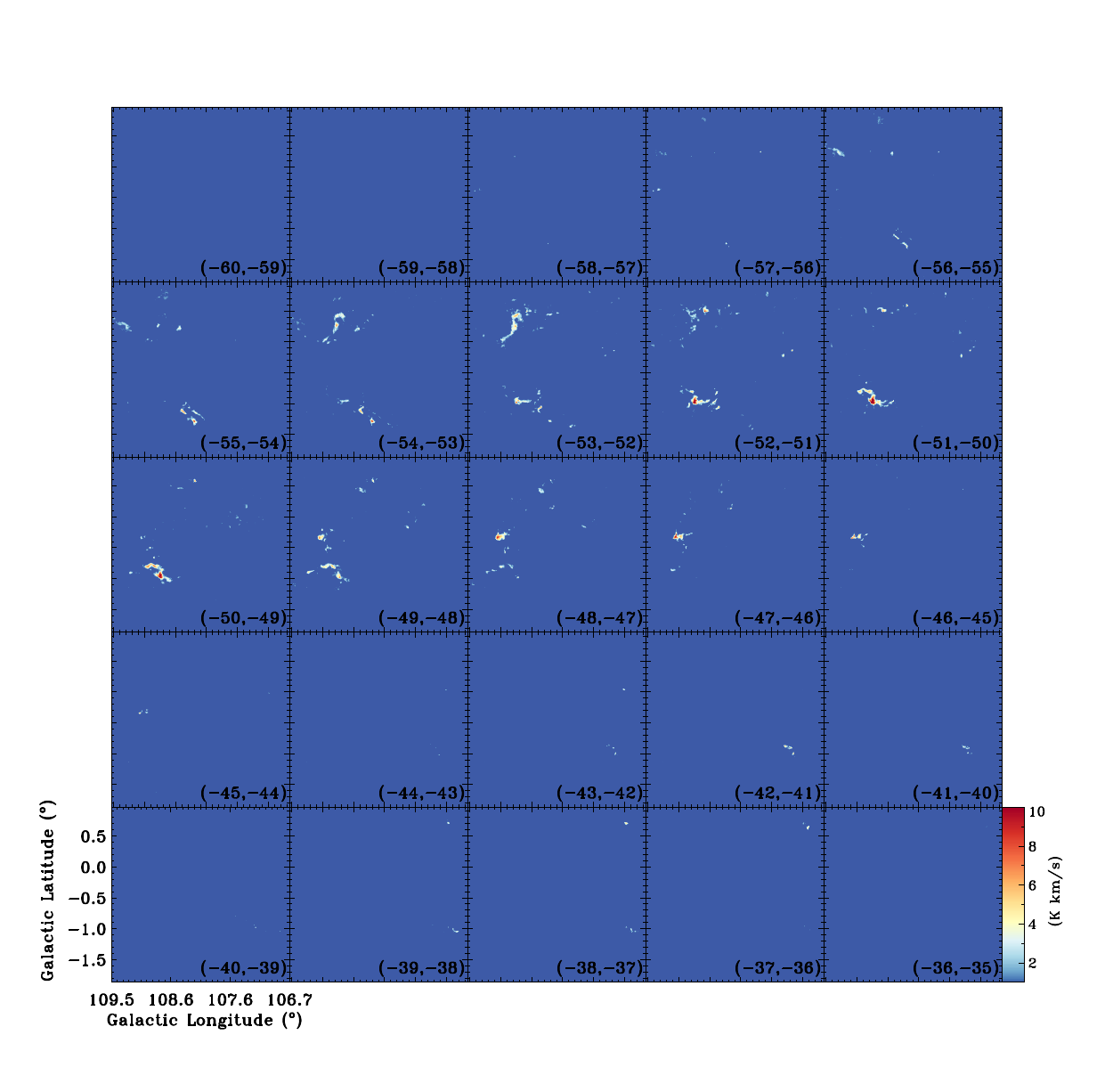}\hfill
    \caption{Velocity channel maps of $\rm {}^{13}{CO}$ from ${-}$60 to ${-}$35 $\rm km~s^{-1}$ of the region. The velocity range of each channel is shown at the bottom-right corner of each panel. }
    \label{FigA.1}

\end{figure}

\begin{figure}[htbp!]
    \centering
    \begin{subfigure}[b]{0.75\linewidth}
    \includegraphics[trim=6cm 1cm 7cm 3cm, width=\linewidth]{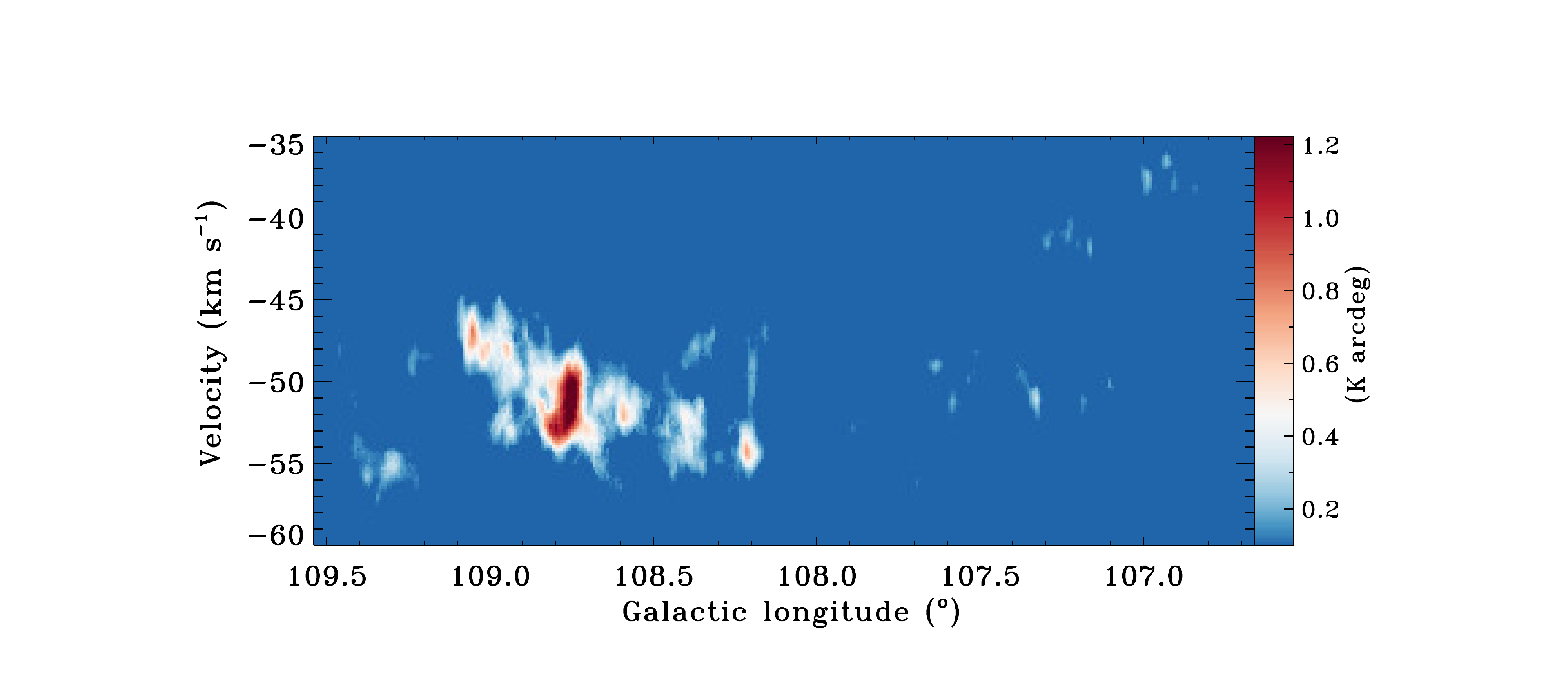}
    \caption{}
   \end{subfigure}
    \begin{subfigure}[b]{0.35\linewidth}
    \includegraphics[trim=5cm 13cm 5cm 10cm,width=\linewidth]{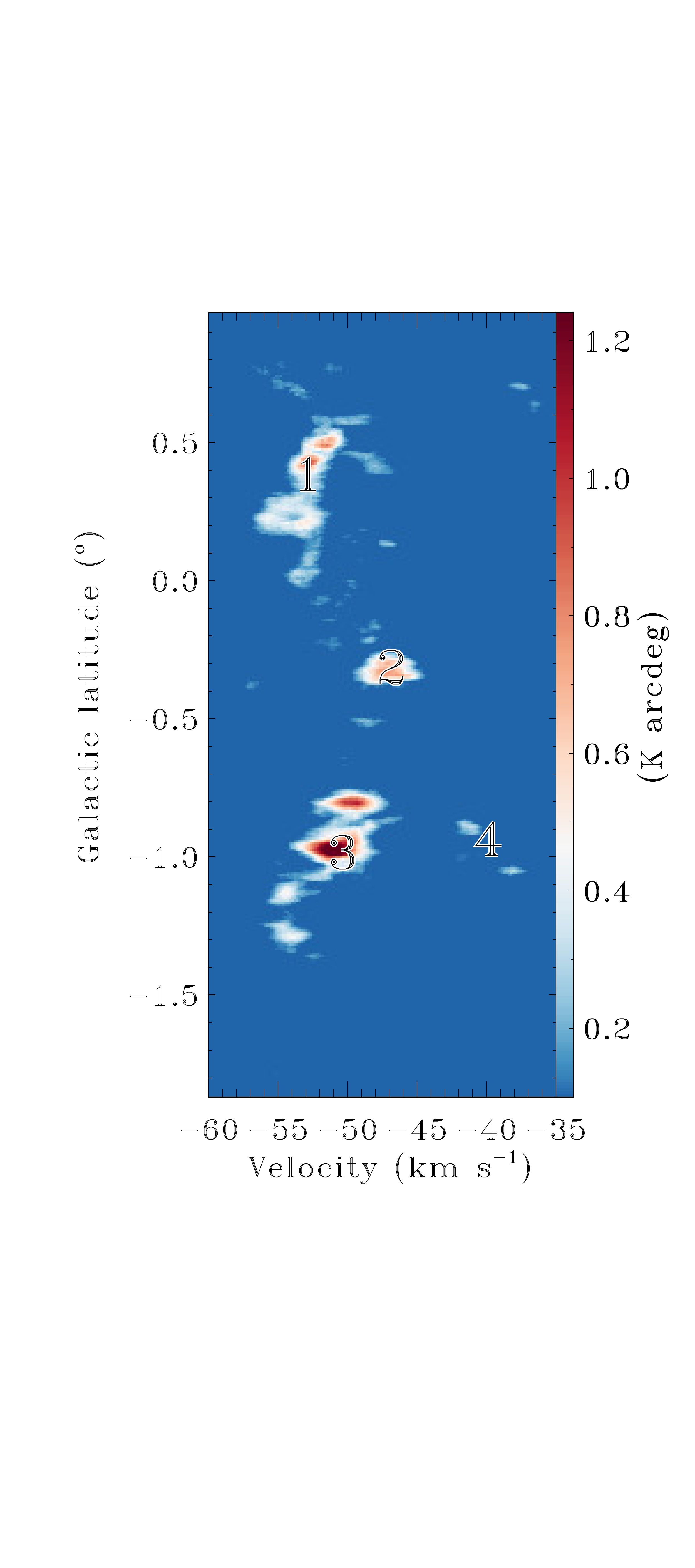}
     \caption{}
    \end{subfigure}
    \caption{Same as Figure~\ref{Fig4} but for $\rm {}^{13}{CO}$ line emission. }
   \label{FigA.2}
  \end{figure}

\begin{figure}[htbp]
    \centering
    \includegraphics[width=7.0cm, trim= 2cm 3cm 1cm 2.5cm]{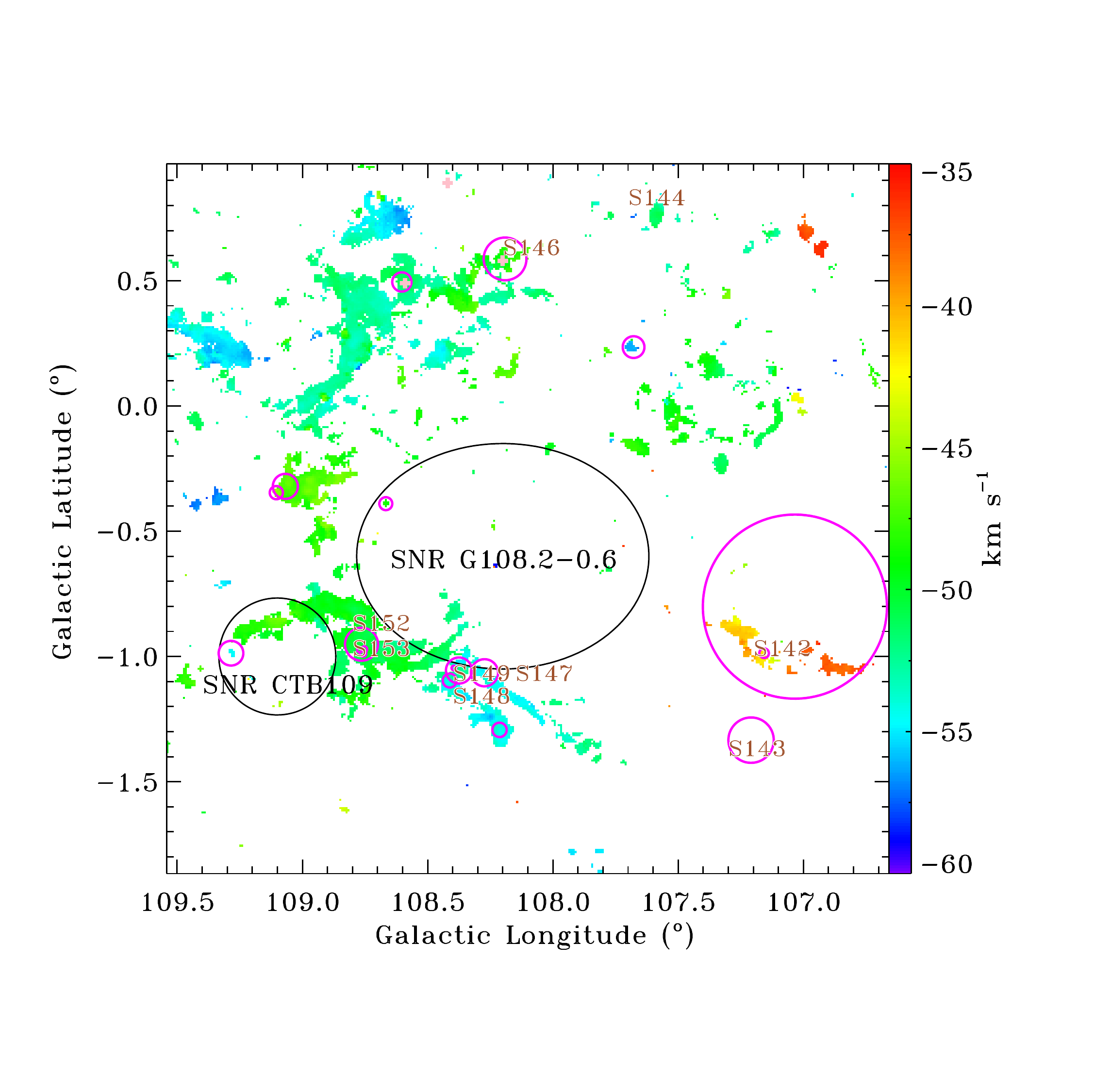}
    \includegraphics[width=7.0cm, trim= 2cm 3cm 1cm 2.5cm]{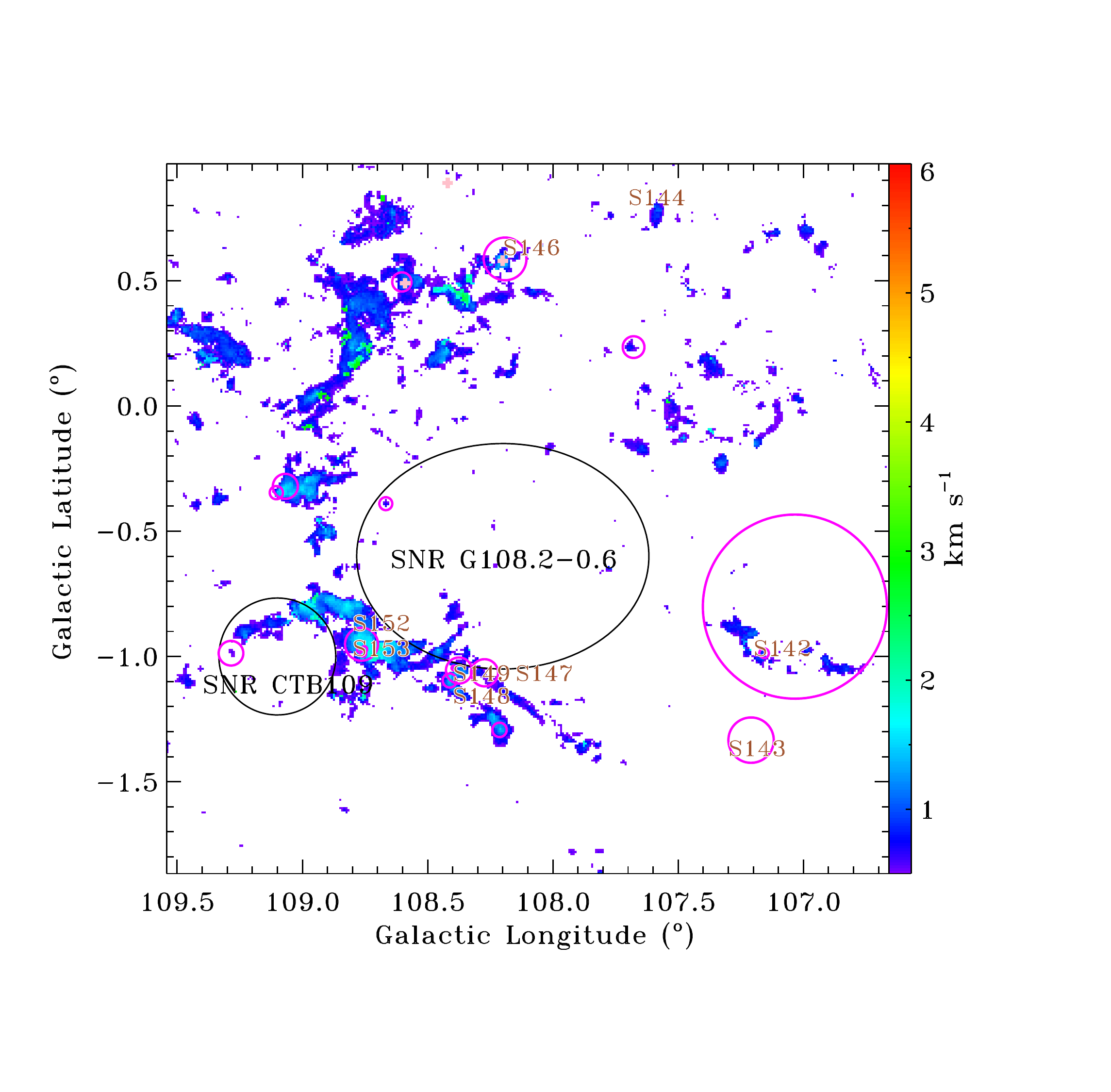}
    \caption{Same as Figure~\ref{Fig5} but for $\rm {}^{13}{CO}$ line emission. }
    \label{FigA.3}
  \end{figure}

\begin{figure}[h!]
    \centering
     \begin{subfigure}[b]{0.4\linewidth}
     \centering
    \includegraphics[trim=15cm 0cm 15cm 0cm, width=.4\textwidth]{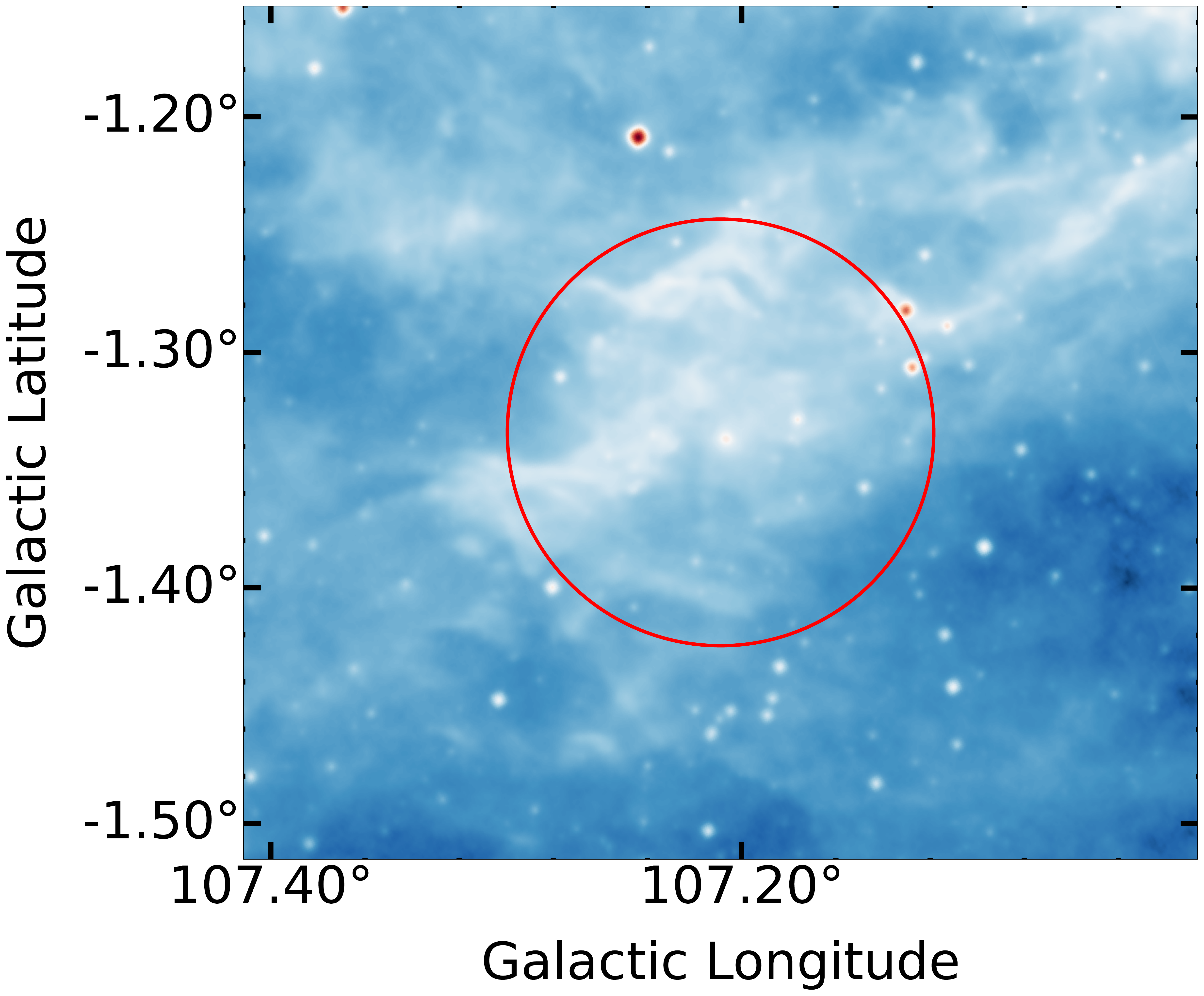}
    \\(a)
   \end{subfigure}
   \begin{subfigure}[b]{0.4\linewidth}
   \centering
    \includegraphics[trim=15cm 0cm 15cm 0cm,width=.4\textwidth]{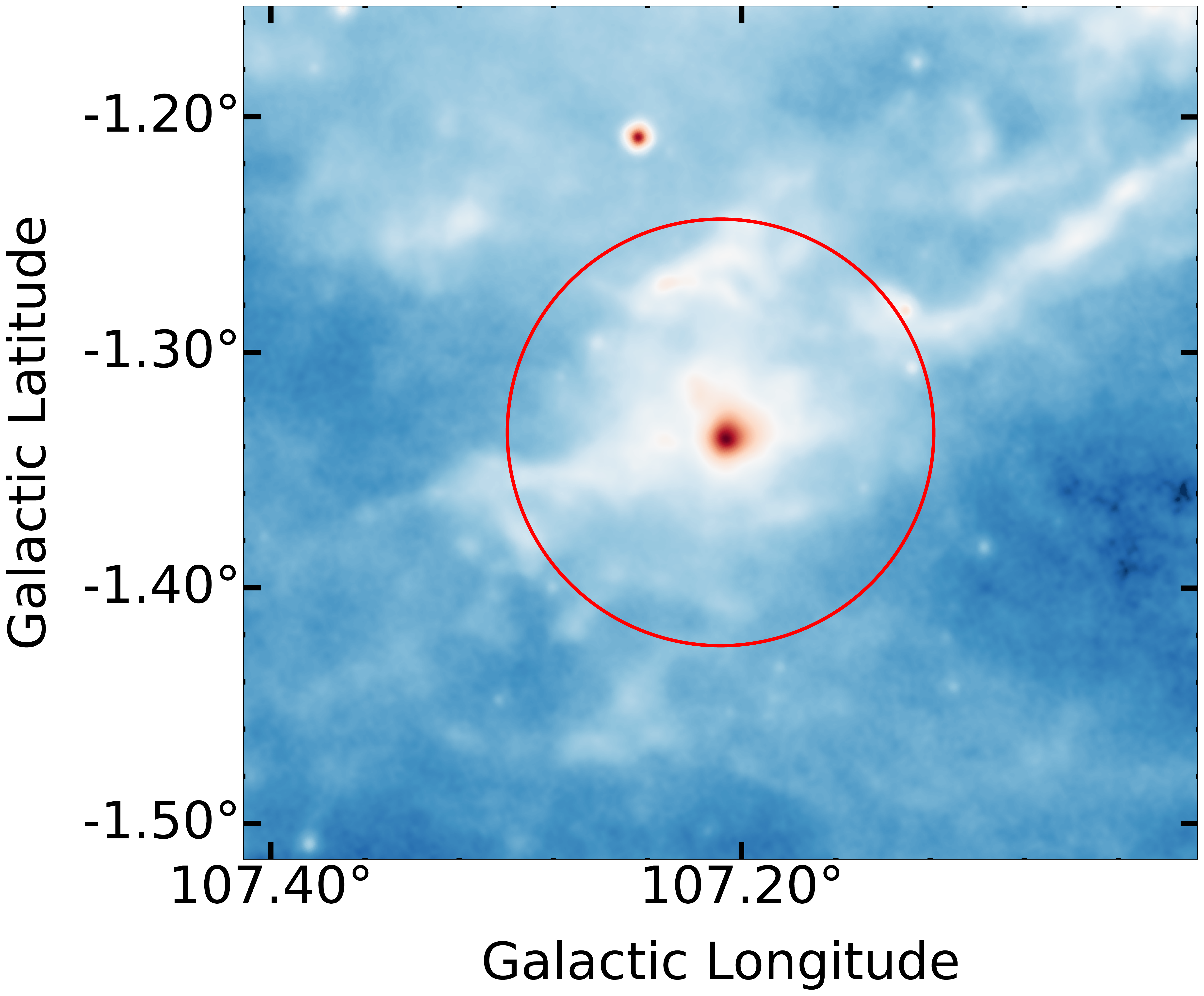}
     \\(b)
    \end{subfigure}
   \begin{subfigure}[b]{0.4\linewidth}
   \centering
    \includegraphics[trim=15cm 0cm 15cm 0cm,width=.4\textwidth]{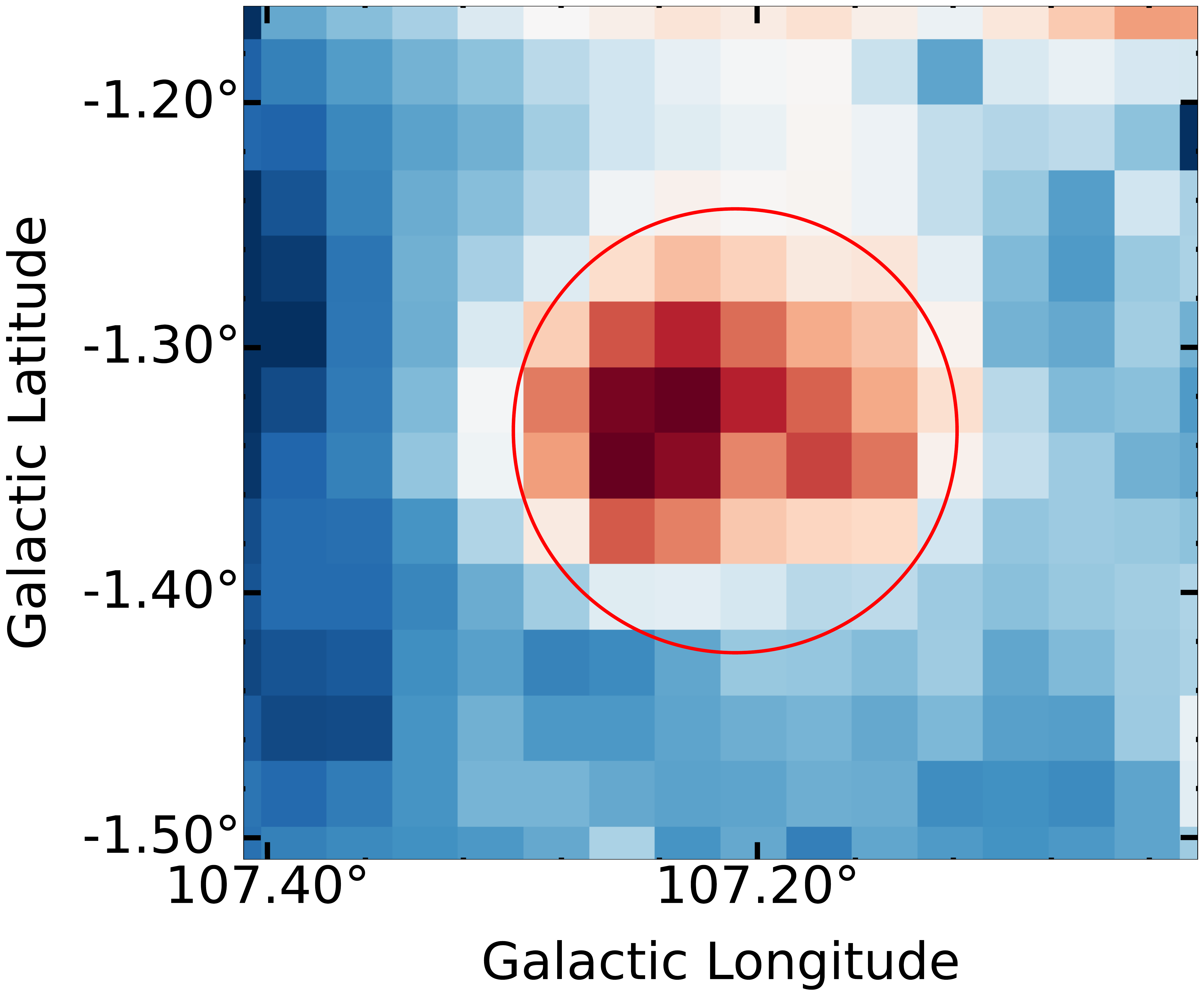}
     \\(c)
    \end{subfigure}
    \begin{subfigure}[b]{0.4\linewidth}
   \centering
    \includegraphics[trim=14.5cm 0cm 21.5cm 0cm,width=.4\textwidth]{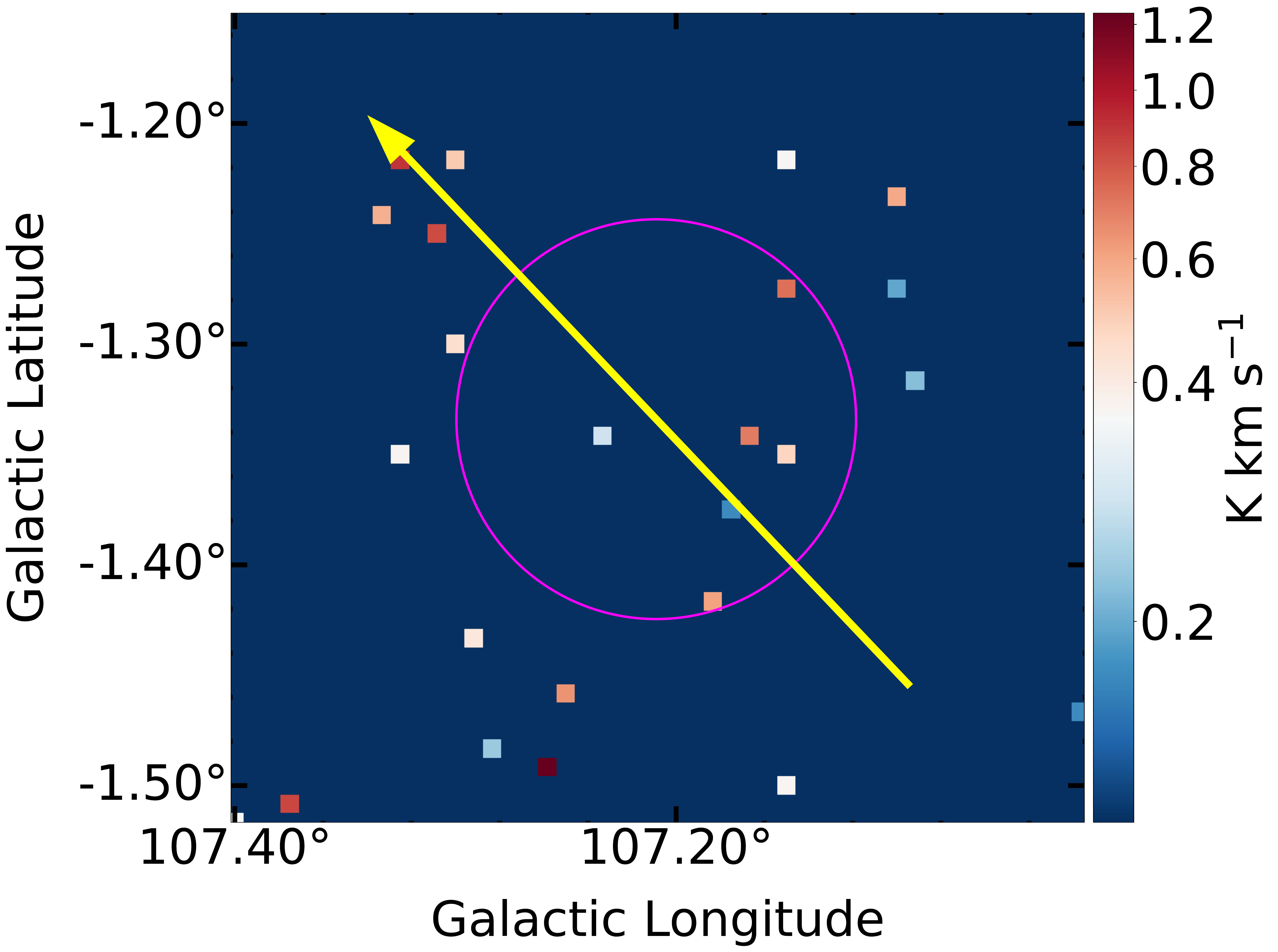}\hfill
    \\(d)
    \end{subfigure}
    \begin{subfigure}[b]{0.4\linewidth}
   \centering
    \includegraphics[trim=15cm 0cm 22.5cm 0cm,width=.4\textwidth]{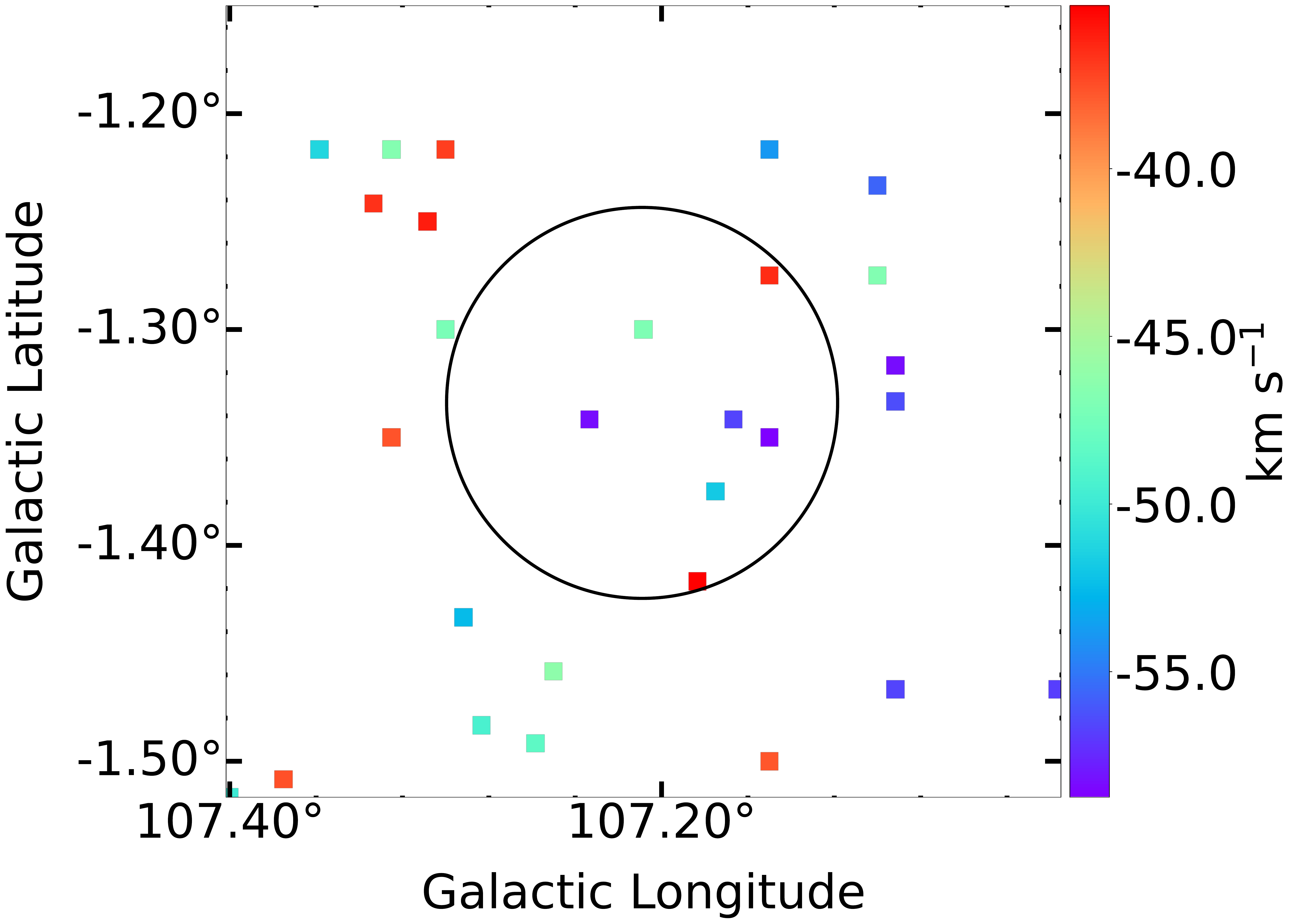}\hfill
    \\(e)
    \end{subfigure}
    \begin{subfigure}[b]{0.4\linewidth}
   \centering
    \includegraphics[trim=9.cm 0cm 26cm 0cm,width=.4\textwidth]{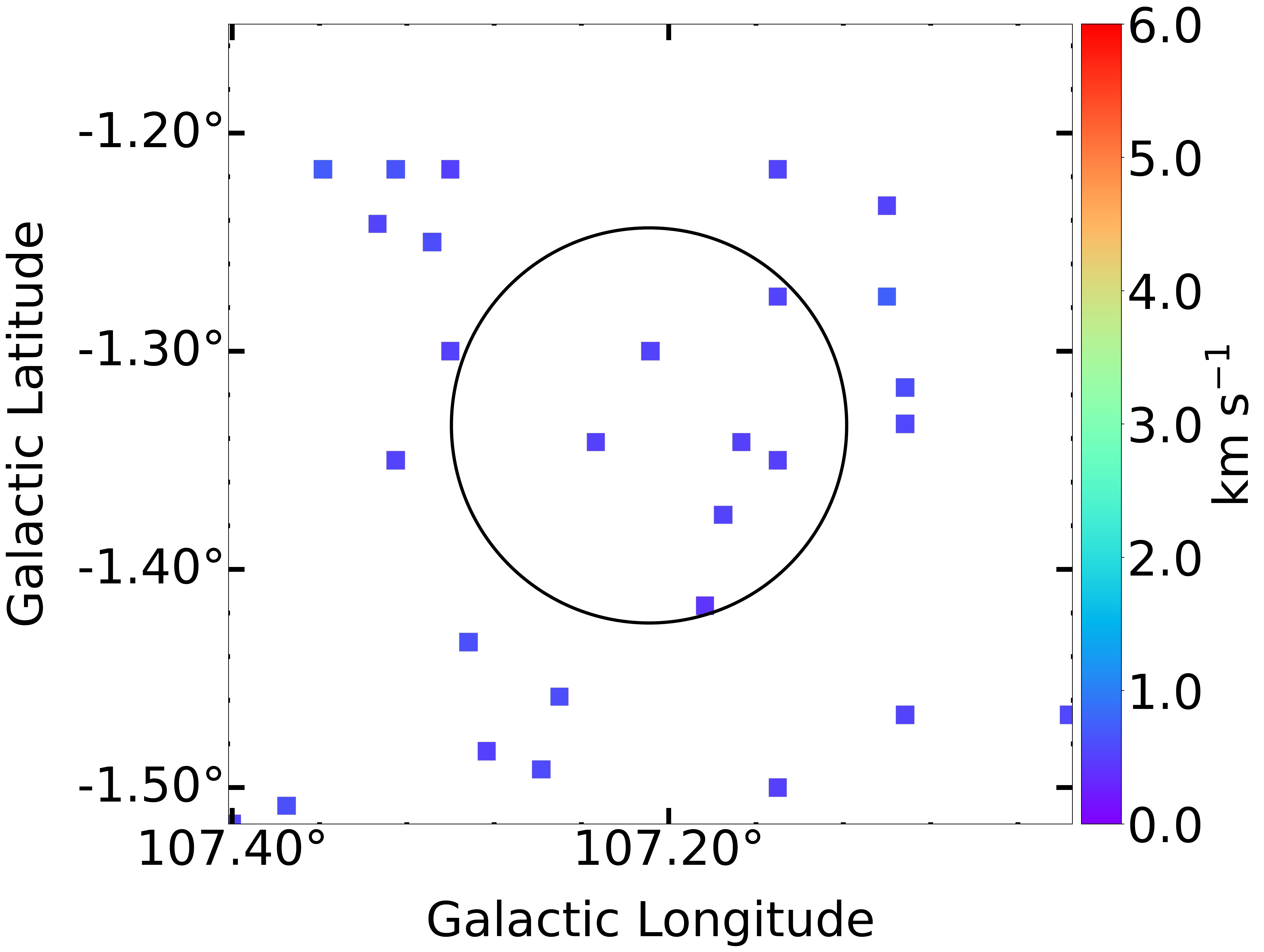}\hfill
    \\(f)
    \end{subfigure}
    \begin{subfigure}[b]{0.4\linewidth}
   \centering
    \includegraphics[trim=14.5cm 0cm 21.3cm 0cm, width=.4\textwidth]{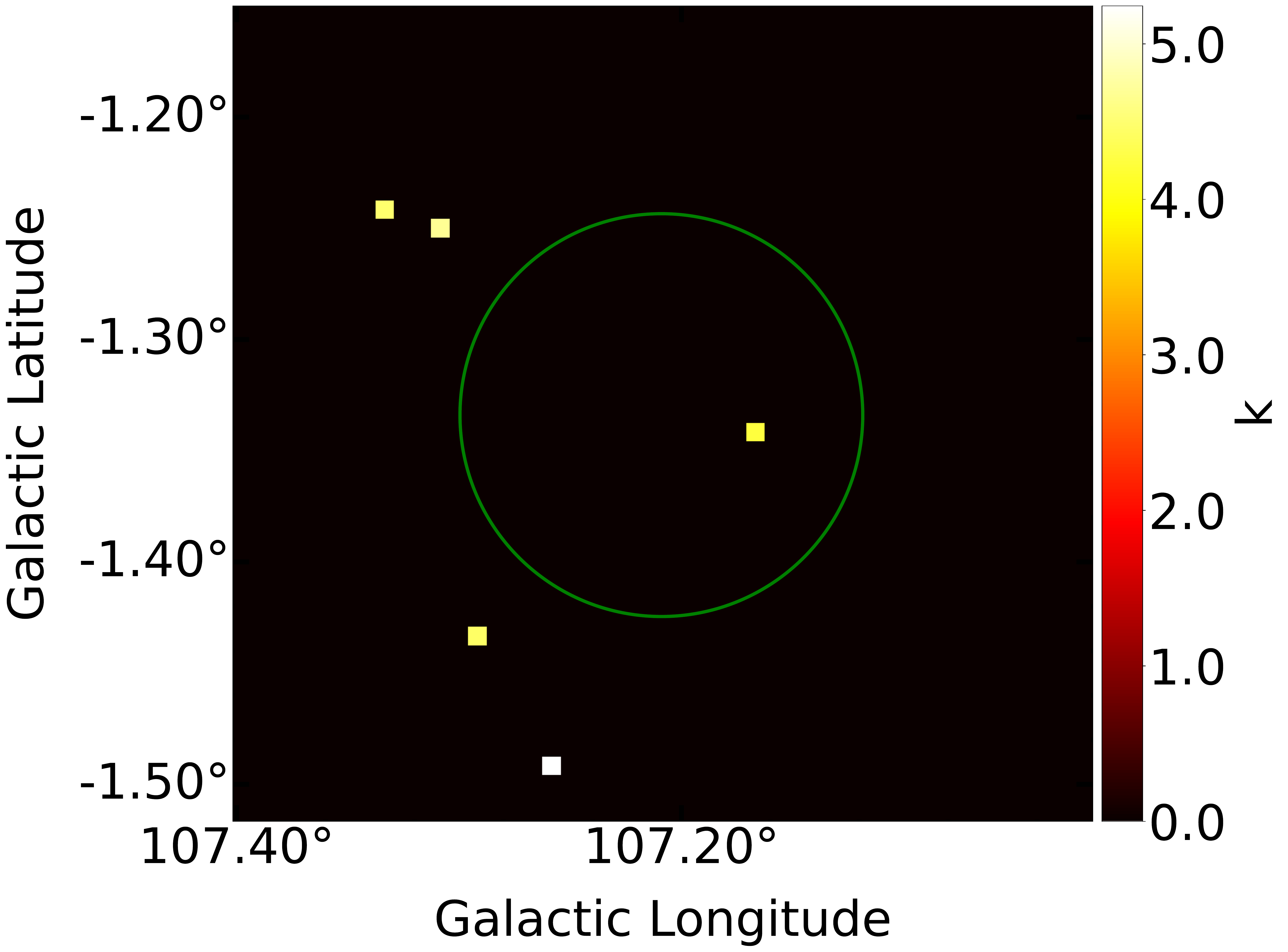}\hfill
    \\(g)
    \end{subfigure}
    \begin{subfigure}[b]{0.4\linewidth}
   \centering
    \includegraphics[trim=1cm -2cm 15cm 0cm, width=.4\textwidth]{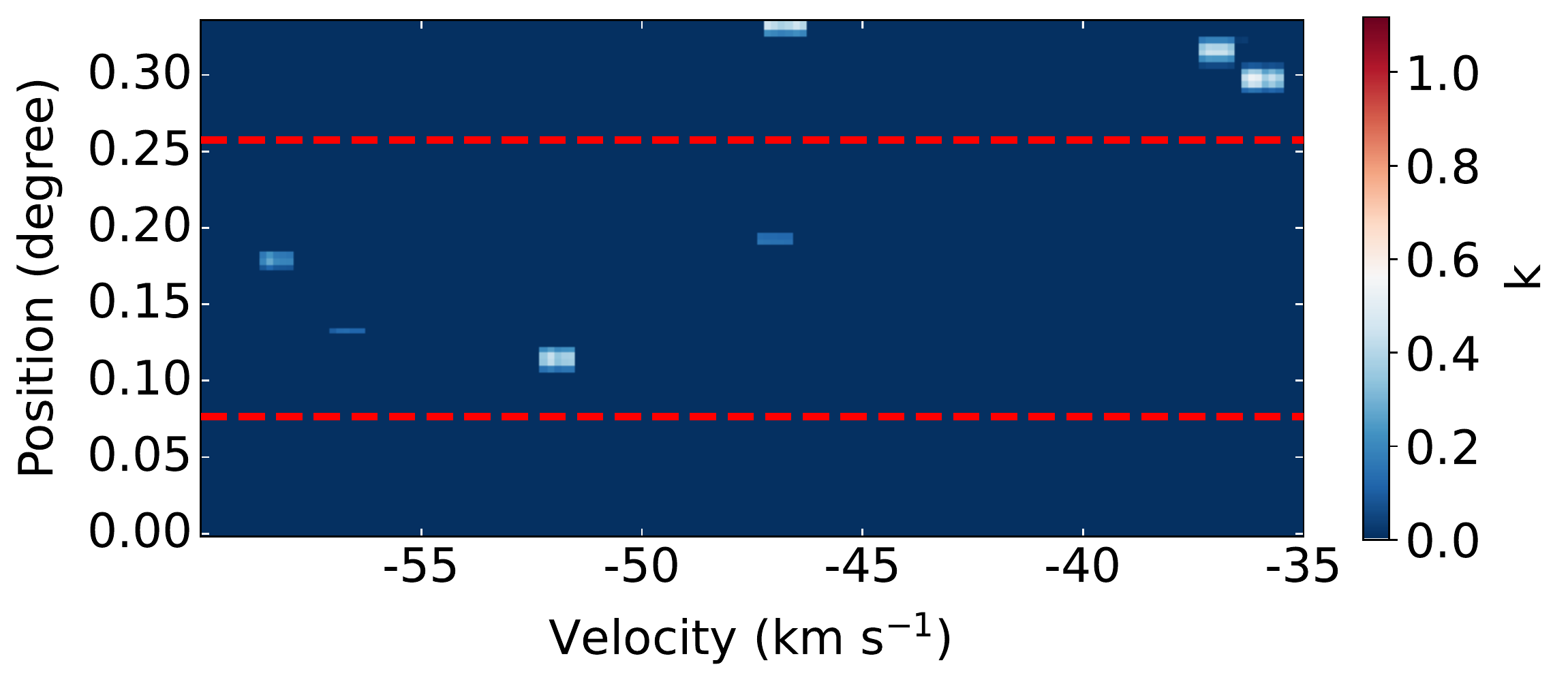}\hfill
    \\(h)
    \end{subfigure}
    \\[\smallskipamount]
    \caption{Same as Figure~\ref{Fig8} but for the S143 region. }
    \label{FigA.4}
\end{figure}

\begin{figure}[h!]
    \centering
     \begin{subfigure}[b]{0.4\linewidth}
     \centering
    \includegraphics[trim=14.5cm 0cm 14.5cm 0cm, width=.4\textwidth]{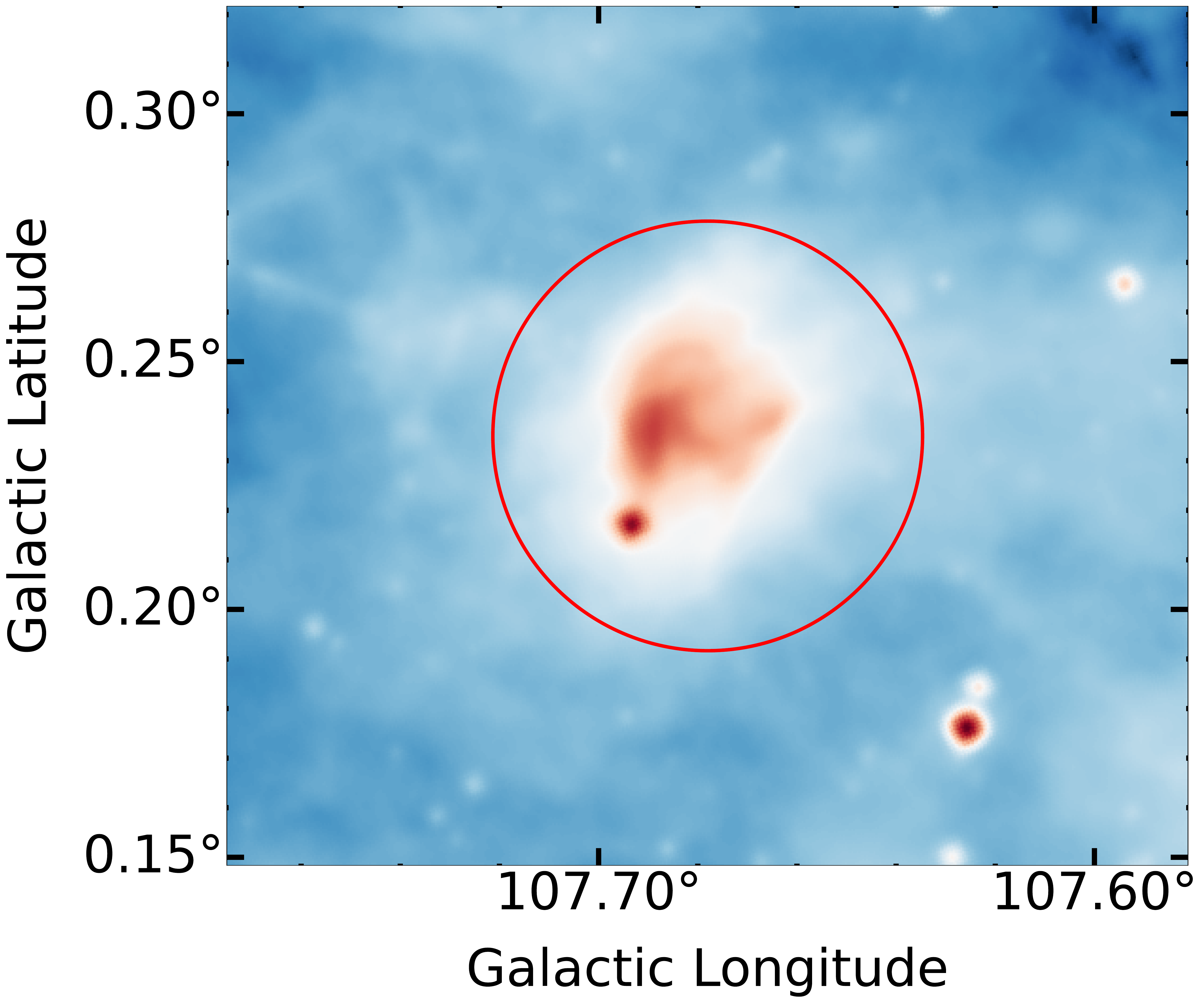}
    \\(a)
   \end{subfigure}
   \begin{subfigure}[b]{0.4\linewidth}
   \centering
    \includegraphics[trim=14.5cm 0cm 14.5cm 0cm,width=.4\textwidth]{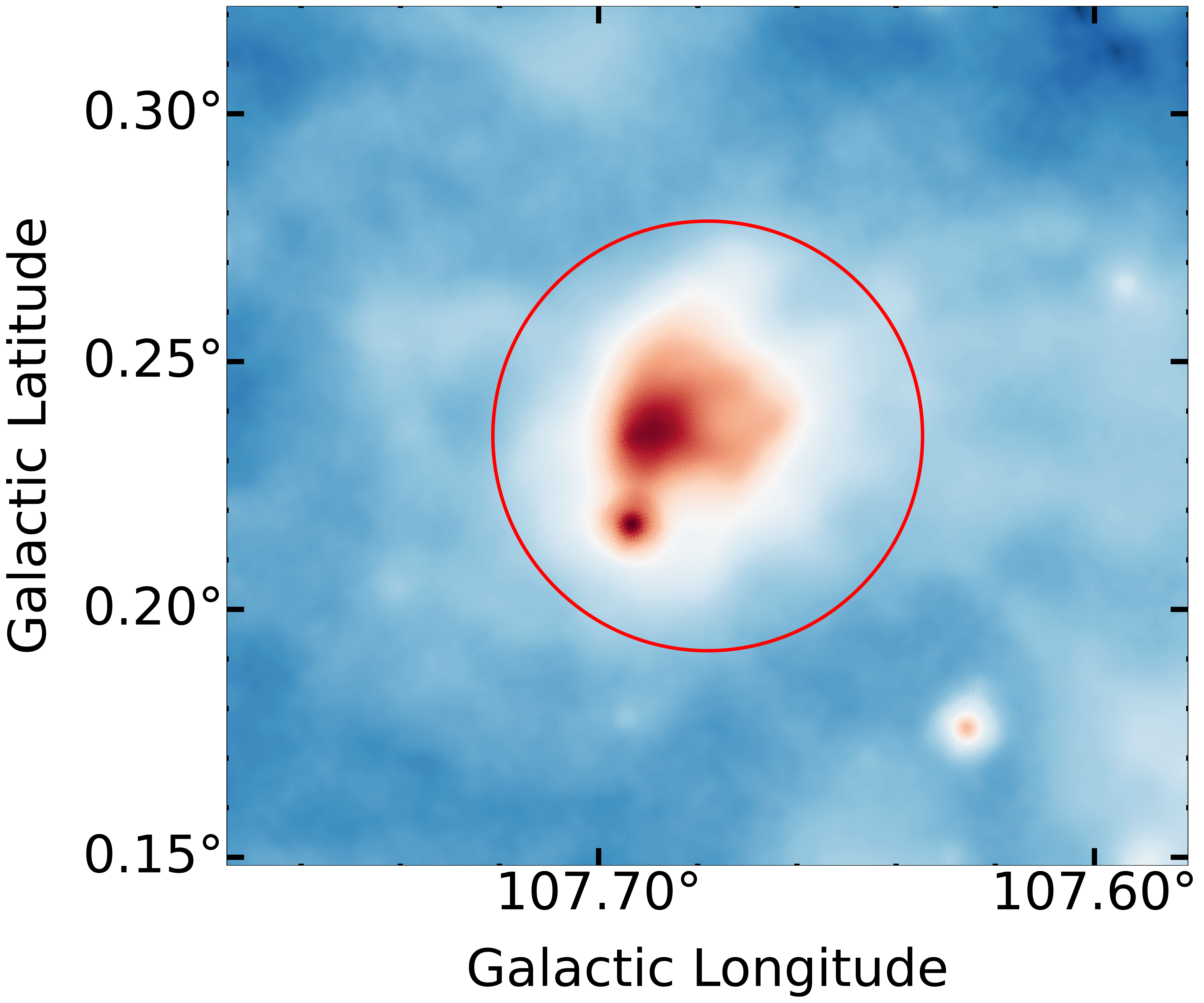}
     \\(b)
    \end{subfigure}
   \begin{subfigure}[b]{0.4\linewidth}
   \centering
    \includegraphics[trim=15cm 0cm 17cm 0cm,width=.4\textwidth]{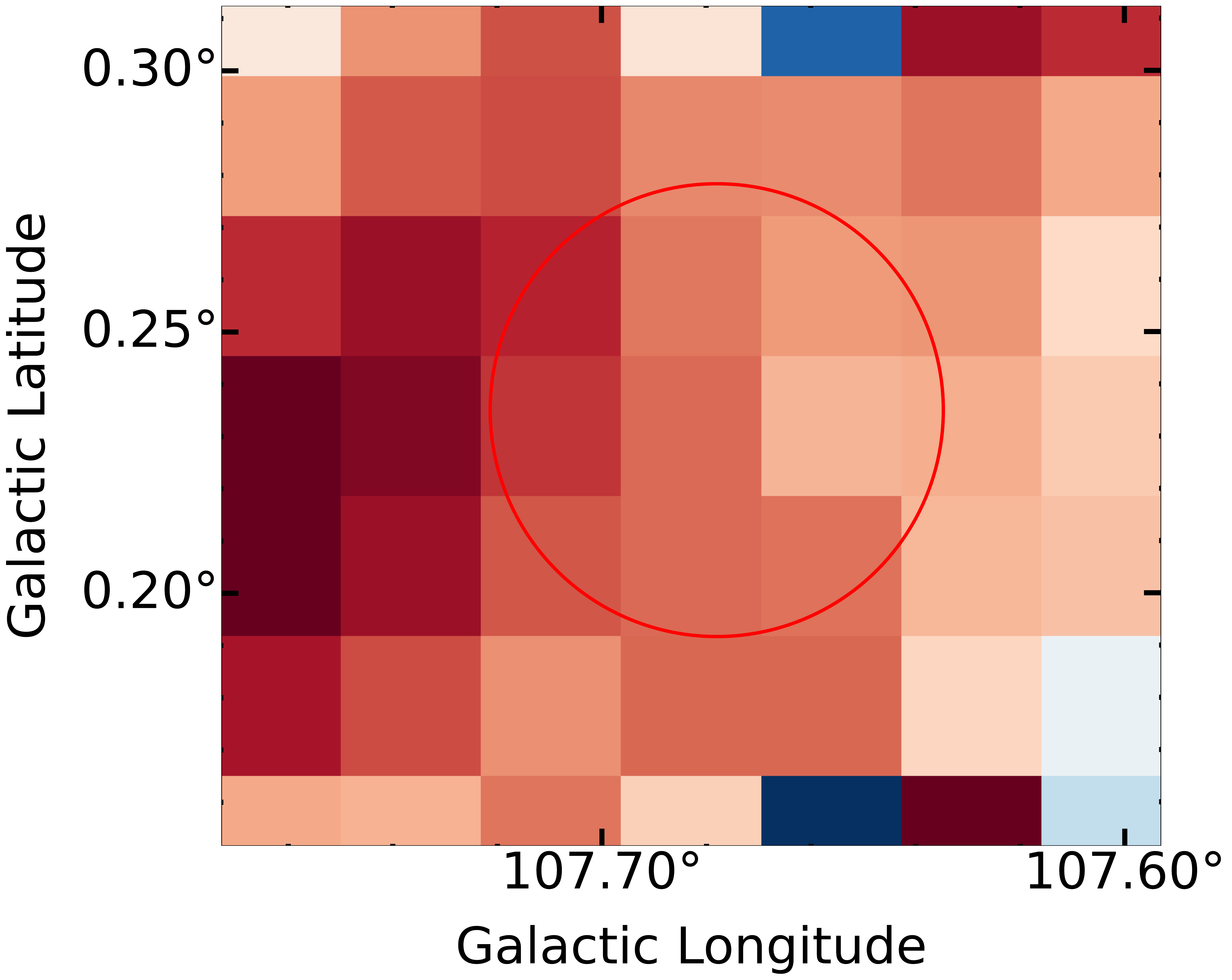}
     \\(c)
    \end{subfigure}
    \begin{subfigure}[b]{0.4\linewidth}
   \centering
    \includegraphics[trim=14.5cm 0cm 22cm 0cm,width=.4\textwidth]{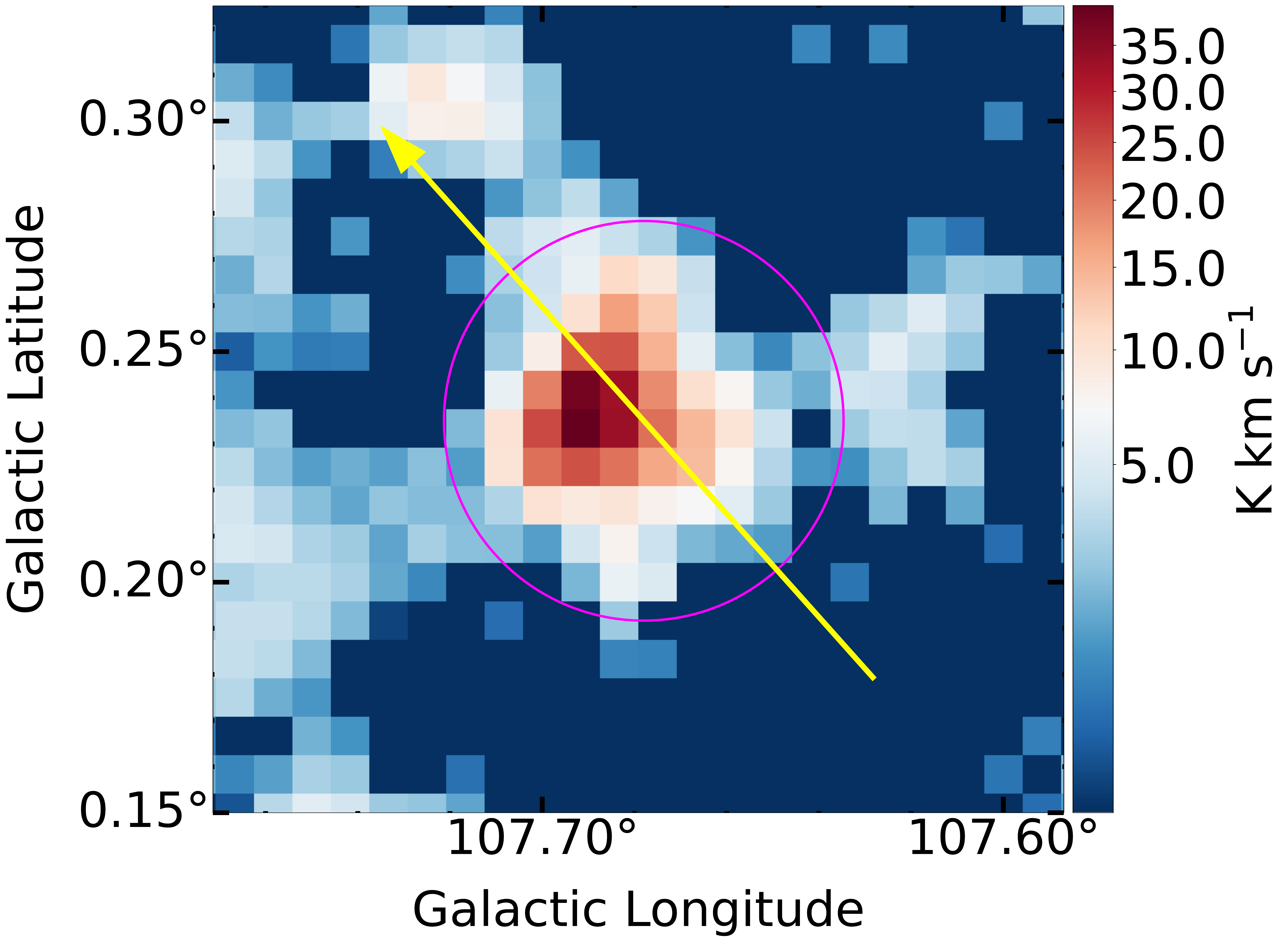}\hfill
    \\(d)
    \end{subfigure}
    \begin{subfigure}[b]{0.4\linewidth}
   \centering
    \includegraphics[trim=14cm 0cm 22.5cm 0cm,width=.4\textwidth]{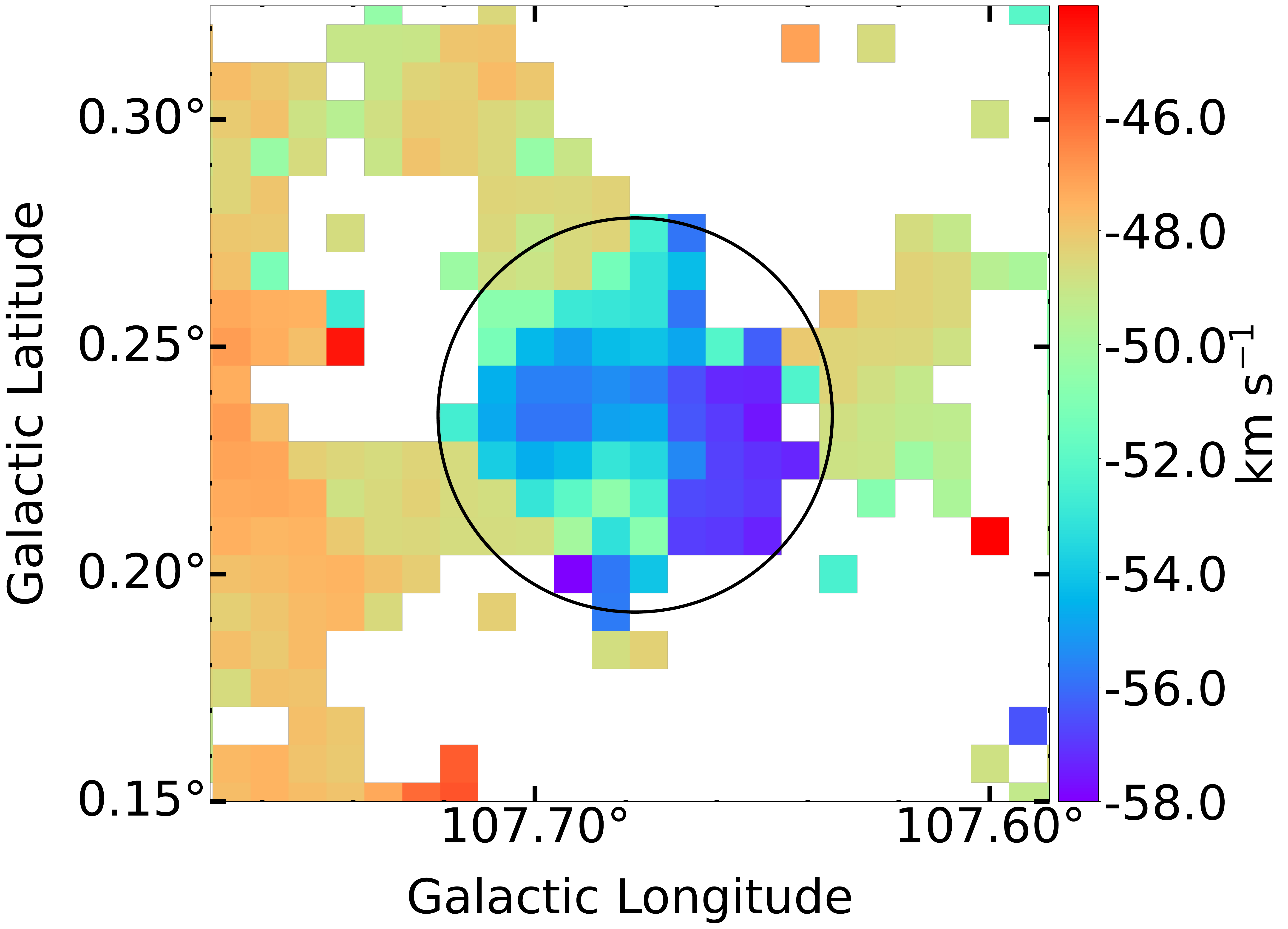}\hfill
    \\(e)
    \end{subfigure}
    \begin{subfigure}[b]{0.4\linewidth}
   \centering
    \includegraphics[trim=9.cm 0cm 25cm 0.7cm,width=.4\textwidth]{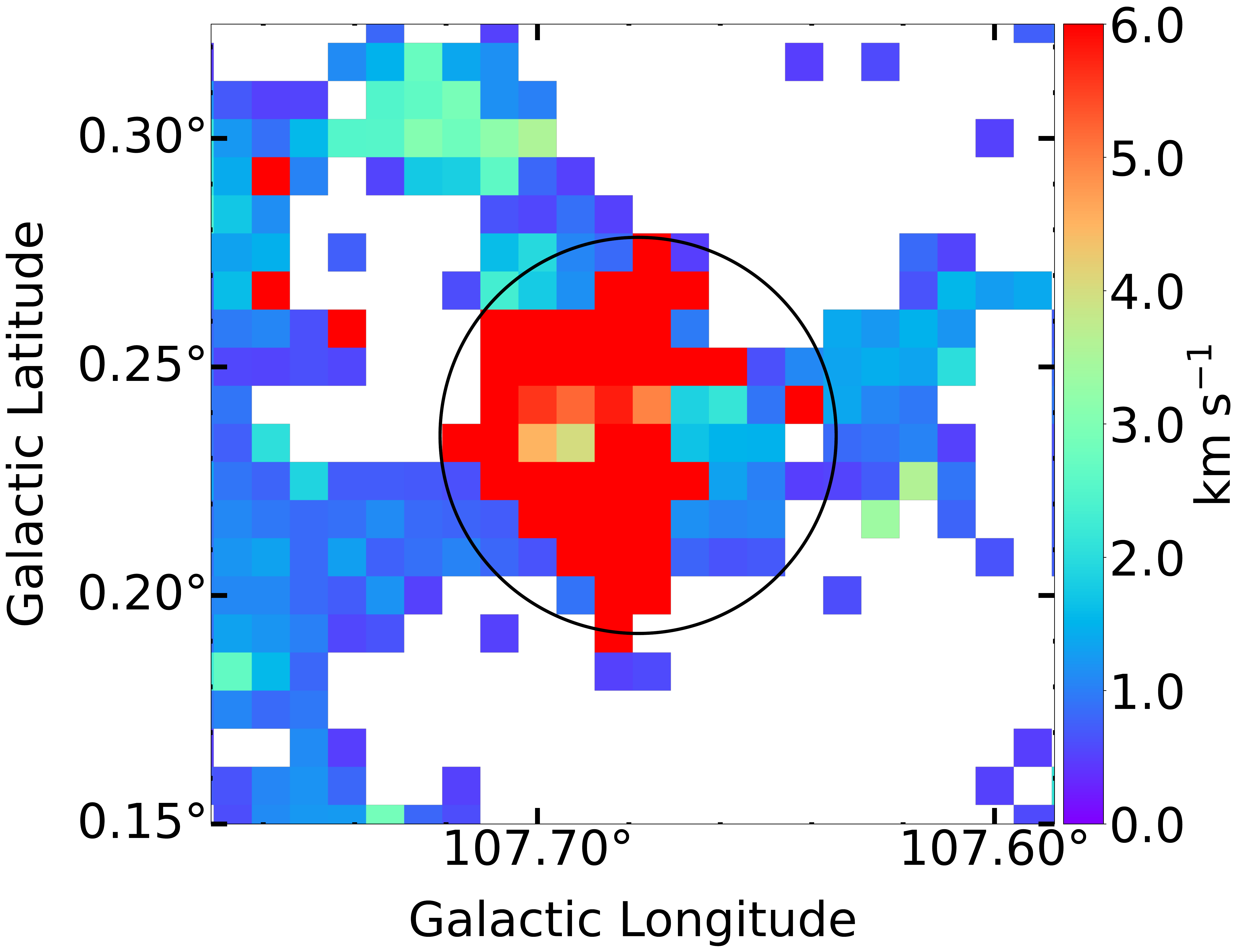}\hfill
    \\(f)
    \end{subfigure}
    \begin{subfigure}[b]{0.4\linewidth}
   \centering
    \includegraphics[trim=14.5cm 0cm 22cm 0cm, width=.4\textwidth]{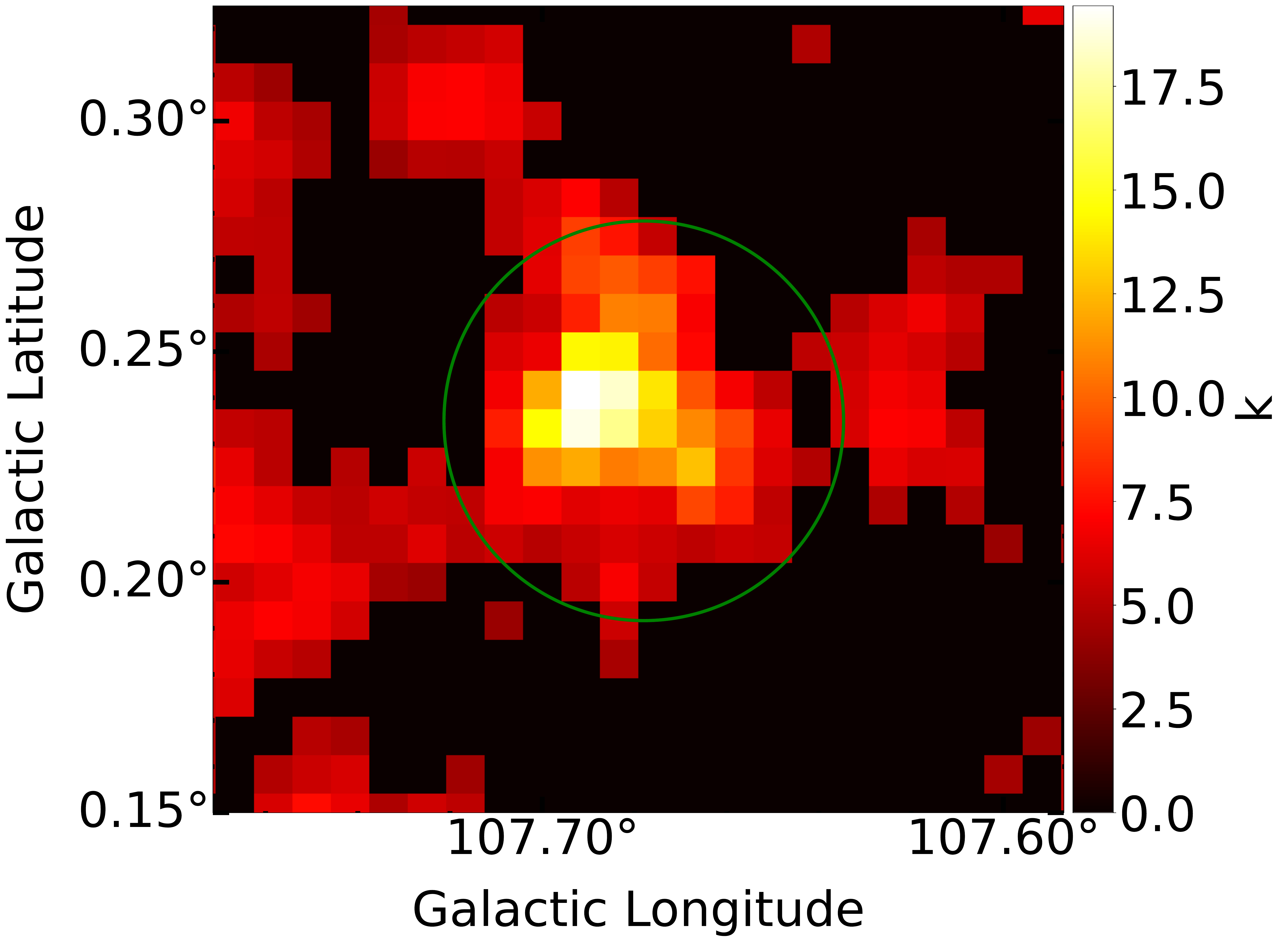}\hfill
    \\(g)
    \end{subfigure}
    \begin{subfigure}[b]{0.4\linewidth}
   \centering
    \includegraphics[trim=2cm -2cm 17cm 0cm, width=.4\textwidth]{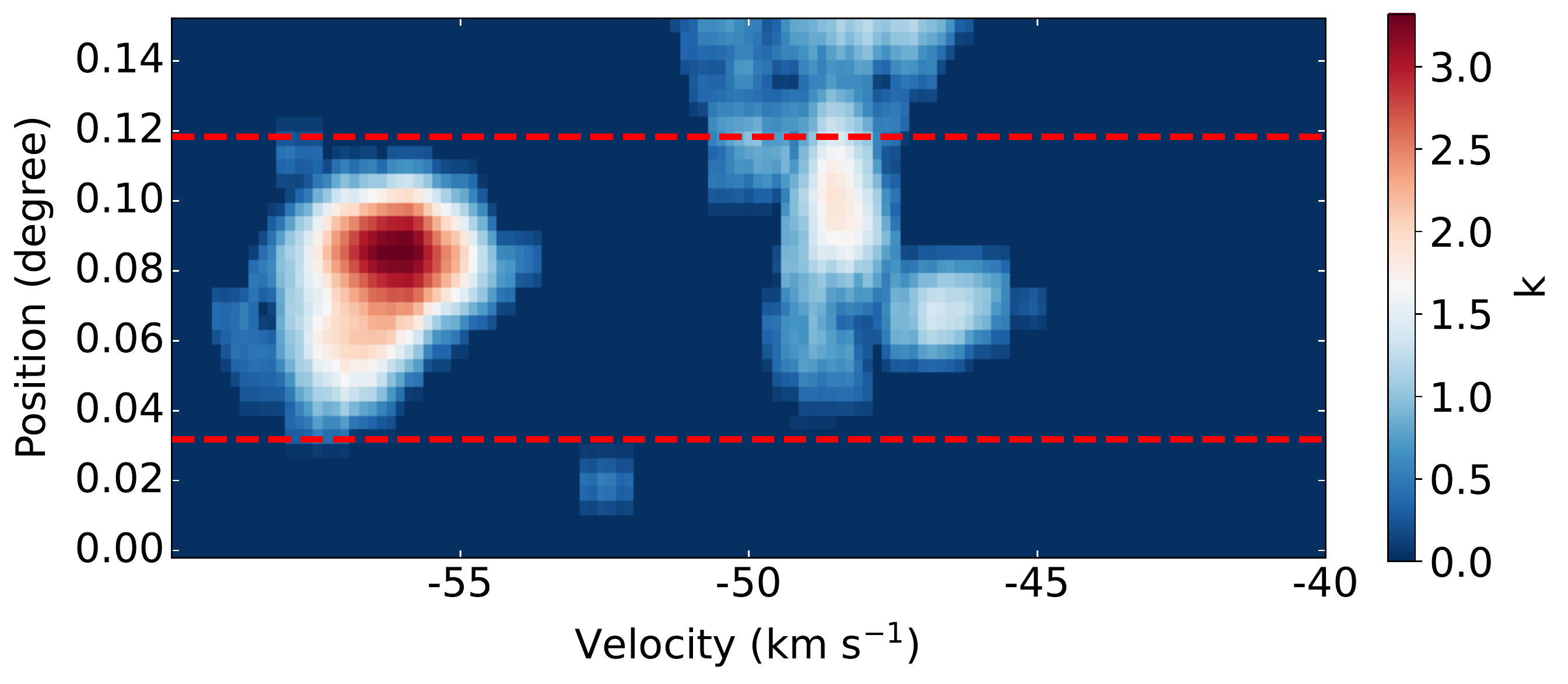}\hfill
    \\(h)
    \end{subfigure}
    \\[\smallskipamount]
    \caption{Same as Figure~\ref{Fig8} but for the G107.678${+}$00.235 region. }
    \label{FigA.5}
\end{figure}

\begin{figure}[h!]
    \centering
     \begin{subfigure}[b]{0.4\linewidth}
     \centering
    \includegraphics[trim=14.5cm 0cm 14.5cm 0cm, width=.4\textwidth]{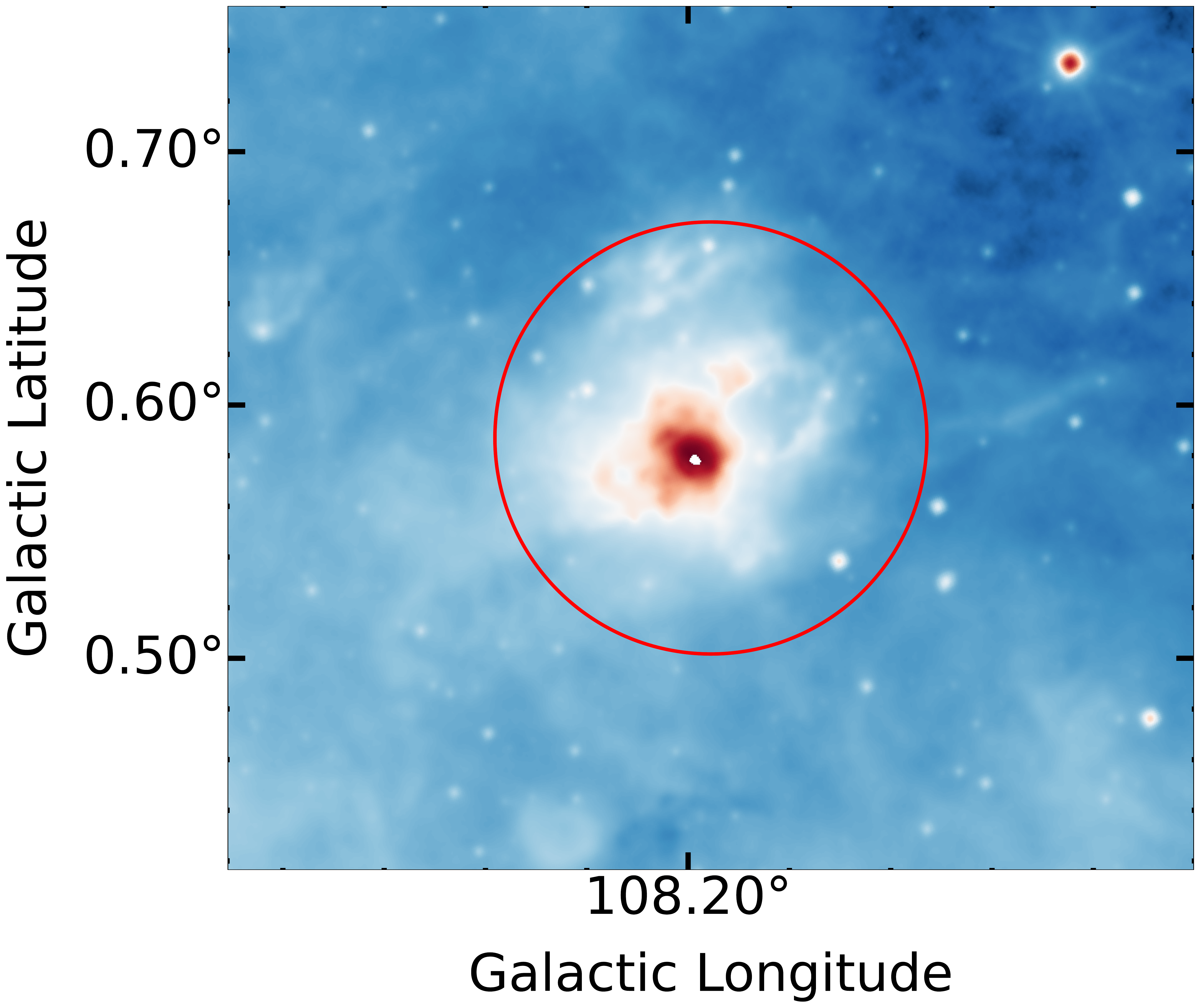}
    \\(a)
   \end{subfigure}
   \begin{subfigure}[b]{0.4\linewidth}
   \centering
    \includegraphics[trim=14.5cm 0cm 14.5cm 0cm,width=.4\textwidth]{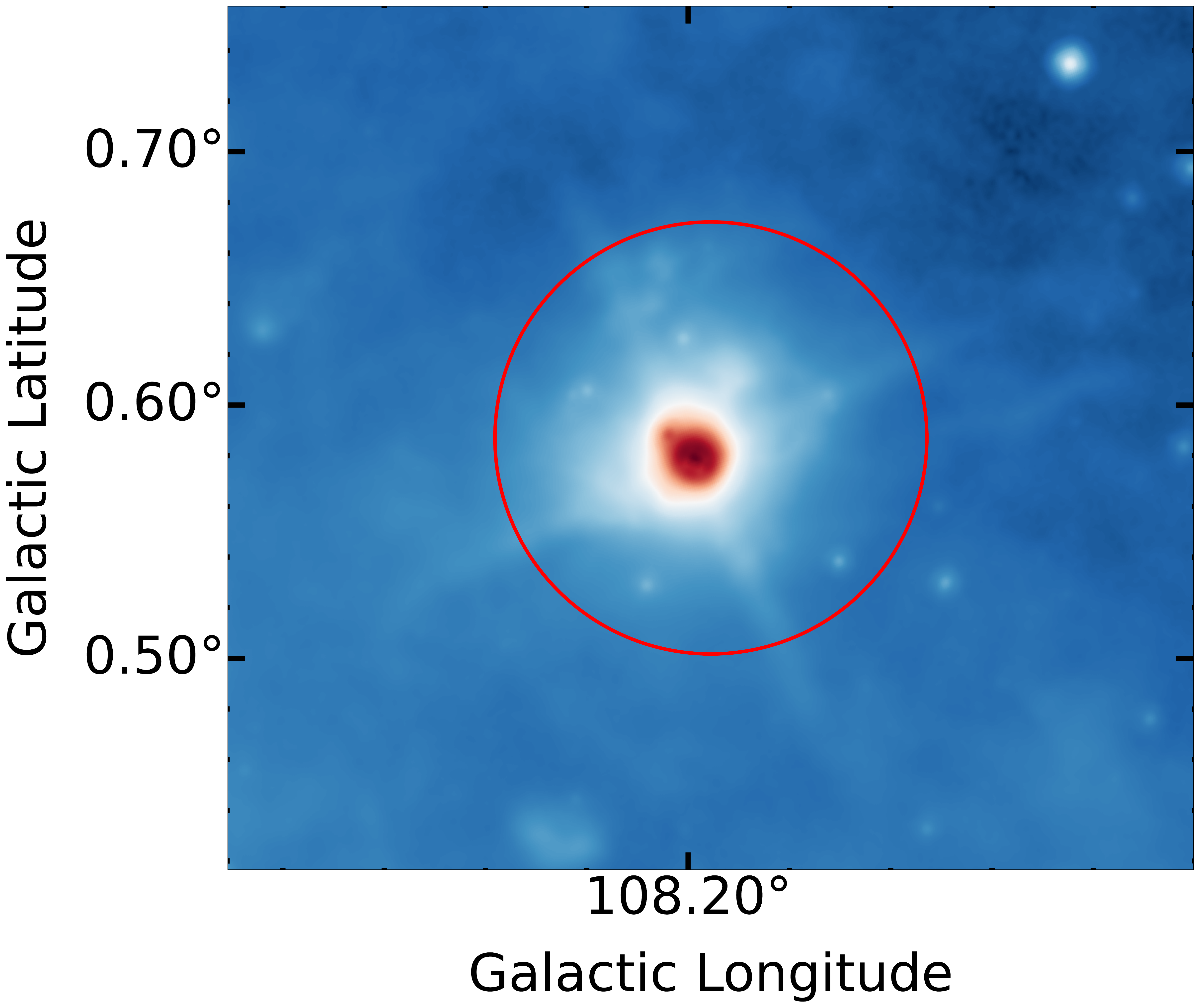}
     \\(b)
    \end{subfigure}
   \begin{subfigure}[b]{0.4\linewidth}
   \centering
    \includegraphics[trim=15cm 0cm 18cm 0cm,width=.4\textwidth]{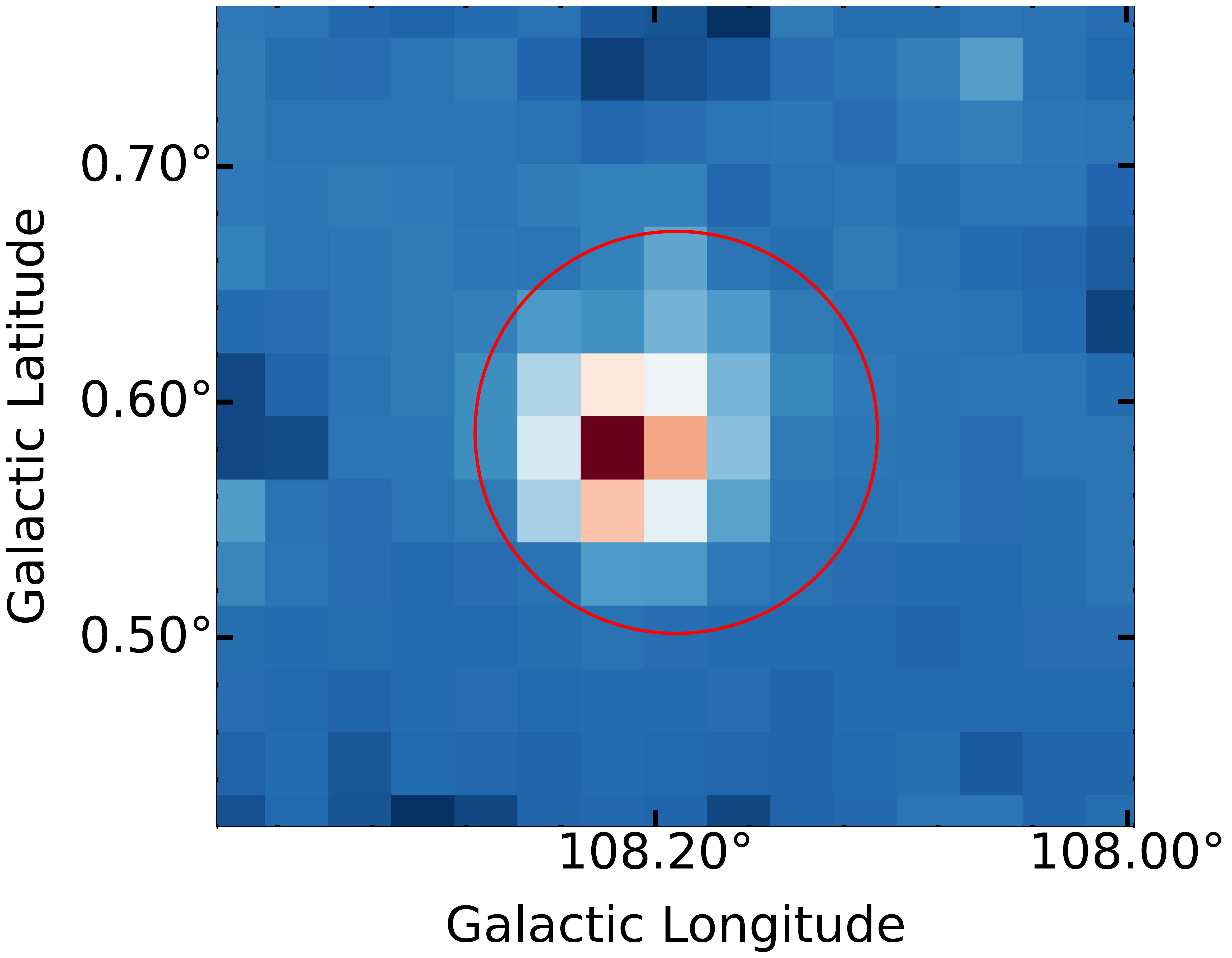}
     \\(c)
    \end{subfigure}
    \begin{subfigure}[b]{0.4\linewidth}
   \centering
    \includegraphics[trim=15.5cm 0cm 22cm 0cm,width=.4\textwidth]{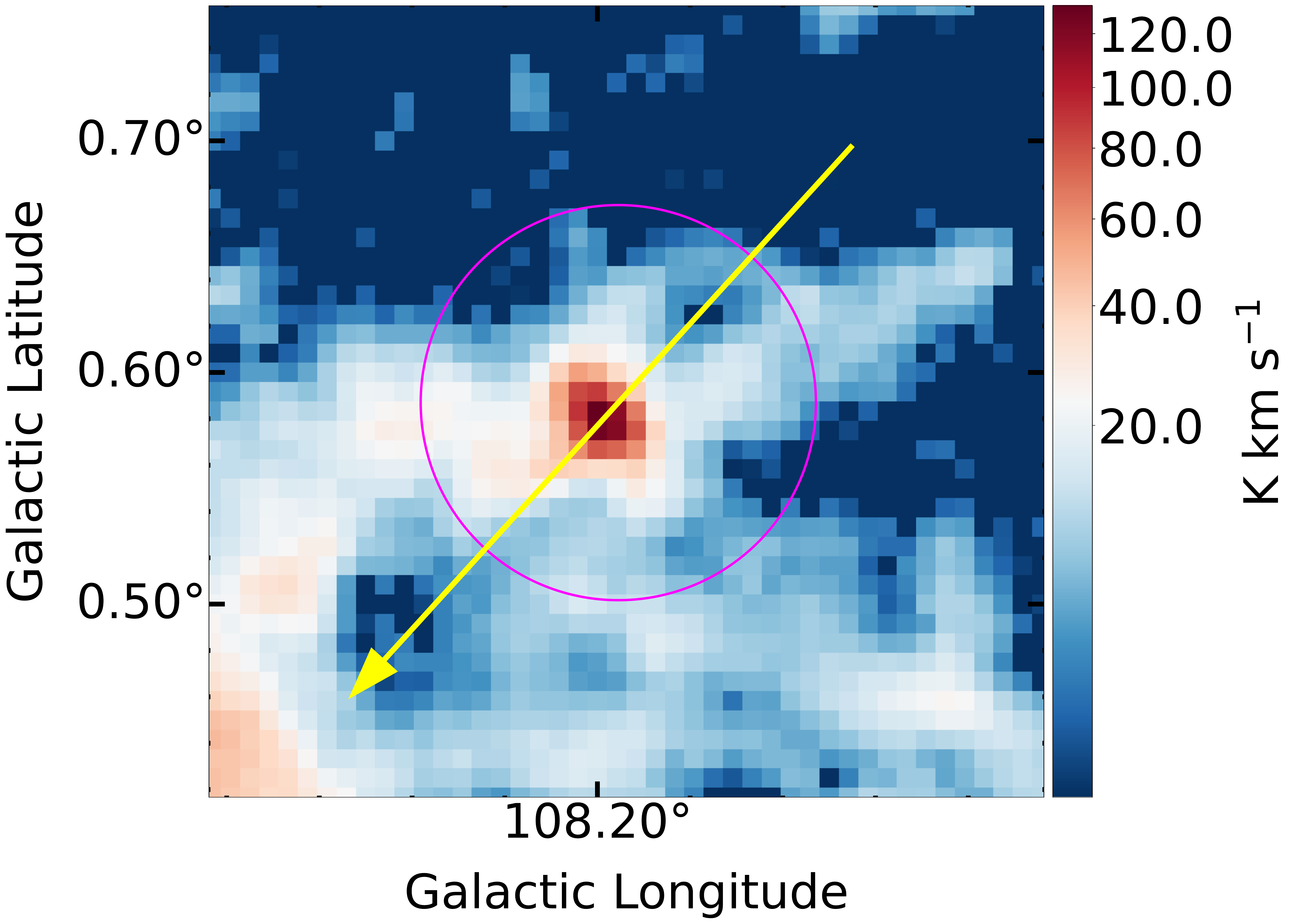}\hfill
    \\(d)
    \end{subfigure}
    \begin{subfigure}[b]{0.4\linewidth}
   \centering
    \includegraphics[trim=14cm 0cm 23cm 0cm,width=.4\textwidth]{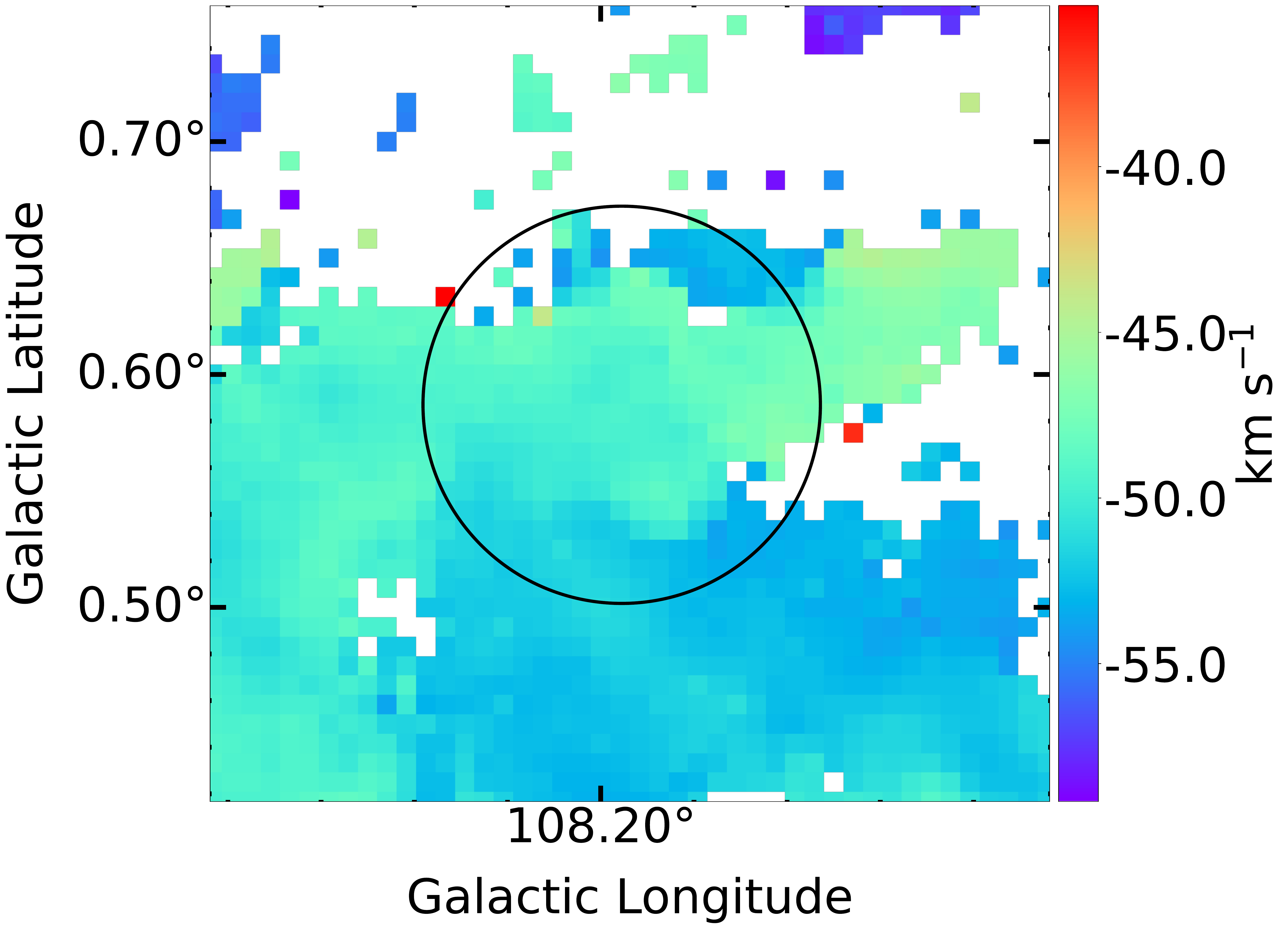}\hfill
    \\(e)
    \end{subfigure}
    \begin{subfigure}[b]{0.4\linewidth}
   \centering
    \includegraphics[trim=8.5cm 0cm 26cm 0cm,width=.4\textwidth]{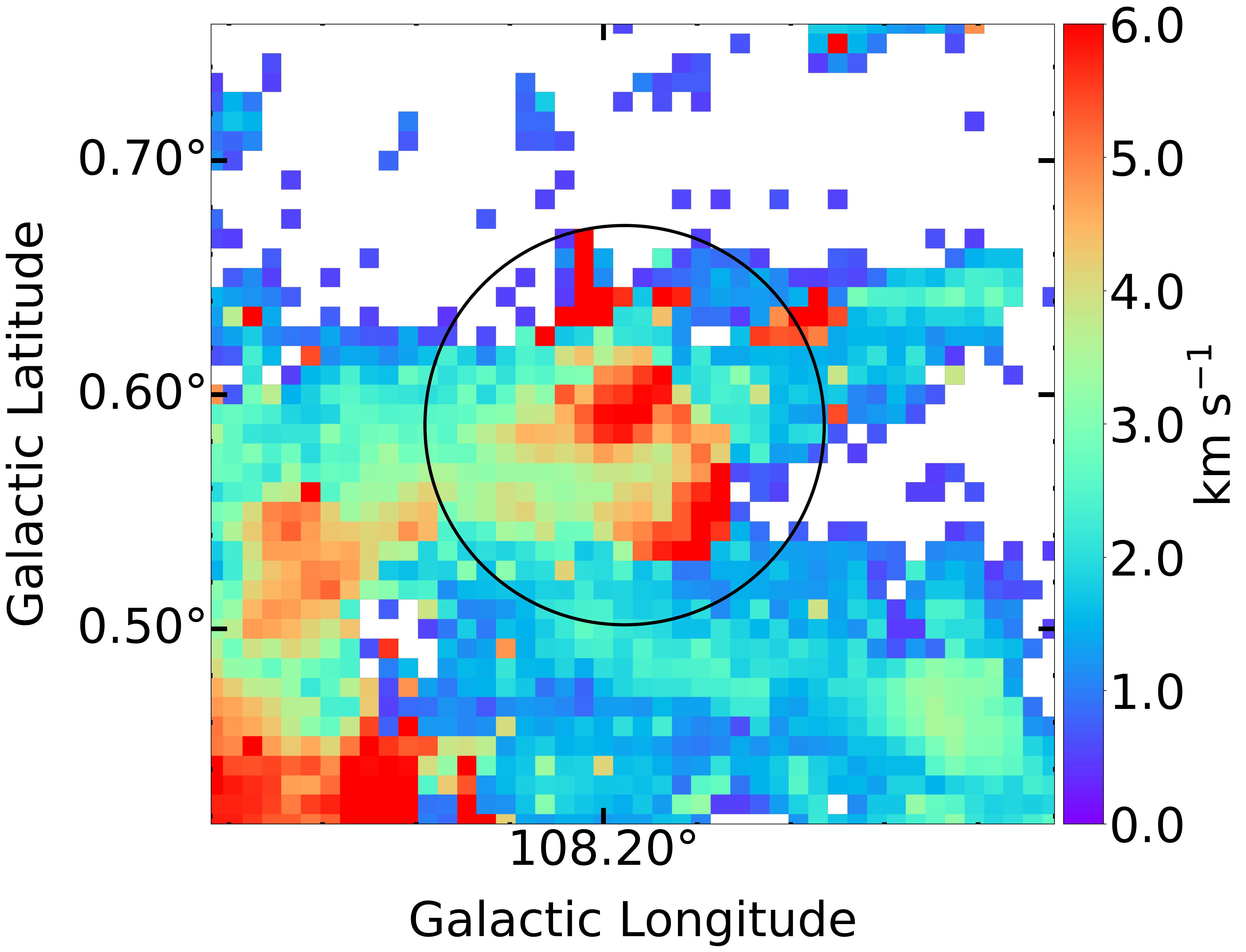}\hfill
    \\(f)
    \end{subfigure}
    \begin{subfigure}[b]{0.4\linewidth}
   \centering
    \includegraphics[trim=14.5cm 0cm 22cm 0cm, width=.4\textwidth]{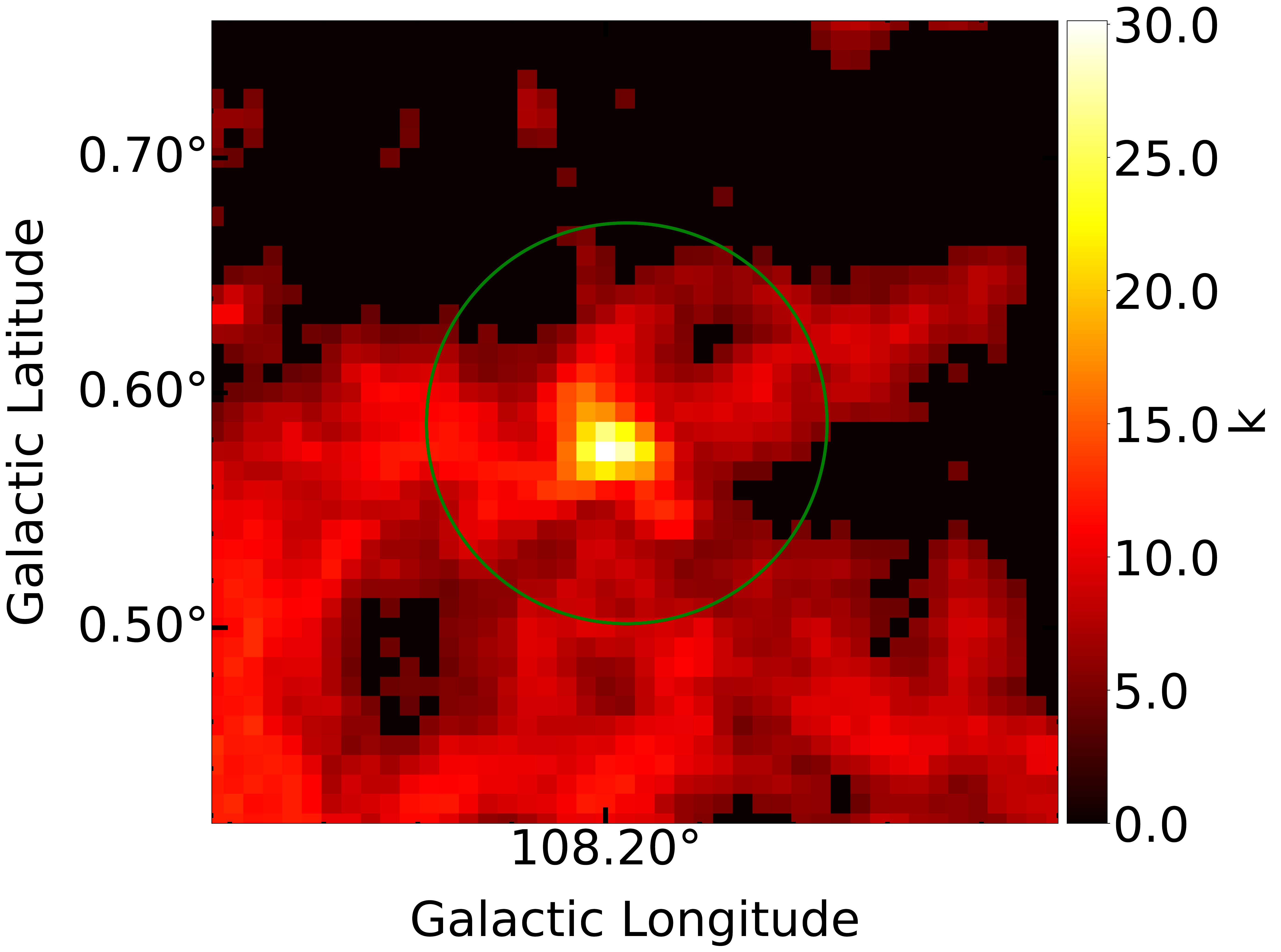}\hfill
    \\(g)
    \end{subfigure}
    \begin{subfigure}[b]{0.4\linewidth}
   \centering
    \includegraphics[trim=1cm -2cm 15cm 0cm, width=.4\textwidth]{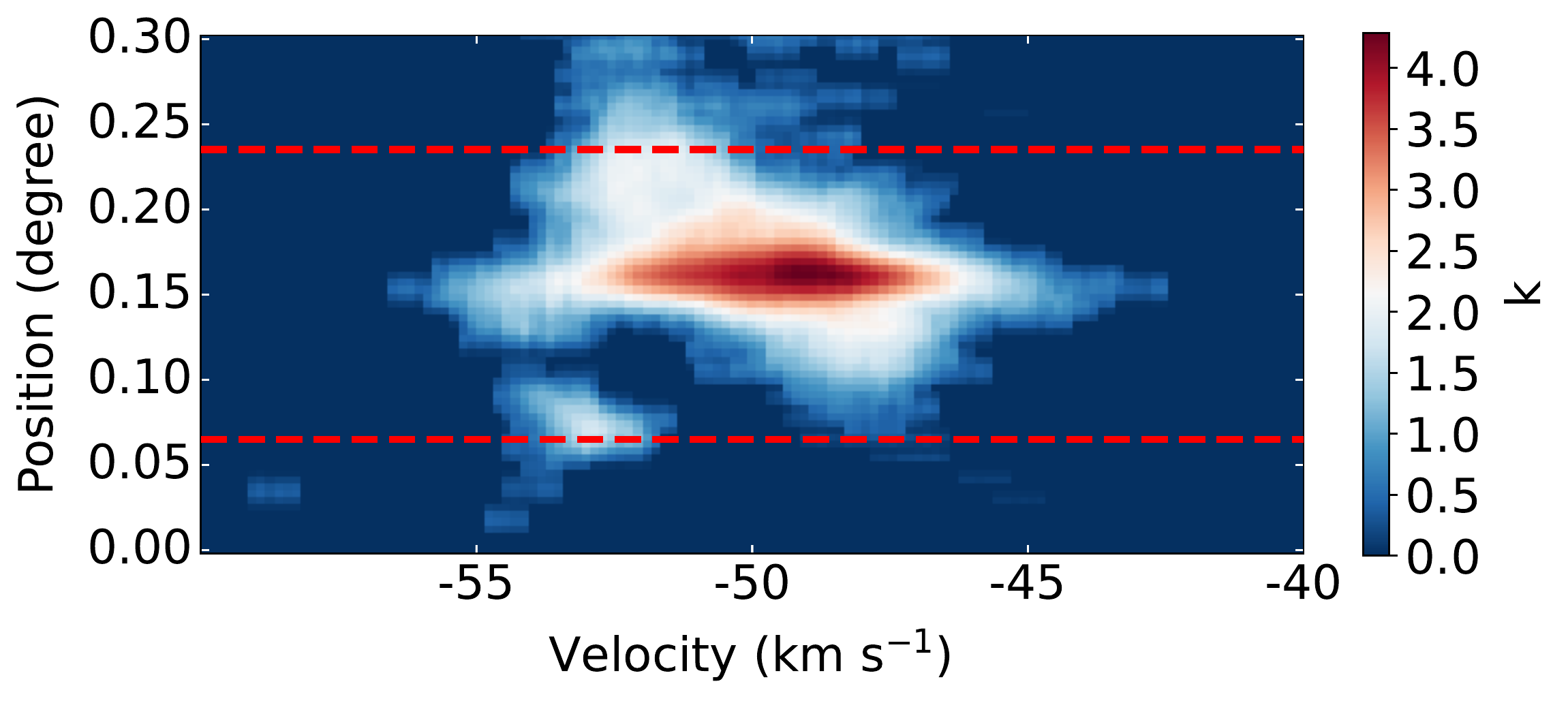}\hfill
    \\(h)
    \end{subfigure}
    \\[\smallskipamount]
    \caption{Same as Figure~\ref{Fig8} but for the S146 region. }
    \label{FigA.6}
\end{figure}

\begin{figure}[h!]
    \centering
     \begin{subfigure}[b]{0.4\linewidth}
     \centering
    \includegraphics[trim=15.5cm 0cm 16.5cm 0cm, width=.4\textwidth]{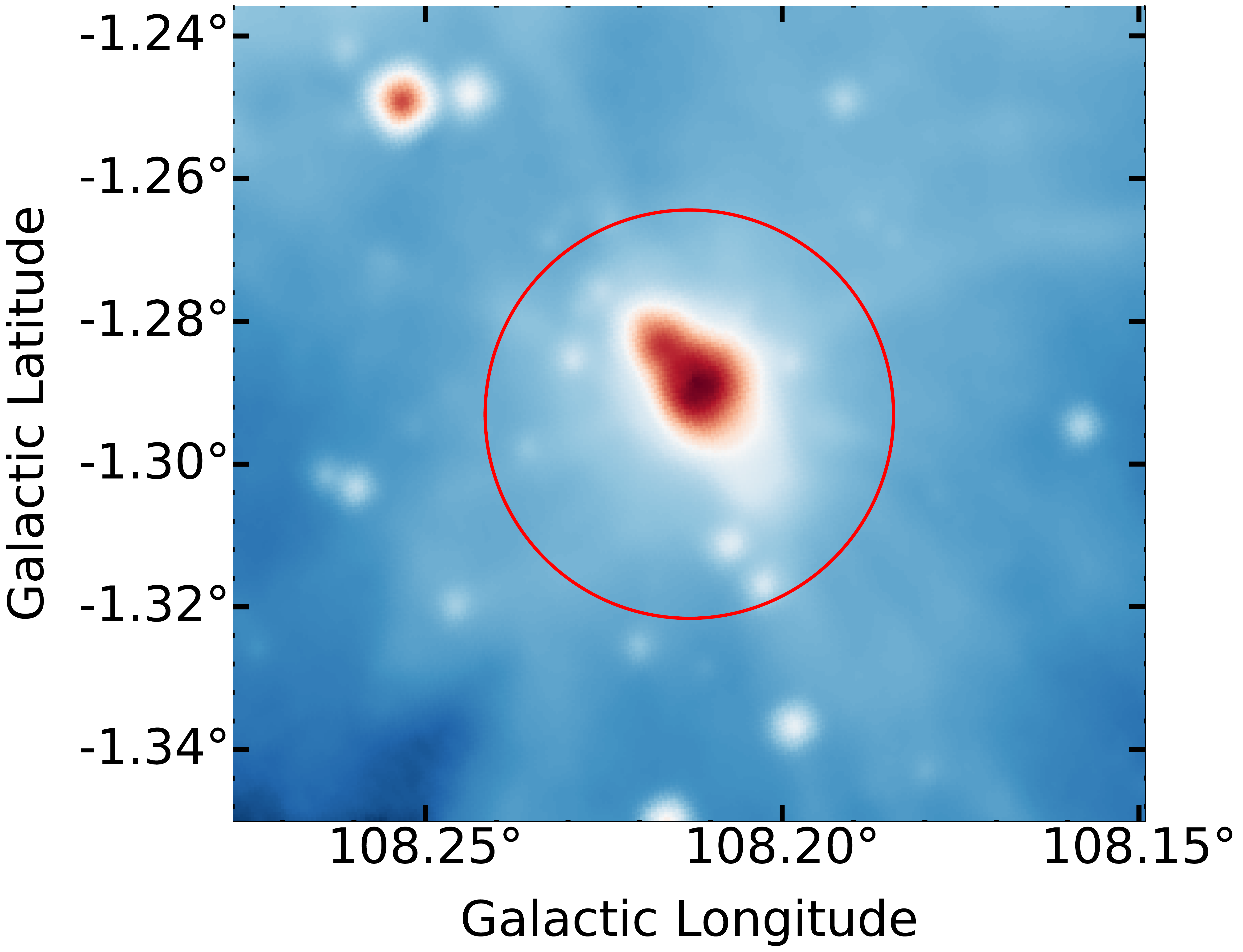}
    \\(a)
   \end{subfigure}
   \begin{subfigure}[b]{0.4\linewidth}
   \centering
    \includegraphics[trim=15.5cm 0cm 17cm 0cm,width=.4\textwidth]{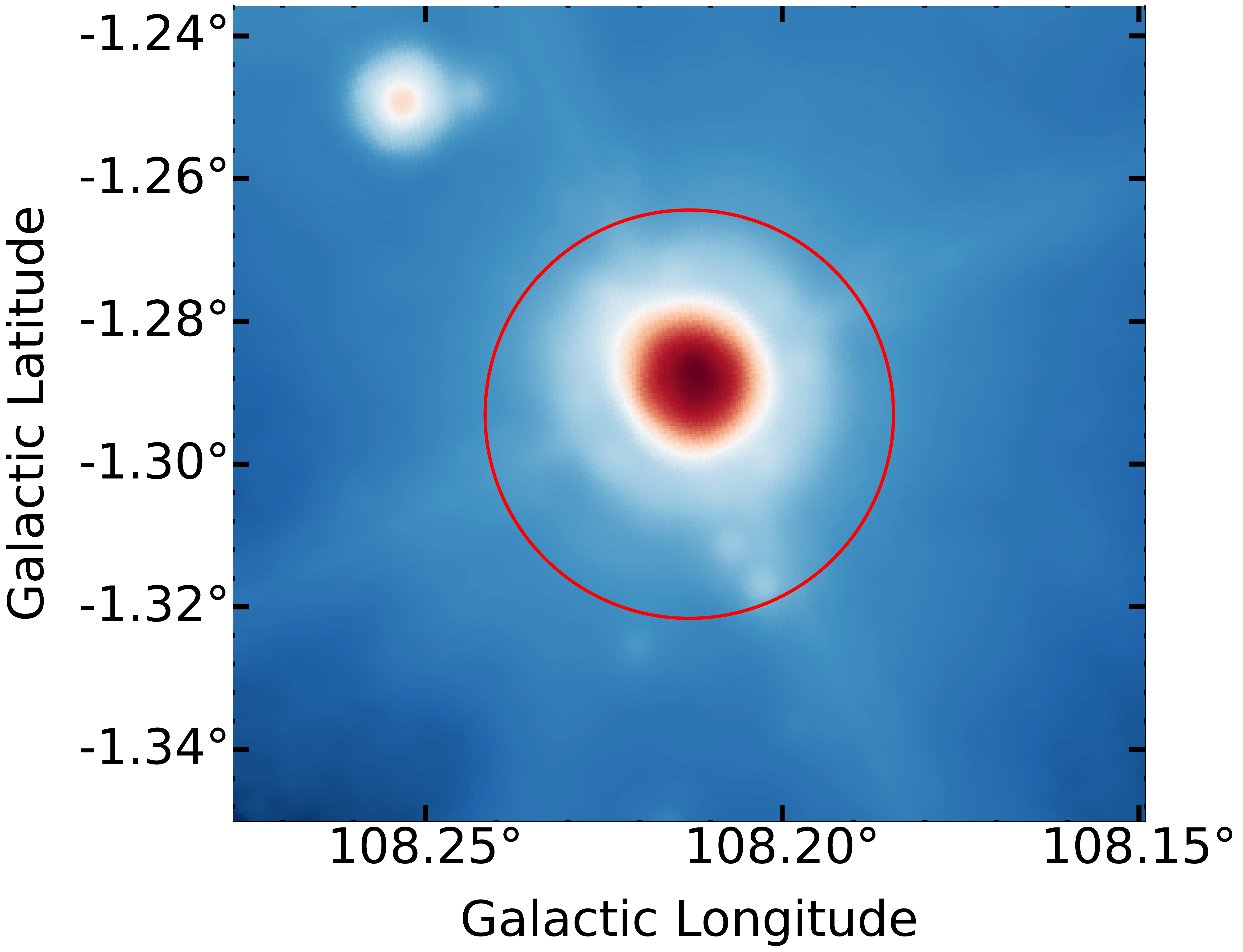}
     \\(b)
    \end{subfigure}
   \begin{subfigure}[b]{0.4\linewidth}
   \centering
    \includegraphics[trim=14.5cm 0cm 14.5cm 0cm,width=.4\textwidth]{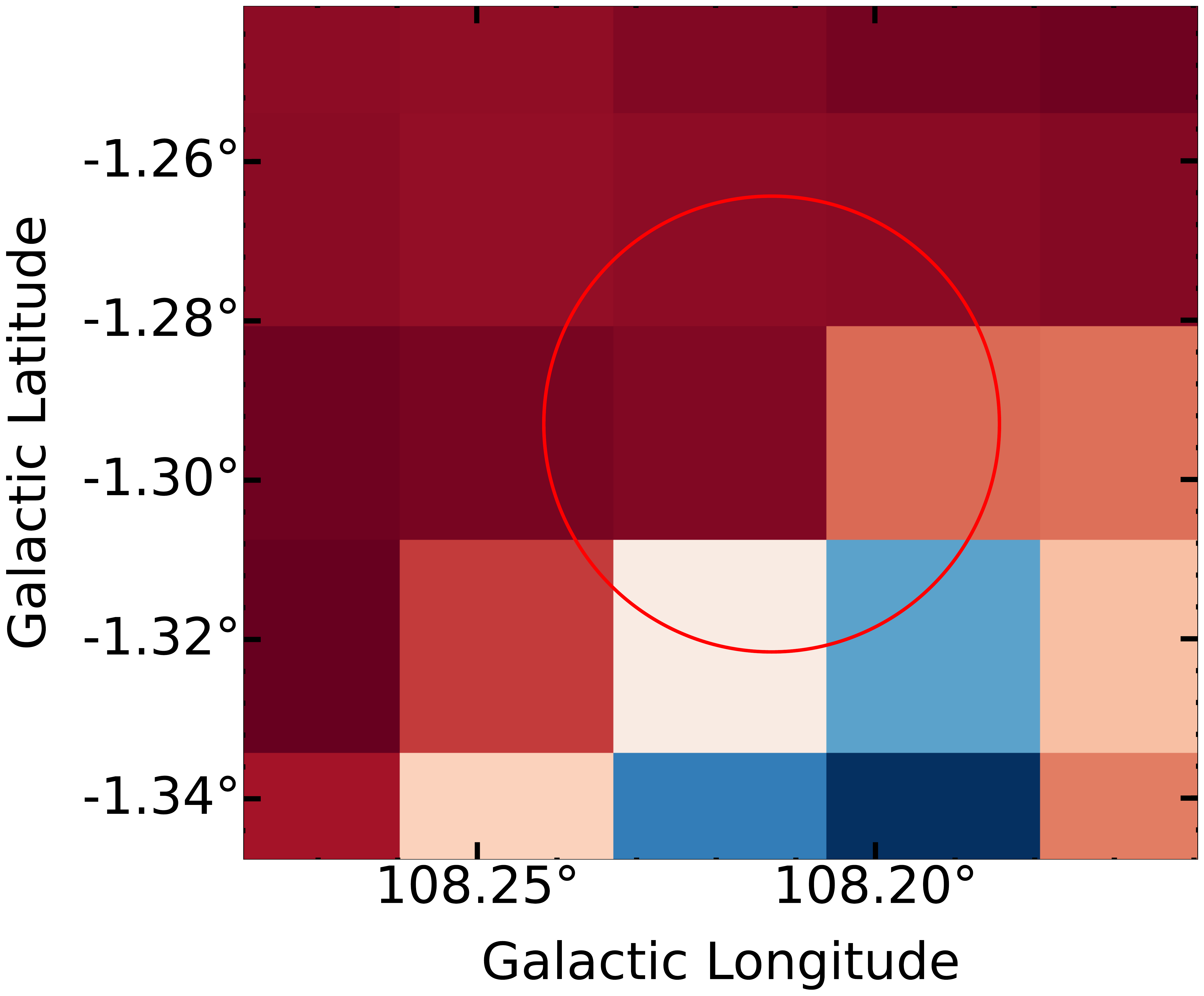}
     \\(c)
    \end{subfigure}
    \begin{subfigure}[b]{0.4\linewidth}
   \centering
    \includegraphics[trim=15cm 0cm 21.5cm 0cm,width=.4\textwidth]{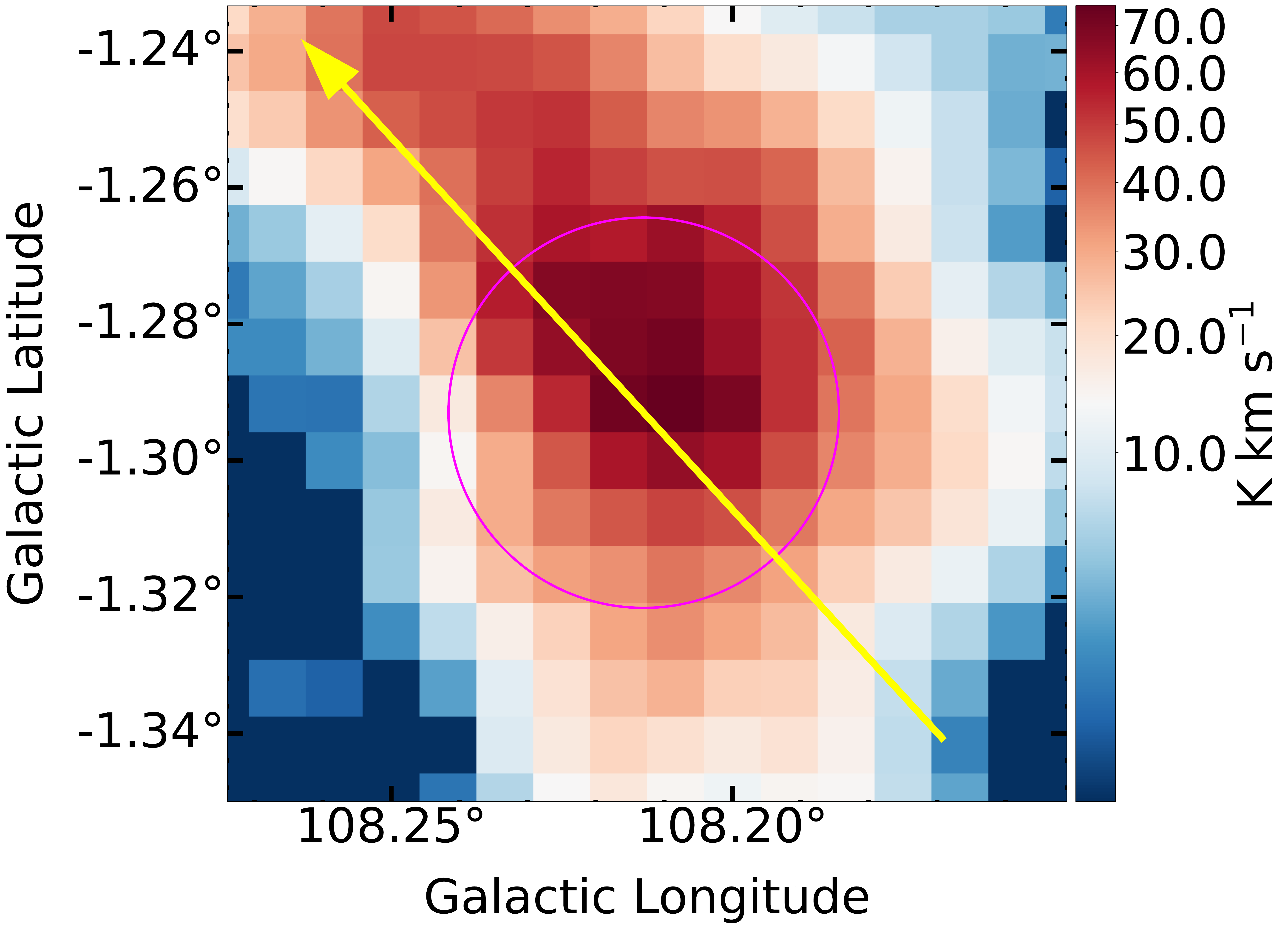}\hfill
    \\(d)
    \end{subfigure}
    \begin{subfigure}[b]{0.4\linewidth}
   \centering
    \includegraphics[trim=14cm 0cm 23cm 0cm,width=.4\textwidth]{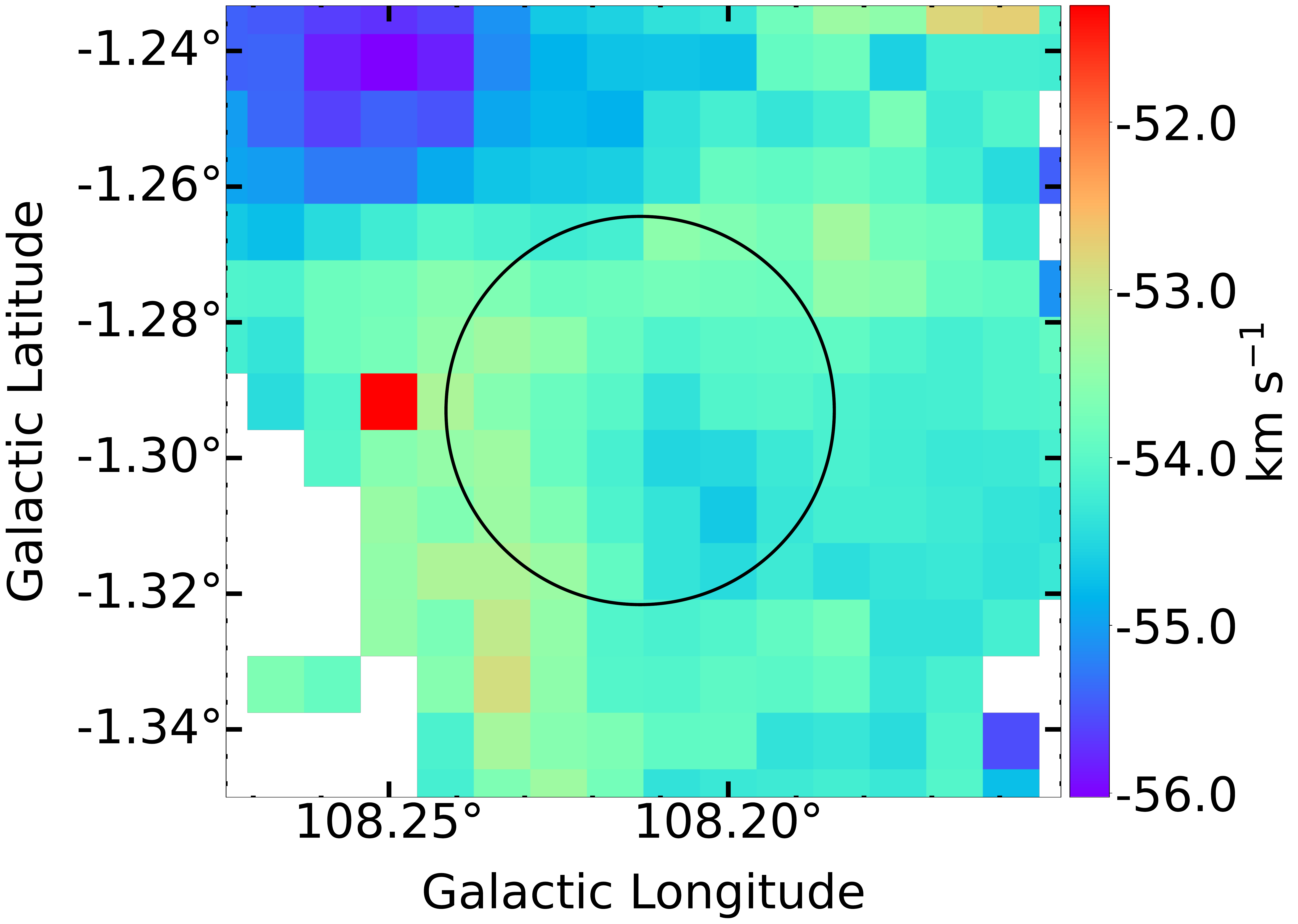}\hfill
    \\(e)
    \end{subfigure}
    \begin{subfigure}[b]{0.4\linewidth}
   \centering
    \includegraphics[trim=8.5cm 0cm 26cm 0cm,width=.4\textwidth]{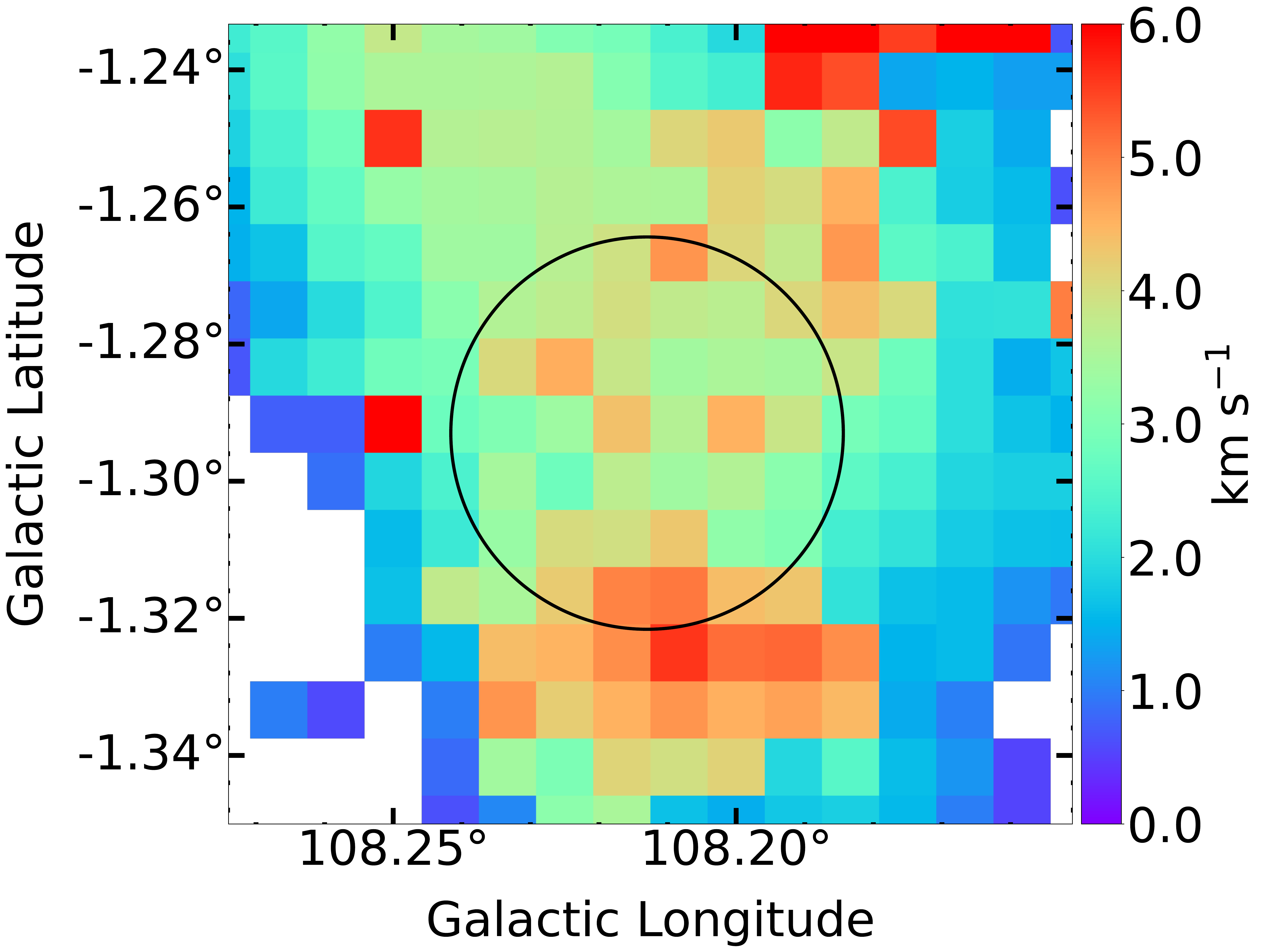}\hfill
    \\(f)
    \end{subfigure}
    \begin{subfigure}[b]{0.4\linewidth}
   \centering
    \includegraphics[trim=14.5cm 0cm 22cm 0cm, width=.4\textwidth]{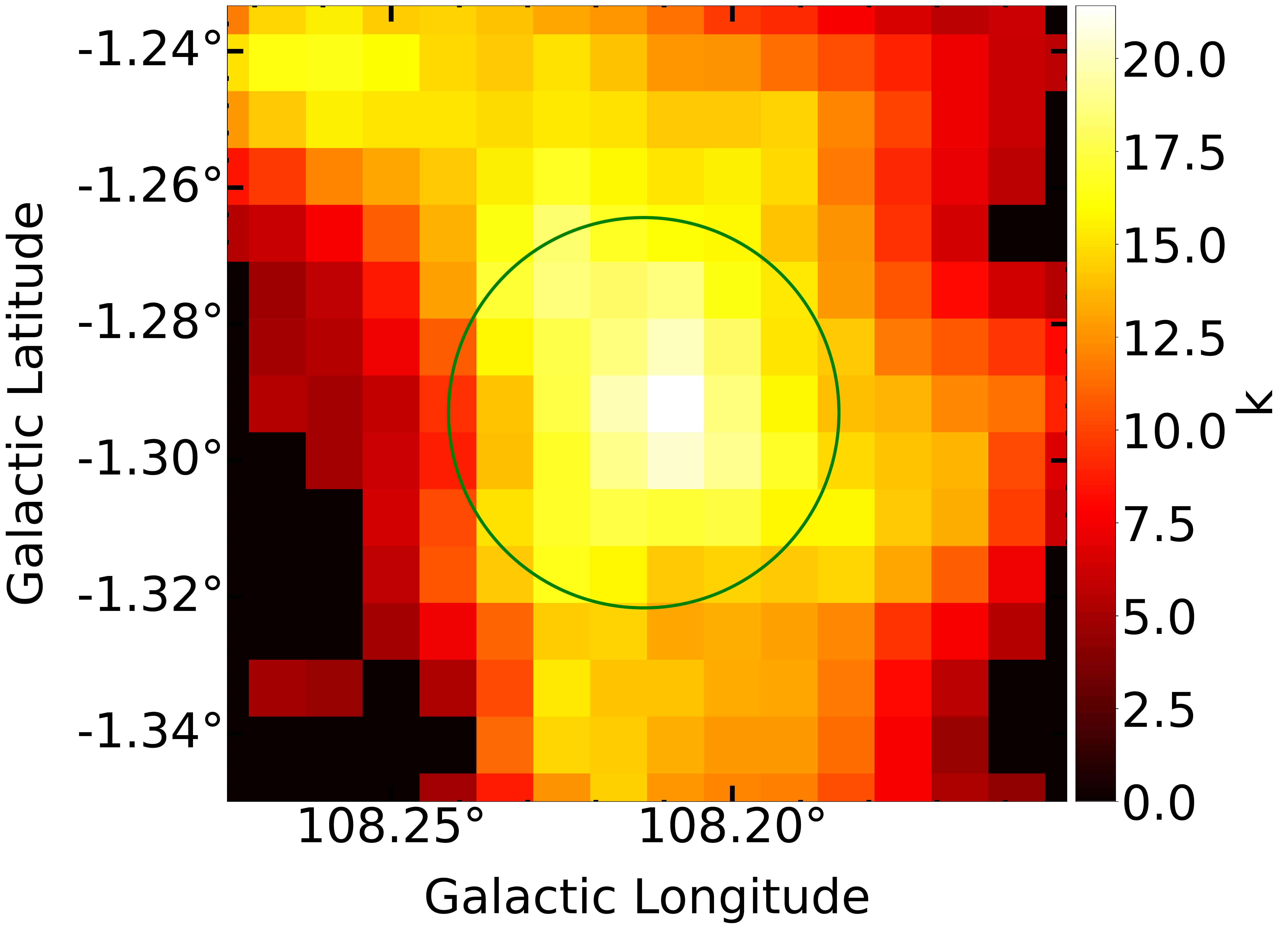}\hfill
    \\(g)
    \end{subfigure}
    \begin{subfigure}[b]{0.4\linewidth}
   \centering
    \includegraphics[trim=1cm -2cm 16cm 0cm, width=.4\textwidth]{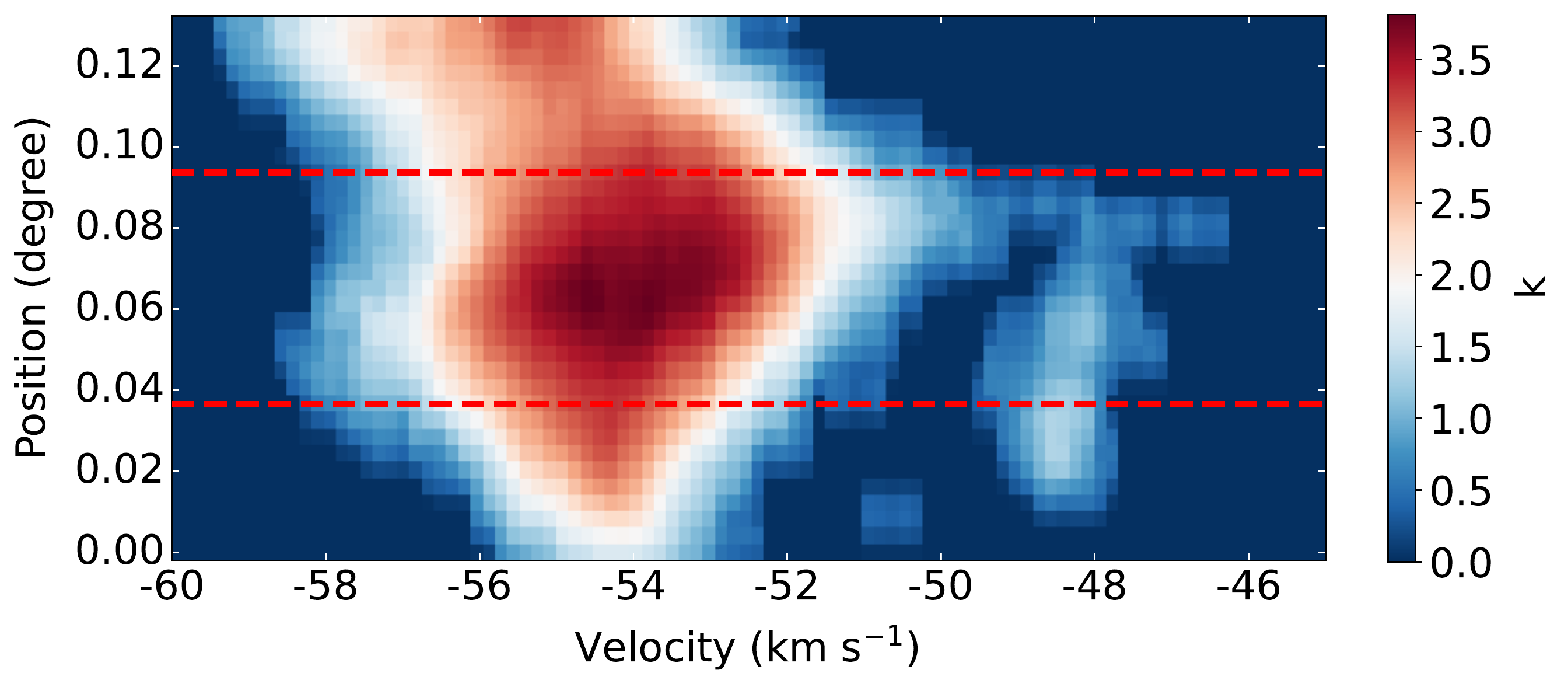}\hfill
    \\(h)
    \end{subfigure}
    \\[\smallskipamount]
    \caption{Same as Figure~\ref{Fig8} but for the G108.213${-}$01.293 region. }
    \label{FigA.7}
\end{figure}

\begin{figure}[h!]
    \centering
     \begin{subfigure}[b]{0.4\linewidth}
     \centering
    \includegraphics[trim=15cm 0cm 15cm 0cm, width=.4\textwidth]{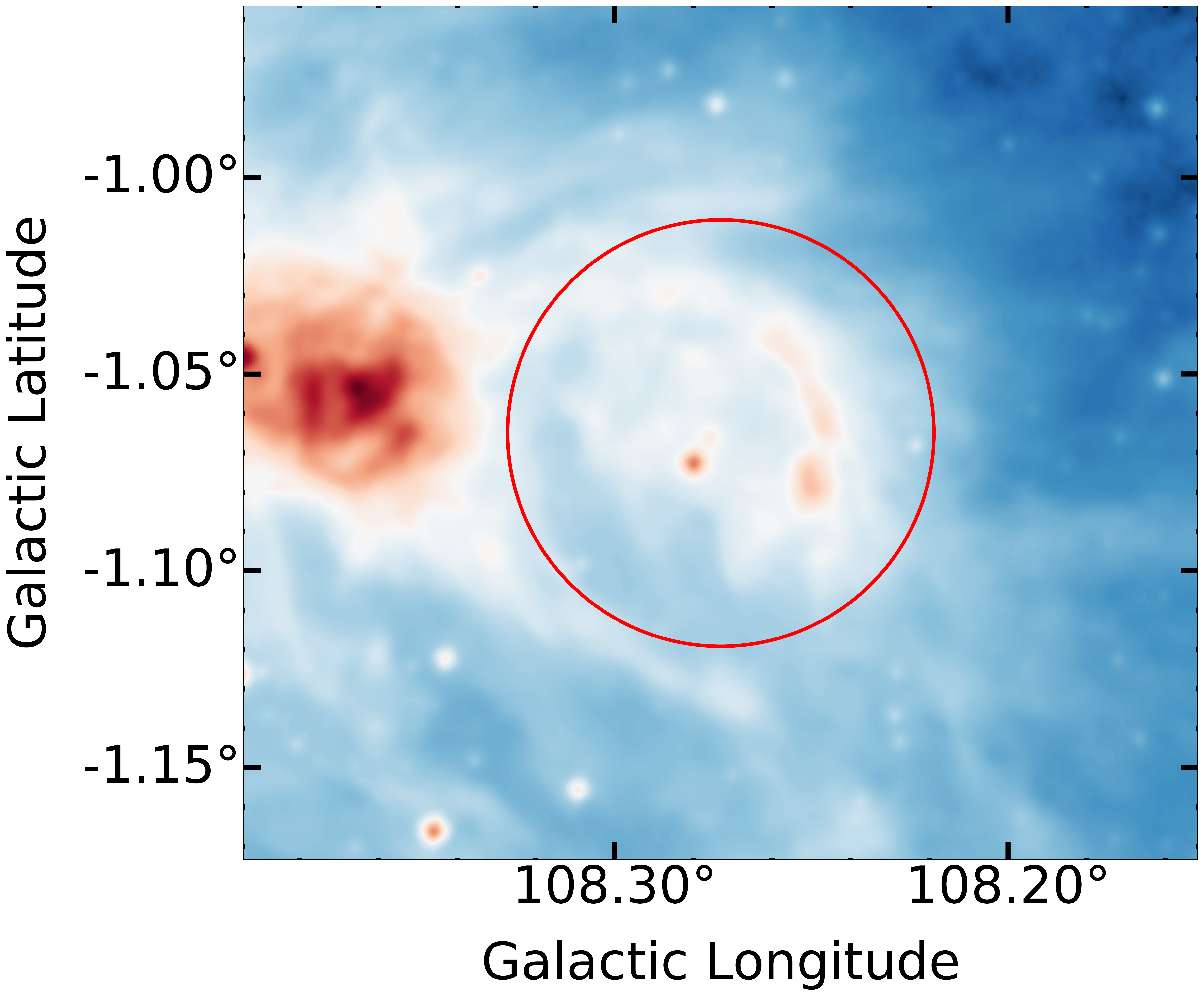}
    \\(a)
   \end{subfigure}
   \begin{subfigure}[b]{0.4\linewidth}
   \centering
    \includegraphics[trim=15cm 0cm 15cm 0cm,width=.4\textwidth]{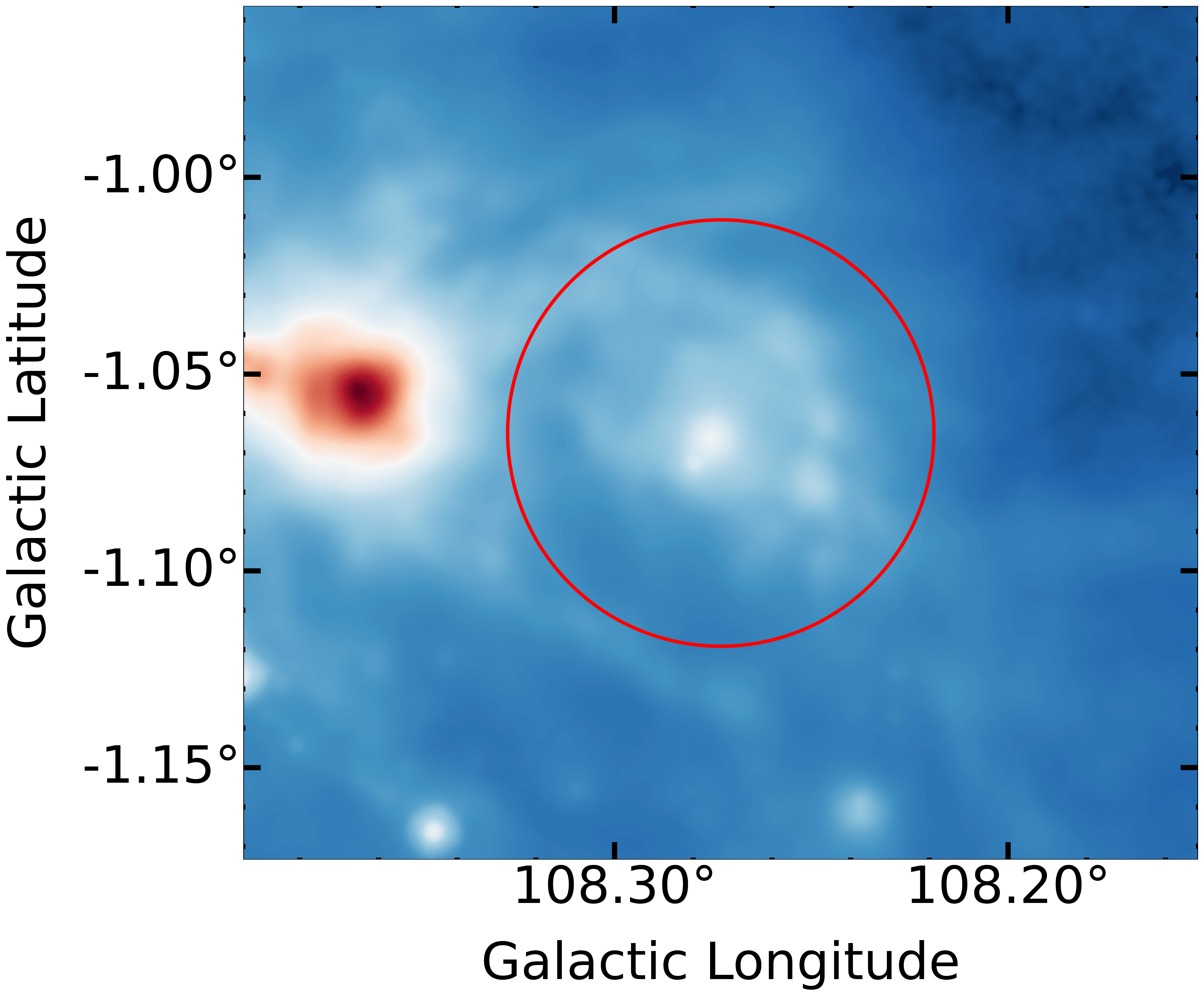}
     \\(b)
    \end{subfigure}
   \begin{subfigure}[b]{0.4\linewidth}
   \centering
    \includegraphics[trim=15cm 0cm 15cm 0cm,width=.4\textwidth]{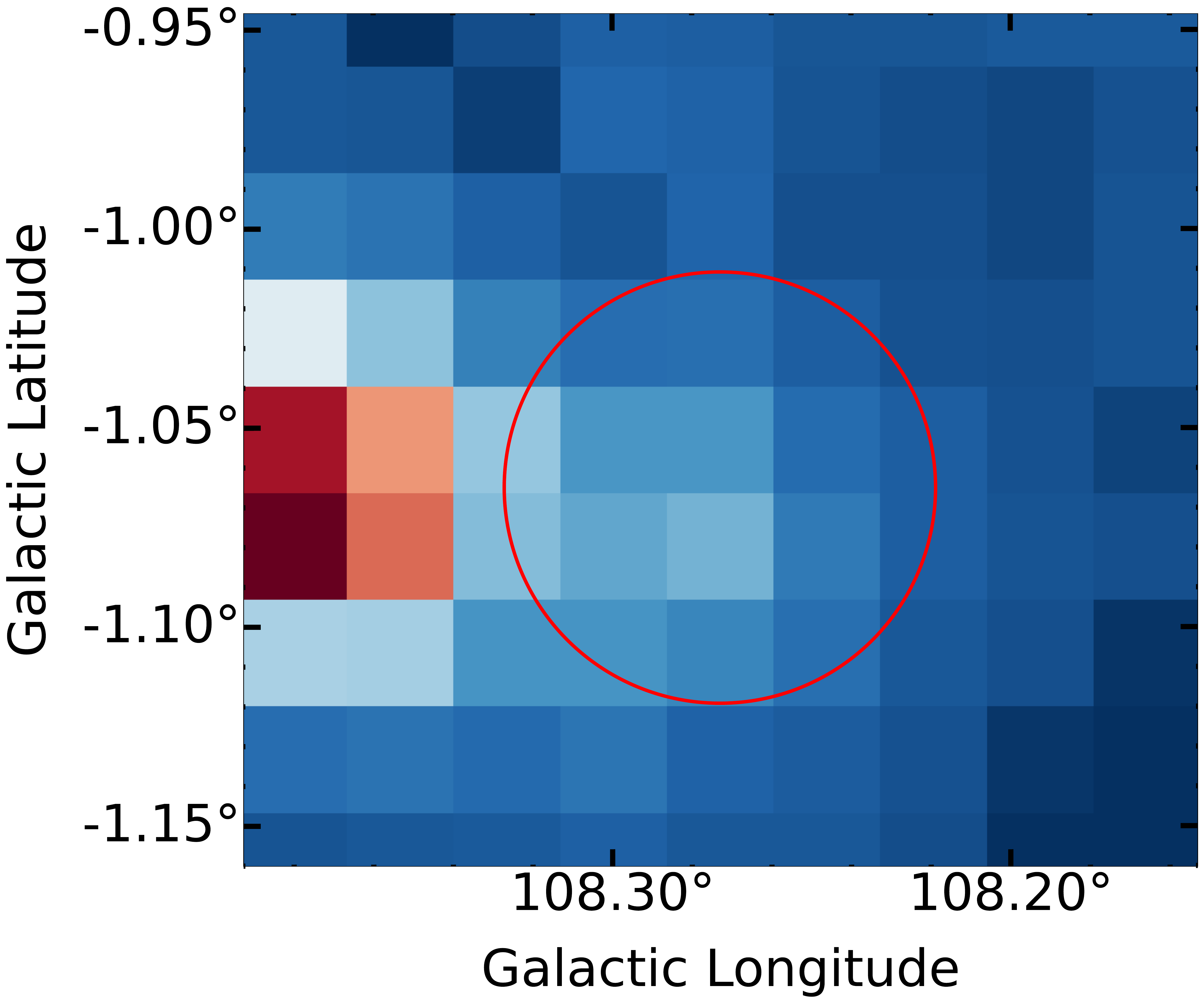}
     \\(c)
    \end{subfigure}
    \begin{subfigure}[b]{0.4\linewidth}
   \centering
    \includegraphics[trim=15cm 0cm 22.5cm 0cm,width=.4\textwidth]{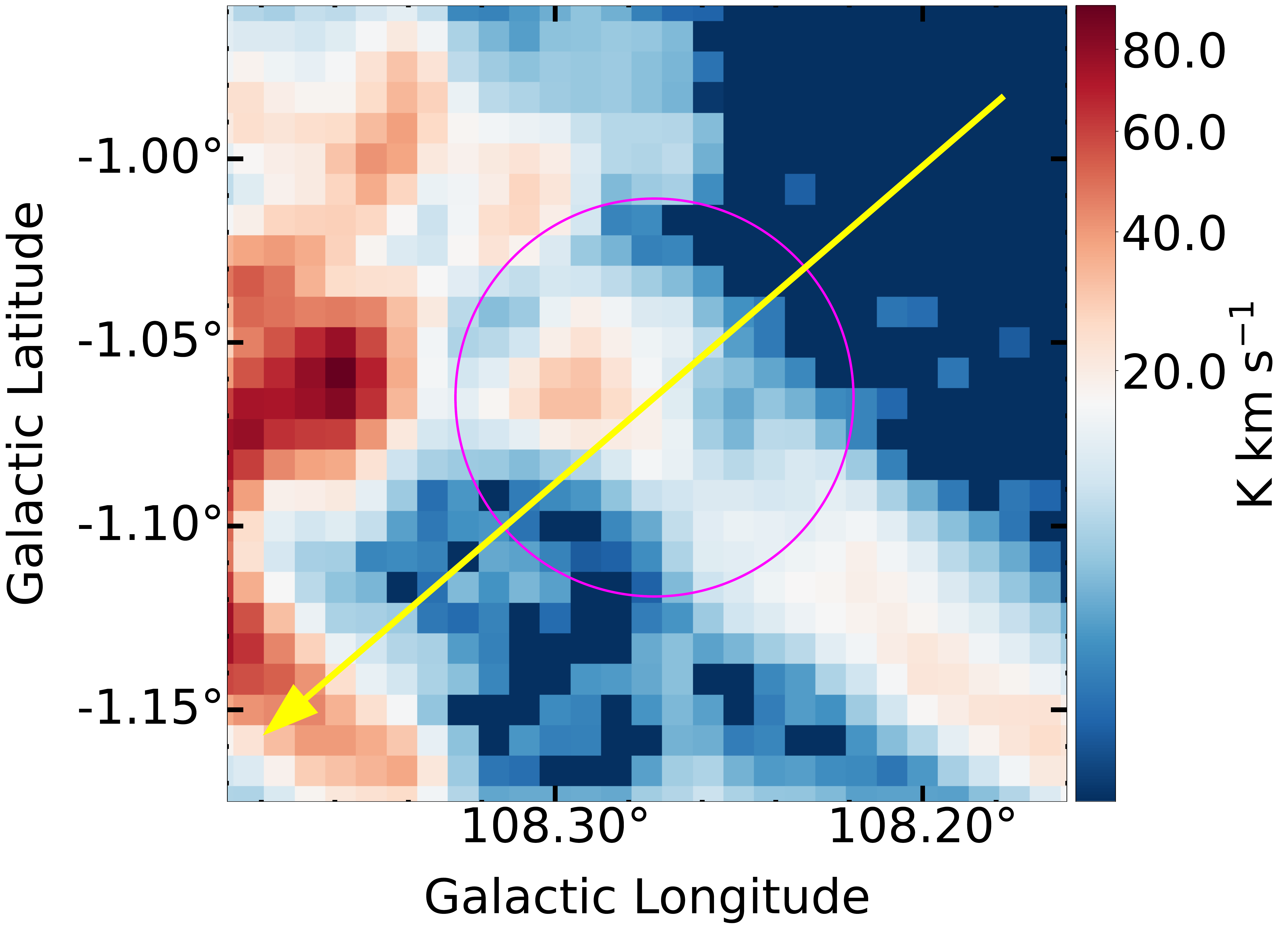}\hfill
    \\(d)
    \end{subfigure}
    \begin{subfigure}[b]{0.4\linewidth}
   \centering
    \includegraphics[trim=14cm 0cm 23cm 0cm,width=.4\textwidth]{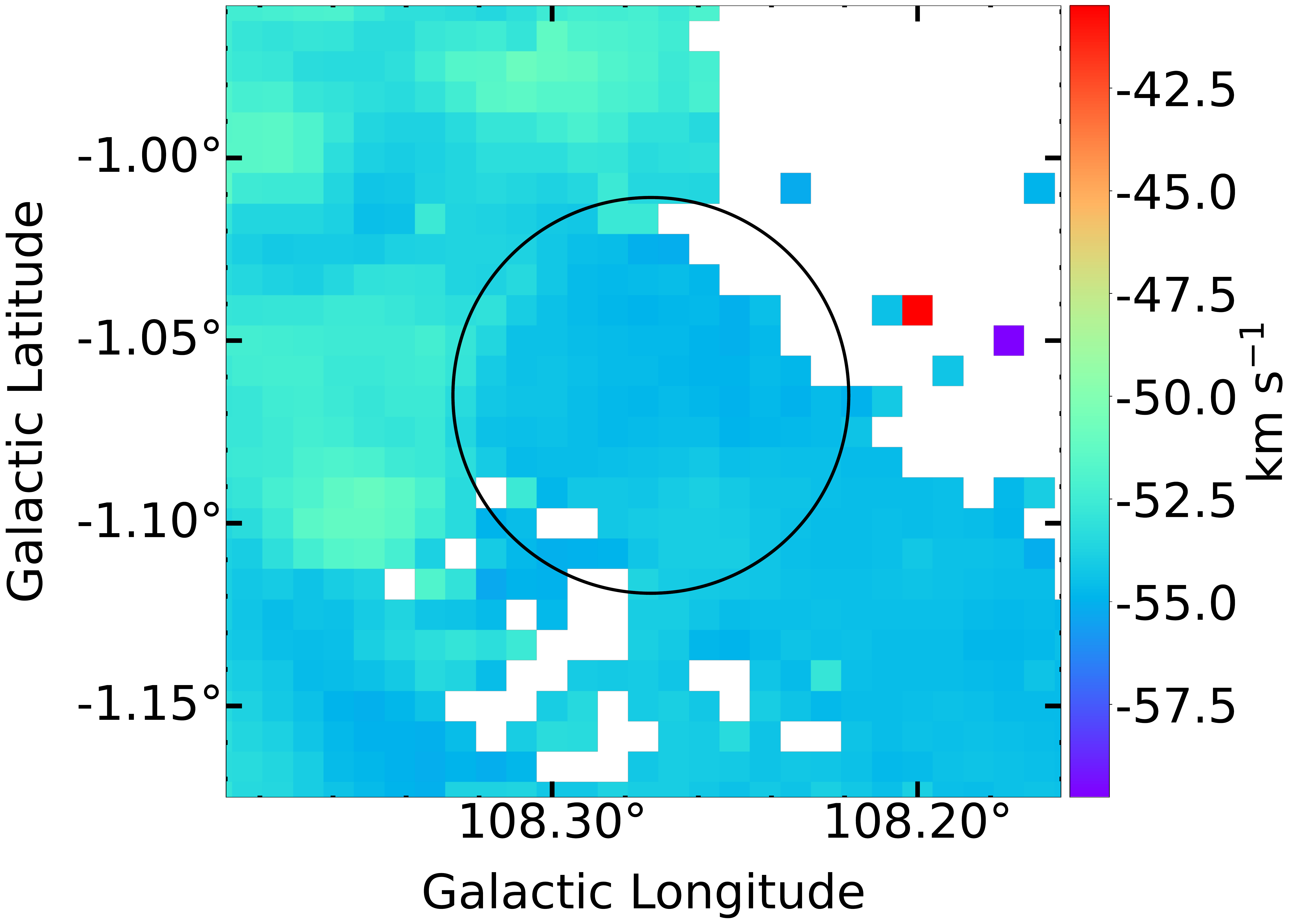}\hfill
    \\(e)
    \end{subfigure}
    \begin{subfigure}[b]{0.4\linewidth}
   \centering
    \includegraphics[trim=8.5cm 0cm 26cm 0cm,width=.4\textwidth]{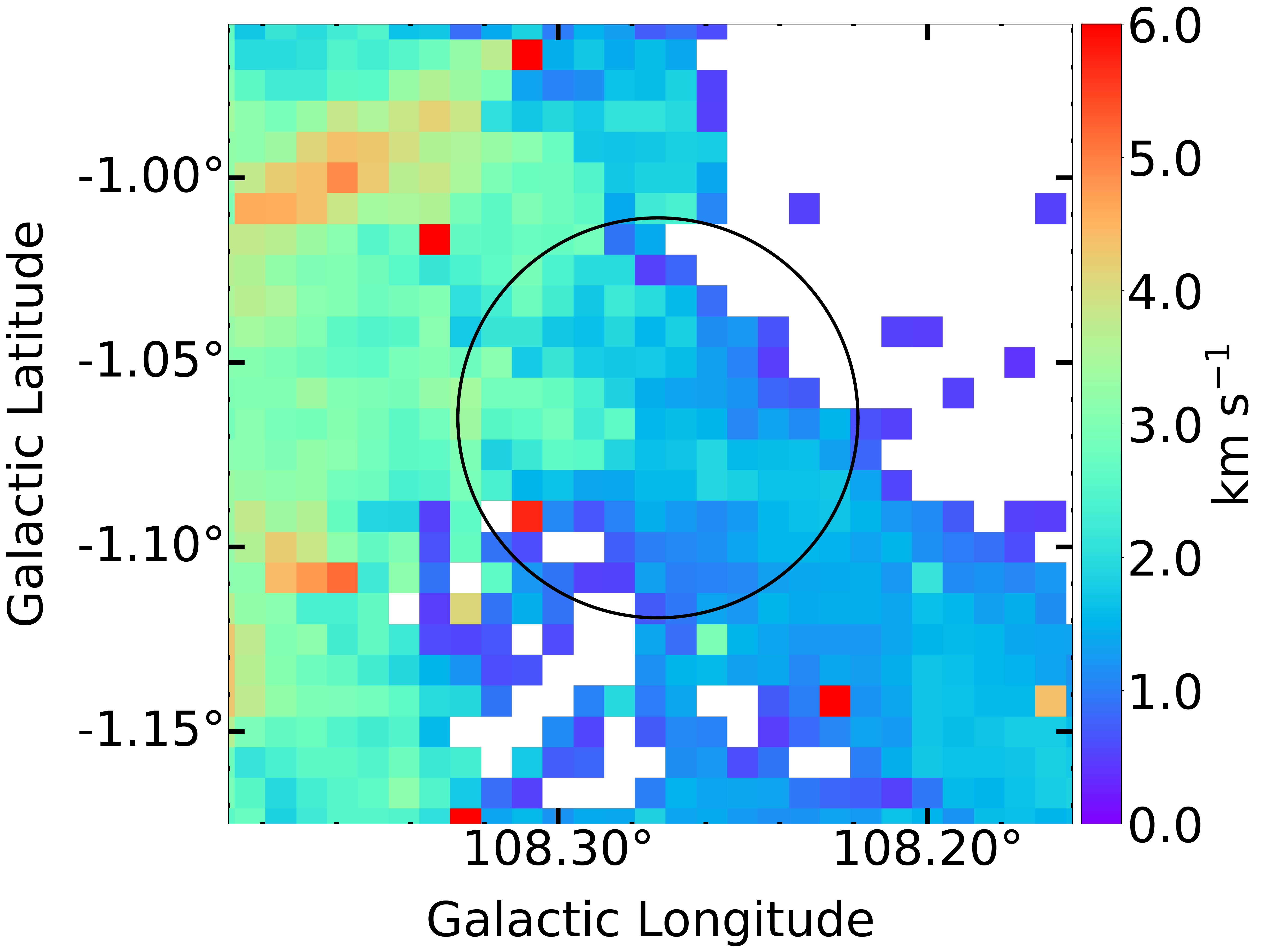}\hfill
    \\(f)
    \end{subfigure}
    \begin{subfigure}[b]{0.4\linewidth}
   \centering
    \includegraphics[trim=14.5cm 0cm 22cm 0cm, width=.4\textwidth]{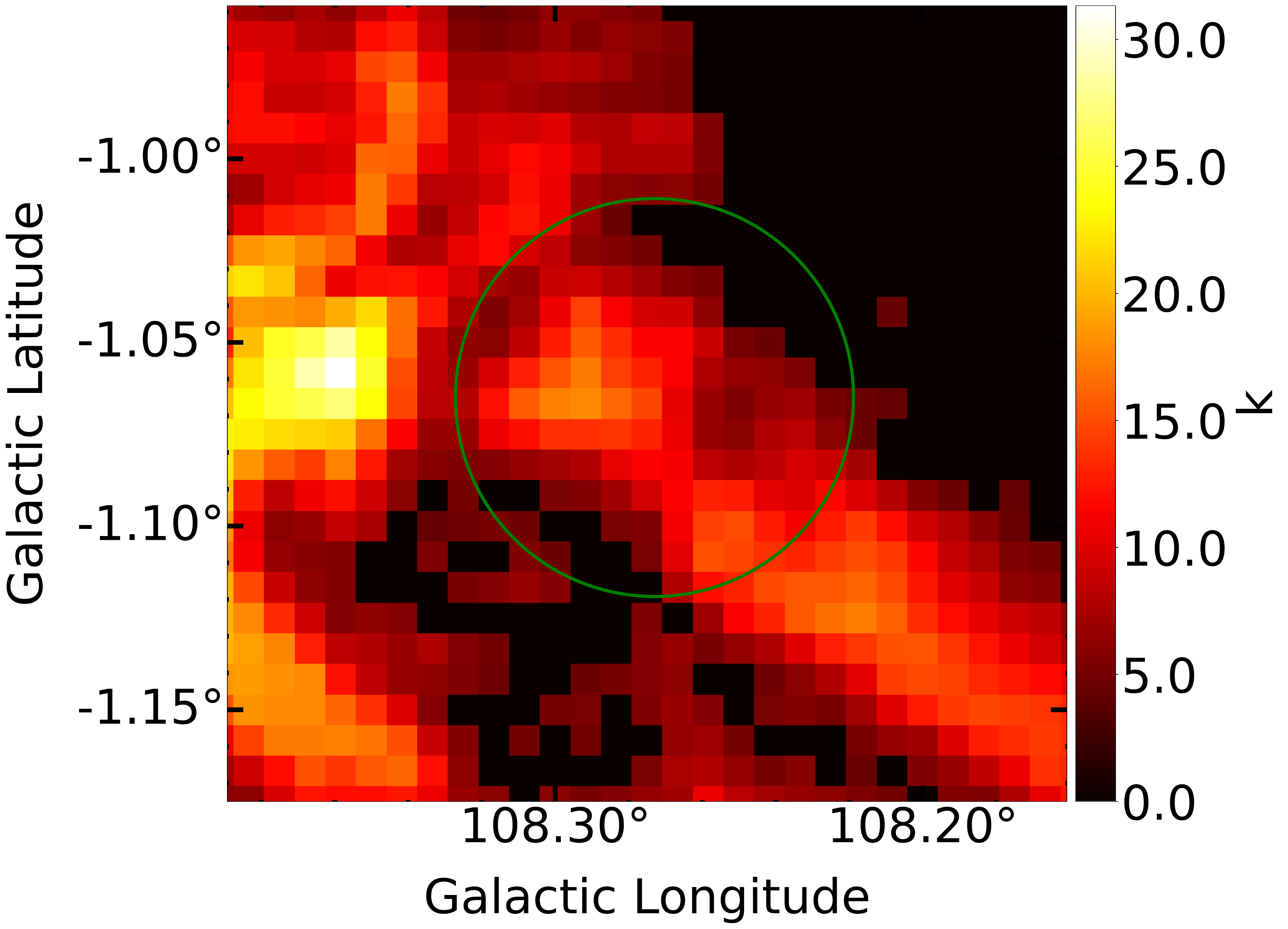}\hfill
    \\(g)
    \end{subfigure}
    \begin{subfigure}[b]{0.4\linewidth}
   \centering
    \includegraphics[trim=1cm -2cm 16cm 0cm, width=.4\textwidth]{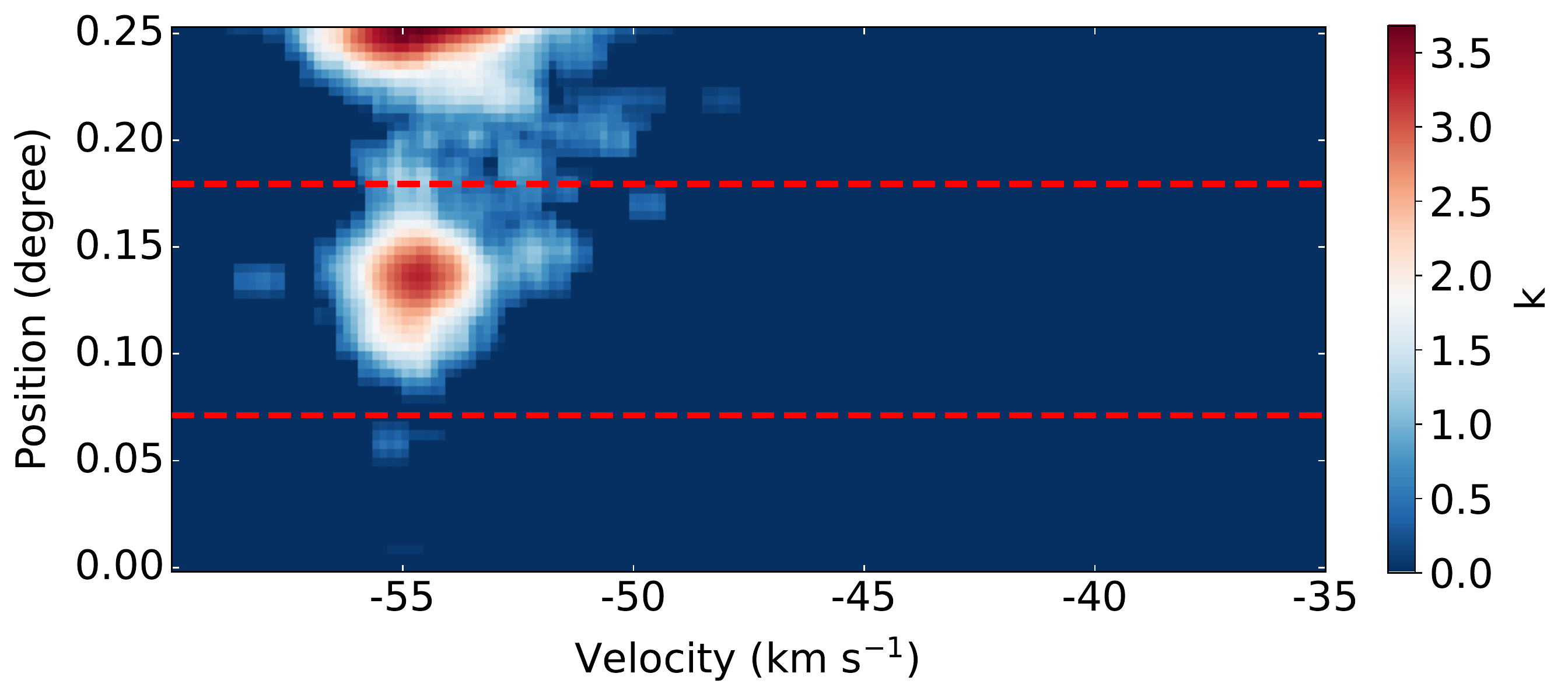}\hfill
    \\(h)
    \end{subfigure}
    \\[\smallskipamount]
    \caption{Same as Figure~\ref{Fig8} but for the S148 region. }
    \label{FigA.8}
\end{figure}

\begin{figure}[h!]
    \centering
     \begin{subfigure}[b]{0.4\linewidth}
     \centering
    \includegraphics[trim=15cm 0cm 15cm 0cm, width=.4\textwidth]{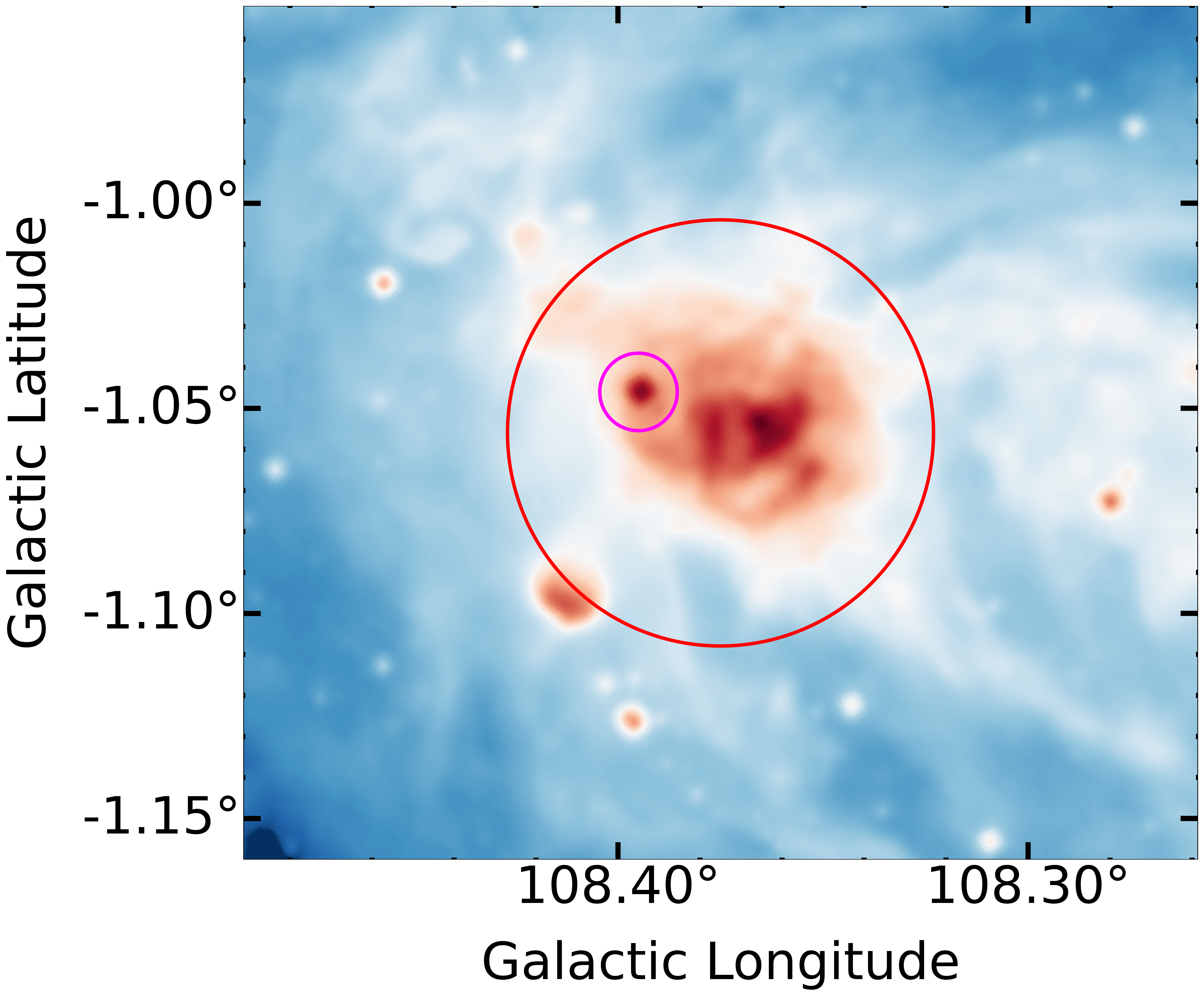}
    \\(a)
   \end{subfigure}
   \begin{subfigure}[b]{0.4\linewidth}
   \centering
    \includegraphics[trim=15cm 0cm 15cm 0cm,width=.4\textwidth]{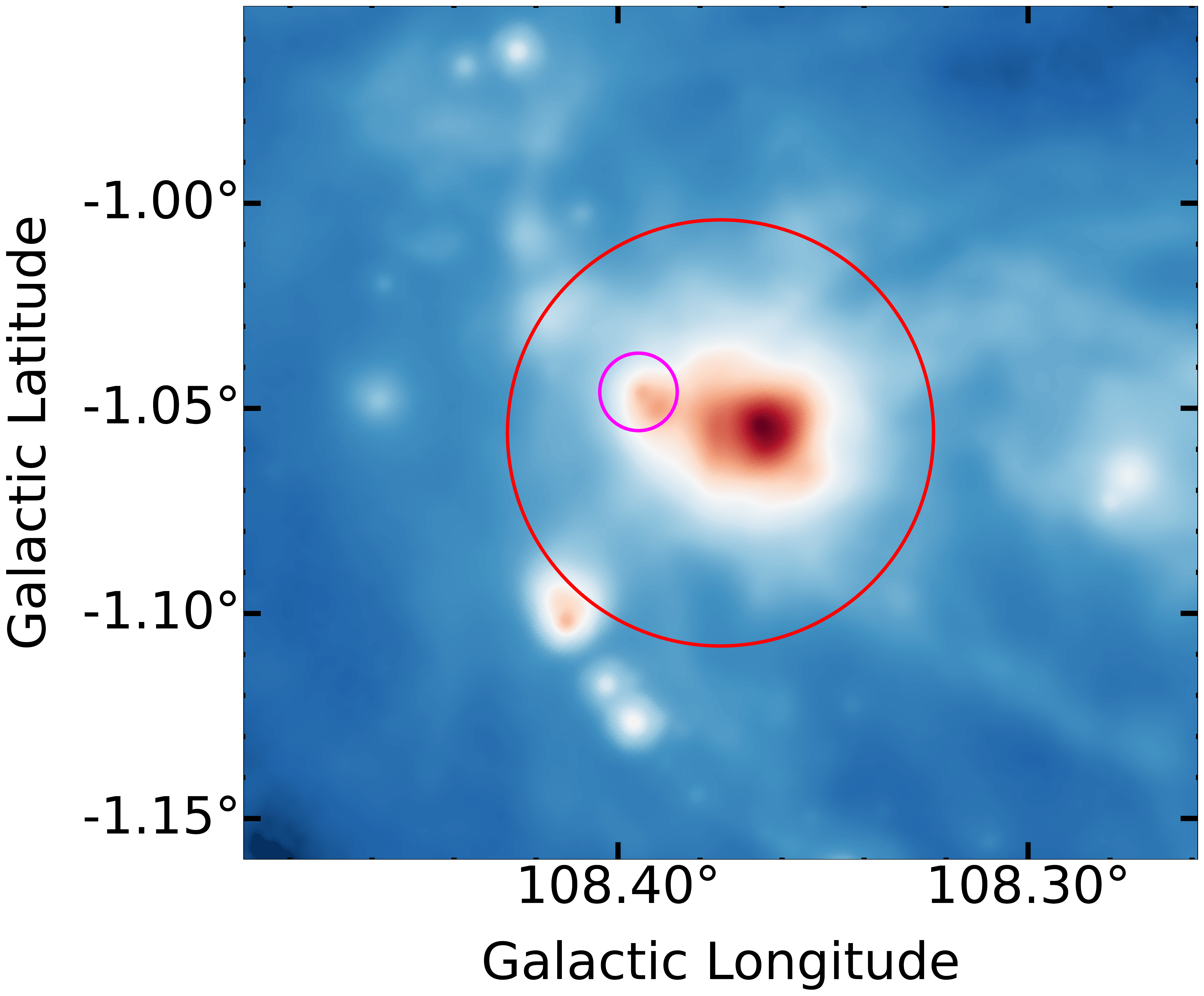}
     \\(b)
    \end{subfigure}
   \begin{subfigure}[b]{0.4\linewidth}
   \centering
    \includegraphics[trim=15cm 0cm 15cm 0cm,width=.4\textwidth]{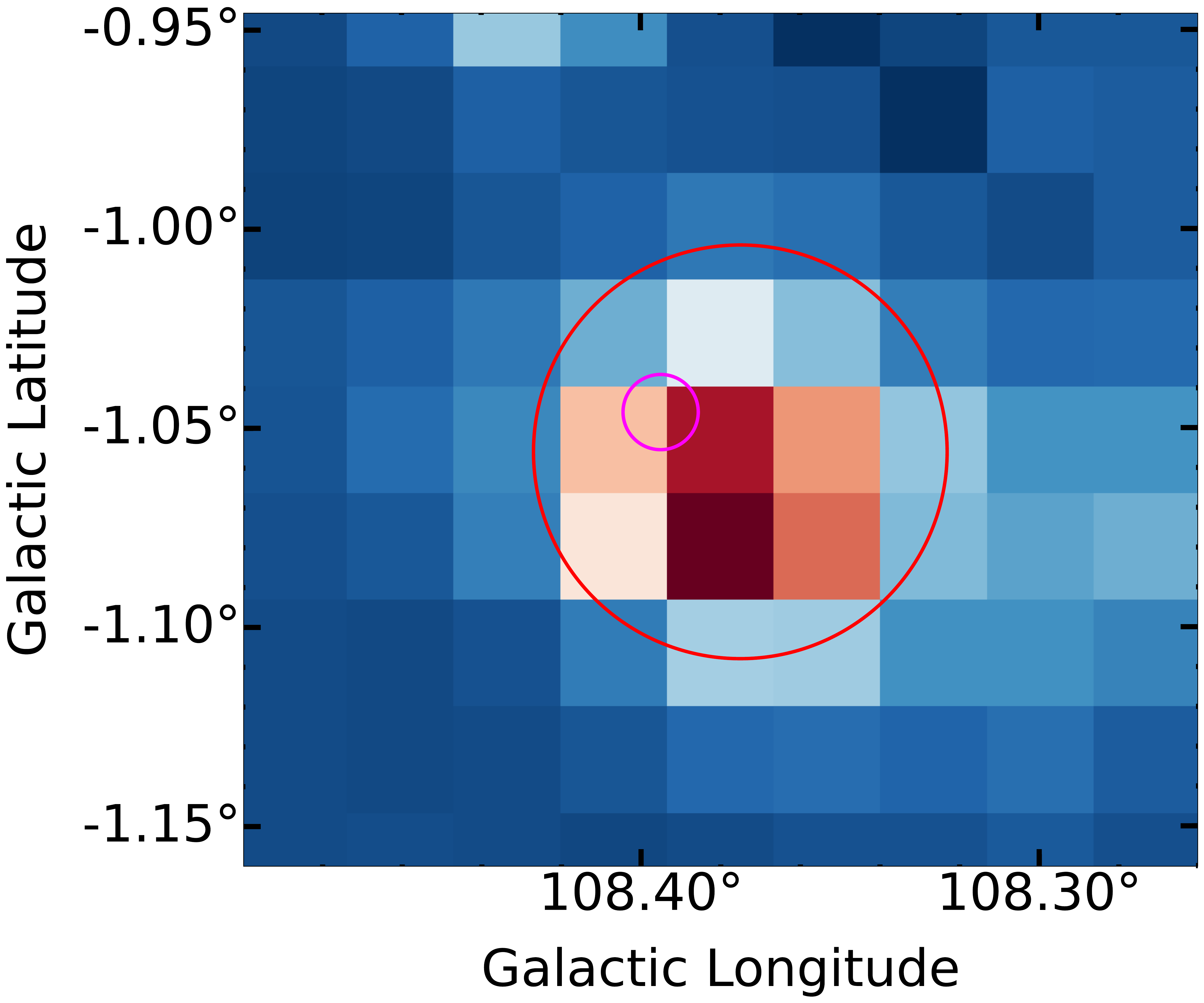}
     \\(c)
    \end{subfigure}
    \begin{subfigure}[b]{0.4\linewidth}
   \centering
    \includegraphics[trim=15cm 0cm 22.5cm 0cm,width=.4\textwidth]{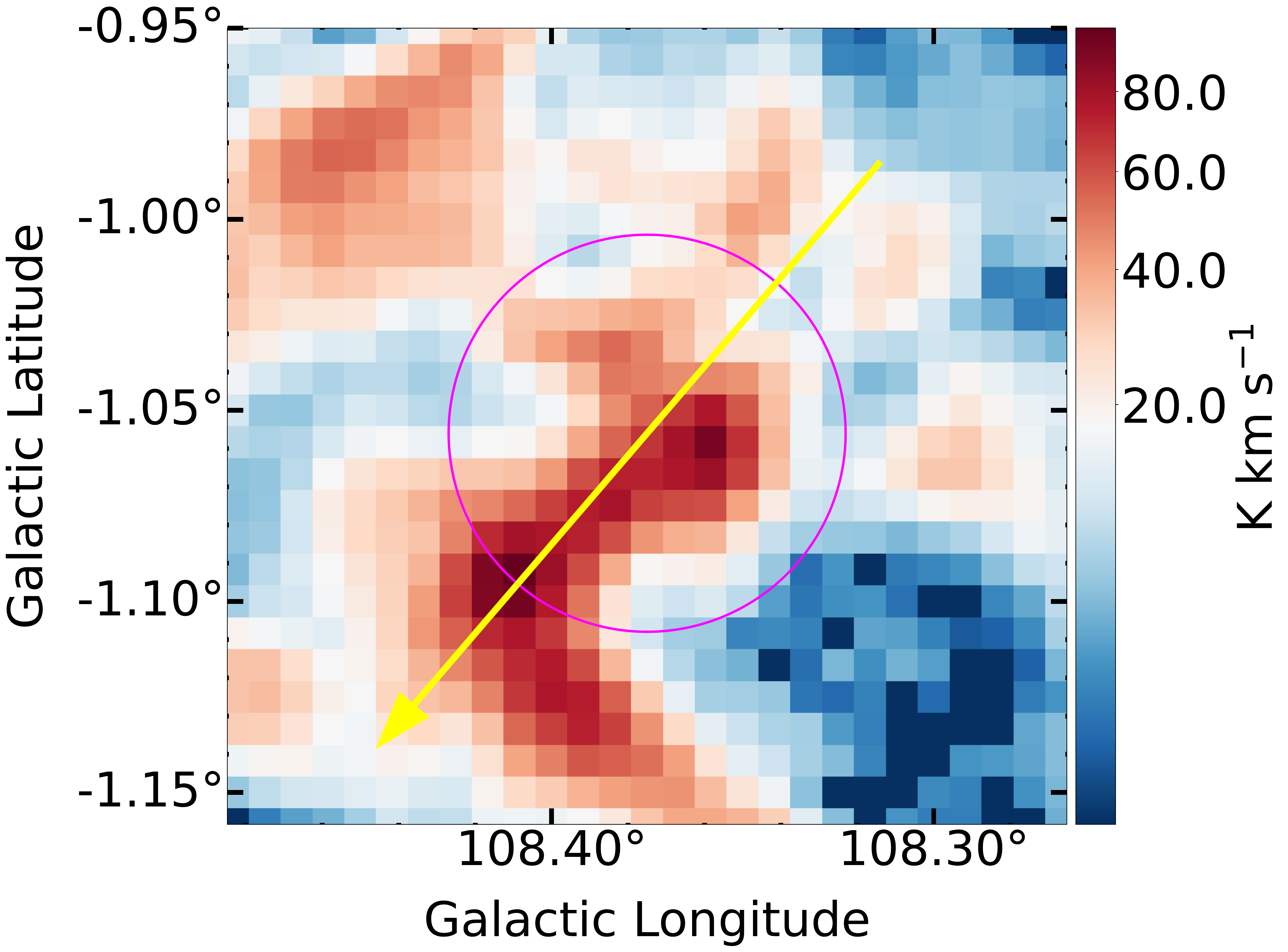}\hfill
    \\(d)
    \end{subfigure}
    \begin{subfigure}[b]{0.4\linewidth}
   \centering
    \includegraphics[trim=14cm 0cm 23cm 0cm,width=.4\textwidth]{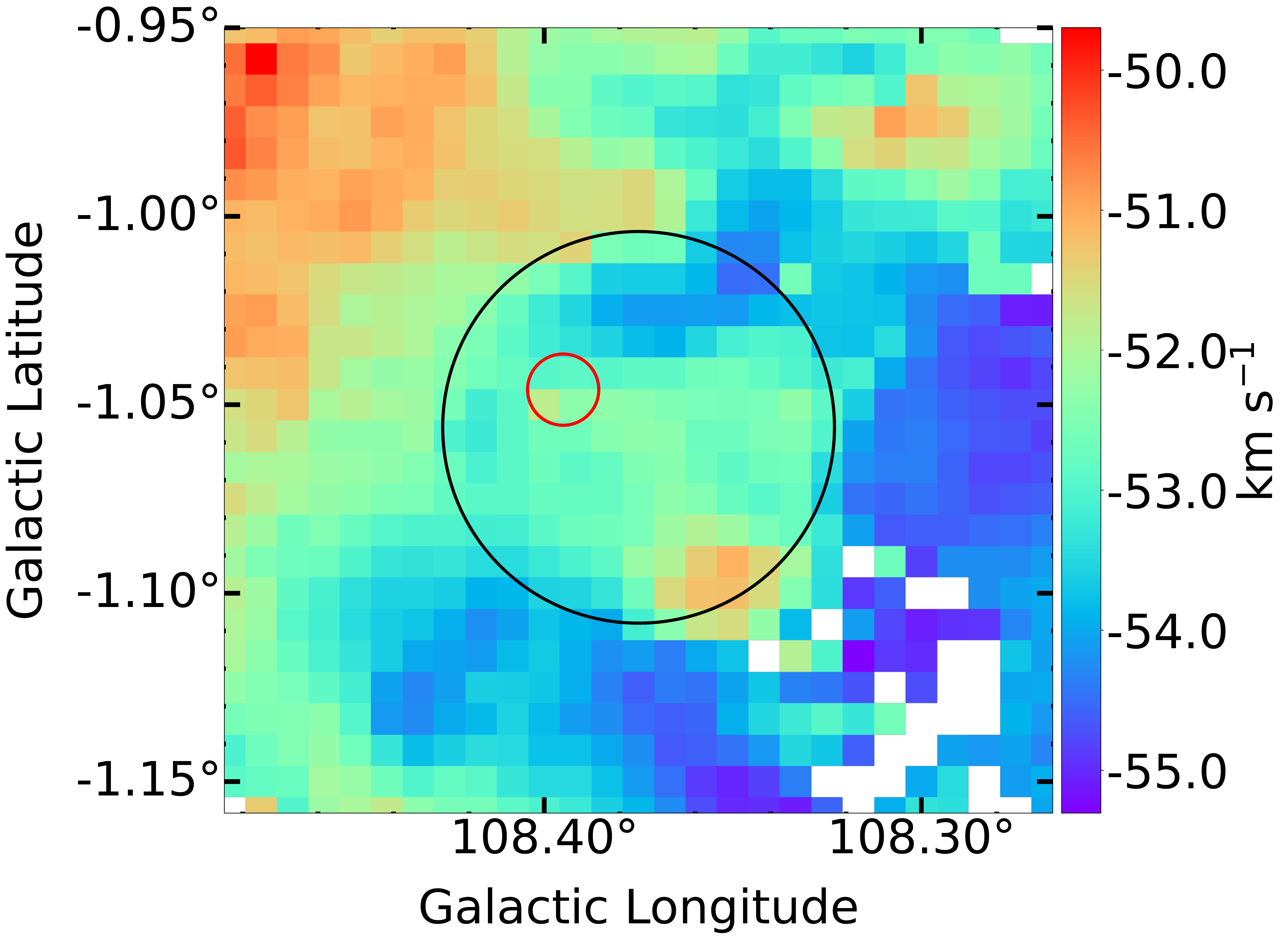}\hfill
    \\(e)
    \end{subfigure}
    \begin{subfigure}[b]{0.4\linewidth}
   \centering
    \includegraphics[trim=8.5cm 0cm 26cm 0cm,width=.4\textwidth]{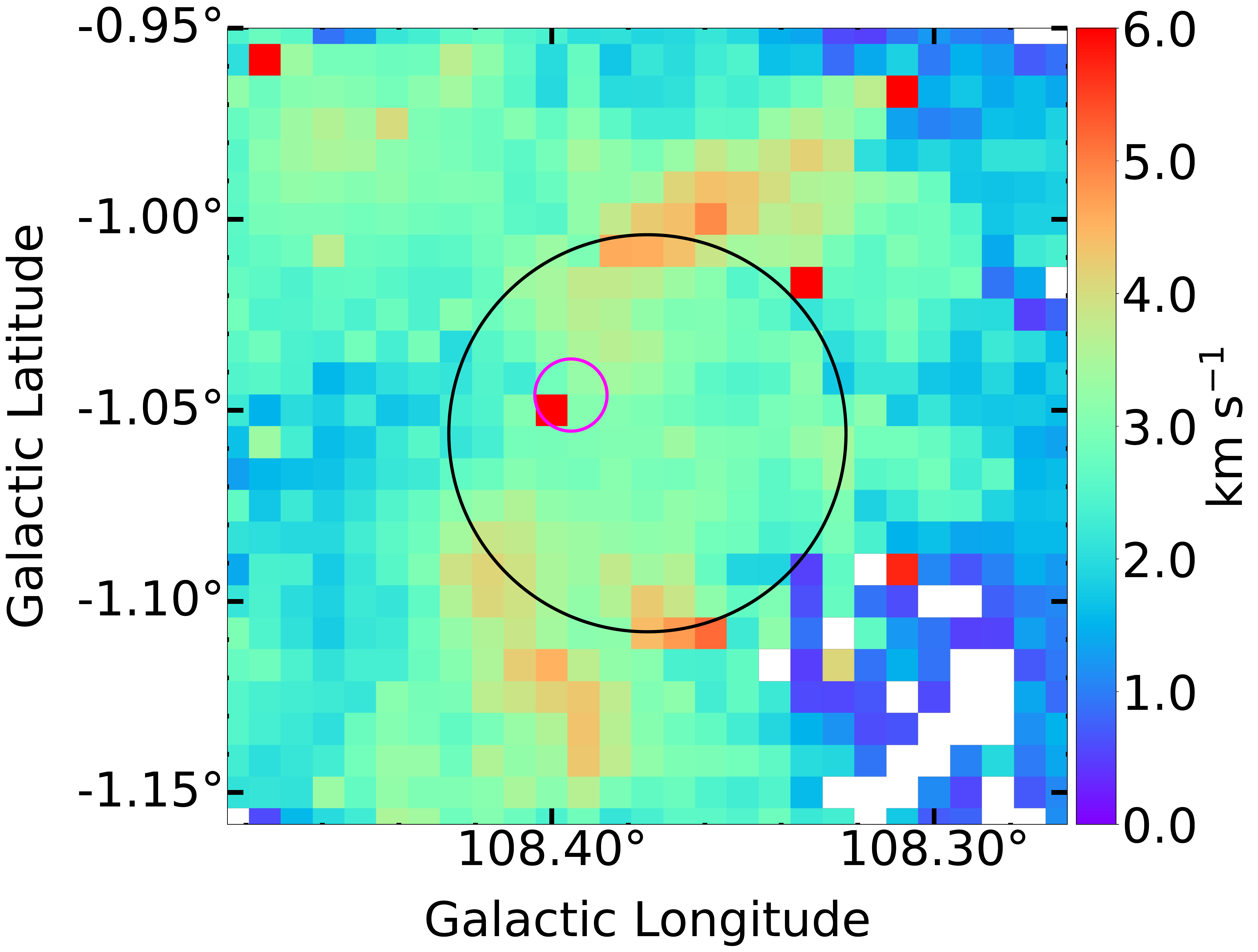}\hfill
    \\(f)
    \end{subfigure}
    \begin{subfigure}[b]{0.4\linewidth}
   \centering
    \includegraphics[trim=14.5cm 0cm 22cm 0cm, width=.4\textwidth]{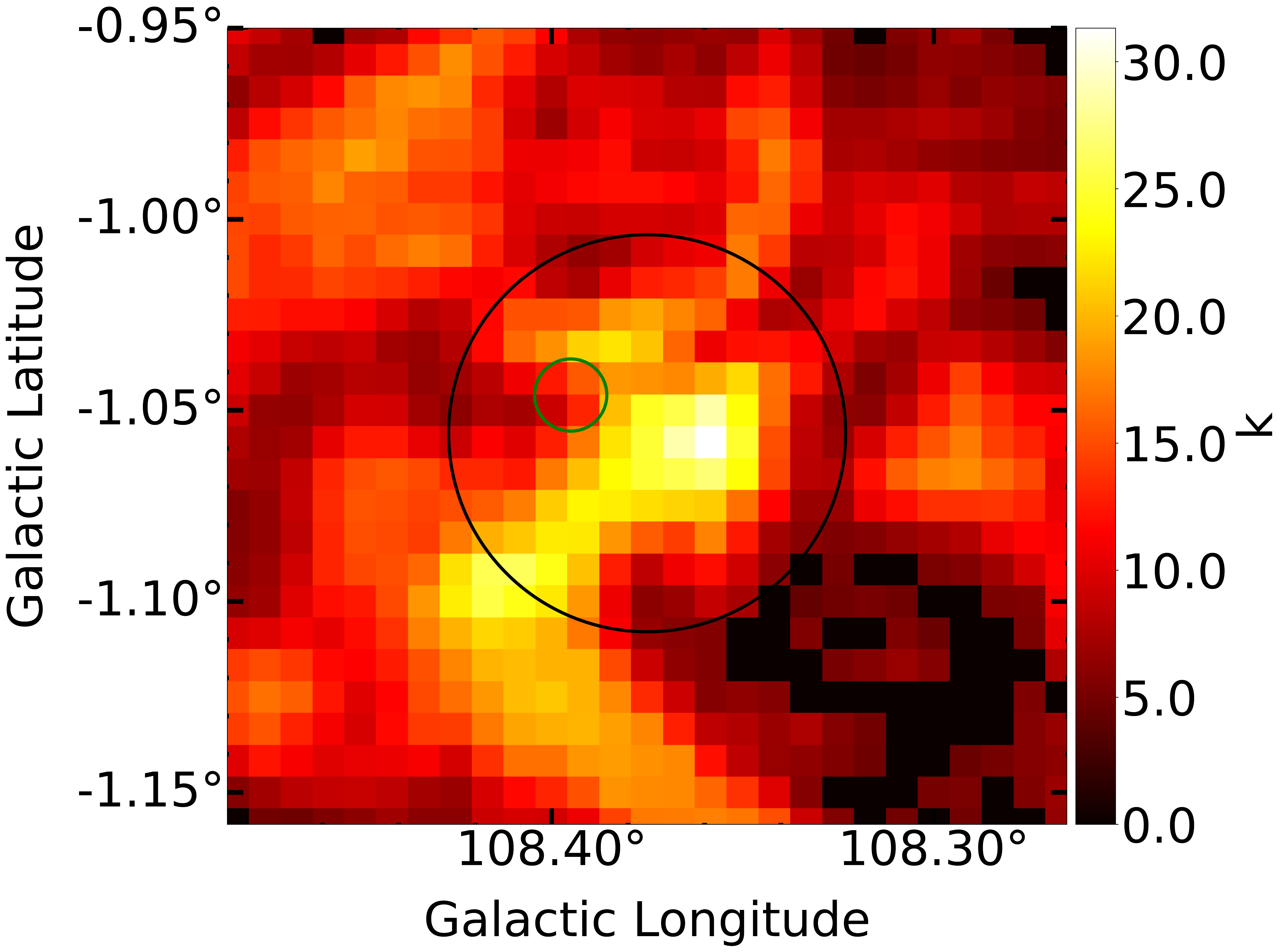}\hfill
    \\(g)
    \end{subfigure}
    \begin{subfigure}[b]{0.4\linewidth}
   \centering
    \includegraphics[trim=1cm -2cm 15cm 0cm, width=.4\textwidth]{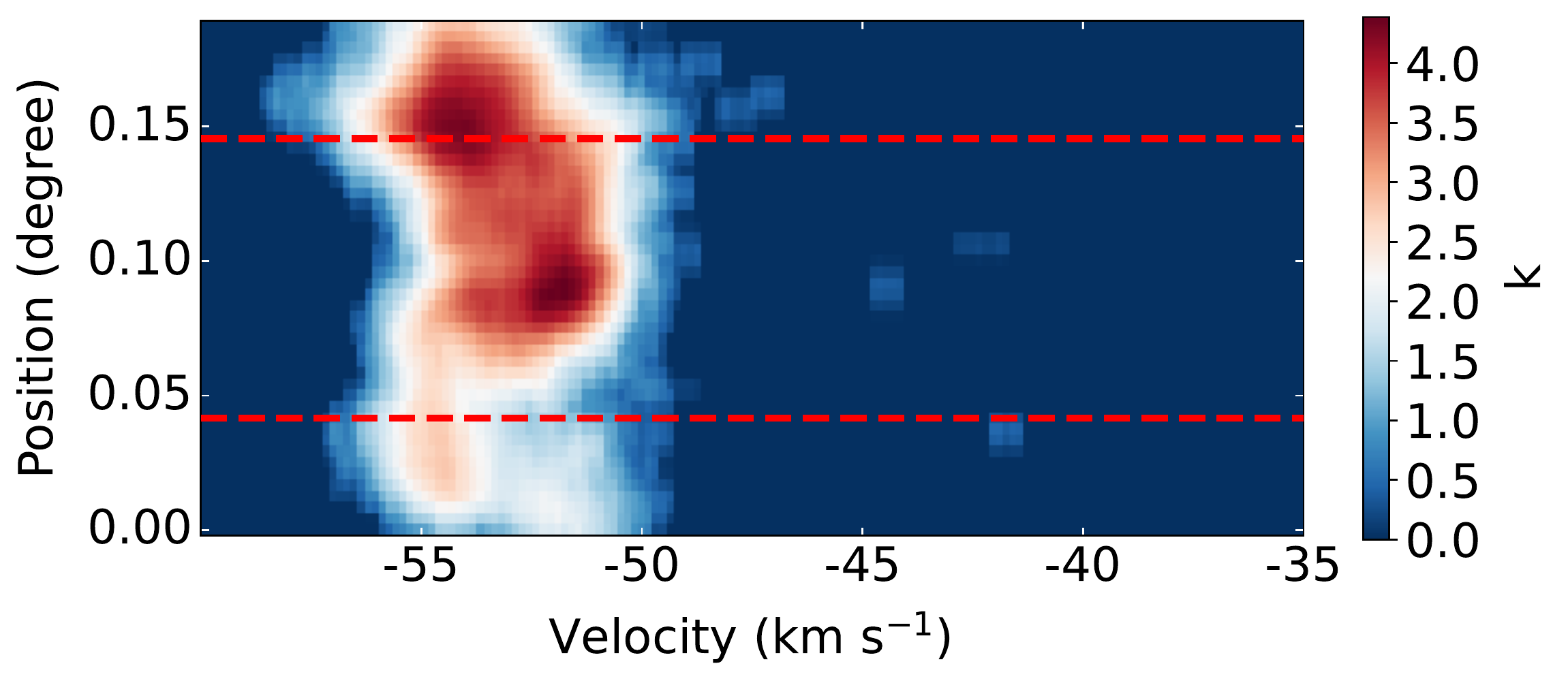}\hfill
    \\(h)
    \end{subfigure}
    \\[\smallskipamount]
    \caption{Same as Figure~\ref{Fig8} but for the S149 and G108.394${-}$01.046 {H \small {II}} regions/candidates. The large and the small circles show the radius of S149 and G108.394${-}$01.046, respectively. }
    \label{FigA.9}
\end{figure}

\begin{figure}[h!]
    \centering
     \begin{subfigure}[b]{0.4\linewidth}
     \centering
    \includegraphics[trim=15cm 0cm 15cm 0cm, width=.4\textwidth]{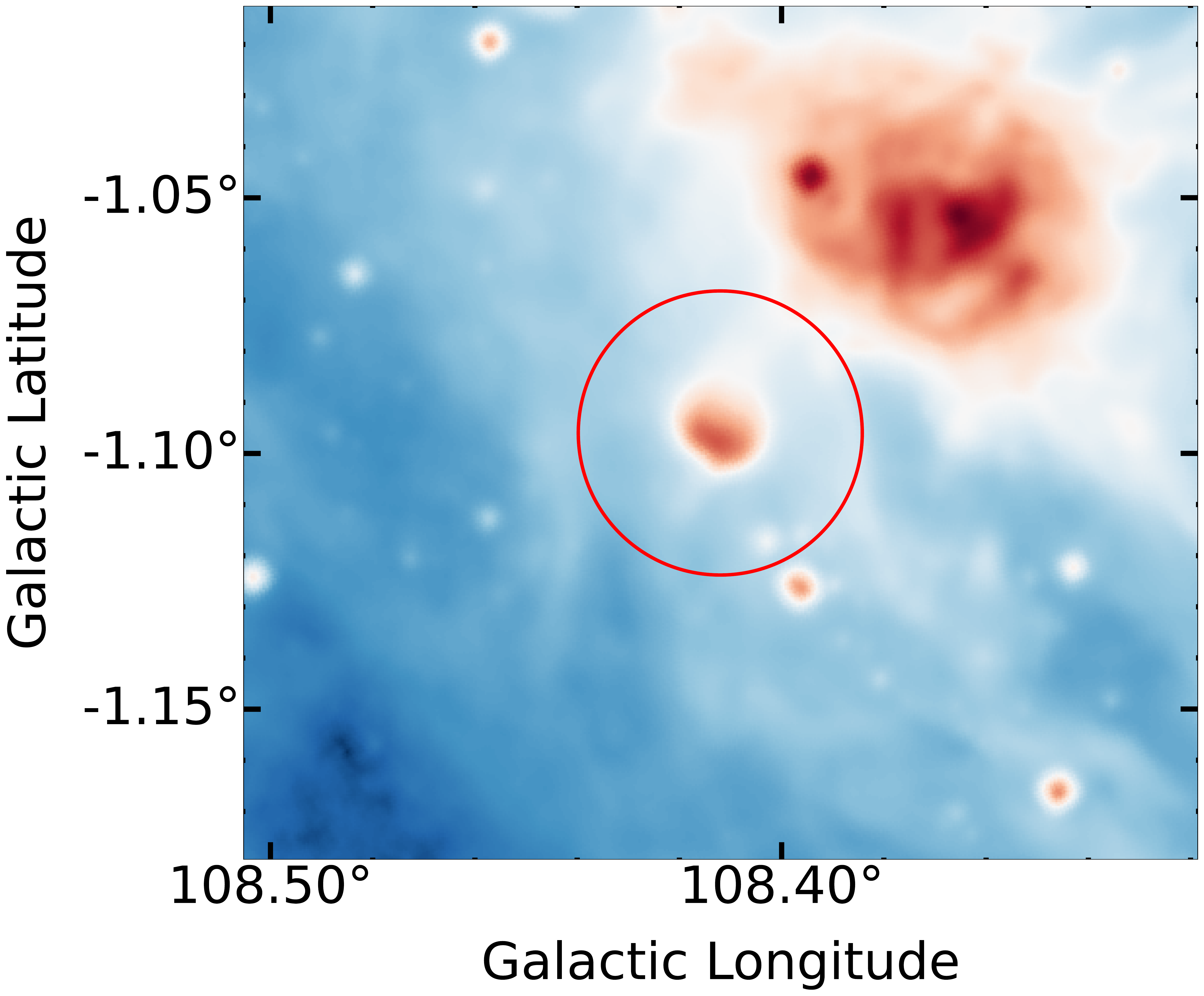}
    \\(a)
   \end{subfigure}
   \begin{subfigure}[b]{0.4\linewidth}
   \centering
    \includegraphics[trim=15cm 0cm 15cm 0cm,width=.4\textwidth]{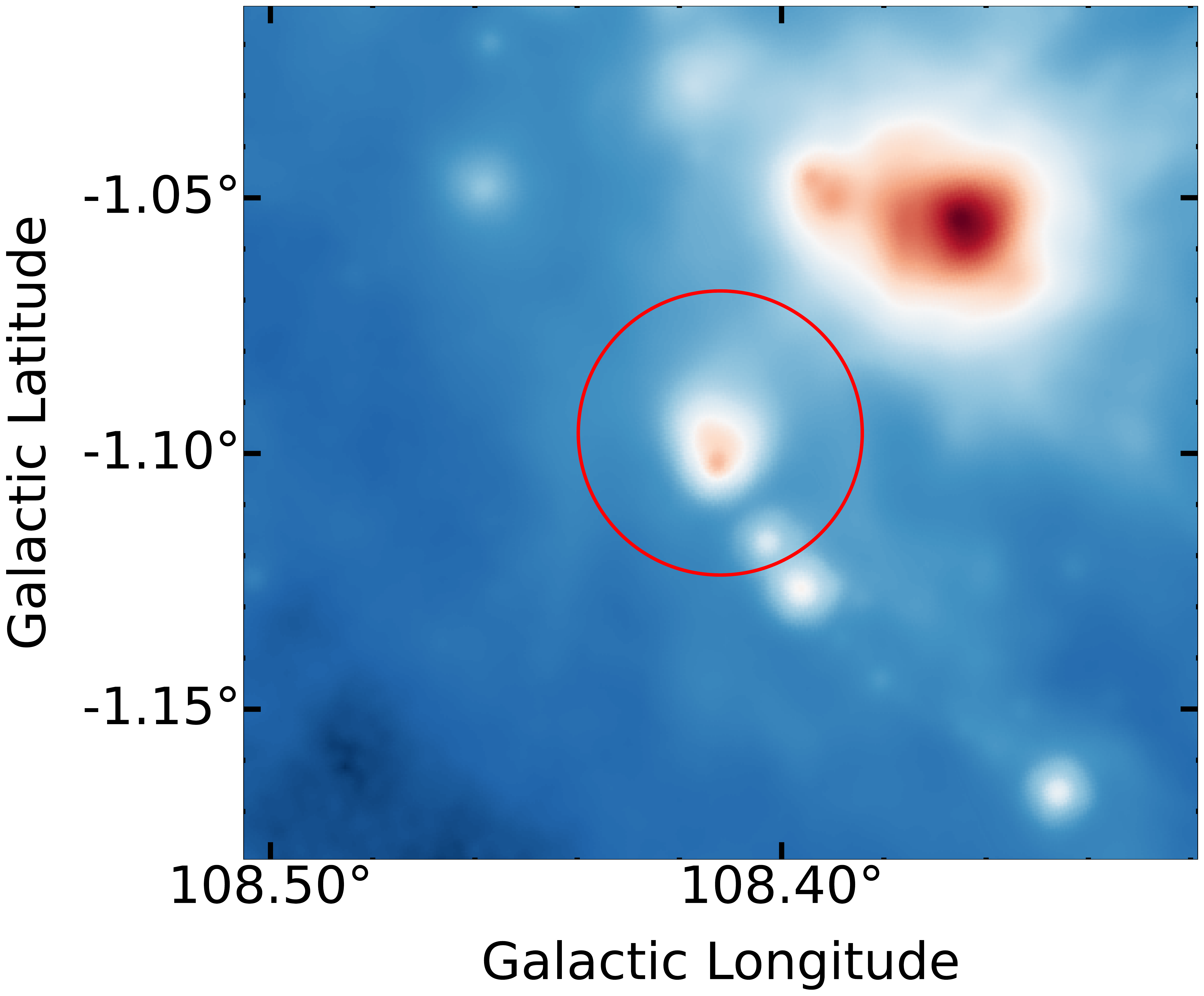}
     \\(b)
    \end{subfigure}
   \begin{subfigure}[b]{0.4\linewidth}
   \centering
    \includegraphics[trim=15cm 0cm 15cm 0cm,width=.4\textwidth]{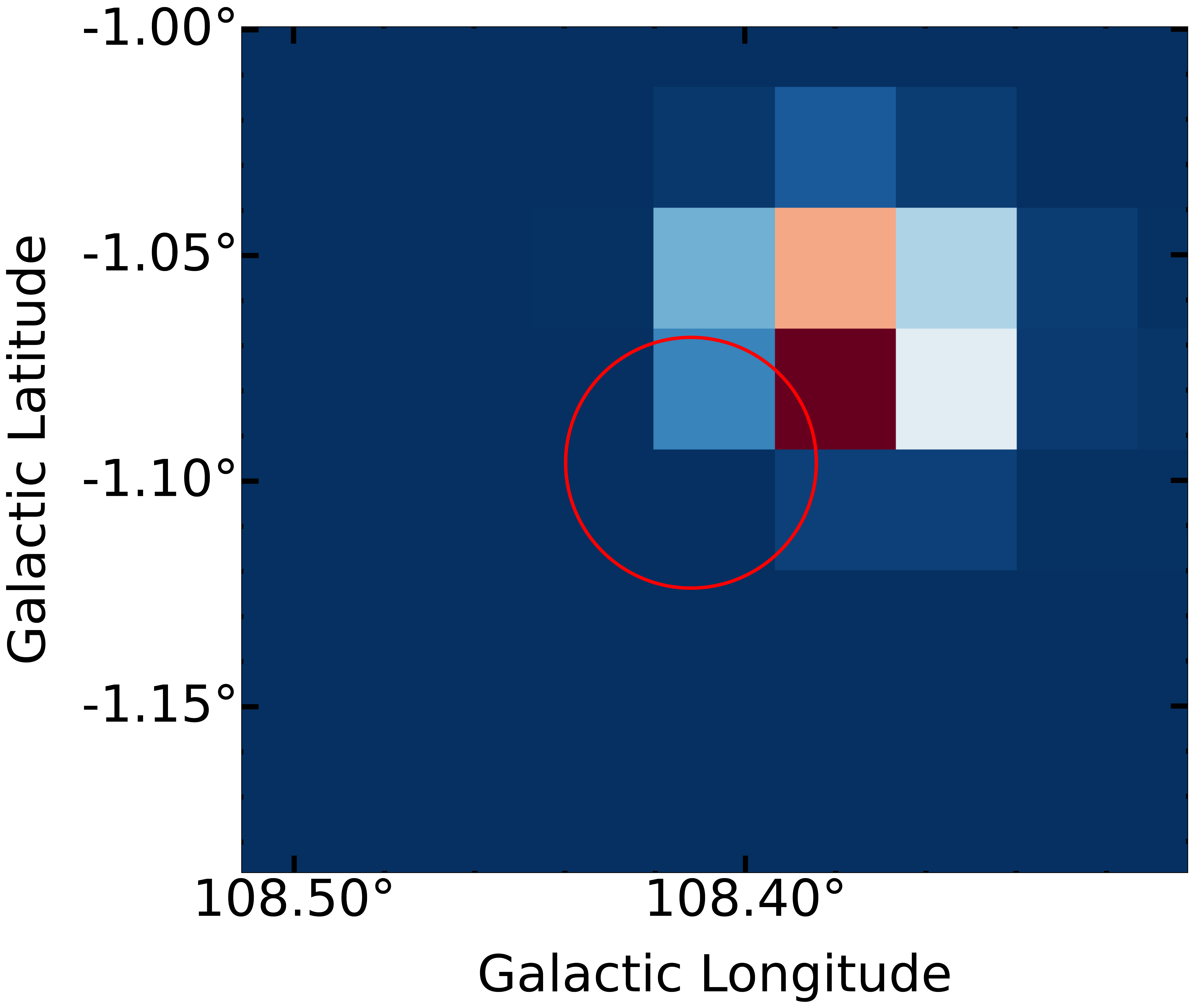}
     \\(c)
    \end{subfigure}
    \begin{subfigure}[b]{0.4\linewidth}
   \centering
    \includegraphics[trim=15cm 0cm 22.5cm 0cm,width=.4\textwidth]{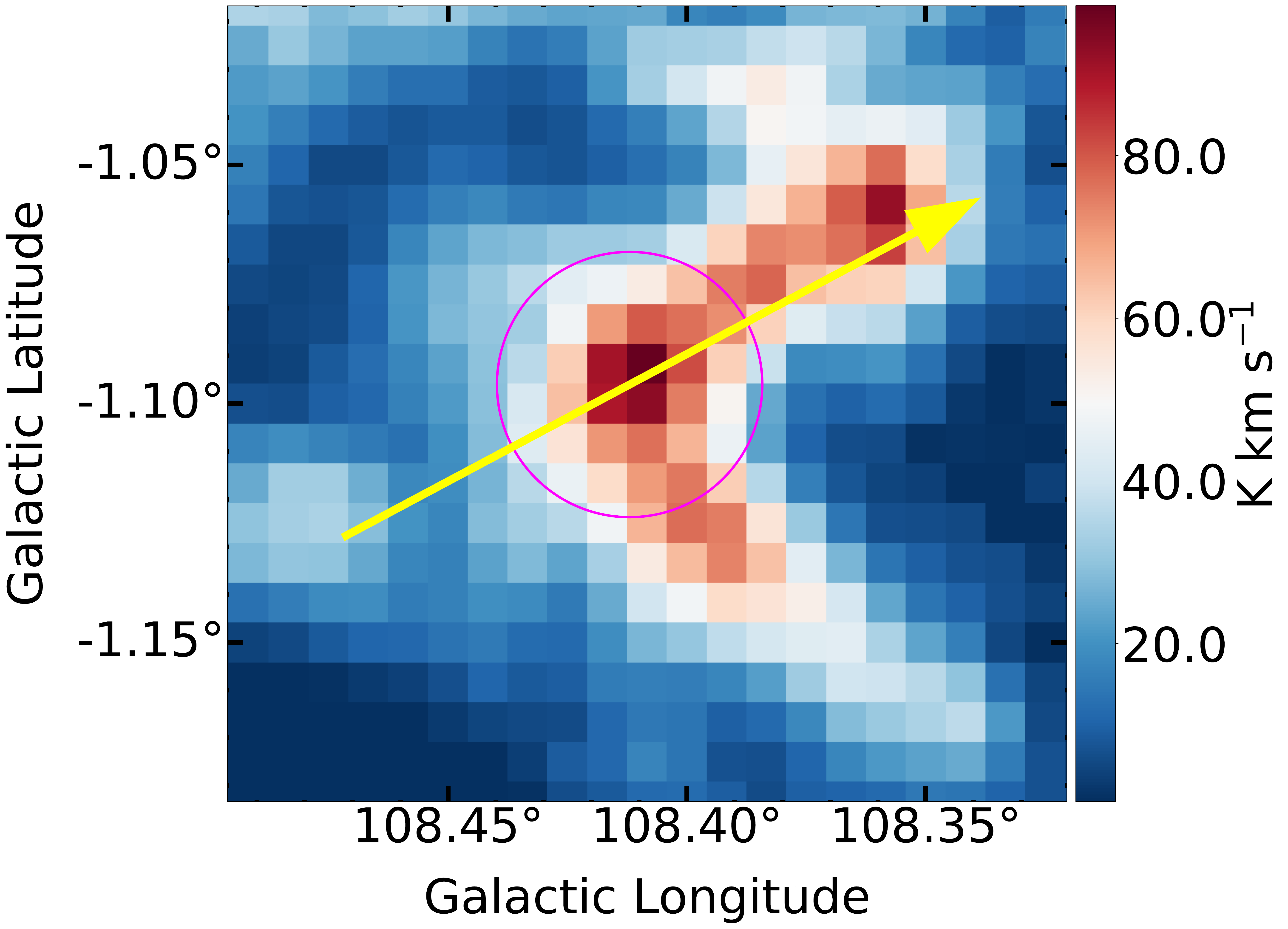}\hfill
    \\(d)
    \end{subfigure}
    \begin{subfigure}[b]{0.4\linewidth}
   \centering
    \includegraphics[trim=14cm 0cm 23cm 0cm,width=.4\textwidth]{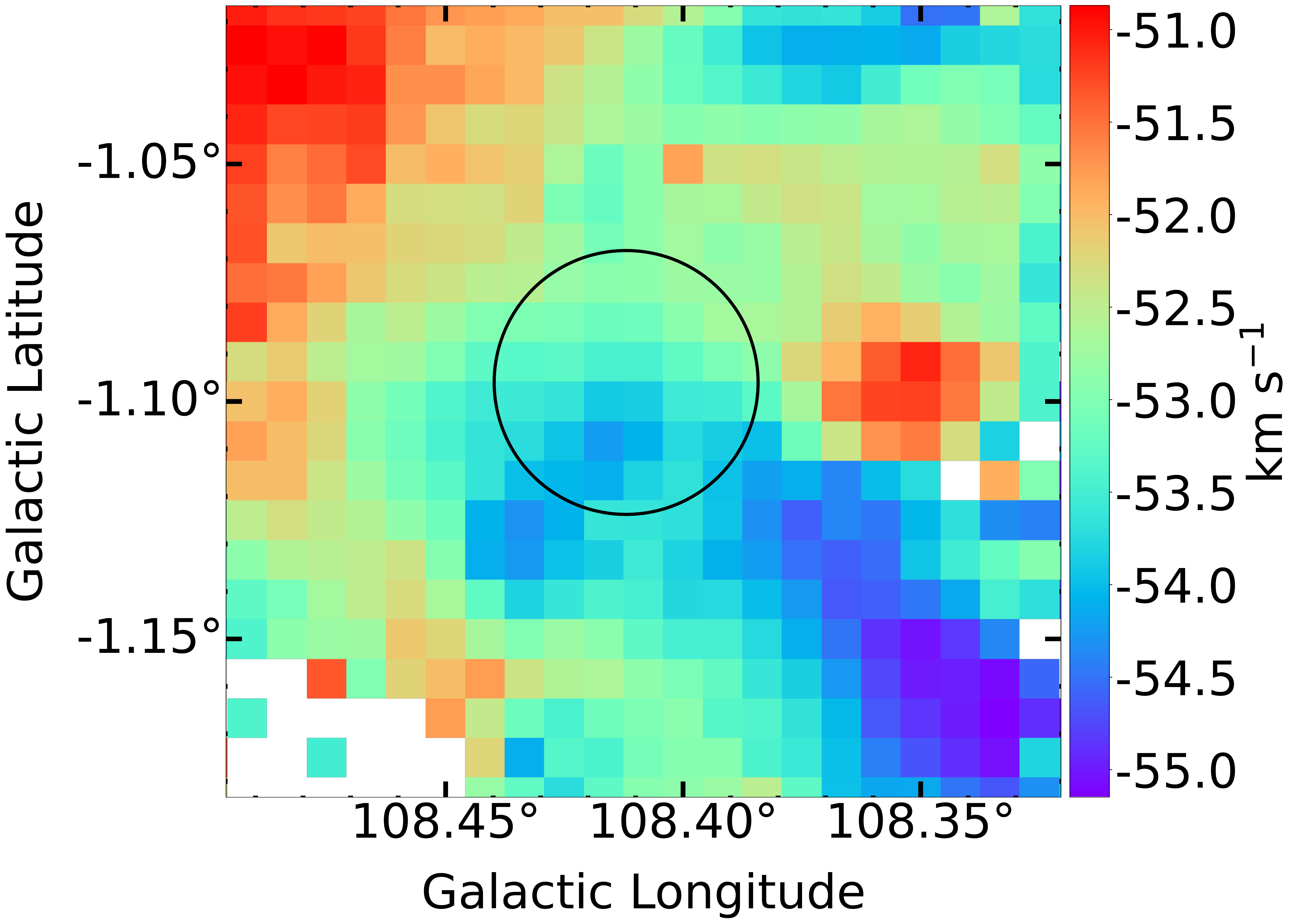}\hfill
    \\(e)
    \end{subfigure}
    \begin{subfigure}[b]{0.4\linewidth}
   \centering
    \includegraphics[trim=8.5cm 0cm 26cm 0cm,width=.4\textwidth]{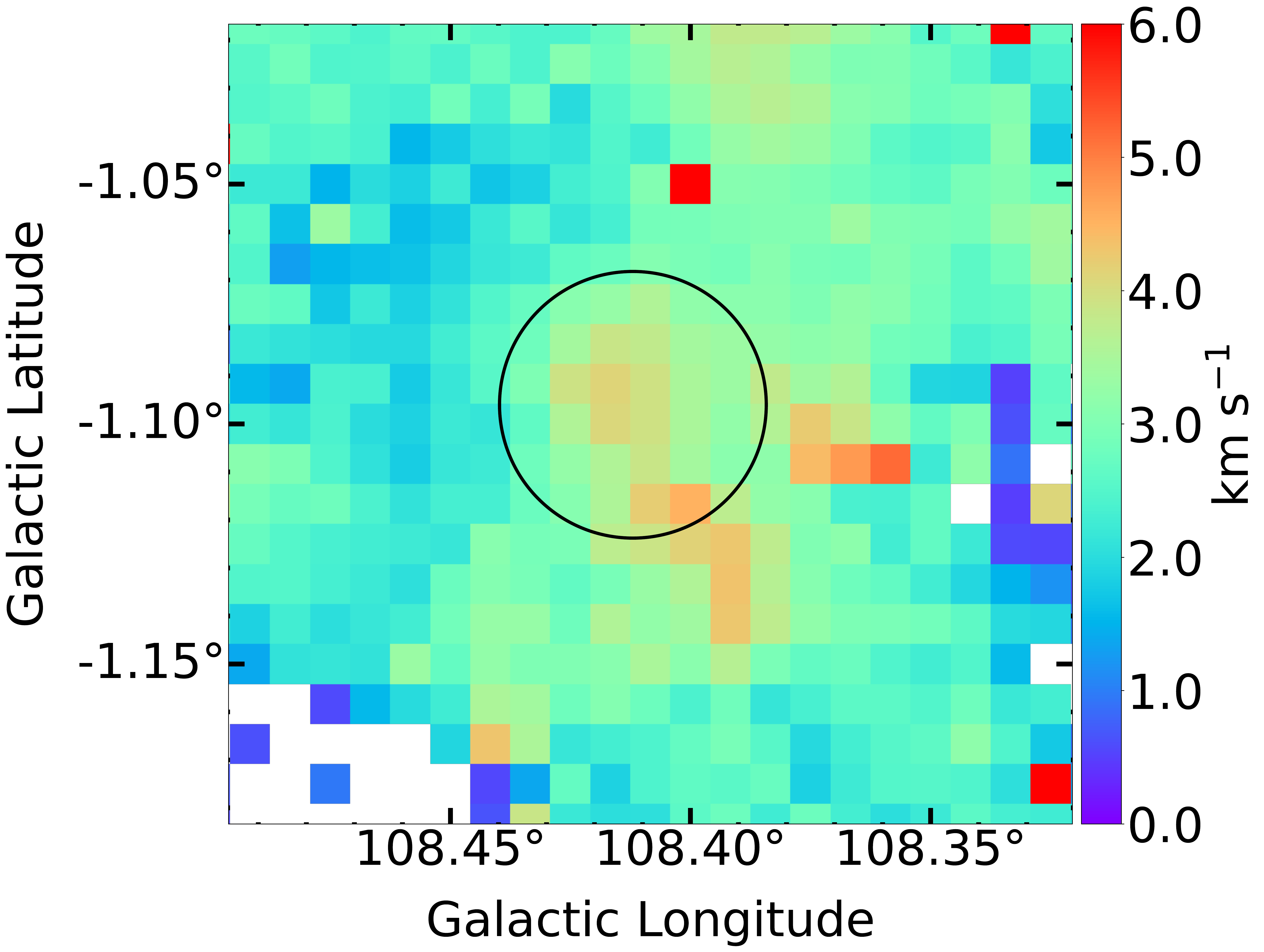}\hfill
    \\(f)
    \end{subfigure}
    \begin{subfigure}[b]{0.4\linewidth}
   \centering
    \includegraphics[trim=14.5cm 0cm 22cm 0cm, width=.4\textwidth]{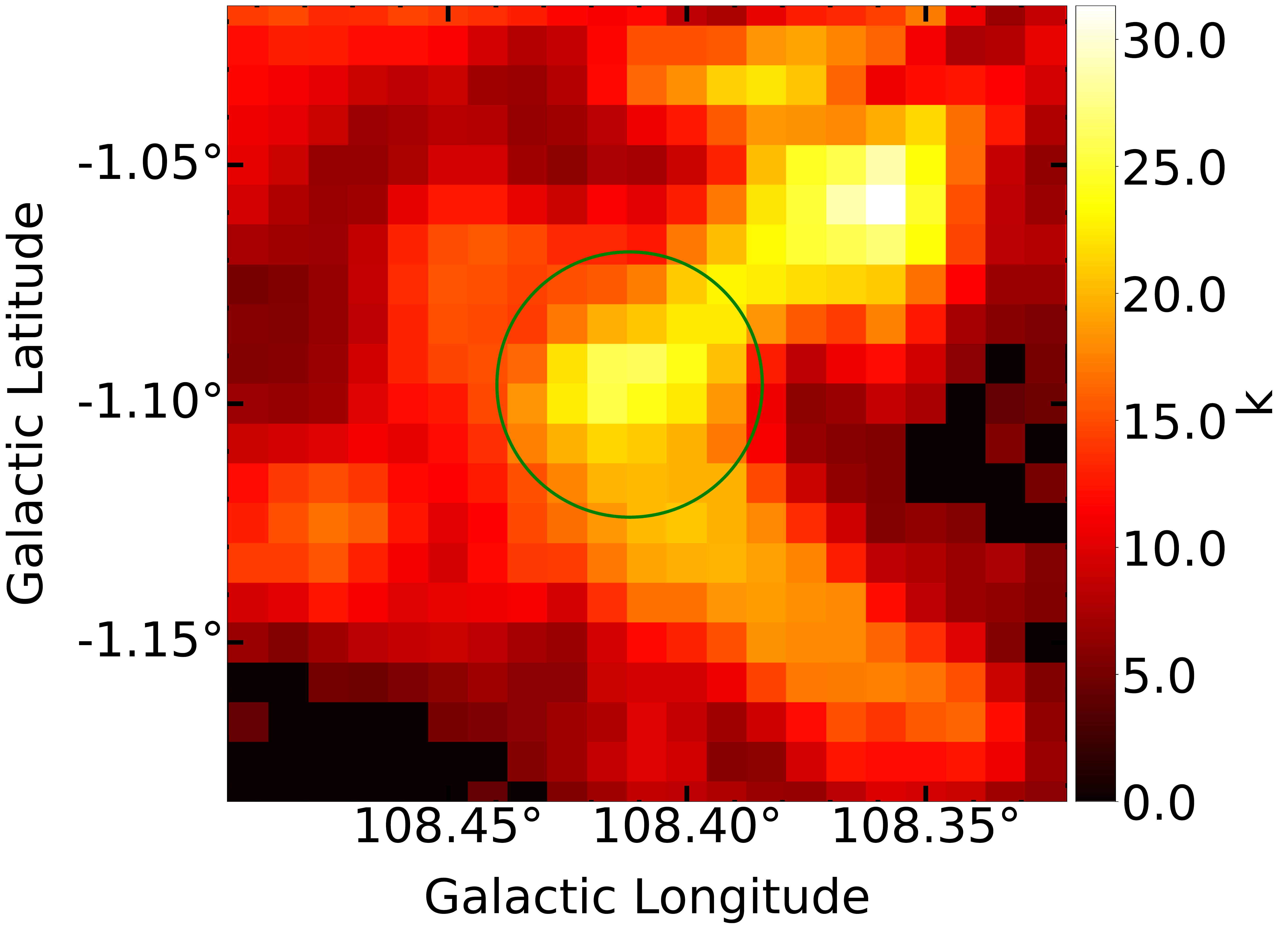}\hfill
    \\(g)
    \end{subfigure}
    \begin{subfigure}[b]{0.4\linewidth}
   \centering
    \includegraphics[trim=1cm -2cm 15cm 0cm, width=.4\textwidth]{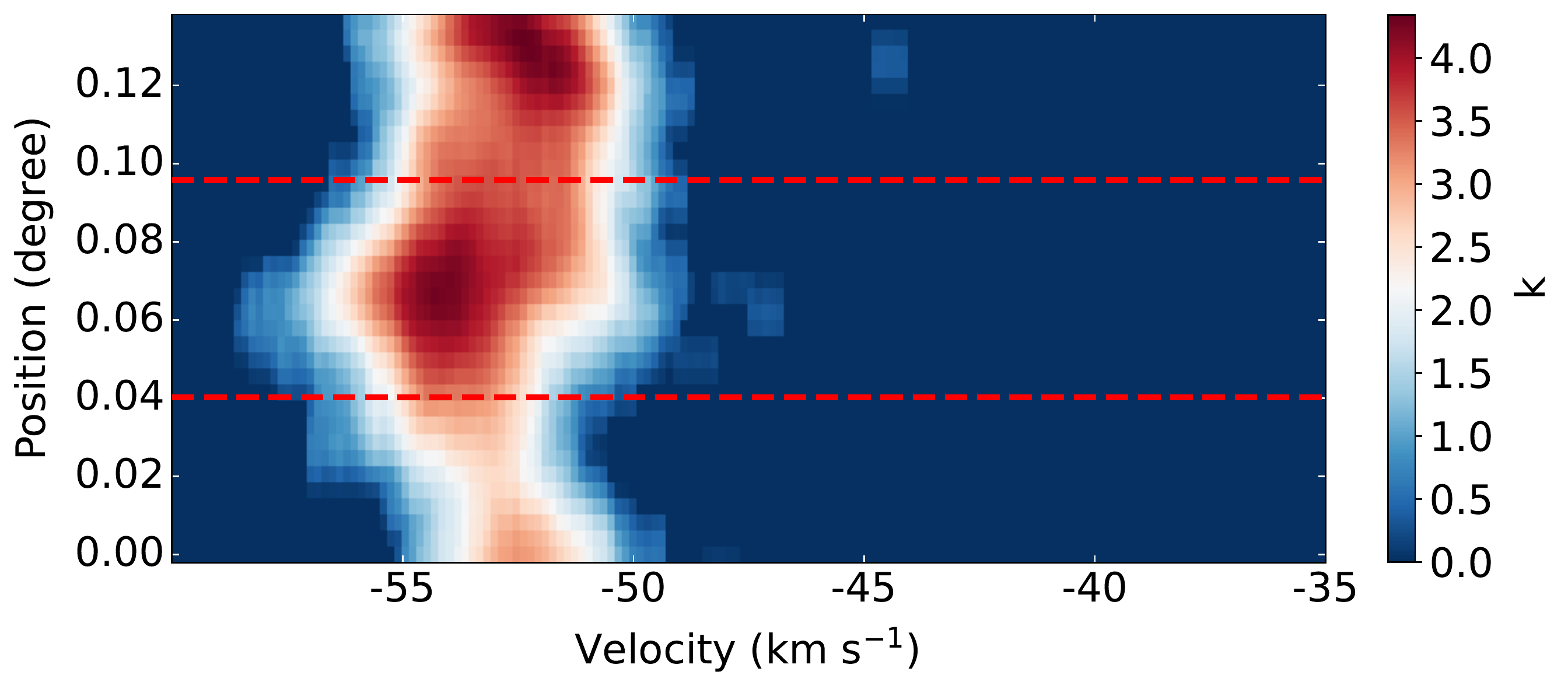}\hfill
    \\(h)
    \end{subfigure}
    \\[\smallskipamount]
    \caption{Same as Figure~\ref{Fig8} but for the G108.412${-}$01.097 region. }
    \label{FigA.10}
\end{figure}

\begin{figure}[h!]
    \centering
     \begin{subfigure}[b]{0.4\linewidth}
     \centering
    \includegraphics[trim=15cm 0cm 15cm 0cm, width=.4\textwidth]{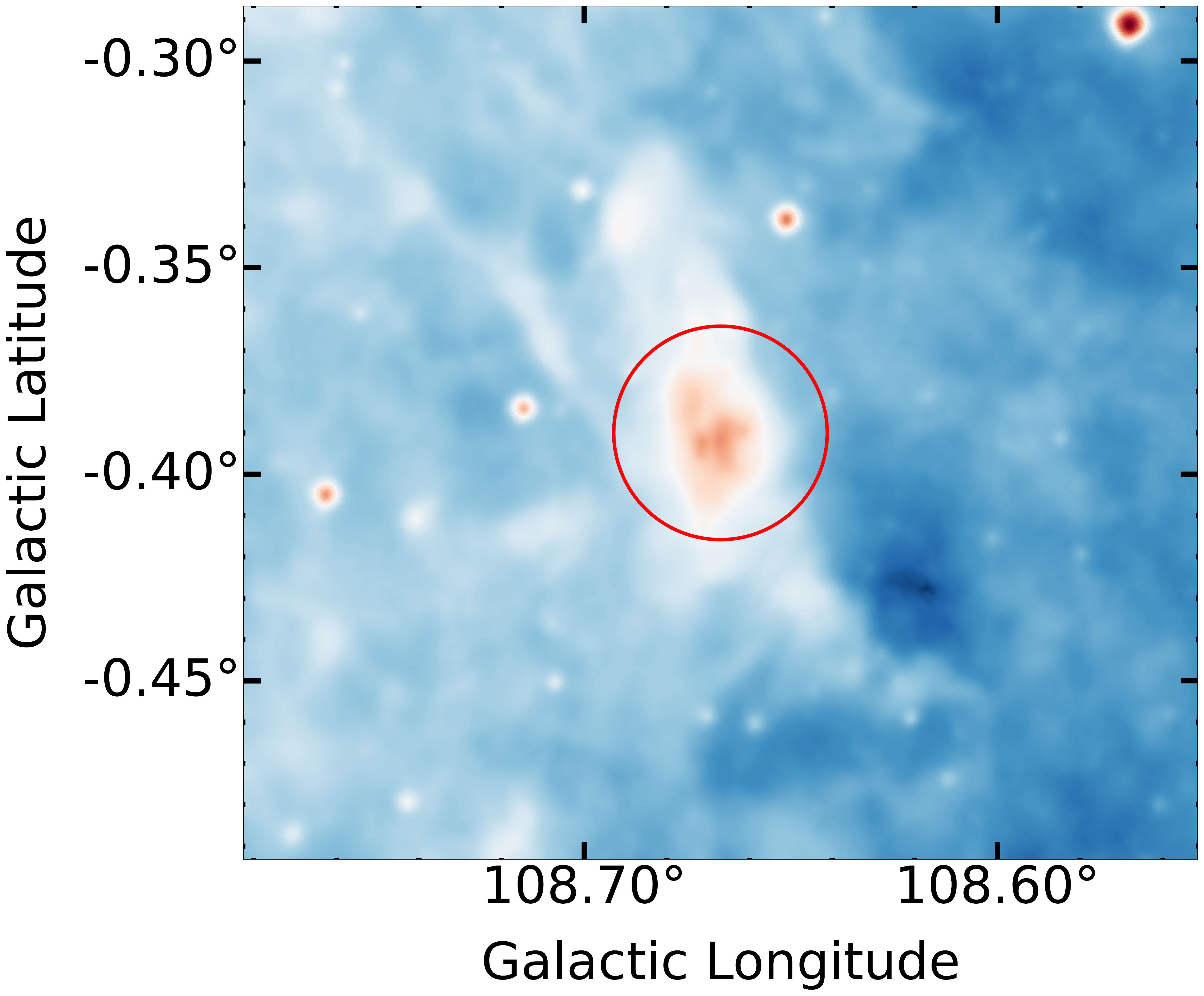}
    \\(a)
   \end{subfigure}
   \begin{subfigure}[b]{0.4\linewidth}
   \centering
    \includegraphics[trim=15cm 0cm 15cm 0cm,width=.4\textwidth]{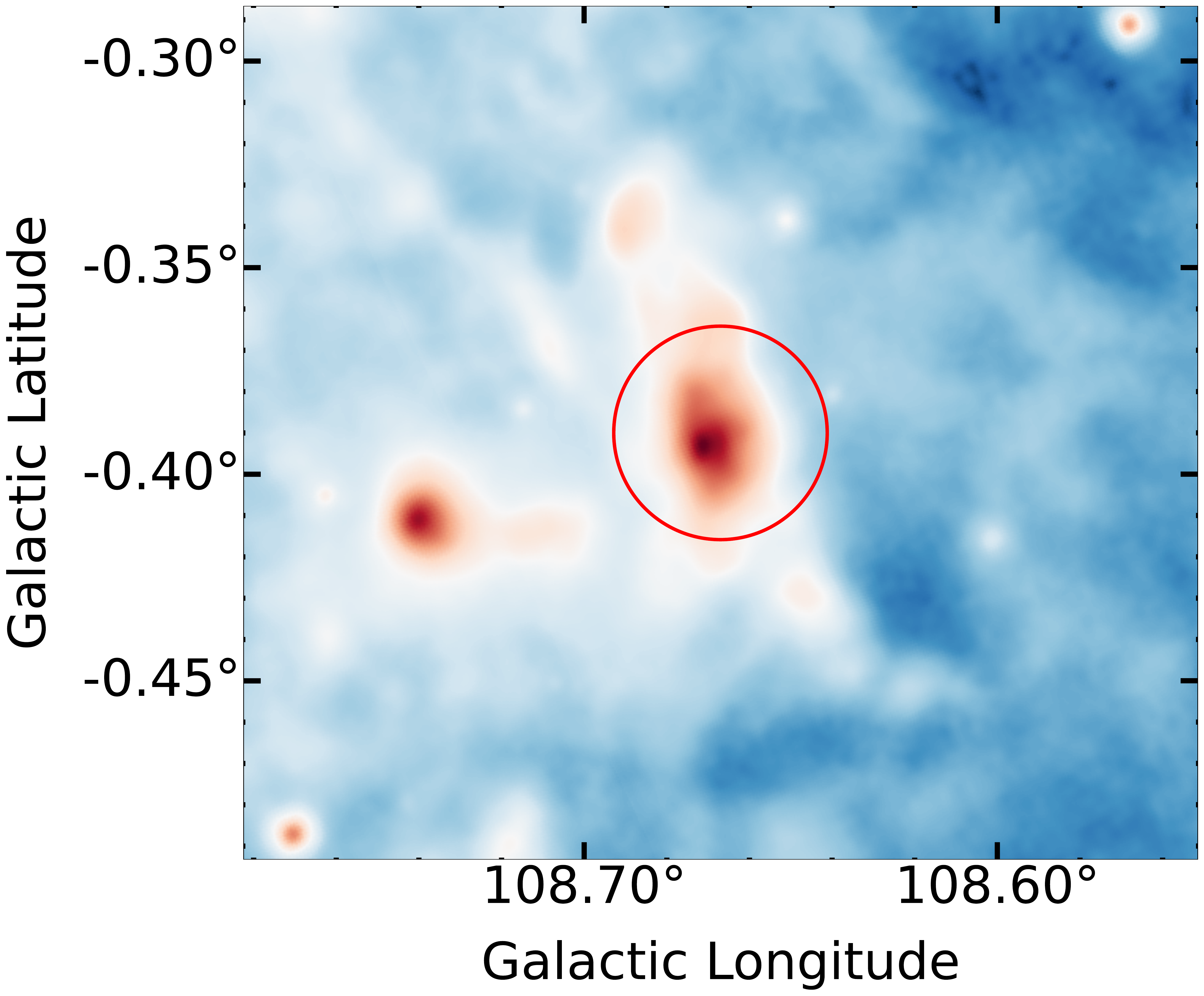}
     \\(b)
    \end{subfigure}
   \begin{subfigure}[b]{0.4\linewidth}
   \centering
    \includegraphics[trim=15cm 0cm 15cm 0cm,width=.4\textwidth]{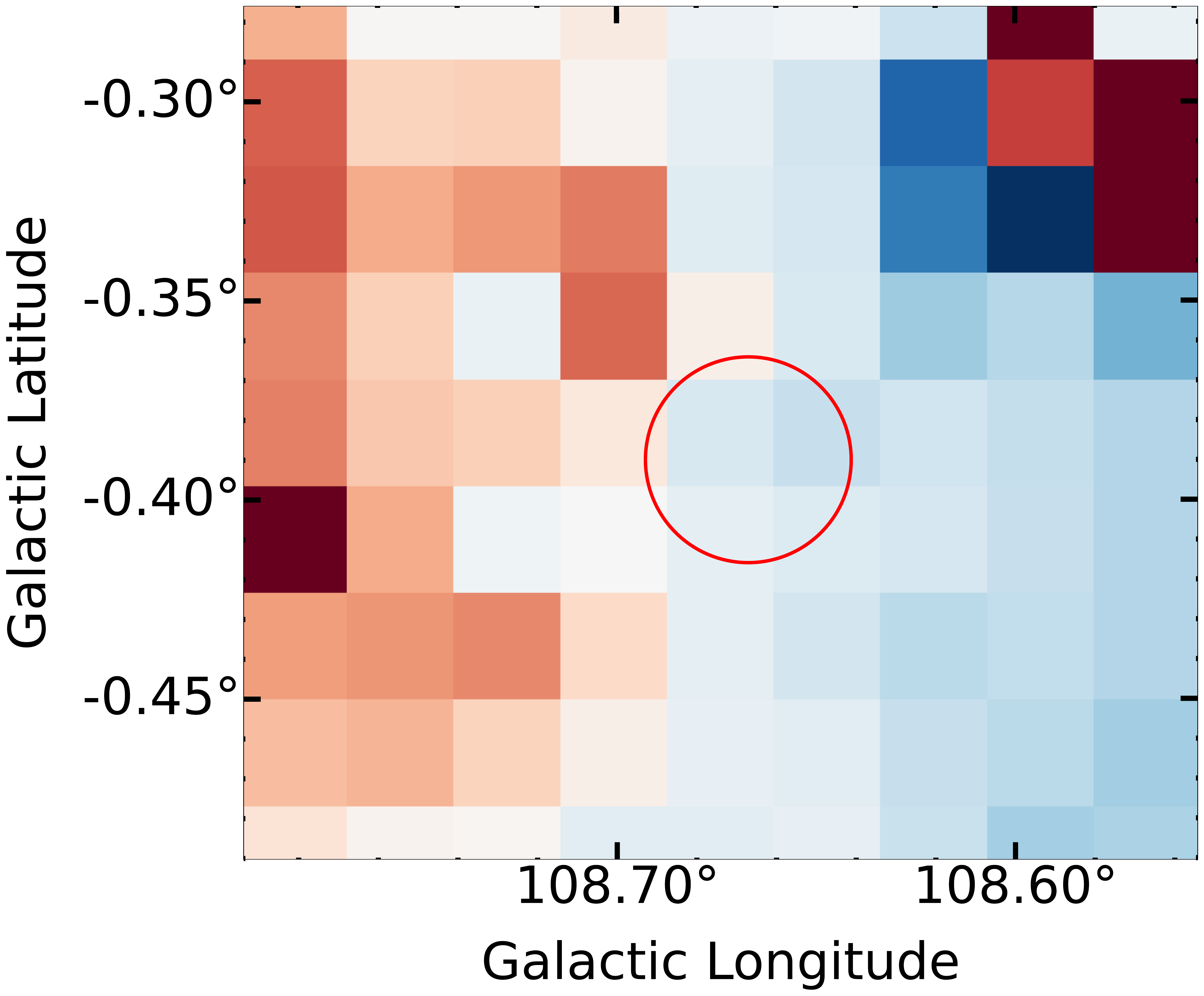}
     \\(c)
    \end{subfigure}
    \begin{subfigure}[b]{0.4\linewidth}
   \centering
    \includegraphics[trim=15cm 0cm 22.5cm 0cm,width=.4\textwidth]{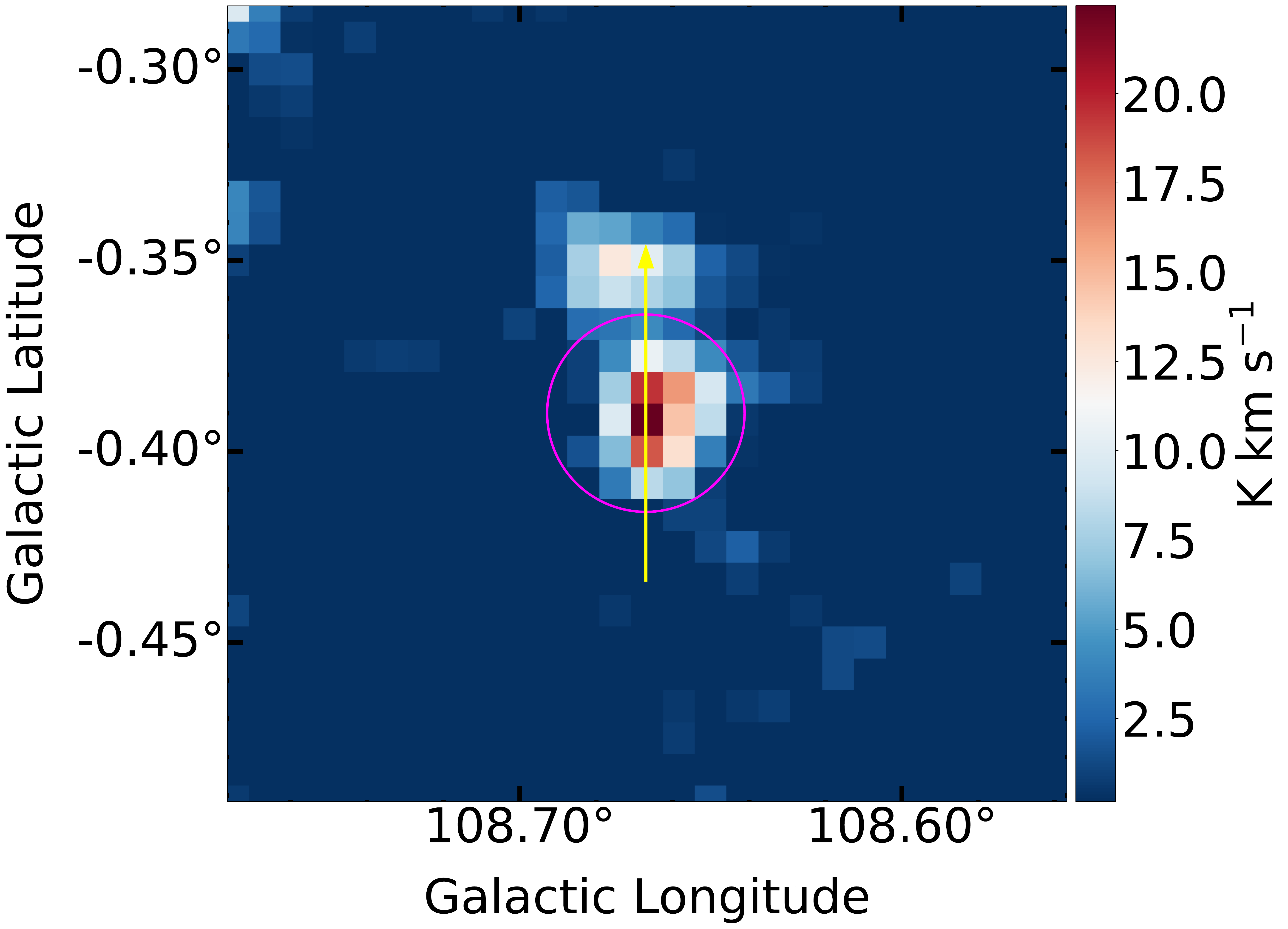}\hfill
    \\(d)
    \end{subfigure}
    \begin{subfigure}[b]{0.4\linewidth}
   \centering
    \includegraphics[trim=14cm 0cm 23cm 0cm,width=.4\textwidth]{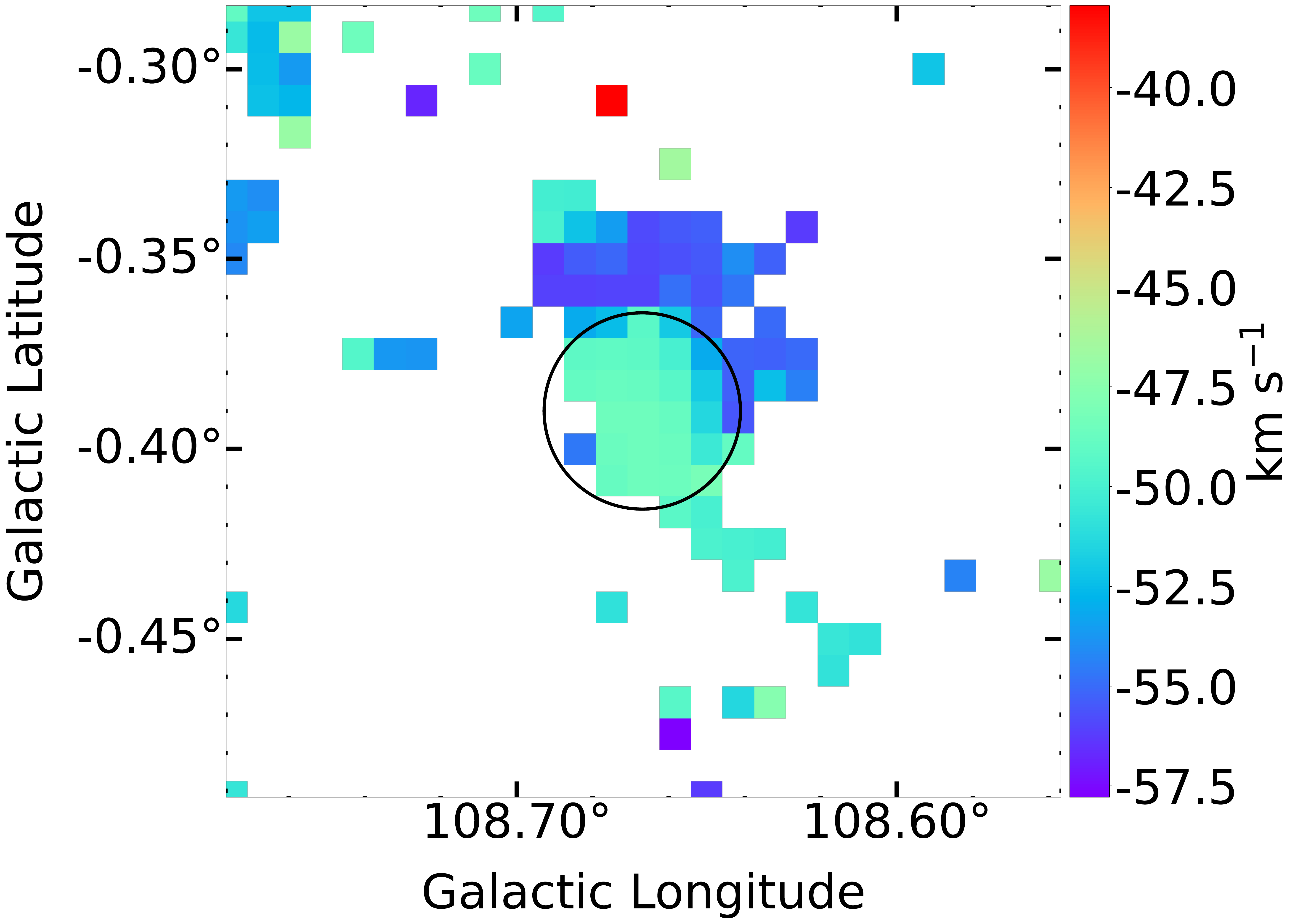}\hfill
    \\(e)
    \end{subfigure}
    \begin{subfigure}[b]{0.4\linewidth}
   \centering
    \includegraphics[trim=8.5cm 0cm 26cm 0cm,width=.4\textwidth]{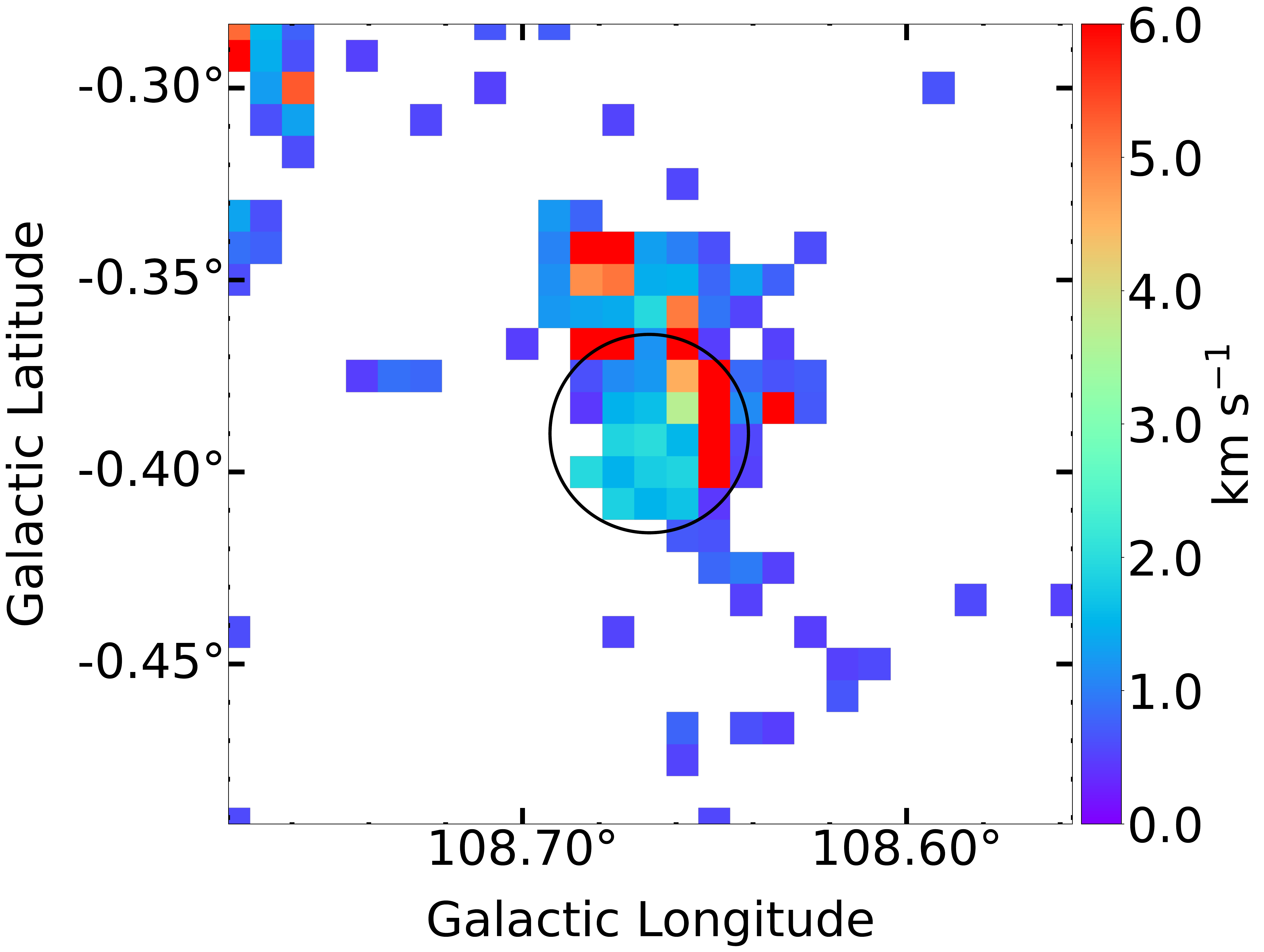}\hfill
    \\(f)
    \end{subfigure}
    \begin{subfigure}[b]{0.4\linewidth}
   \centering
    \includegraphics[trim=14.5cm 0cm 22cm 0cm, width=.4\textwidth]{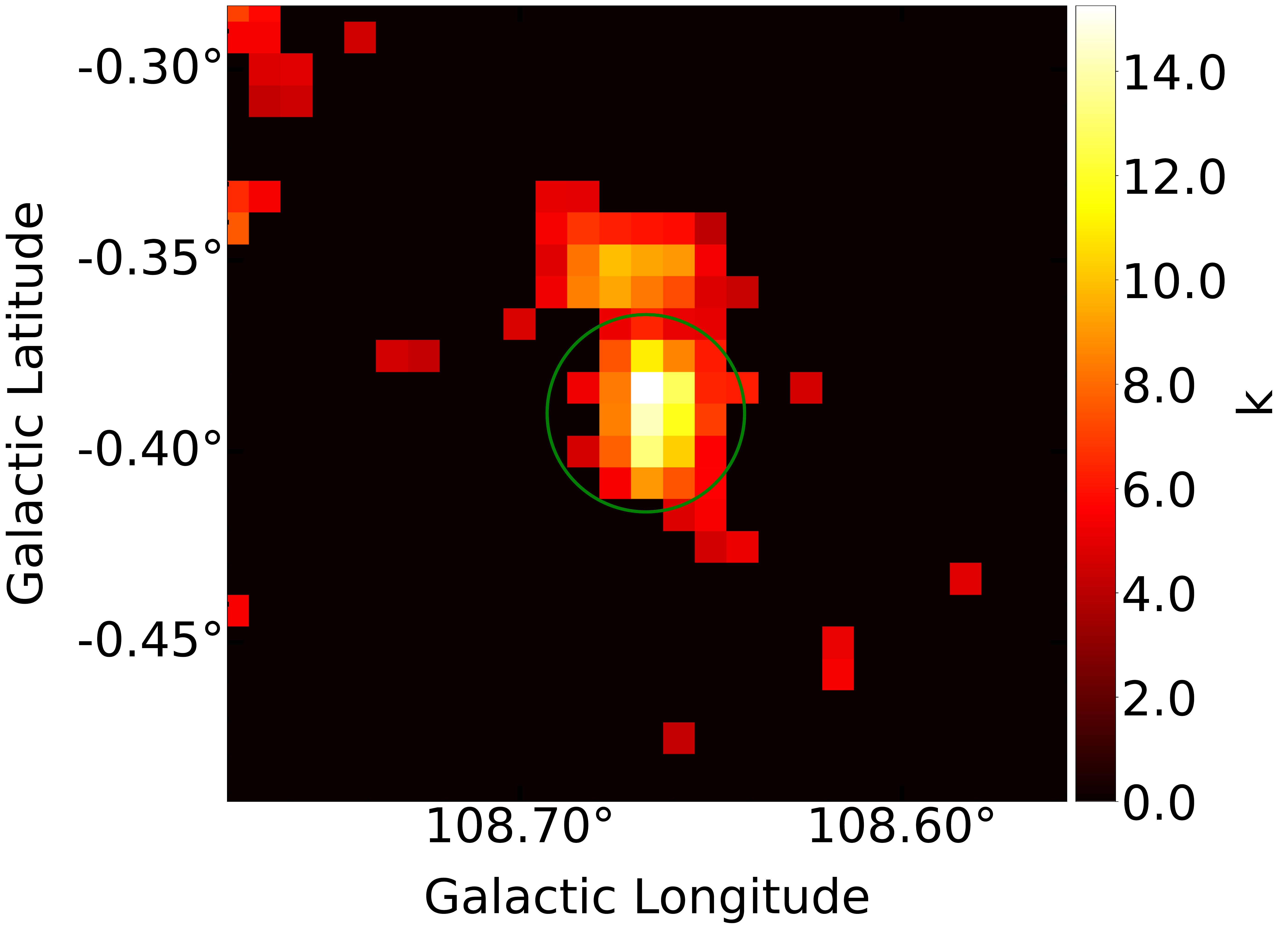}\hfill
    \\(g)
    \end{subfigure}
    \begin{subfigure}[b]{0.4\linewidth}
   \centering
    \includegraphics[trim=1cm -2cm 16cm 0cm, width=.4\textwidth]{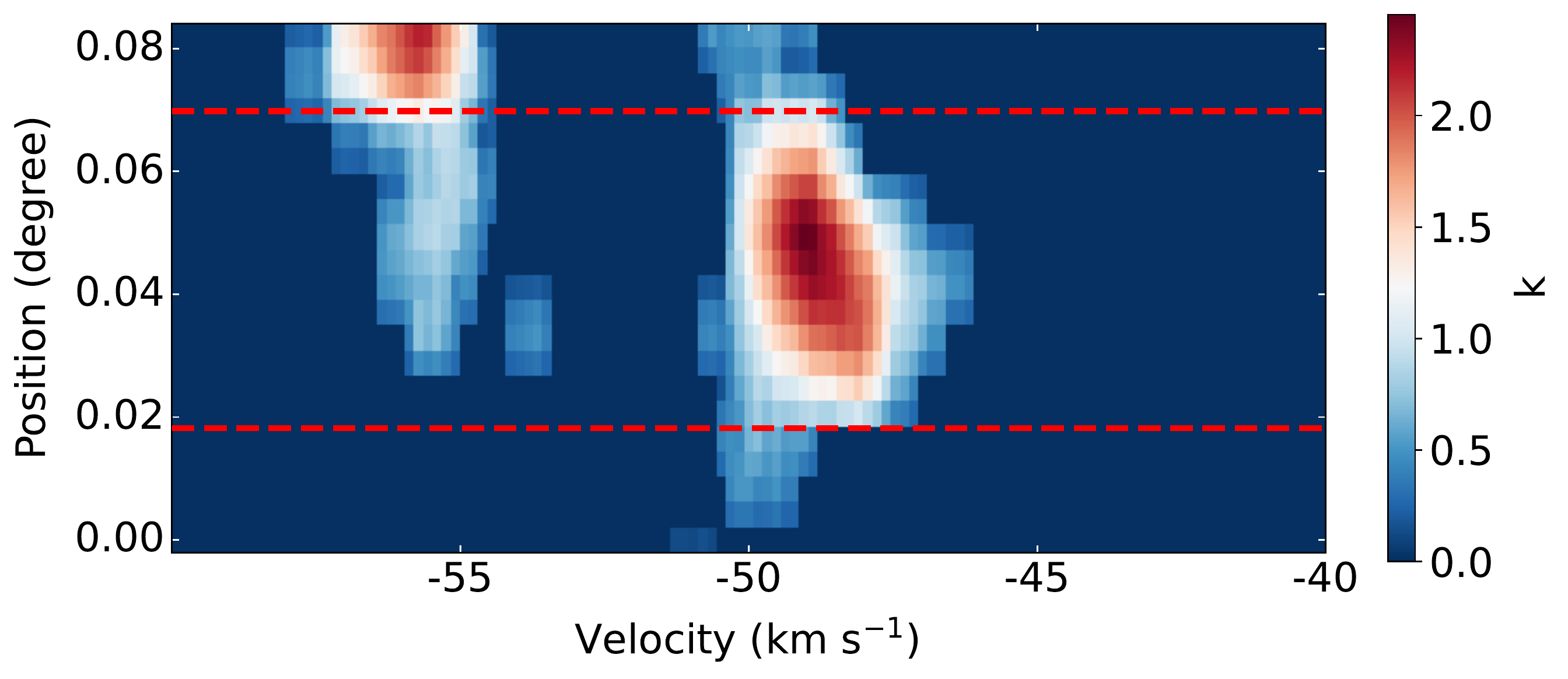}\hfill
    \\(h)
    \end{subfigure}
    \\[\smallskipamount]
    \caption{Same as Figure~\ref{Fig8} but for the G108.666${-}$00.391 region. }
    \label{FigA.11}
\end{figure}

\begin{figure}[h!]
    \centering
     \begin{subfigure}[b]{0.4\linewidth}
     \centering
    \includegraphics[trim=14.5cm 0cm 14.5cm 0cm, width=.4\textwidth]{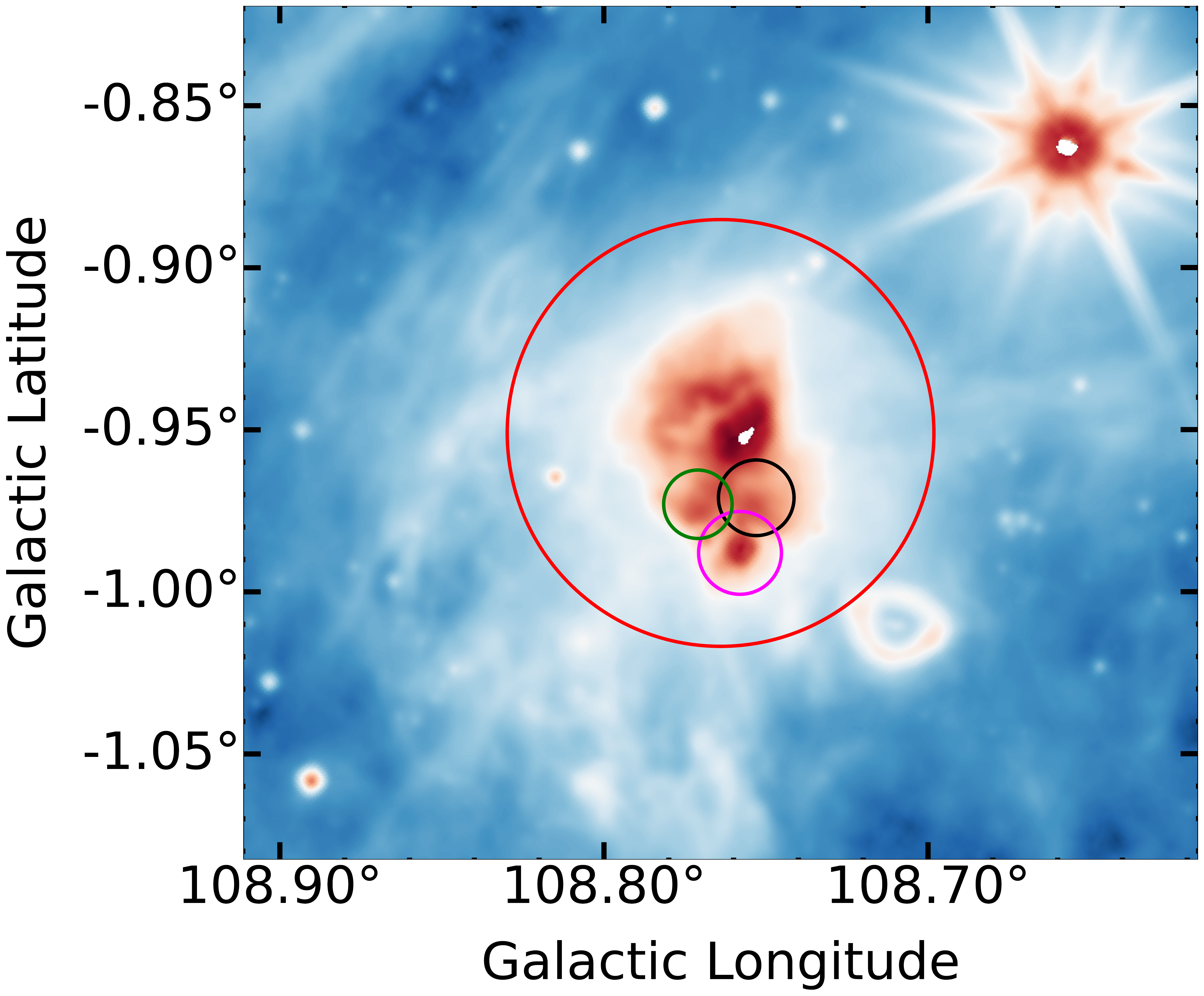}
    \\(a)
   \end{subfigure}
   \begin{subfigure}[b]{0.4\linewidth}
   \centering
    \includegraphics[trim=14.5cm 0cm 14.5cm 0cm,width=.4\textwidth]{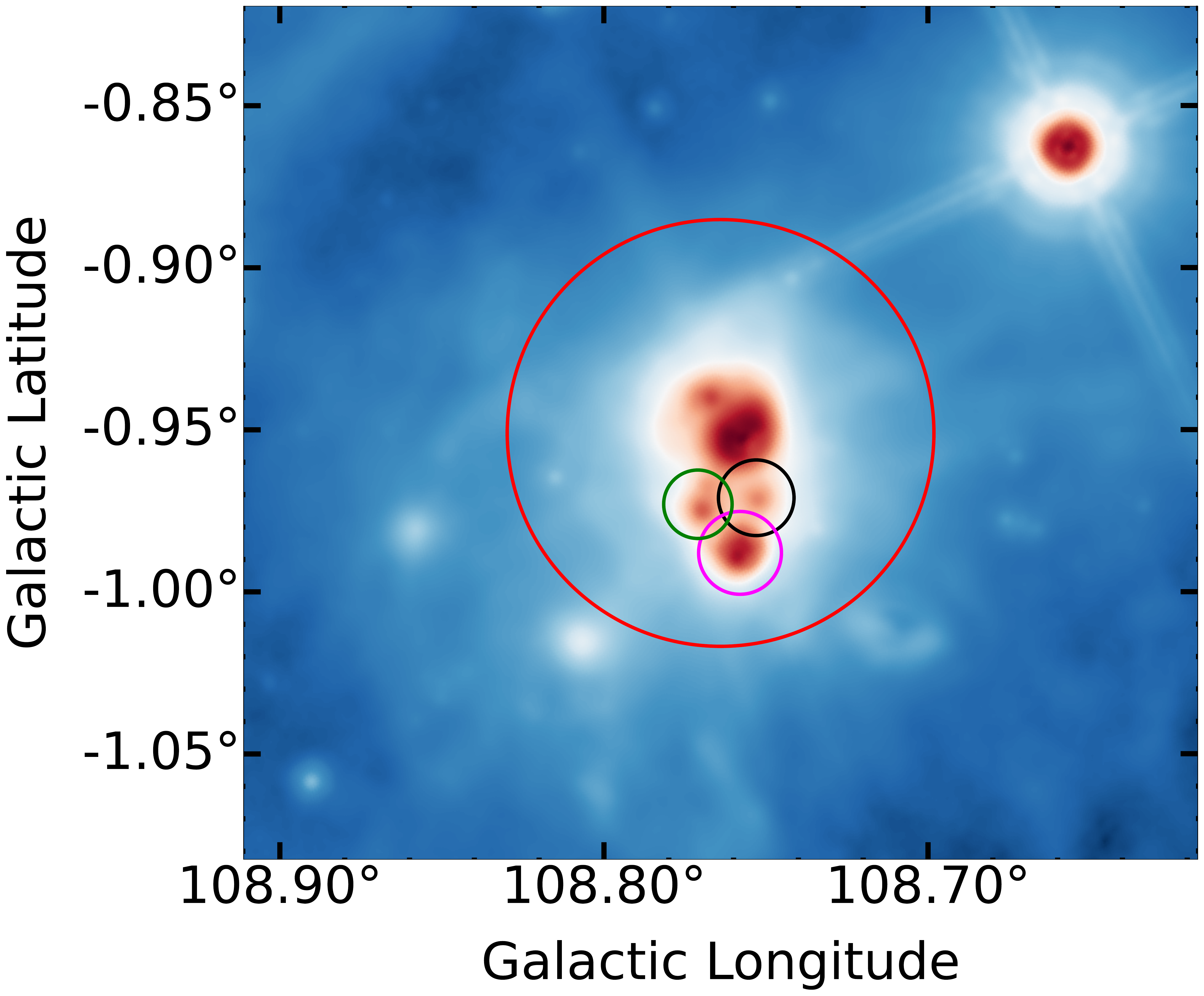}
     \\(b)
    \end{subfigure}
   \begin{subfigure}[b]{0.4\linewidth}
   \centering
    \includegraphics[trim=14.5cm 0cm 14.5cm 0cm,width=.4\textwidth]{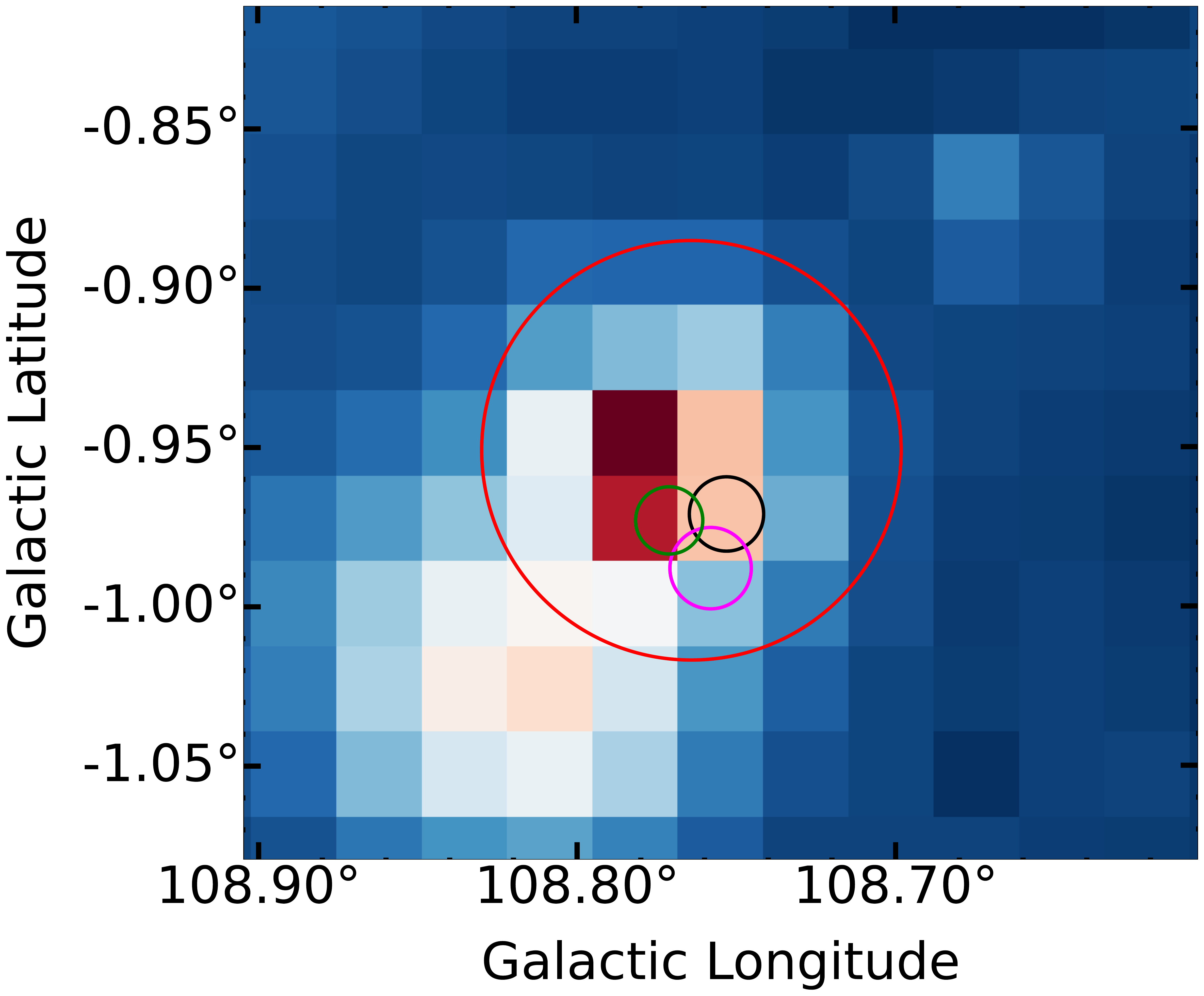}
     \\(c)
    \end{subfigure}
    \begin{subfigure}[b]{0.4\linewidth}
   \centering
    \includegraphics[trim=15cm 0cm 22.5cm 0cm,width=.4\textwidth]{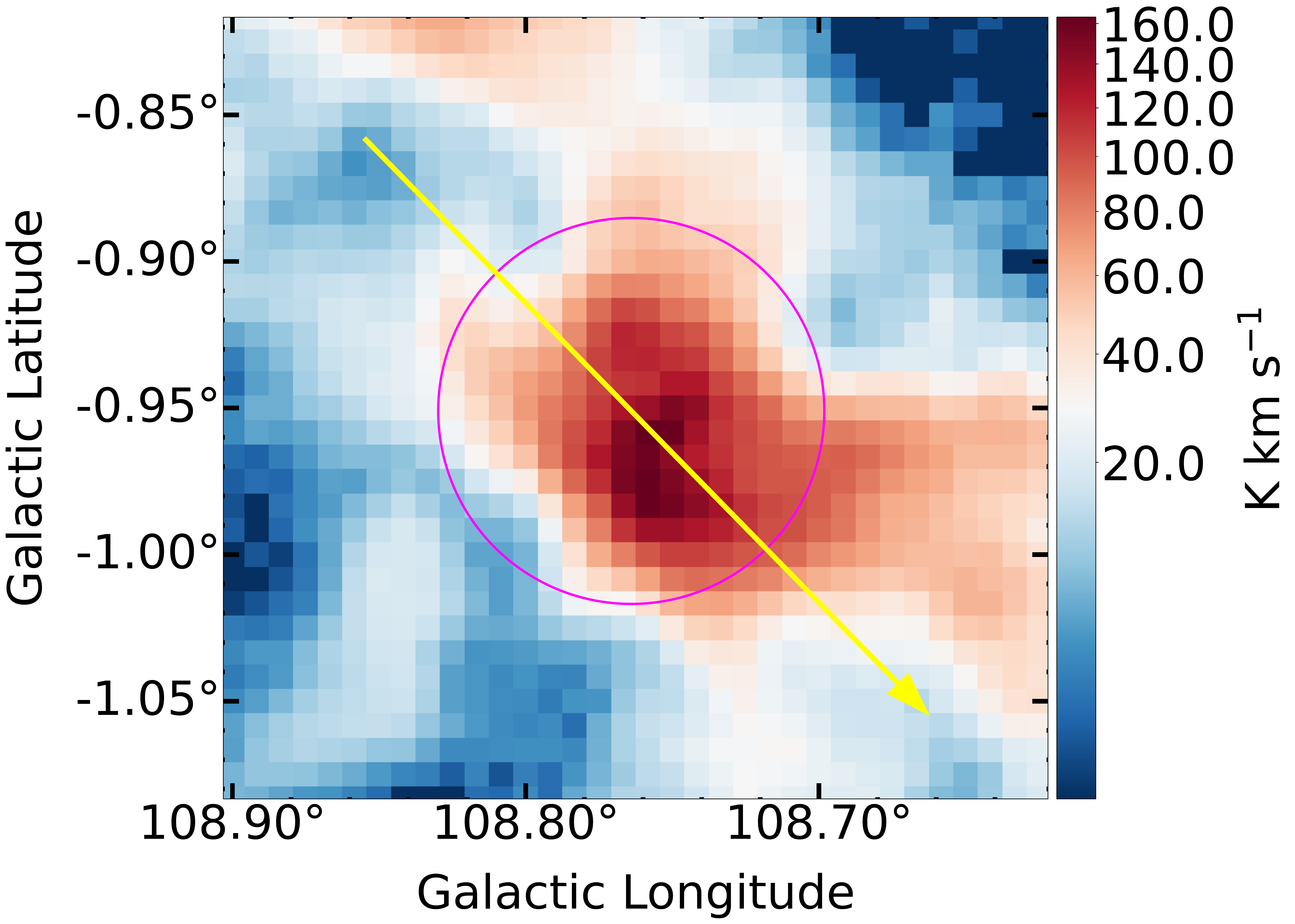}\hfill
    \\(d)
    \end{subfigure}
    \begin{subfigure}[b]{0.4\linewidth}
   \centering
    \includegraphics[trim=14cm 0cm 23cm 0cm,width=.4\textwidth]{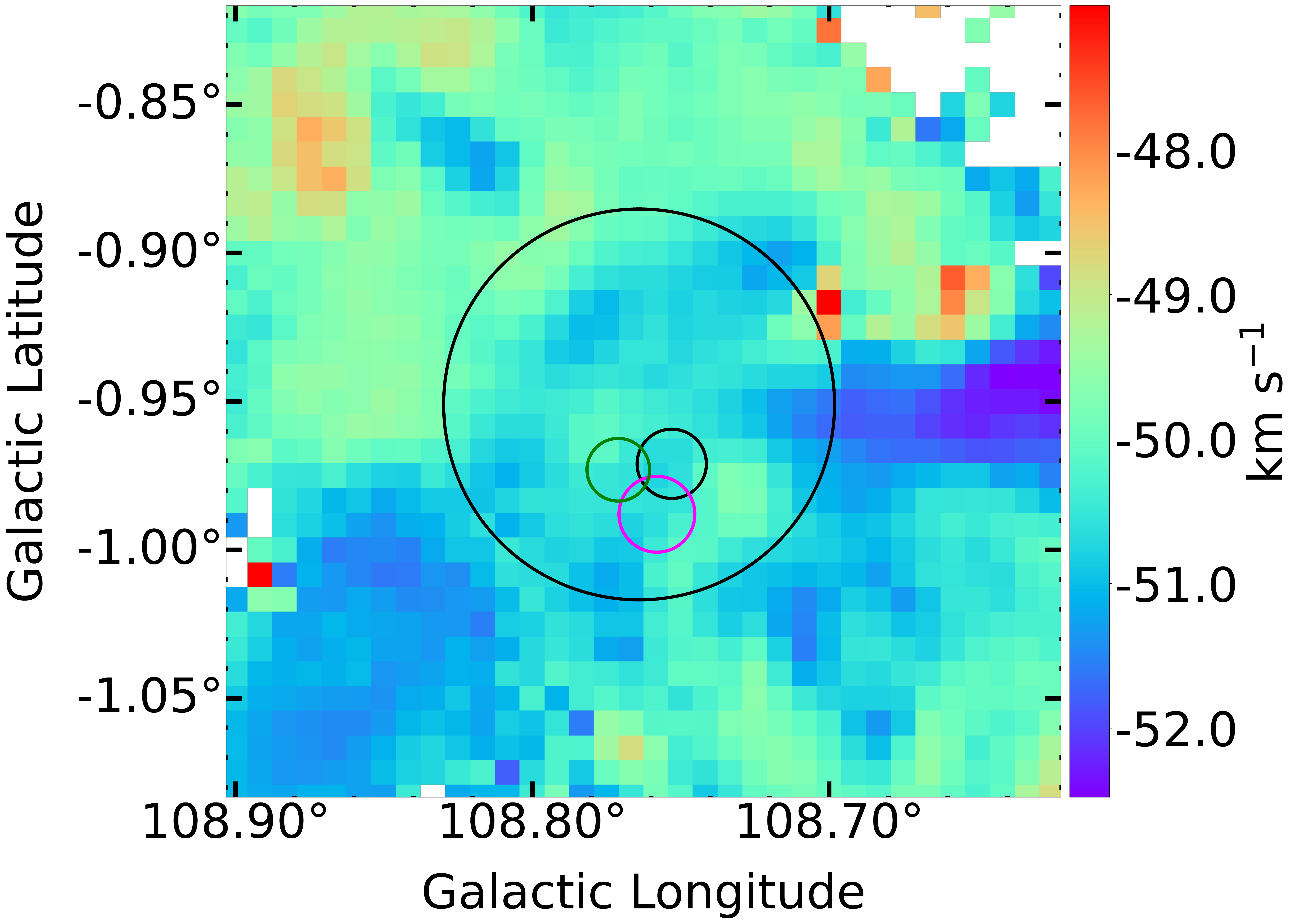}\hfill
    \\(e)
    \end{subfigure}
    \begin{subfigure}[b]{0.4\linewidth}
   \centering
    \includegraphics[trim=8.5cm 0cm 26cm 0cm,width=.4\textwidth]{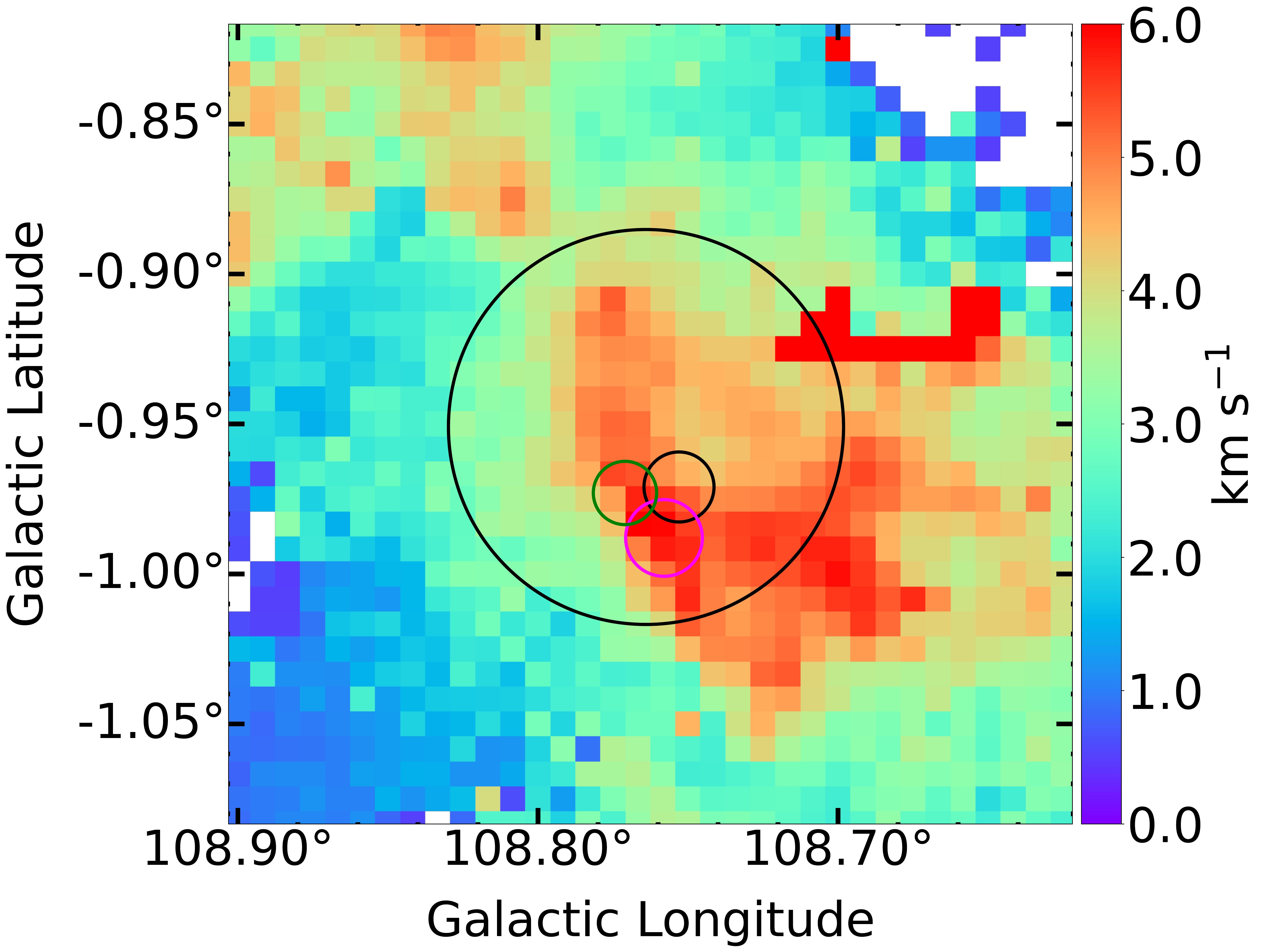}\hfill
    \\(f)
    \end{subfigure}
    \begin{subfigure}[b]{0.4\linewidth}
   \centering
    \includegraphics[trim=14.5cm 0cm 22cm 0cm, width=.4\textwidth]{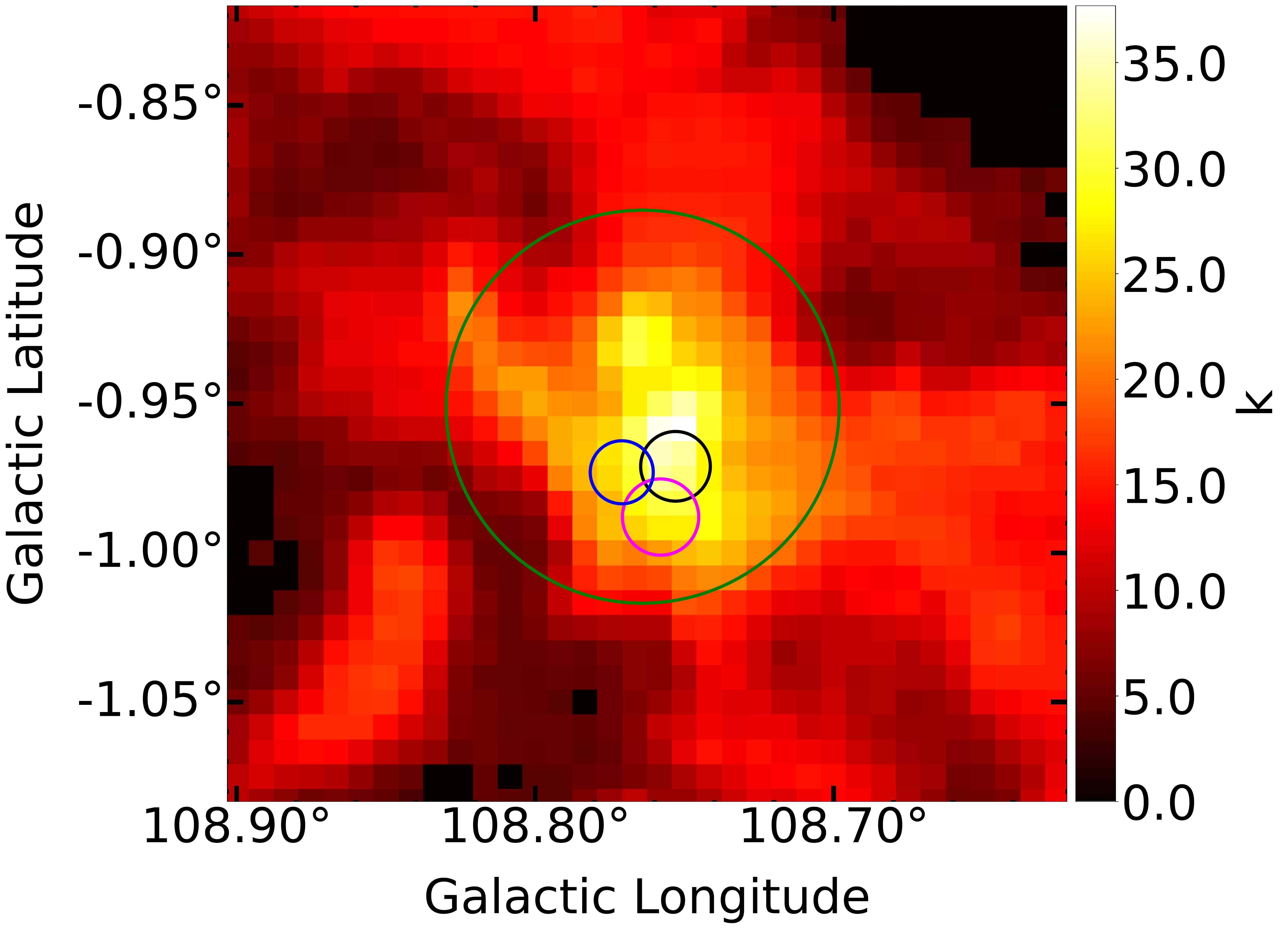}\hfill
    \\(g)
    \end{subfigure}
    \begin{subfigure}[b]{0.4\linewidth}
   \centering
    \includegraphics[trim=1cm -2cm 15cm 0cm, width=.4\textwidth]{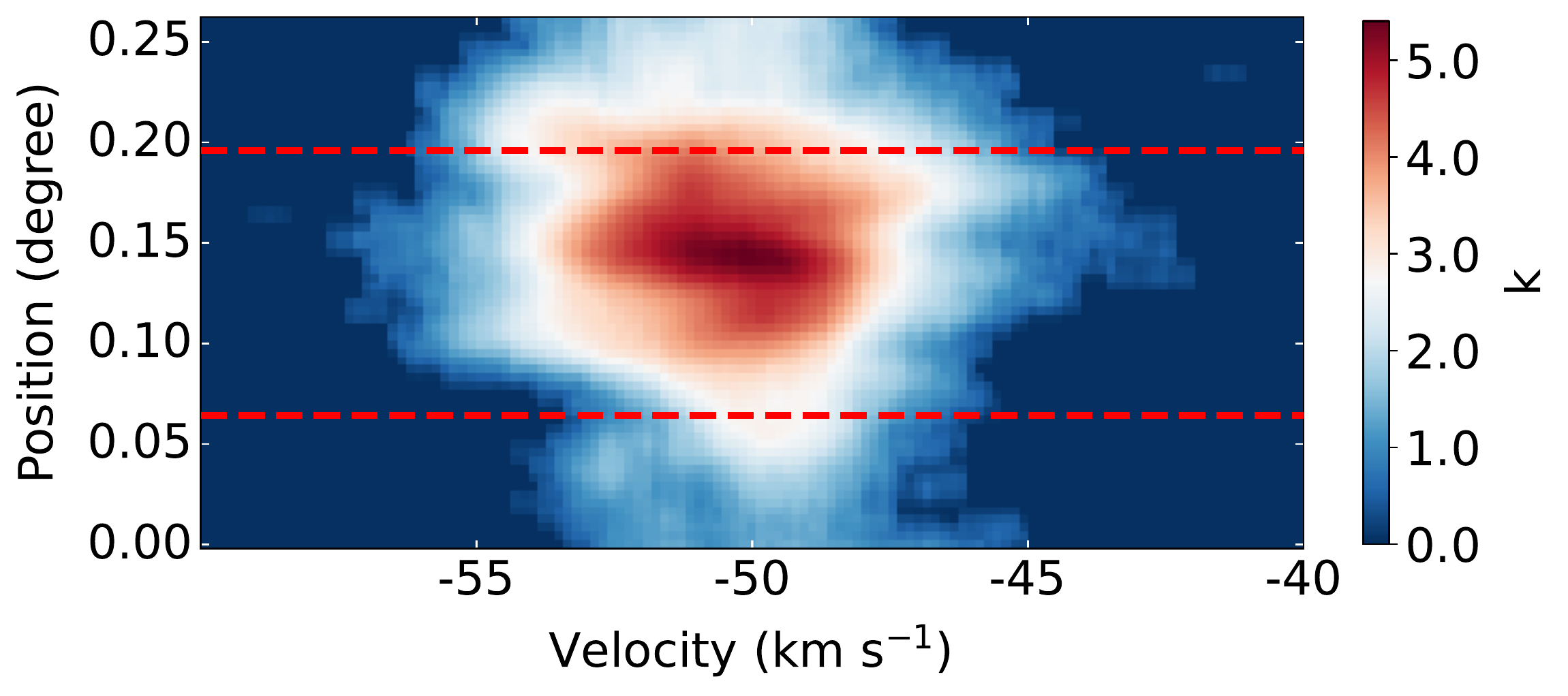}\hfill
    \\(h)
    \end{subfigure}
    \\[\smallskipamount]
    \caption{Same as Figure~\ref{Fig8} but for the S152, G108.752${-}$00.972, G108.758${-}$00.989, and G108.770${-}$00.974. The large circle shows the radius of S152. The black, magenta, and green circles show the radius of G108.752${-}$00.972, G108.758${-}$00.989, and G108.770${-}$00.974, respectively.}
    \label{FigA.12}
\end{figure}

\begin{figure}[h!]
    \centering
     \begin{subfigure}[b]{0.4\linewidth}
     \centering
    \includegraphics[trim=14.5cm 0cm 14.5cm 0cm, width=.4\textwidth]{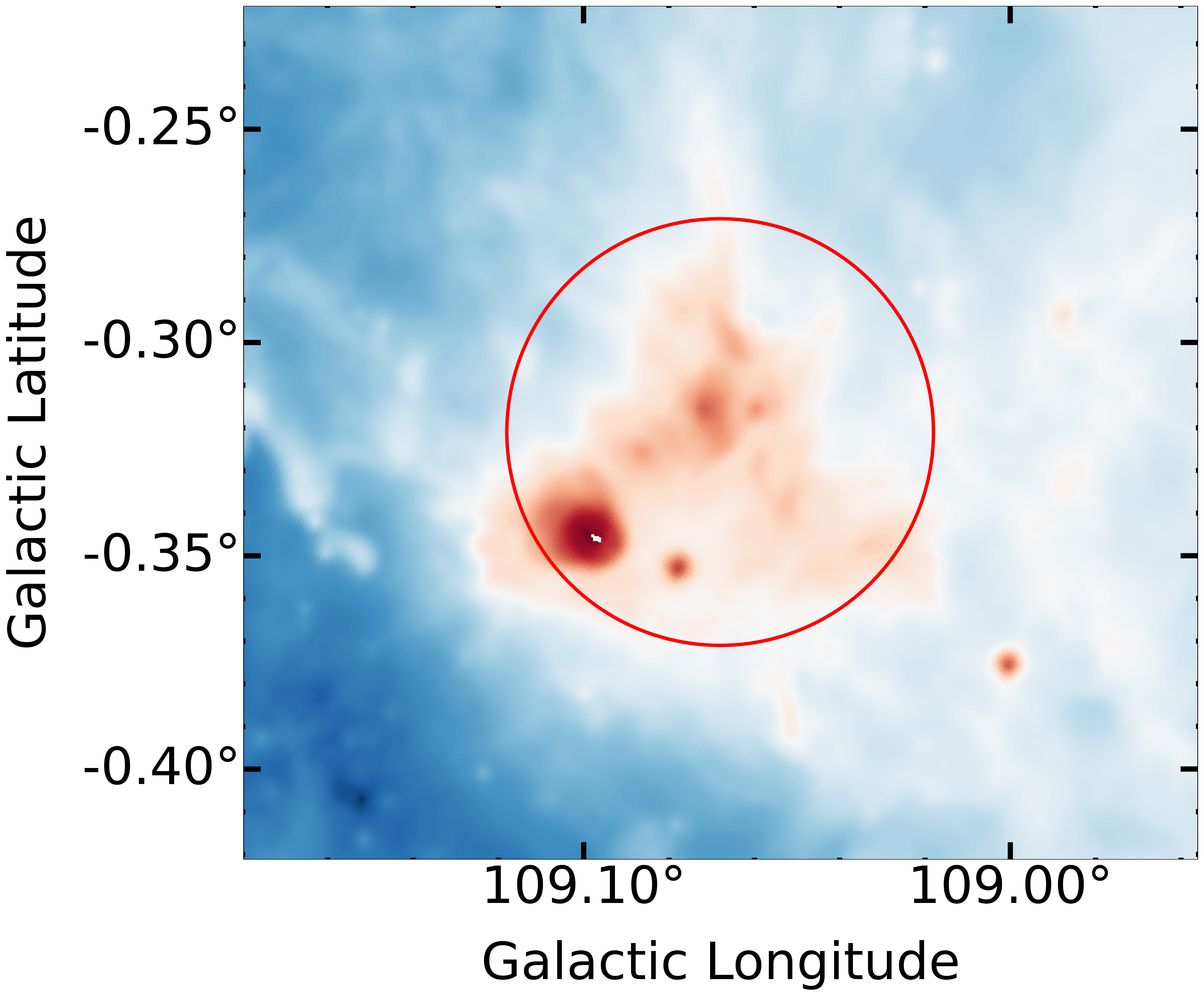}
    \\(a)
   \end{subfigure}
   \begin{subfigure}[b]{0.4\linewidth}
   \centering
    \includegraphics[trim=14.5cm 0cm 14.5cm 0cm,width=.4\textwidth]{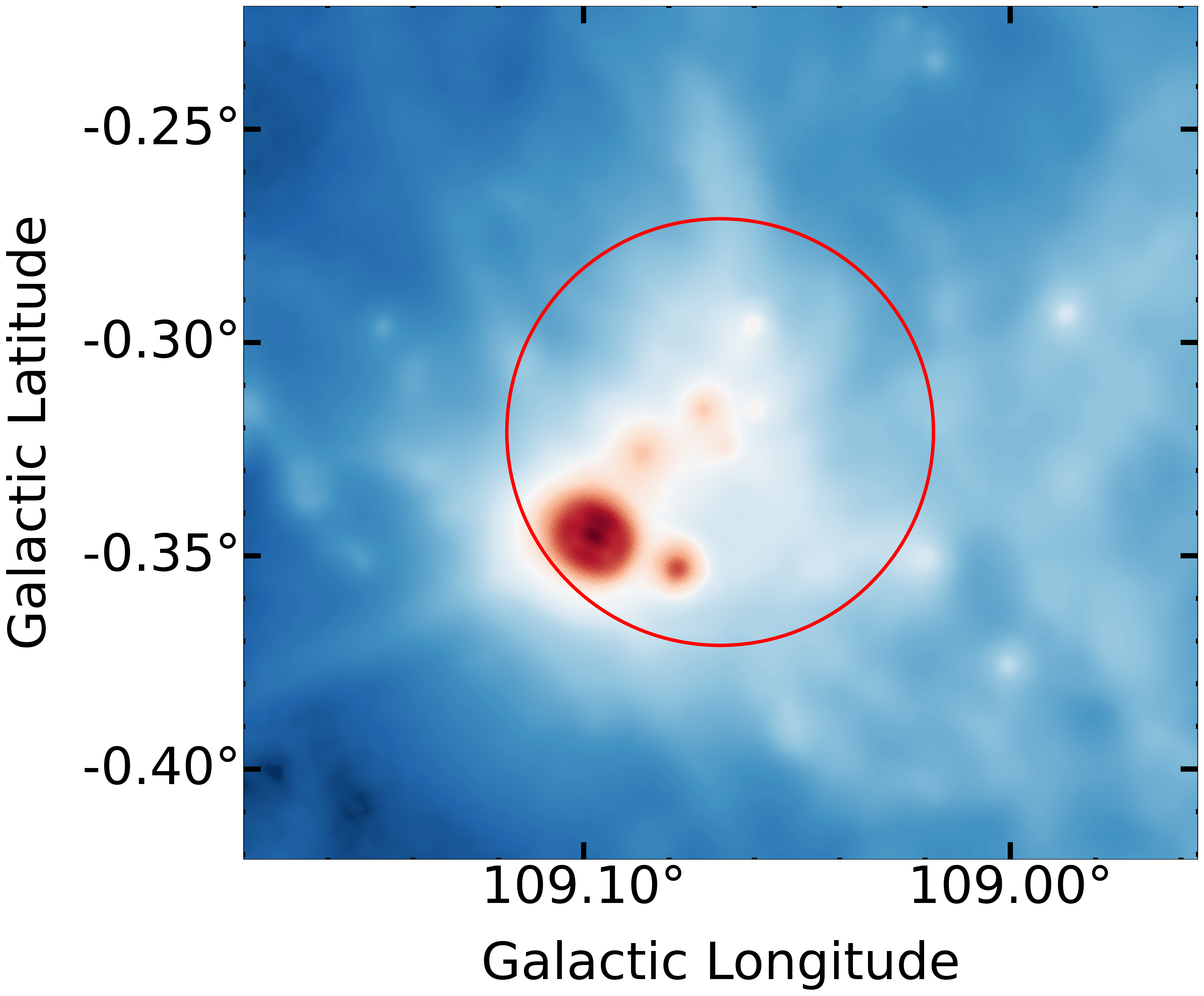}
     \\(b)
    \end{subfigure}
   \begin{subfigure}[b]{0.4\linewidth}
   \centering
    \includegraphics[trim=14.5cm 0cm 14.5cm 0cm,width=.4\textwidth]{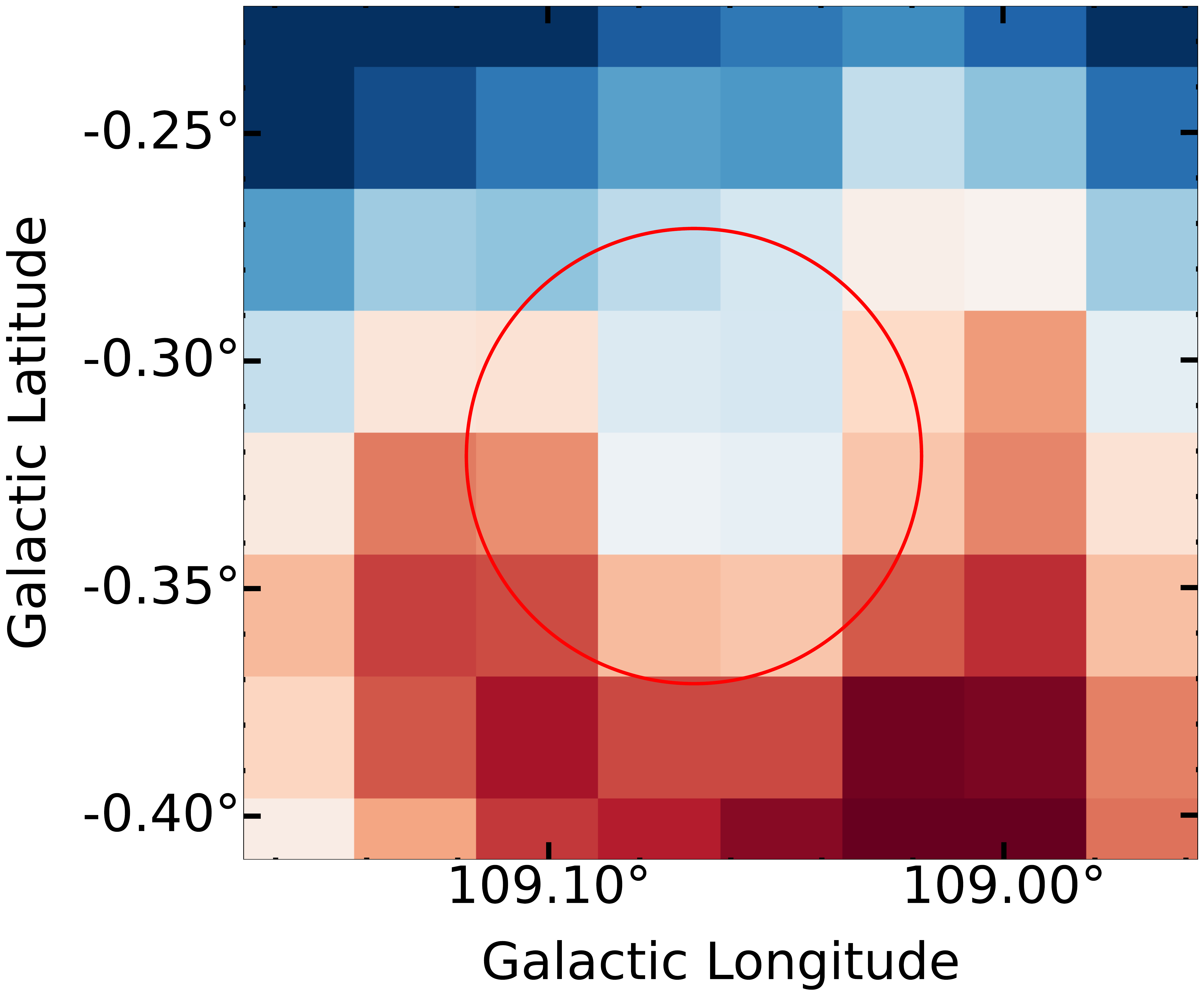}
     \\(c)
    \end{subfigure}
    \begin{subfigure}[b]{0.4\linewidth}
   \centering
    \includegraphics[trim=15cm 0cm 22.5cm 0cm,width=.4\textwidth]{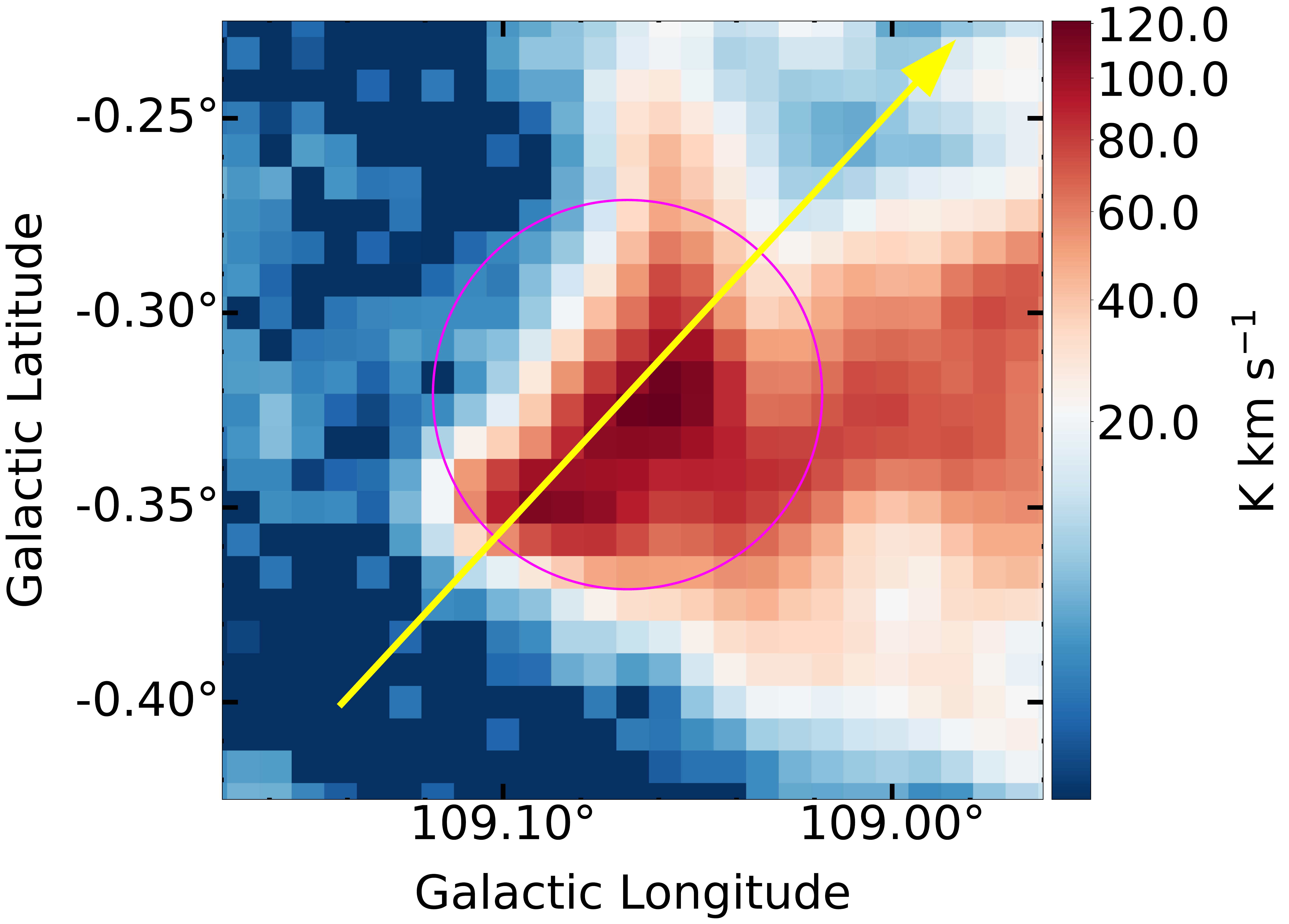}\hfill
    \\(d)
    \end{subfigure}
    \begin{subfigure}[b]{0.4\linewidth}
   \centering
    \includegraphics[trim=14cm 0cm 23cm 0cm,width=.4\textwidth]{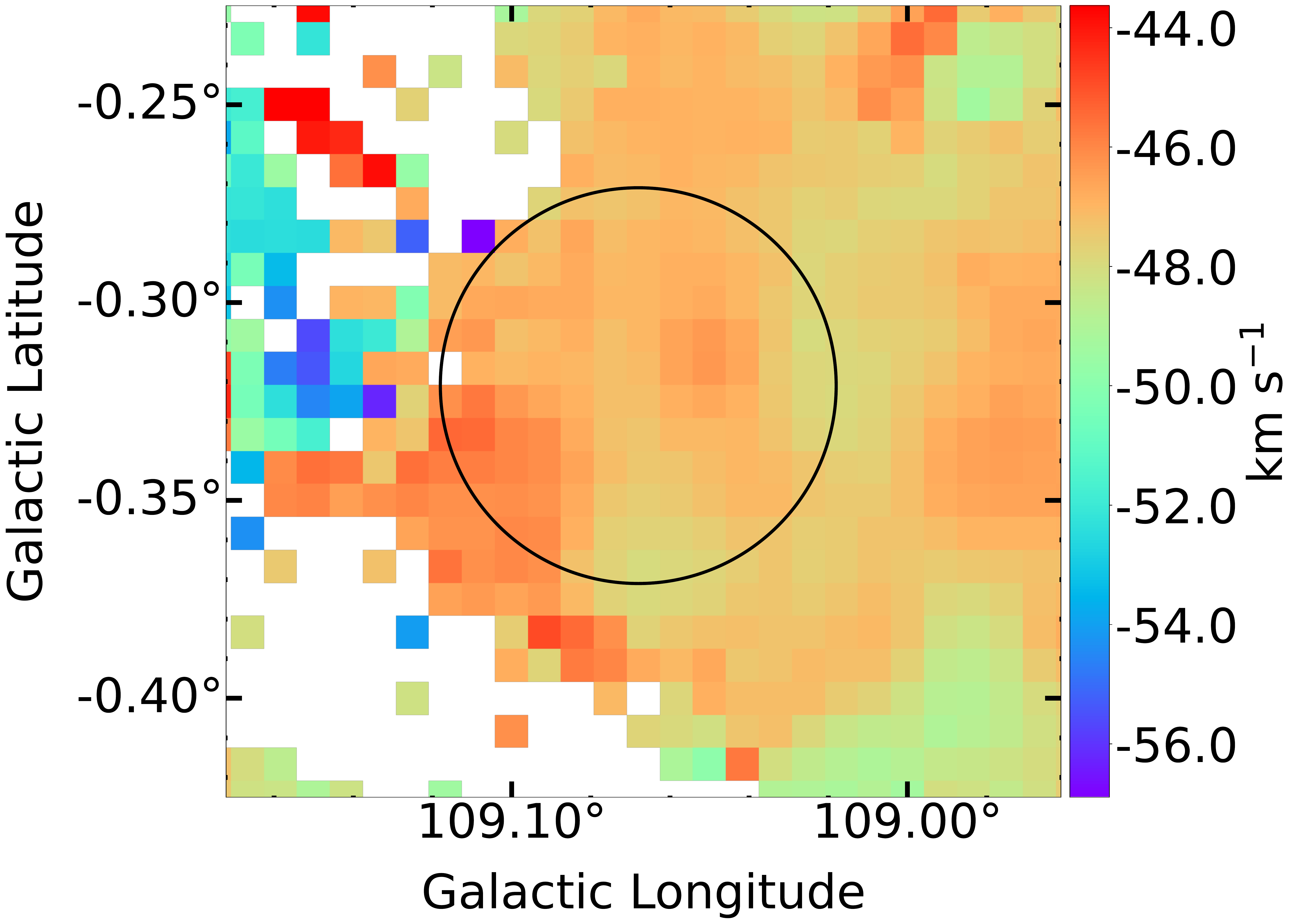}\hfill
    \\(e)
    \end{subfigure}
    \begin{subfigure}[b]{0.4\linewidth}
   \centering
    \includegraphics[trim=8.5cm 0cm 26cm 0cm,width=.4\textwidth]{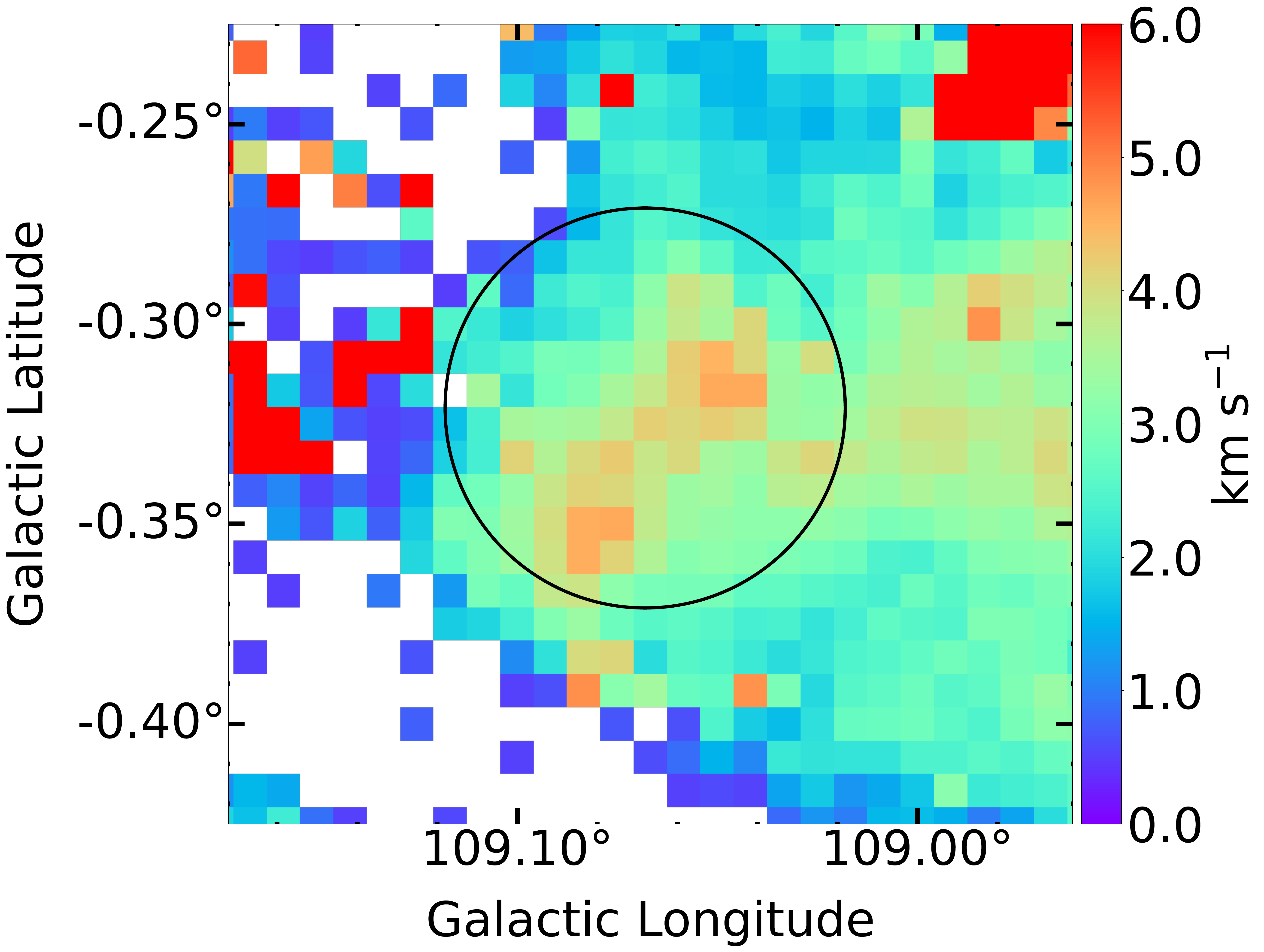}\hfill
    \\(f)
    \end{subfigure}
    \begin{subfigure}[b]{0.4\linewidth}
   \centering
    \includegraphics[trim=14.5cm 0cm 22cm 0cm, width=.4\textwidth]{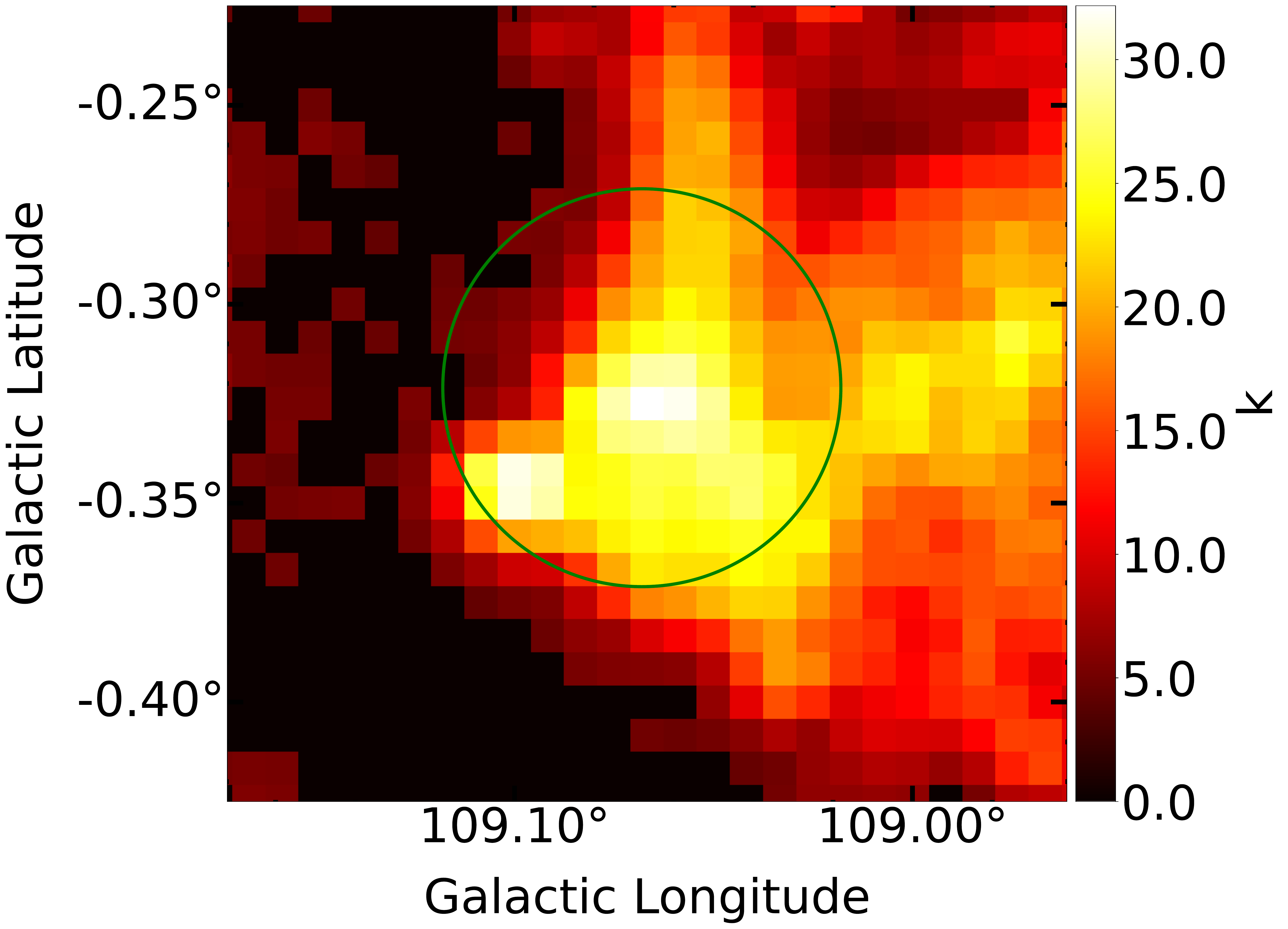}\hfill
    \\(g)
    \end{subfigure}
    \begin{subfigure}[b]{0.4\linewidth}
   \centering
    \includegraphics[trim=1cm -2cm 15cm 0cm, width=.4\textwidth]{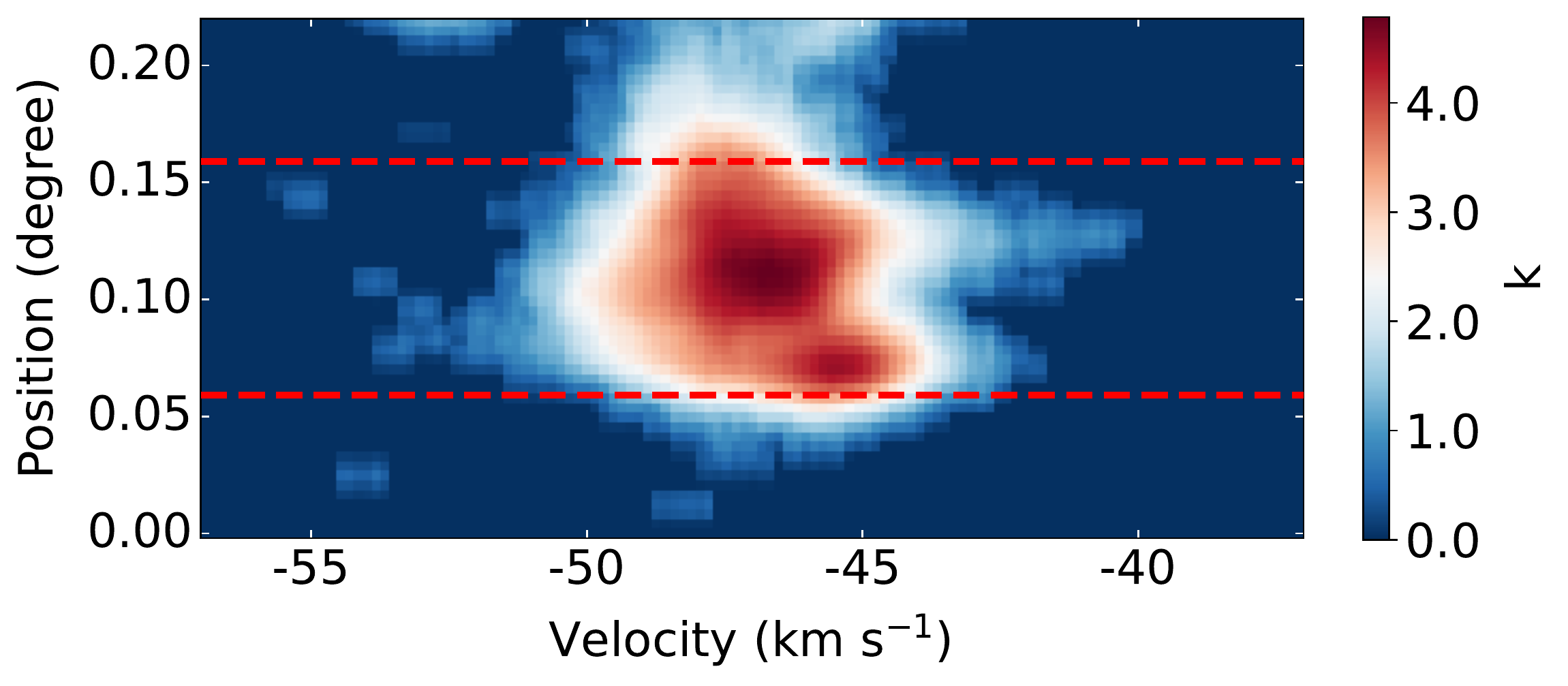}\hfill
    \\(h)
    \end{subfigure}
    \\[\smallskipamount]
    \caption{Same as Figure~\ref{Fig8} but for the G109.068${-}$00.322 region. }
    \label{FigA.13}
\end{figure}

\begin{figure}[h!]
    \centering
     \begin{subfigure}[b]{0.4\linewidth}
     \centering
    \includegraphics[trim=15cm 0cm 15cm 0cm, width=.4\textwidth]{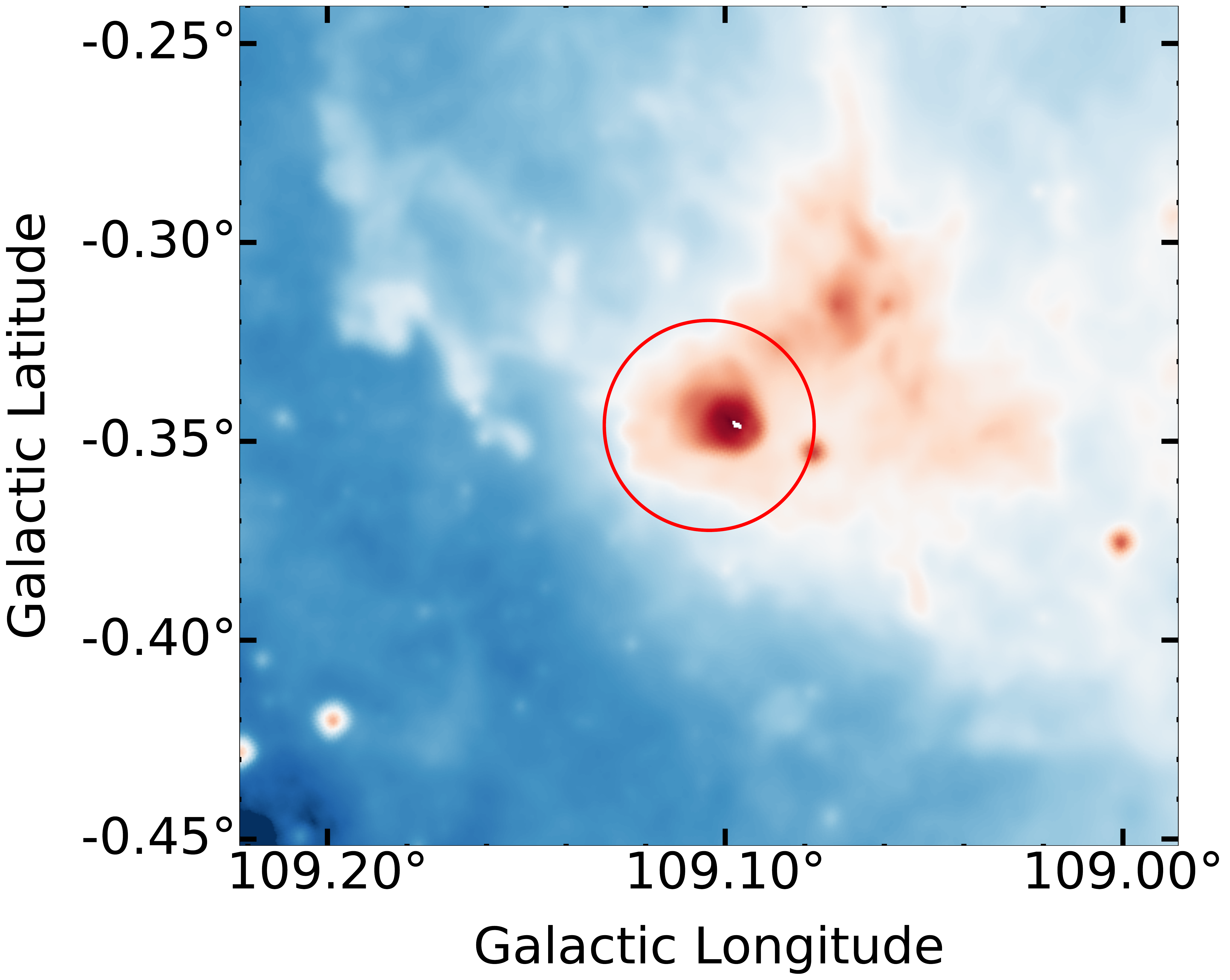}
    \\(a)
   \end{subfigure}
   \begin{subfigure}[b]{0.4\linewidth}
   \centering
    \includegraphics[trim=15cm 0cm 15cm 0cm,width=.4\textwidth]{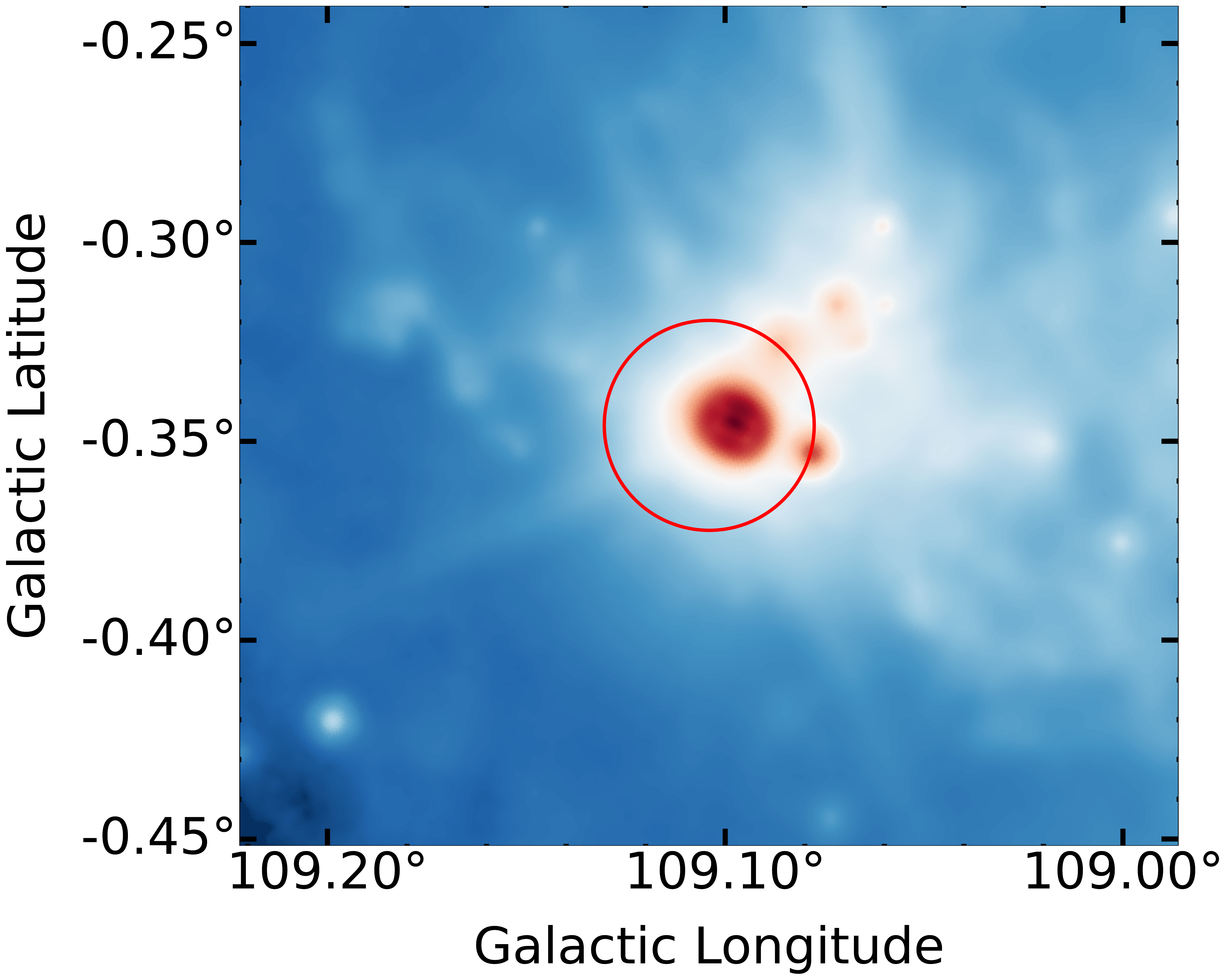}
     \\(b)
    \end{subfigure}
   \begin{subfigure}[b]{0.4\linewidth}
   \centering
    \includegraphics[trim=15cm 0cm 15cm 0cm,width=.4\textwidth]{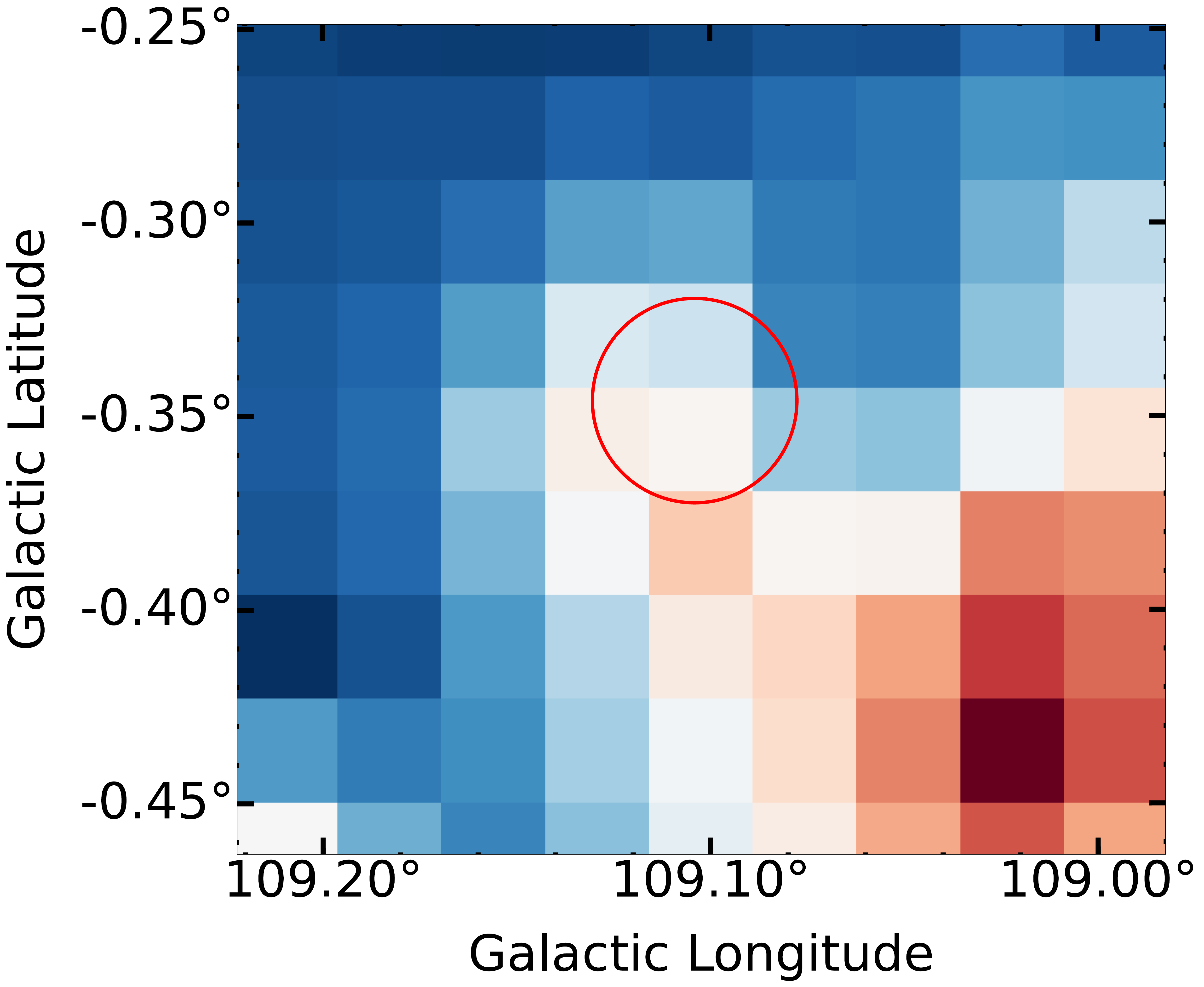}
     \\(c)
    \end{subfigure}
    \begin{subfigure}[b]{0.4\linewidth}
   \centering
    \includegraphics[trim=15cm 0cm 22.5cm 0cm,width=.4\textwidth]{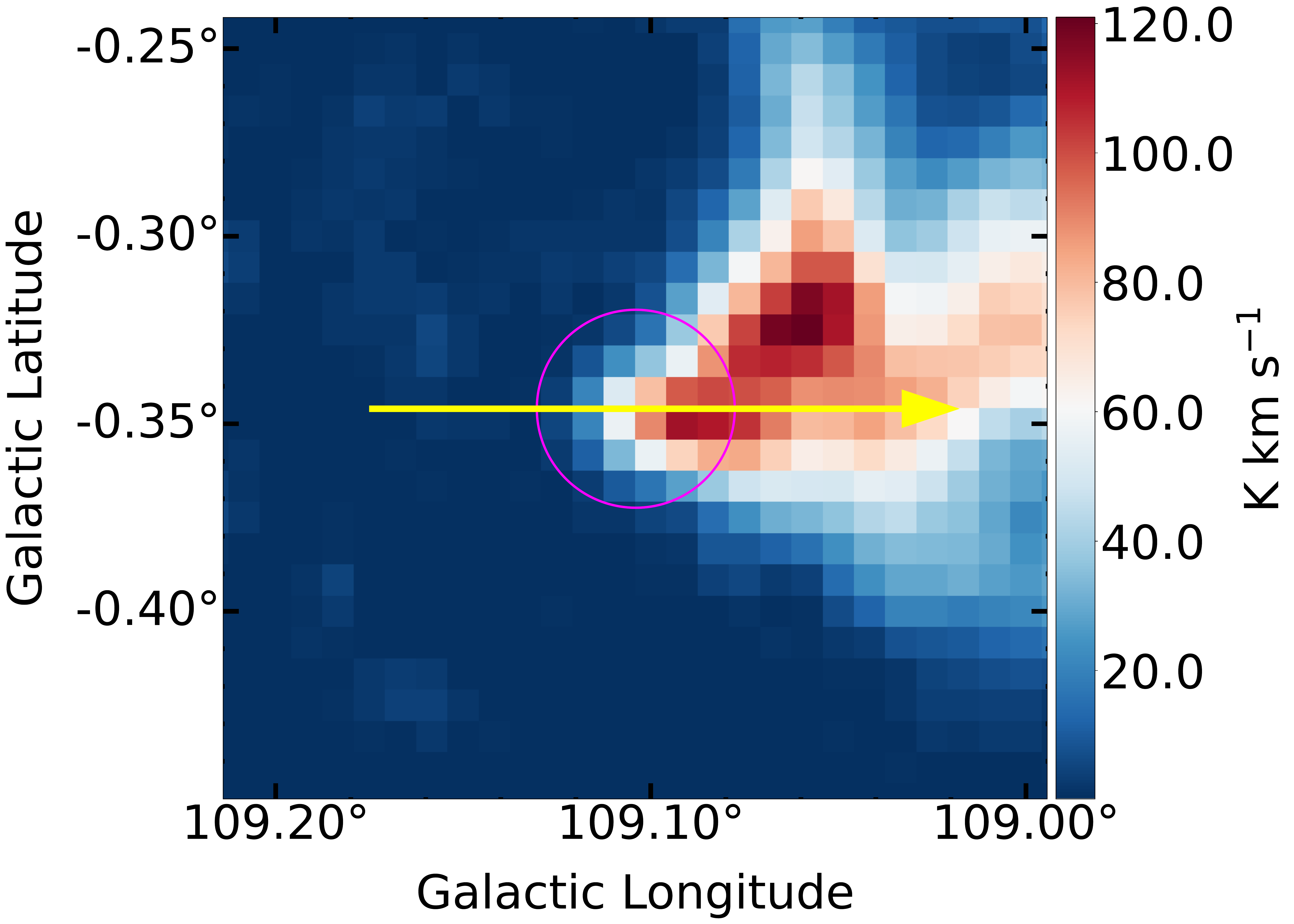}\hfill
    \\(d)
    \end{subfigure}
    \begin{subfigure}[b]{0.4\linewidth}
   \centering
    \includegraphics[trim=14cm 0cm 23cm 0cm,width=.4\textwidth]{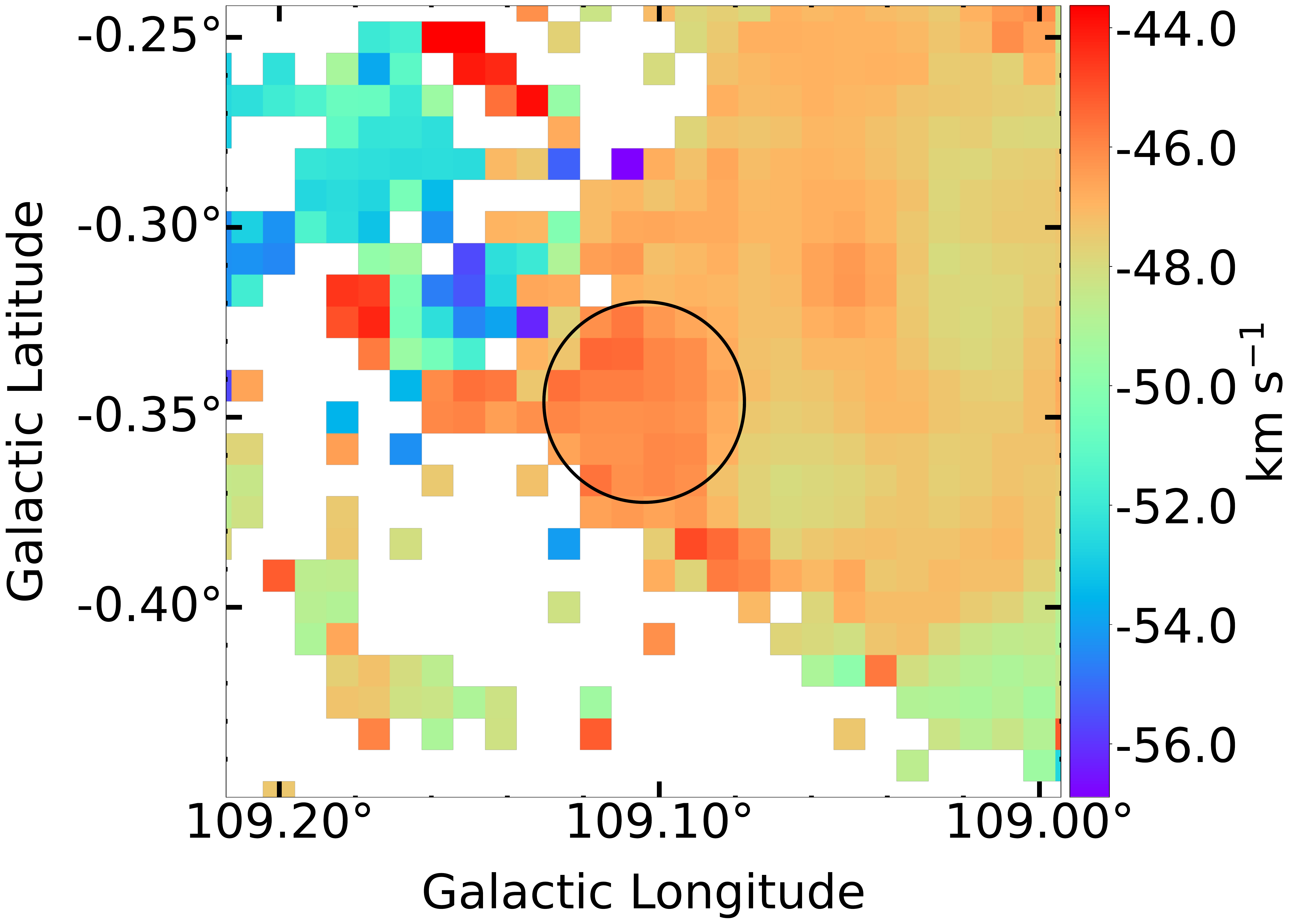}\hfill
    \\(e)
    \end{subfigure}
    \begin{subfigure}[b]{0.4\linewidth}
   \centering
    \includegraphics[trim=8.5cm 0cm 26cm 0cm,width=.4\textwidth]{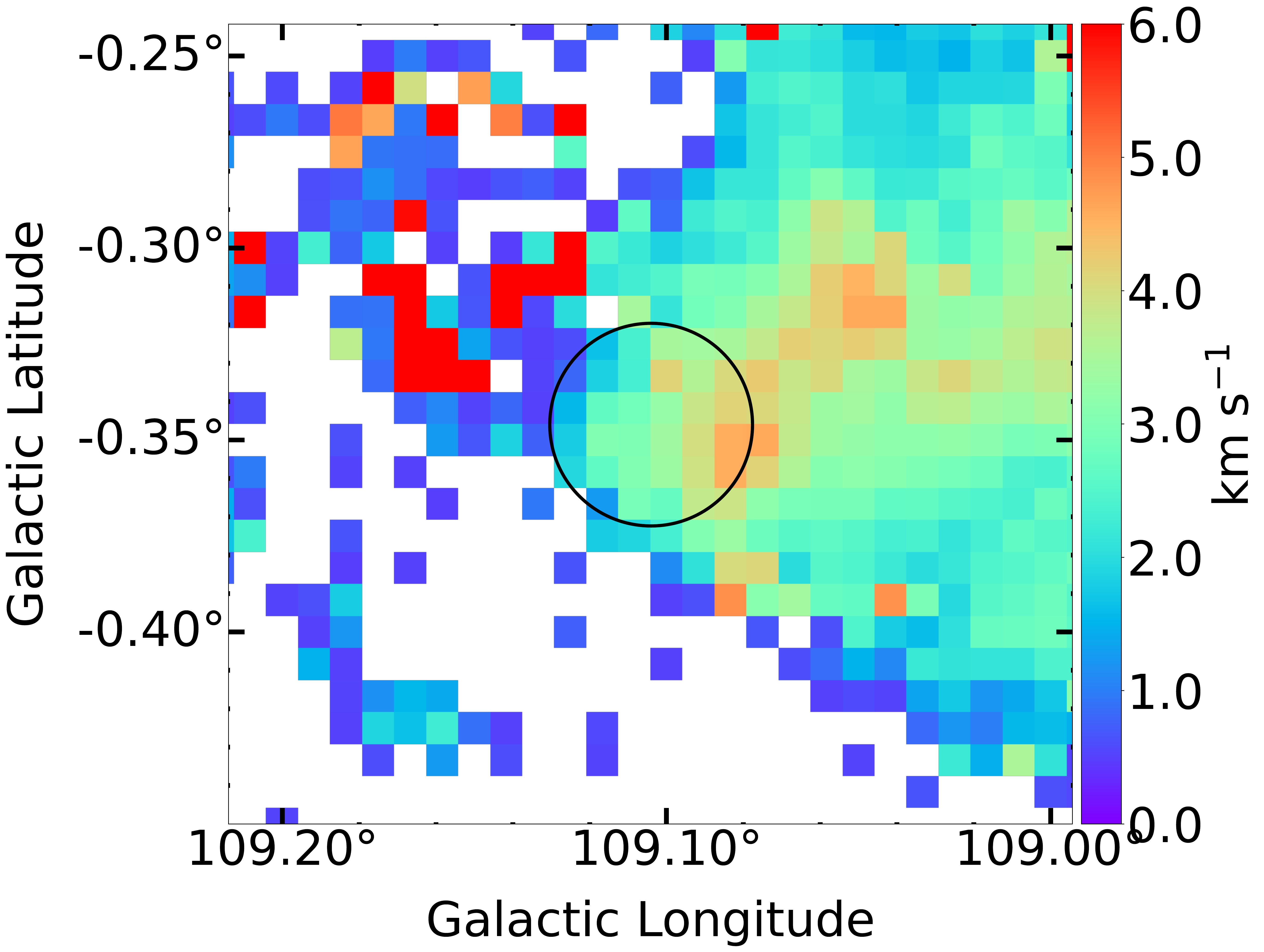}\hfill
    \\(f)
    \end{subfigure}
    \begin{subfigure}[b]{0.4\linewidth}
   \centering
    \includegraphics[trim=14.5cm 0cm 22cm 0cm, width=.4\textwidth]{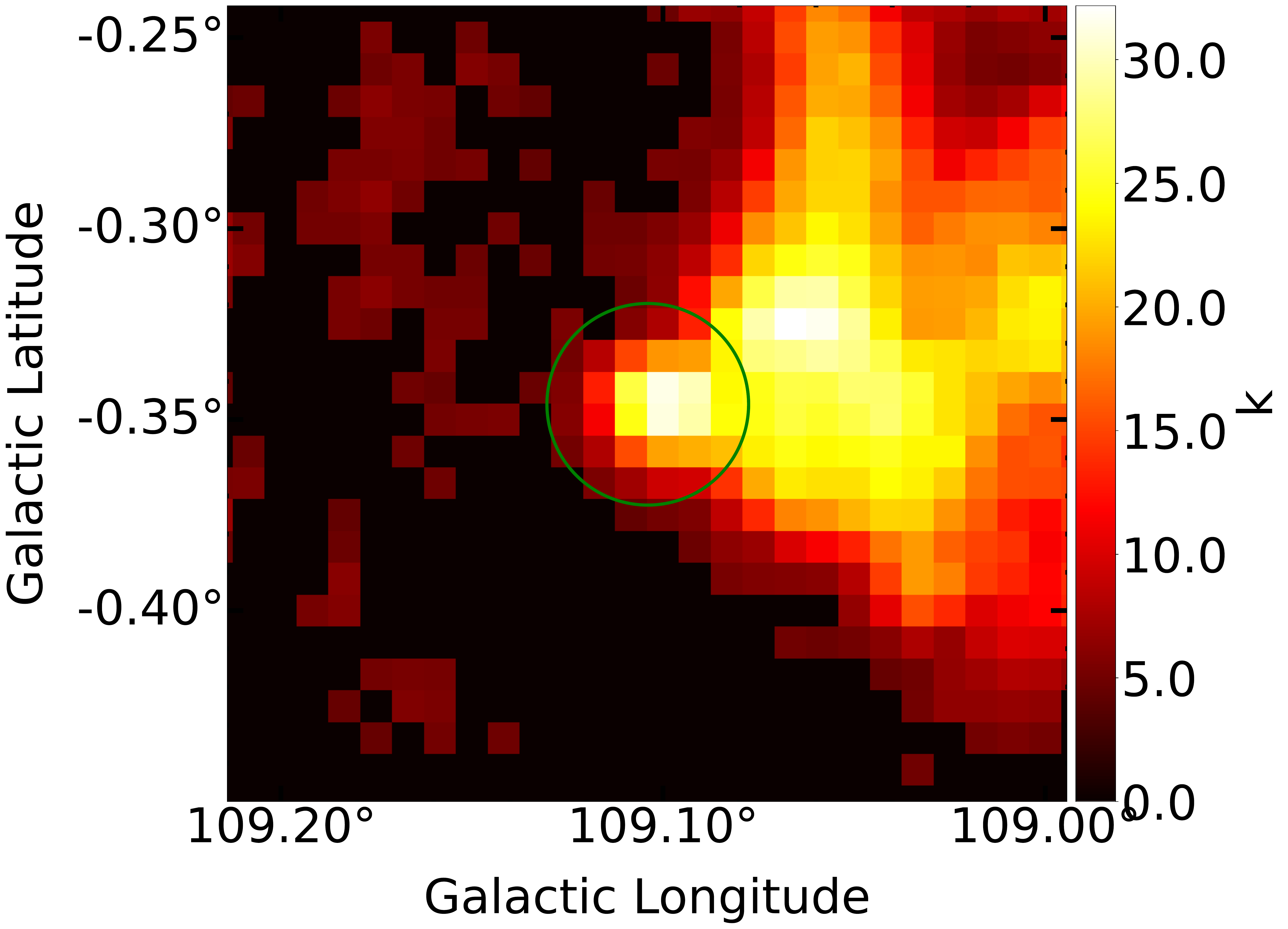}\hfill
    \\(g)
    \end{subfigure}
    \begin{subfigure}[b]{0.4\linewidth}
   \centering
    \includegraphics[trim=1cm -2cm 15cm 0cm, width=.4\textwidth]{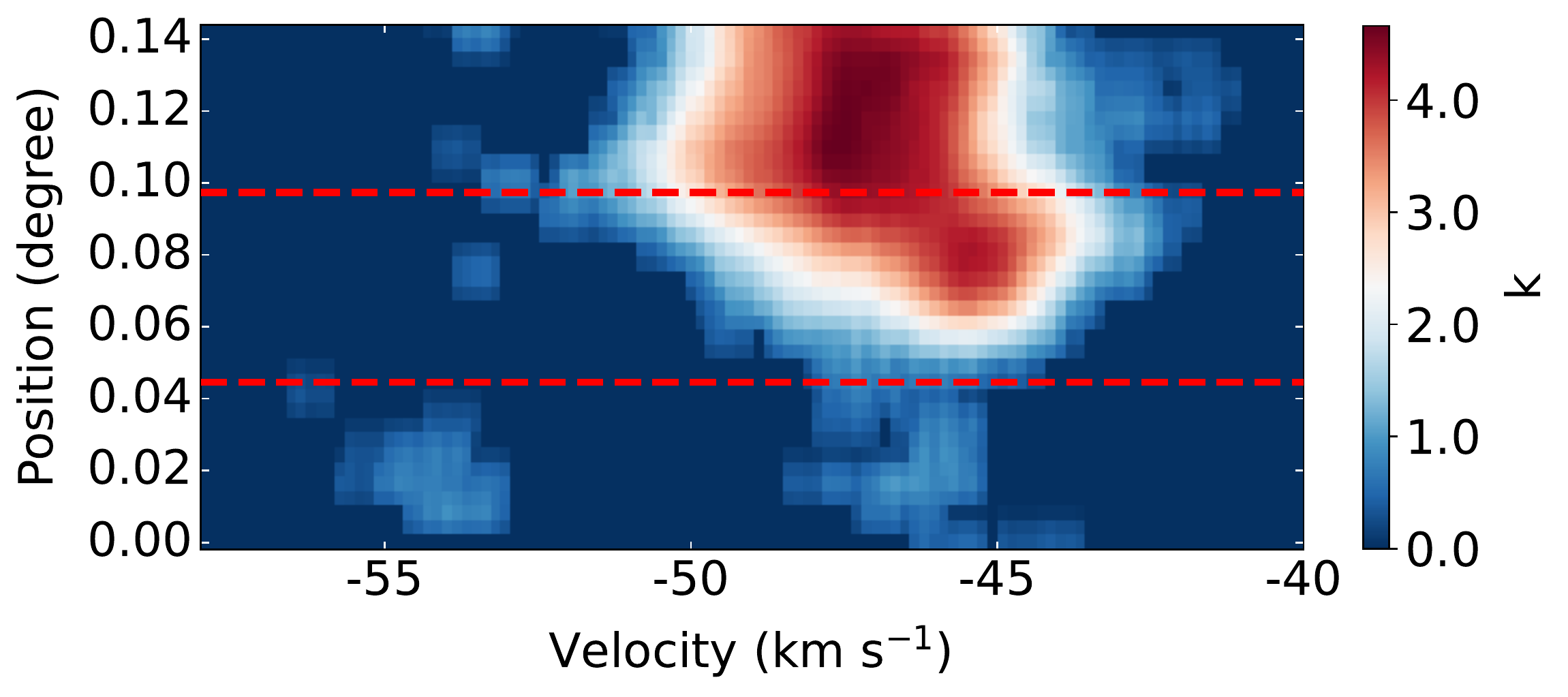}\hfill
    \\(h)
    \end{subfigure}
    \\[\smallskipamount]
    \caption{Same as Figure~\ref{Fig8} but for the G109.104${-}$00.347 region. }
    \label{FigA.14}
\end{figure}

\begin{figure}[h!]
    \centering
     \begin{subfigure}[b]{0.4\linewidth}
     \centering
    \includegraphics[trim=14.5cm 0cm 14.5cm 0cm, width=.4\textwidth]{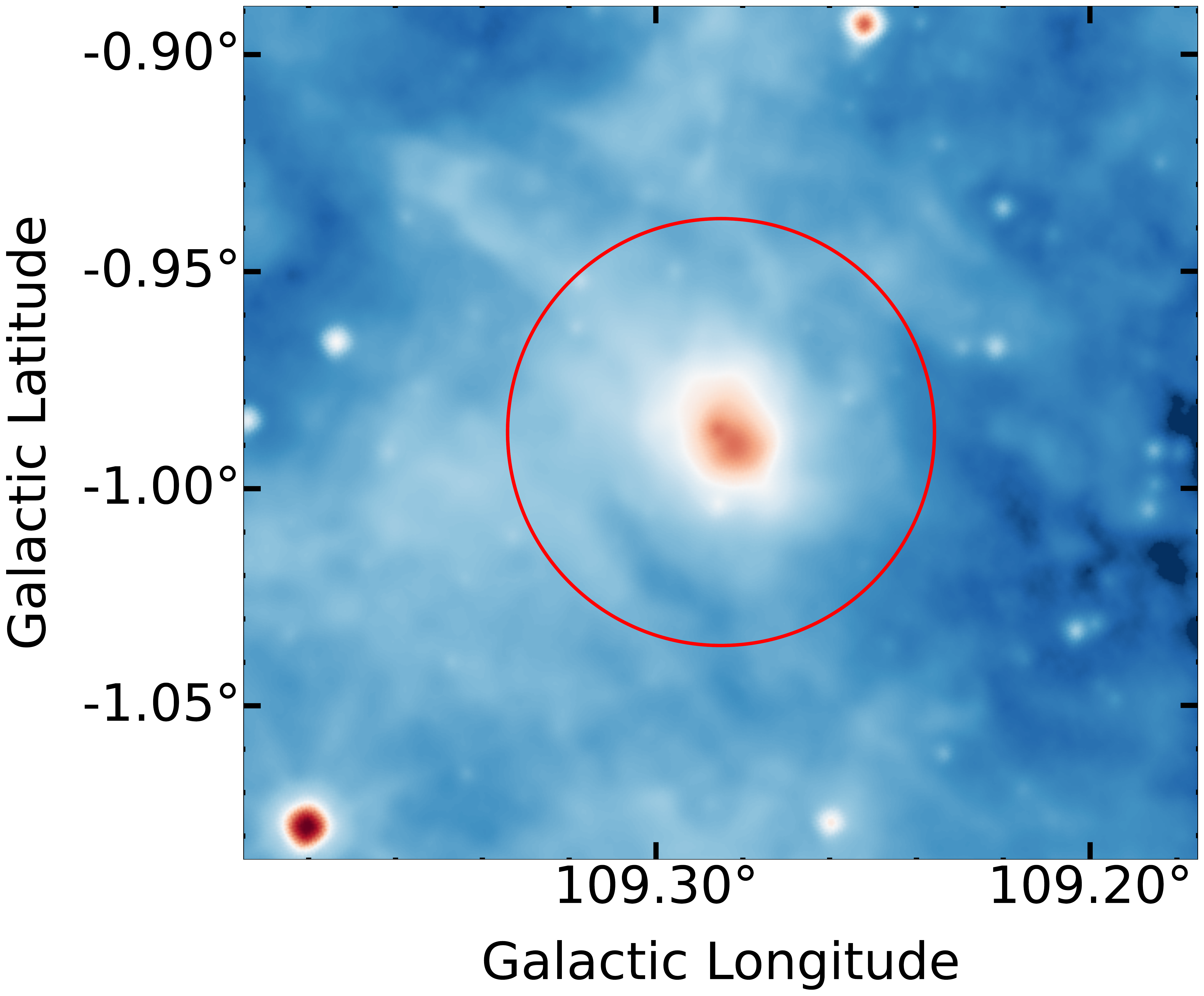}
    \\(a)
   \end{subfigure}
   \begin{subfigure}[b]{0.4\linewidth}
   \centering
    \includegraphics[trim=14.5cm 0cm 14.5cm 0cm,width=.4\textwidth]{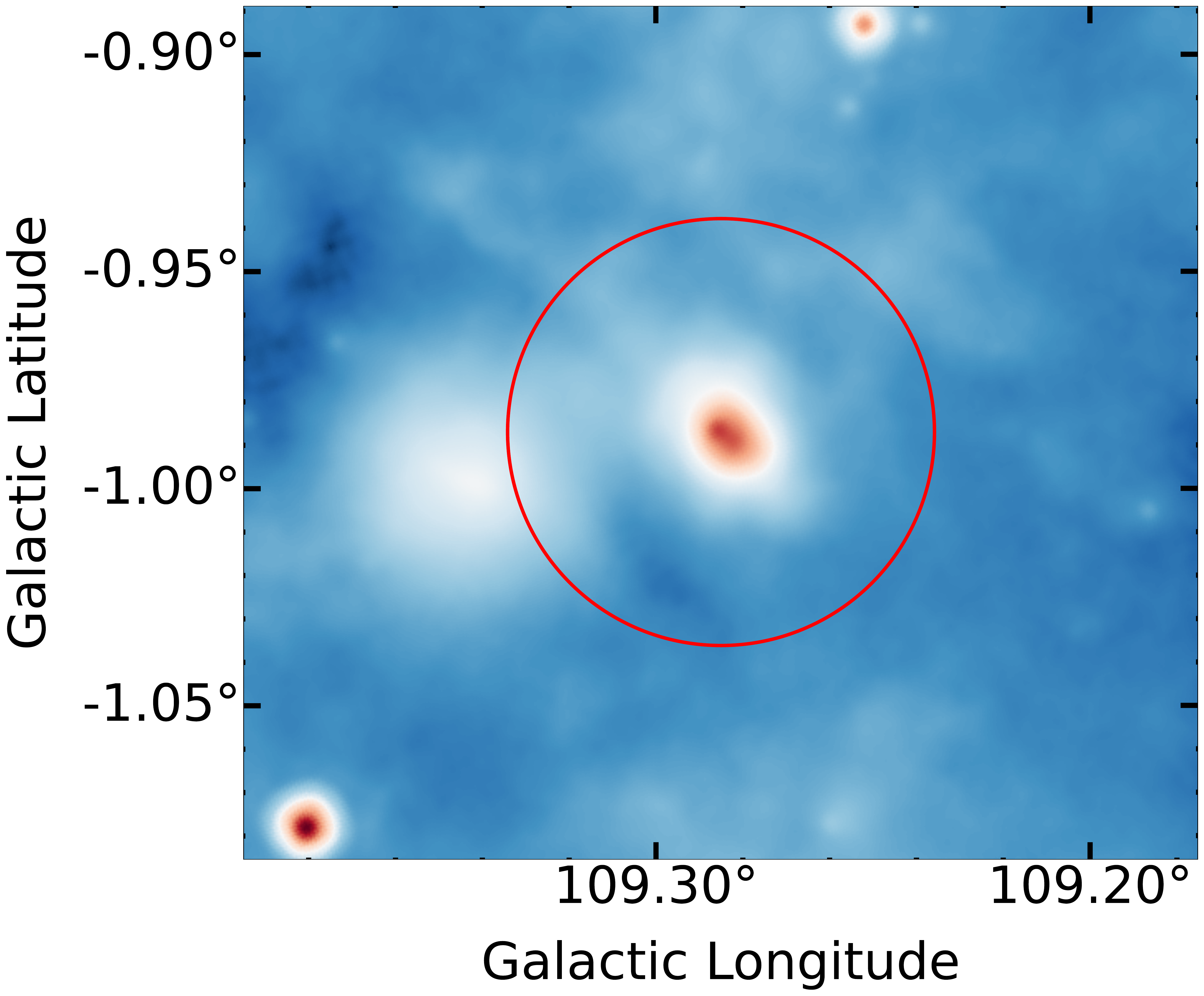}
     \\(b)
    \end{subfigure}
   \begin{subfigure}[b]{0.4\linewidth}
   \centering
    \includegraphics[trim=14.5cm 0cm 14.5cm 0cm,width=.4\textwidth]{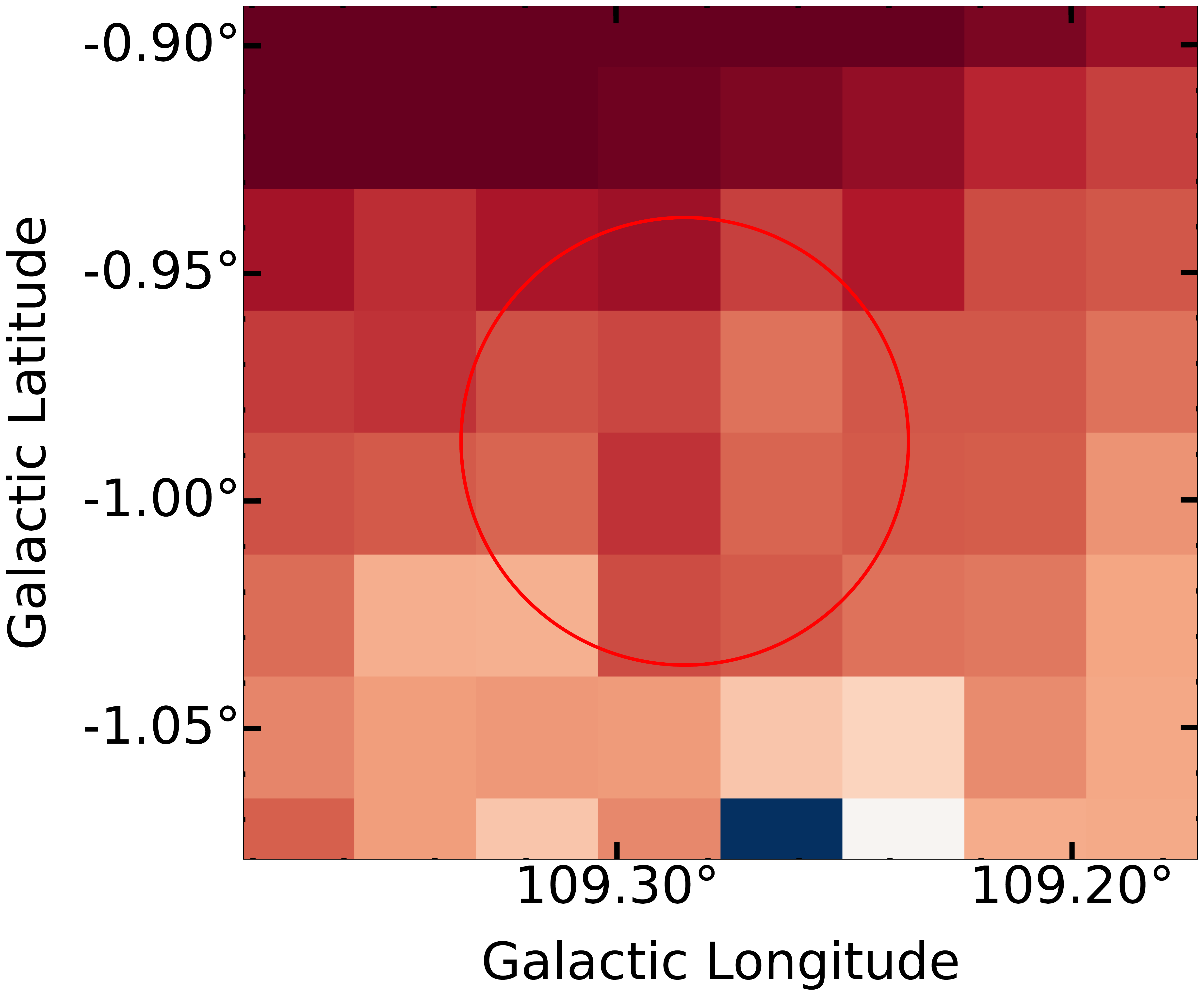}
     \\(c)
    \end{subfigure}
    \begin{subfigure}[b]{0.4\linewidth}
   \centering
    \includegraphics[trim=15cm 0cm 22cm 0cm,width=.4\textwidth]{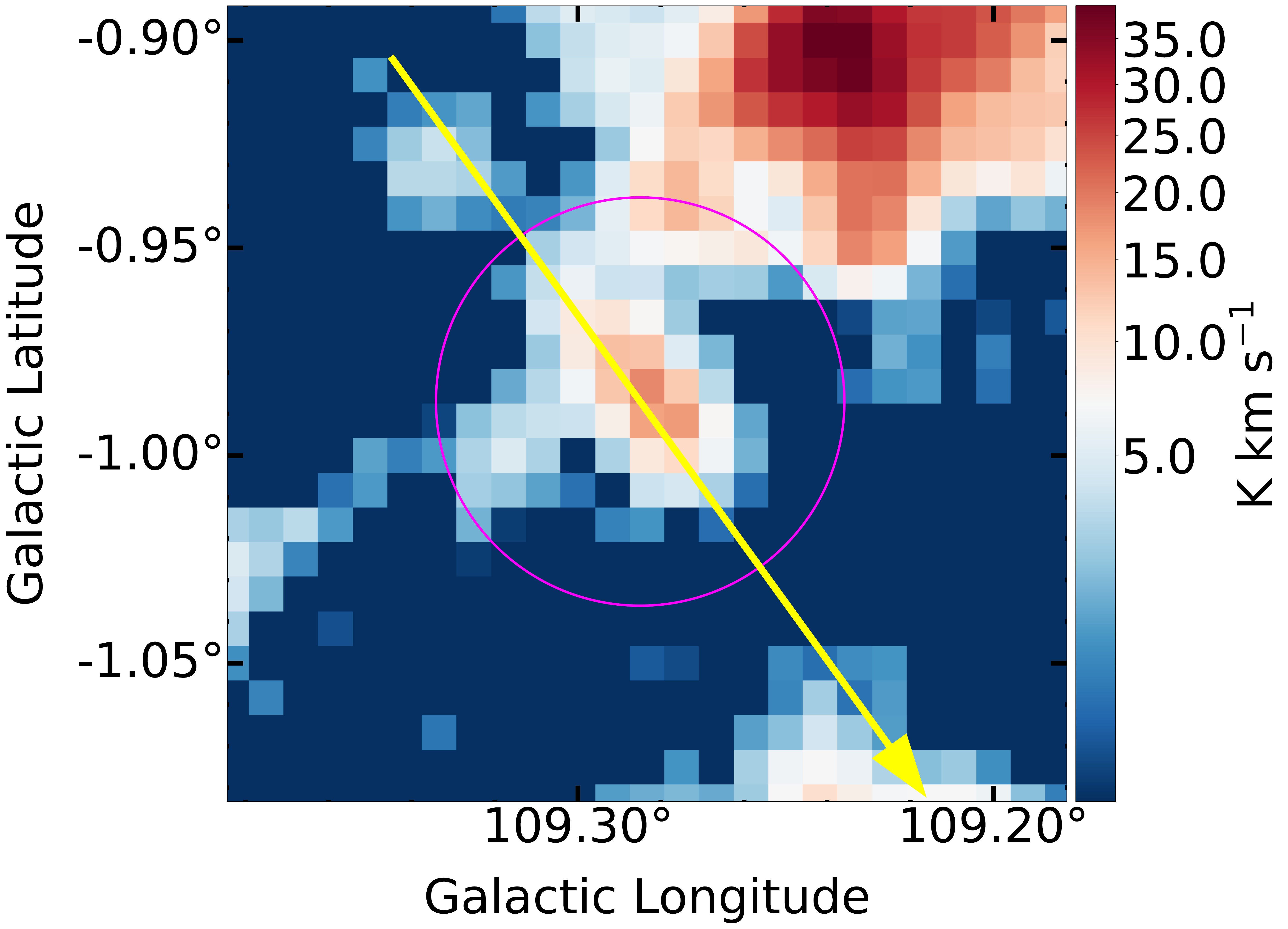}\hfill
    \\(d)
    \end{subfigure}
    \begin{subfigure}[b]{0.4\linewidth}
   \centering
    \includegraphics[trim=14cm 0cm 23cm 0cm,width=.4\textwidth]{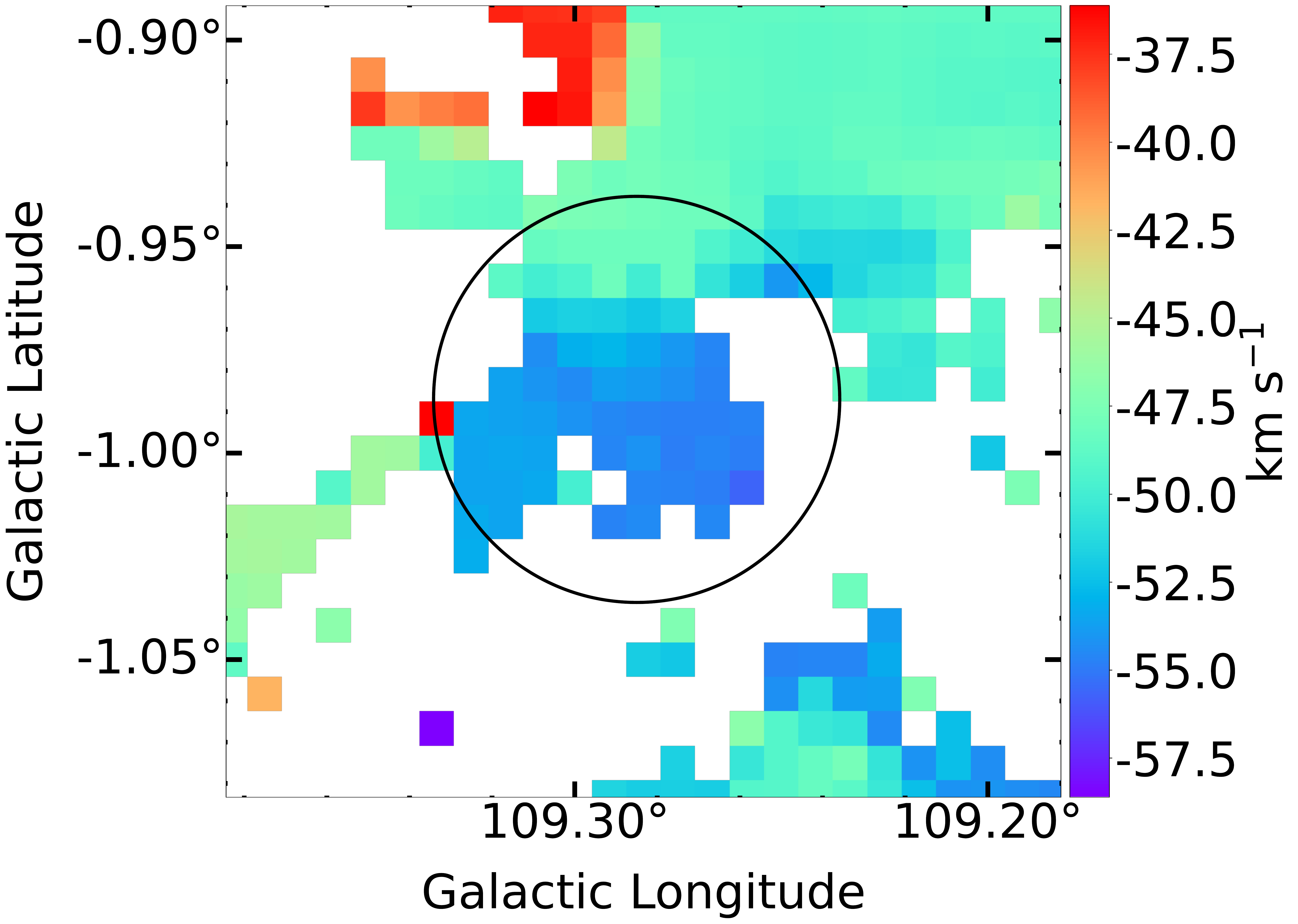}\hfill
    \\(e)
    \end{subfigure}
    \begin{subfigure}[b]{0.4\linewidth}
   \centering
    \includegraphics[trim=8.5cm 0cm 26cm 0cm,width=.4\textwidth]{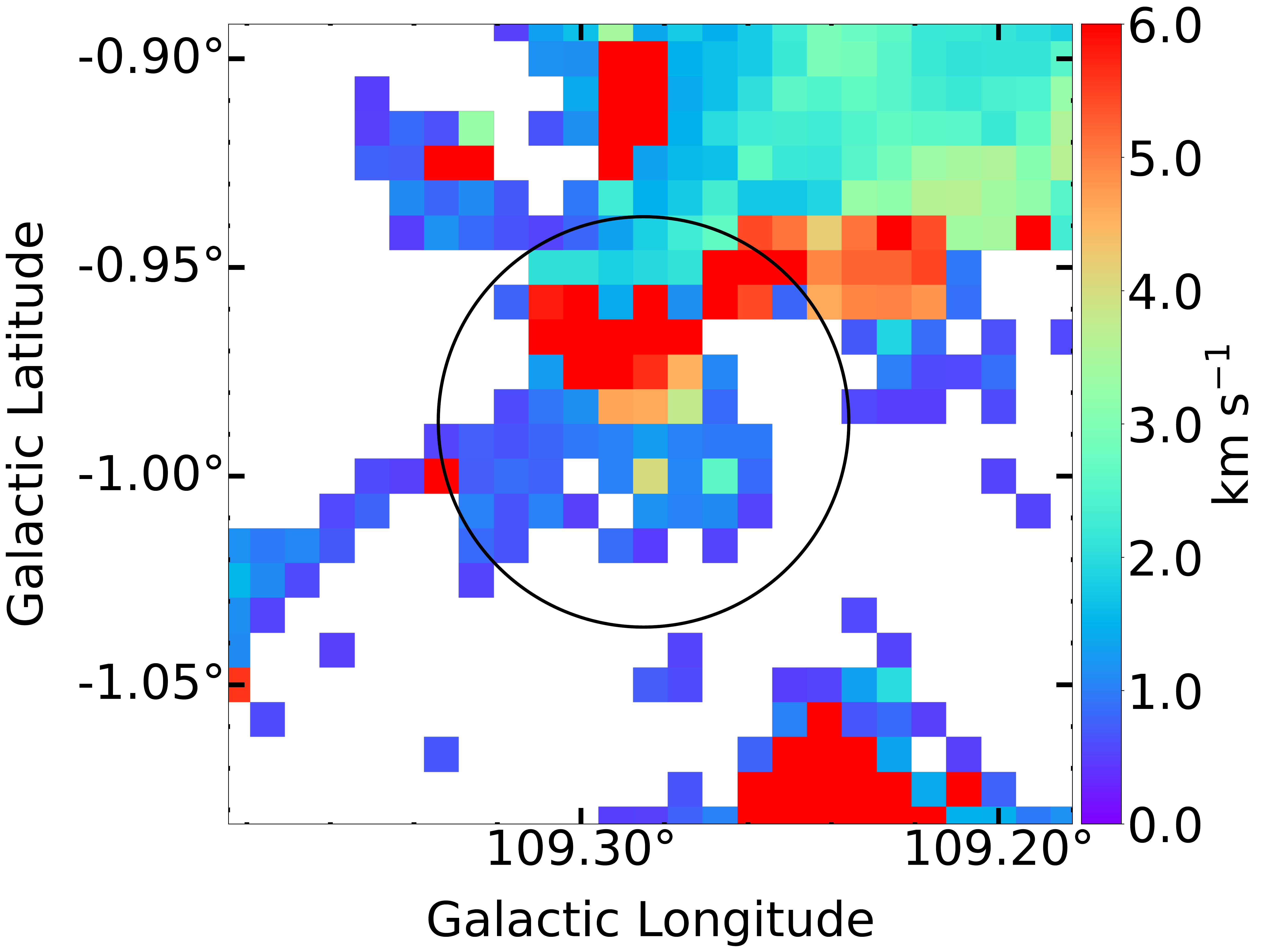}\hfill
    \\(f)
    \end{subfigure}
    \begin{subfigure}[b]{0.4\linewidth}
   \centering
    \includegraphics[trim=14.5cm 0cm 22cm 0cm, width=.4\textwidth]{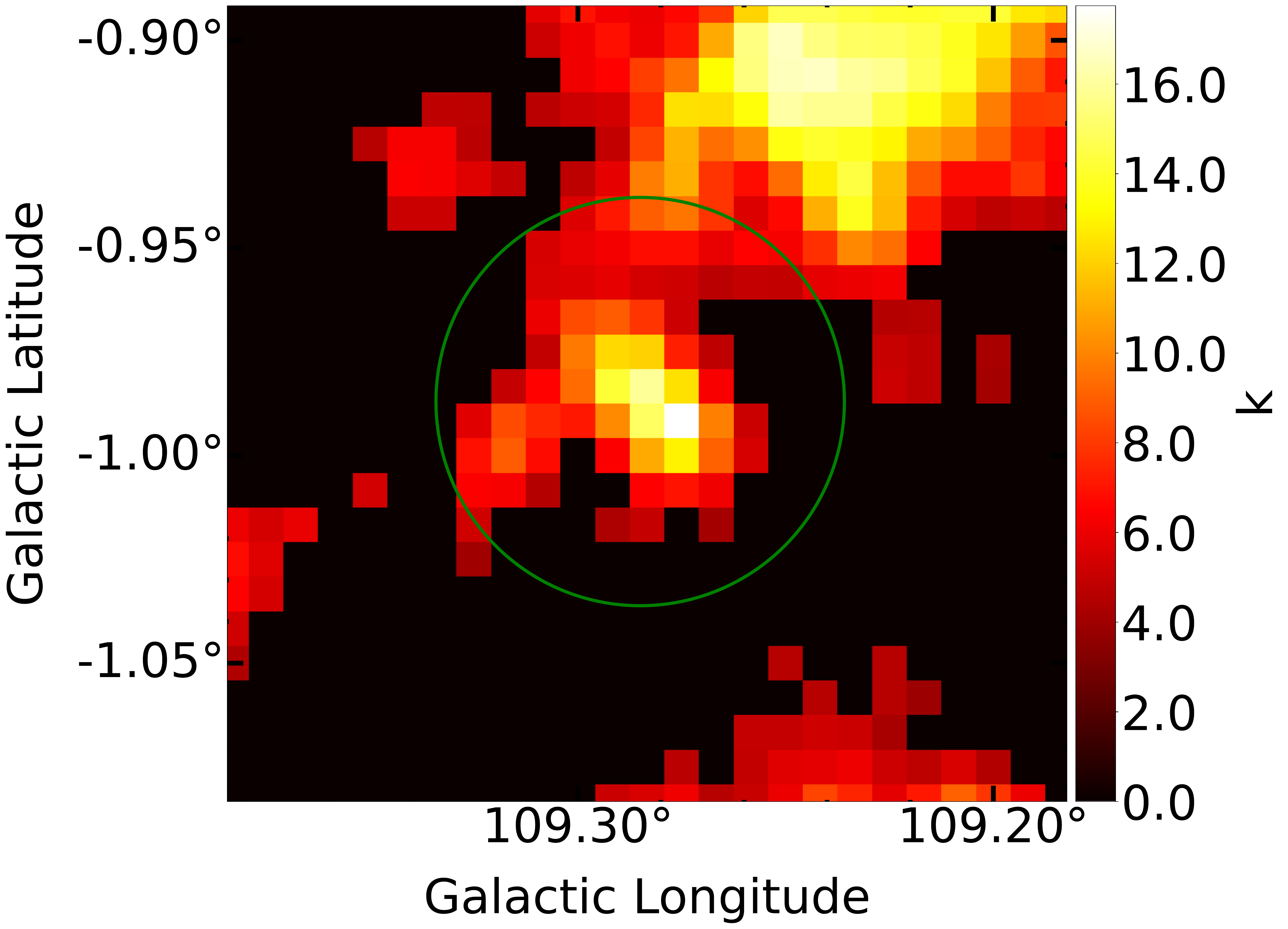}\hfill
    \\(g)
    \end{subfigure}
    \begin{subfigure}[b]{0.4\linewidth}
   \centering
    \includegraphics[trim=1cm -2cm 15cm 0cm, width=.4\textwidth]{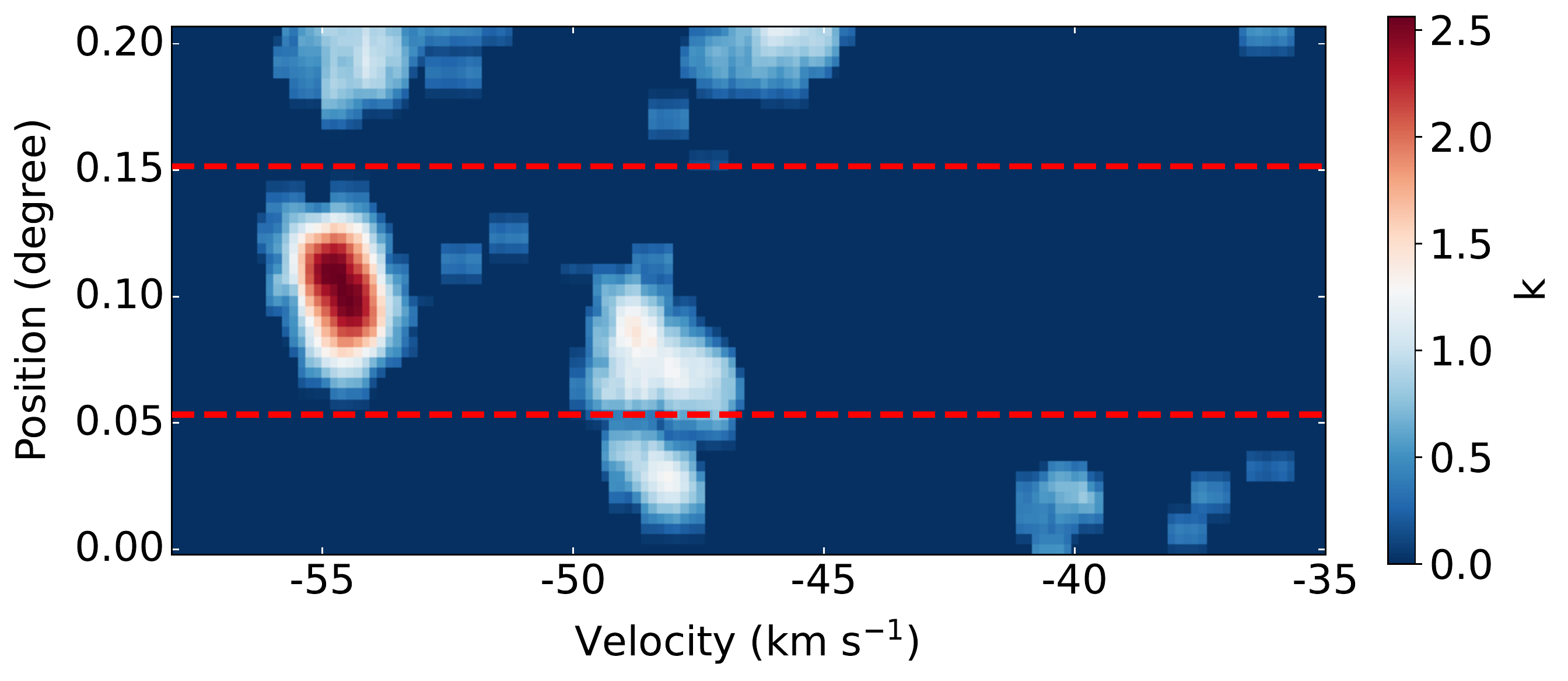}\hfill
    \\(h)
    \end{subfigure}
    \\[\smallskipamount]
    \caption{Same as Figure~\ref{Fig8} but for the G109.285${-}$00.987 region. }
    \label{FigA.15}
\end{figure}

The velocity channel map, position-velocity diagrams, intensity-weighted centroid velocity  and velocity dispersion maps of the $\rm {}^{13}{CO}$ line emission are presented in Figure~\ref{FigA.1}, Figure~\ref{FigA.2}, and Figure~\ref{FigA.3}, respectively. The morphology of {H \small {II}} regions/candidates in various tracers are presented in Figures from ~\ref{FigA.4} to ~\ref{FigA.15}. The properties of the OB stars in this region are listed in Table~\ref{table A.1}

%加3个结束

%\begin{landscape}
\begin{table}
\bc
\begin{minipage}[]{100mm}
\caption[]{OB Stars in the Surveyed Region\label{table A.1}}\end{minipage}
\setlength{\tabcolsep}{8pt}
\small
 \begin{tabular}{cccrcccc}
  \hline\noalign{\smallskip}
  (1)& (2)& (3)&(4)& (5)& (6)&(7)&(8)\\

Number &Glon&Glat& Sptype& Parallax & Parallax${\_}$err & Dist  & Reference\\
 &degree & degree & & mas&mas& kpc&  \\
  \hline\noalign{\smallskip}
1& 106.771 & -1.007 & B0:e& 0.495    &0.030 & 1.917 & 1   \\
  2&106.893& -1.132  &  O9.7V &0.204  & 0.024&4.280&  1    \\
  3& 106.998 & -0.921 & B1.5V&  0.332  &0.020 &2.780 &  1   \\
  4 & 107.066  & -0.898 & B0 & 0.308  &0.034 & 3.011 & 1    \\
  5& 107.091&-0.917 &B1.5V &0.350 &0.036 &2.694 & 3   \\
  6&107.130 &  -0.863 &  B0.5V  & 0.298 &0.032 &3.092& 2     \\
  7& 107.136  & -0.697 & B0Ib  &0.303   &0.035 &3.069& 1   \\
  8& 107.152   & -0.855 & B2  &0.228   &0.031 &3.942& 1    \\
  9& 107.155 & -0.981 & B1V& 0.374    &0.027 & 2.495&  1  \\
  10& 107.166 & -0.902 & B1.5V  &0.355   &0.027 &2.620 &1     \\
  11&107.178 &  -0.953 &  O8V((f)) & 0.411 &0.027 & 2.283 &1      \\
  12&  107.180   & -0.924 & B1V &0.242   &0.029 &3.683 & 1   \\
  13& 107.185 & -0.946& B0.5V&0.380&0.028 &2.396 & 4   \\
  14 & 107.187 & -0.886 & B1/3Ve  & 0.207   &0.019 &4.252& 1  \\
  15& 107.189  & -0.877 & B0.5V  & 0.363  &0.033 & 2.576 & 1  \\
  16& 107.194 & -1.044 &B0.5V &  0.393  &0.036 & 2.396 & 1    \\
  17&107.206&  -1.336 &  O9.5V&   0.420&0.029 &2.235 & 1      \\
  18 & 107.300  & -0.564 & B2  & 0.325  &0.037 &2.883 & 1   \\
  19&108.261&  -1.065& B1& 1.259 &0.312 & 1.090&1     \\
  20&108.275 &  -1.066 &  B2III &0.329  &0.026 & 2.807 & 2   \\
  21&108.365&-1.048 & B0 & 0.331 &0.013 & 2.821&3     \\
  22&108.389 &  -1.049 &  B1V  & 0.294 &0.028 &3.117& 2     \\

  \noalign{\smallskip}\hline
\end{tabular}
\ec
\tablerefs{0.95\textwidth}{(1) \cite{2019MNRAS.487.1400C}; (2) \cite{2015AJ....150..147F}; (3) \cite{2021ApJS..253...22X}; (4)  \cite{2003AJ....125.2531R}}
\end{table}
%\end{landscape}

\label{Appendix}

\clearpage
\bibliographystyle{raa}
\bibliography{ms2021-0363.bbl}

\end{document}